\documentclass[aps,pra,floatfix,reprint,showpacs,superscriptaddress,groupedaddress]{revtex4-1}
\usepackage{latexsym}
\usepackage{graphicx}
\usepackage{newtxmath}
\usepackage[colorlinks]{hyperref}
\usepackage{rotating}
\usepackage[normalem]{ulem}
\usepackage{hyperref}
\usepackage{amsmath,amsfonts}
\usepackage{mathtools}
\usepackage{bm}
\usepackage{color}
\usepackage[dvipsnames]{xcolor}
\usepackage{textgreek}
\usepackage{soul}  
\usepackage{qcircuit}
\usepackage[greek,english]{babel}

\newcommand{\ket}[1]{\ensuremath{| #1 \rangle}}

\usepackage{color}

\newcommand{\lyxmathsym}[1]{\ifmmode\begingroup\def\b@ld{bold}
  \text{\ifx\math@version\b@ld\bfseries\fi#1}\endgroup\else#1\fi}

\newcommand{\sx}[1]{#1}
\newcommand{\ps}[1]{#1}

\usepackage{amsfonts}
\usepackage{mathrsfs}
\usepackage{physics}
\usepackage{multirow}

\usepackage{qcircuit}
\usepackage{braket}
\usepackage{caption}
\usepackage{subcaption}
\hypersetup{colorlinks=true,urlcolor=blue,linkcolor=blue}

\newcommand{\appendixref}[1]{Appendix.~#1}
\newcommand{\Appendices}{Appendices}

\providecommand{\reviewerA}[1]{\color{black}#1\color{black}~}

\usepackage[nolist]{acronym}

\begin{document}
\title{Superconducting qubit readout enhanced by path signature}

\author{Shuxiang Cao$^{1}$}
\email{shuxiang.cao@physics.ox.ac.uk}
\altaffiliation[Present address: ]{%
    NVIDIA Corporation, 2788 San Tomas Expressway, Santa Clara, 95051, CA, USA%
  }%
\author{Zhen Shao$^{2}$}
\author{Jian-Qing Zheng$^{3}$}
\author{Peter A Spring $^{6}$}
\author{Simone D Fasciati$^{1}$}
\author{Mohammed Alghadeer$^{1}$}
\author{Shiyu Wang $^{6}$}
\author{Shuhei Tamate $^{6}$}
\author{Neel Vora$^{4}$}
\author{Yilun Xu$^{4}$}
\author{Gang Huang$^{4}$}
\author{Kasra Nowrouzi$^{4}$}
\author{Michele Piscitelli$^{1}$}
\author{Yasunobu Nakamura $^{6,7}$}
\author{Irfan Siddiqi$^{4,5}$}
\author{Peter Leek$^{1}$}
\author{Terry Lyons$^{2}$}
\author{Mustafa Bakr$^{1}$}

\affiliation{$^1$ Clarendon Laboratory, Department of Physics, University of Oxford, Oxford, OX1 3PU, UK}

\affiliation{$^2$ Mathematical Institute, University of Oxford, Oxford, OX2 6GG, UK}
\affiliation{$^3$ The Kennedy Institute of Rheumatology, University of Oxford, Oxford, OX3 7FY, UK}
\affiliation{$^4$ Lawrence Berkeley National Laboratory, Berkeley, CA 94720, USA}
\affiliation{$^5$ University of California, Berkeley, CA 94720, USA}
\affiliation{$^6$ RIKEN Center for Quantum Computing (RQC), Wako, Saitama 351-0198, Japan}
\affiliation{$^7$ Department of Applied Physics, Graduate School of Engineering, The University of Tokyo, Bunkyo-ku, Tokyo 113-8656, Japan}

\begin{abstract}
Quantum non-demolition measurement plays an essential role in quantum technology, crucial for quantum error correction, metrology, and sensing. Conventionally, the qubit state is classified from the raw or integrated time-domain measurement record. Here, we demonstrate a method to enhance the assignment fidelity of the readout by considering the ``path signature'' of this measurement record, where the path signature is a mathematical tool for analyzing stochastic time series. We evaluate this approach across five different hardware setups, including those with and without readout multiplexing and parametric amplifiers, and demonstrate a significant improvement in assignment fidelity across all setups. Moreover, we show that the path signature of the measurement record \sx{provides an expressive feature set} that can be used to detect and classify state transitions that occurred during the measurement, improving the prediction of the qubit state at the end of the measurement. This method has the potential to become a foundational tool for quantum technology.
\end{abstract}
\maketitle

\begin{figure}[tb!]
     \centering
     \includegraphics[width=\linewidth]{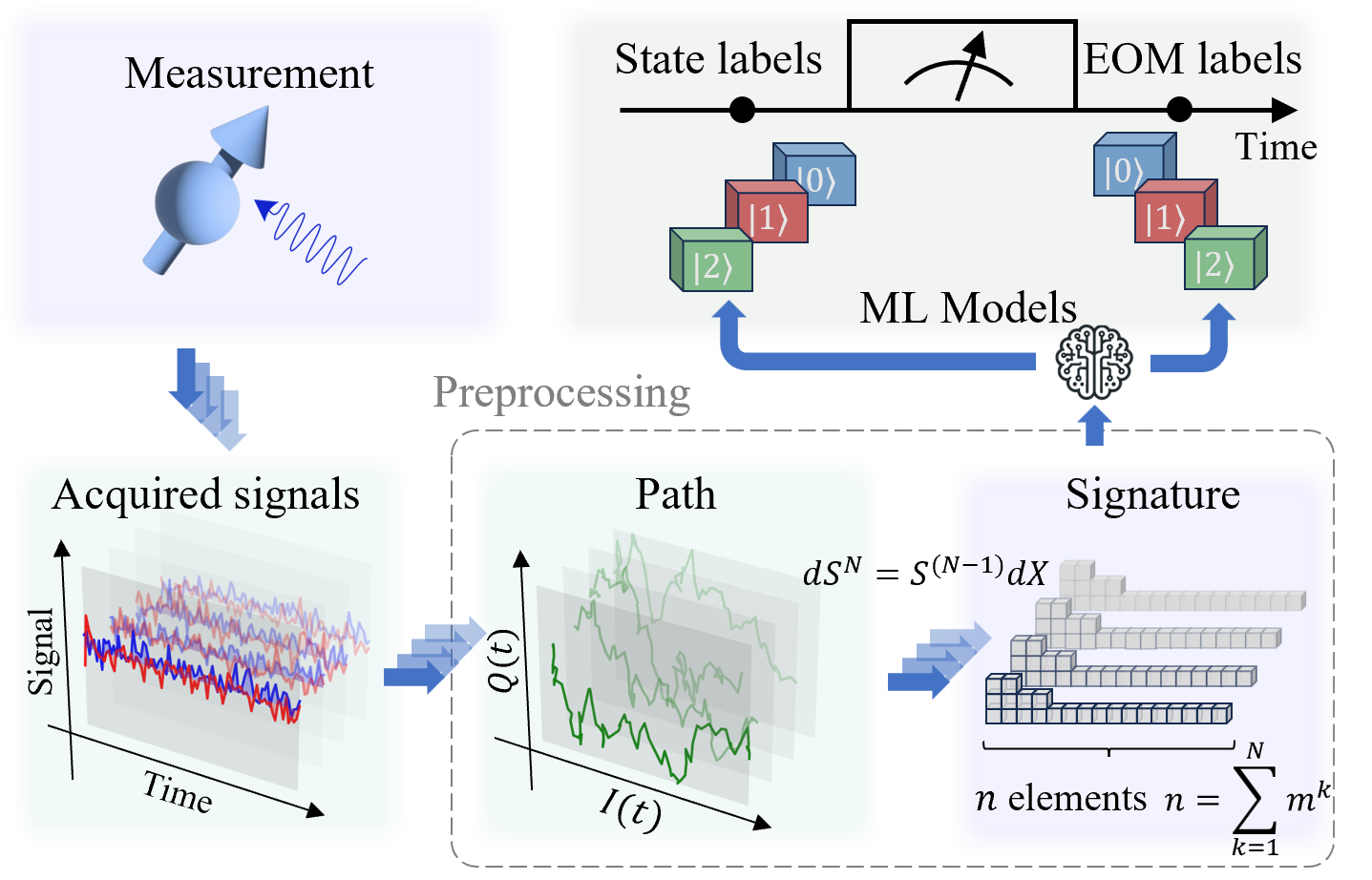}
        \caption{State discrimination using signatures. First, the readout signal record is collected. A weighted cumulative integral is then applied to construct the path of the readout record. The signature is evaluated on this constructed path, and machine learning (ML) models are subsequently applied to determine both the initial state at the start of the measurement and the state at the end of the measurement (EOM). The symbol $m$ denotes the dimension of the acquired signals, $N$ denotes the depth of the signature and $S^{(n)}$ denotes a signature of degree $n$.}
        \label{fig:procedure}
\end{figure}

\begin{figure}[tb!]
     \centering
        \includegraphics[width=\linewidth]{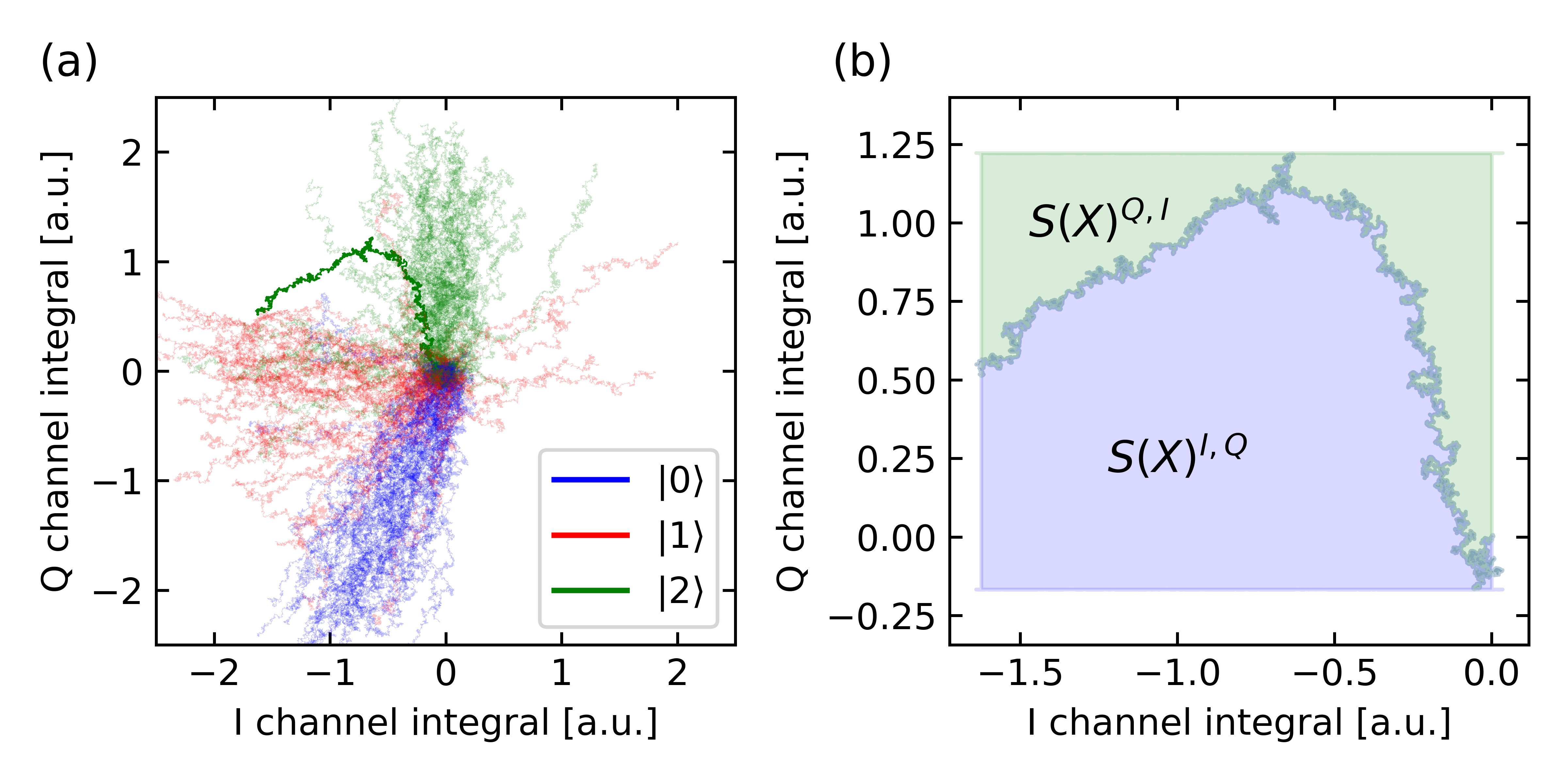}
        \caption{(a) An ensemble of paths constructed by cumulatively summing the acquired readout signal from a qutrit device. The green, red, and blue paths correspond to the states prepared as $\ket{0}$, $\ket{1}$, and $\ket{2}$, respectively. The highlighted green path exhibits an obvious state transition, changing its direction halfway. (b) The values of the degree-2 signatures $S(X)^{I,Q}$ and $S(X)^{Q,I}$ for the highlighted path are given by the areas of the blue and green regions, respectively.}
        \label{fig:state_transition}
\end{figure}

\begin{figure}
     \centering
        \includegraphics[width=\linewidth]{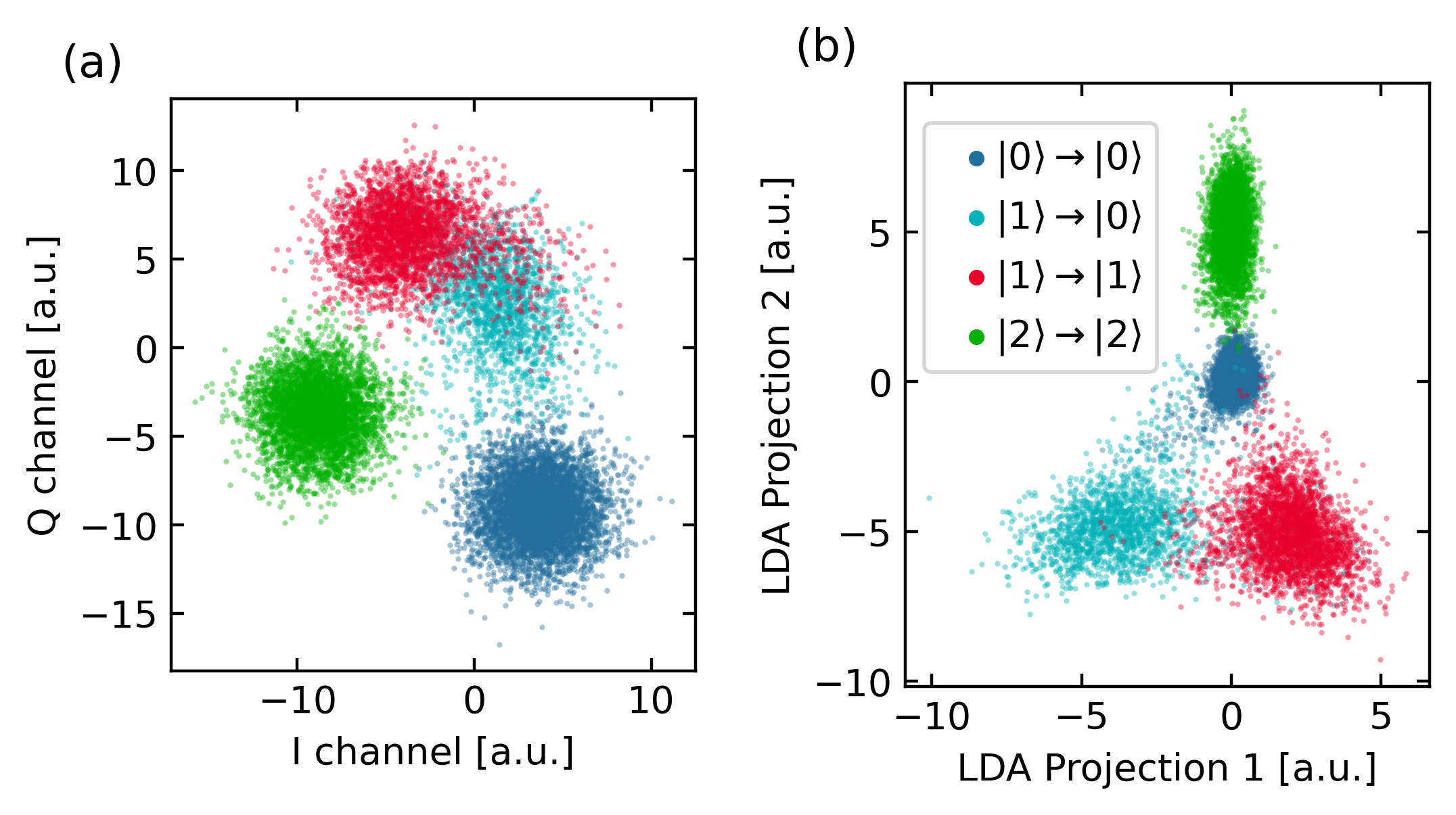}
        \caption{Distribution of single-shot measurement signals on the IQ plane for the AQT Qutrit dataset, which exhibits considerable state transition during the measurement when prepared to the $\ket{1}$ state. (a)~The optimally weighted integrated readout time series data. (b)~The projection of the depth-5 path signature. The labels $\ket{i}\rightarrow\ket{j}$ indicate that the state is expected to start at $\ket{i}$ and end at $\ket{j}$ during the measurement process. For readability, the plot only shows the distribution that indicates the transition from $\ket{1}$ to $\ket{0}$. The more detailed plots indicating more projection directions and the distribution of other state transitions can be found in the \Appendices.}
        \label{fig:visualization}
\end{figure}

\acrodef{FFNN}[FFNN]{Feed-forward Neural Networks}
\acrodef{FPGAs}[FPGAs]{Field-Programmable Gate Arrays}
\acrodef{GMM}[GMM]{Gaussian Mixture Model}
\acrodef{HMM}[HMM]{Hidden Markov Model}
\acrodef{LDA}[LDA]{Linear Discriminant Analysis}
\acrodef{QDA}[QDA]{Quadratic Discriminant Analysis}
\acrodef{RF}[RF]{Random Forest}
\acrodef{SVC}[SVC]{Support Vector Classifier}
\acrodef{SVMs}[SVMs]{Support Vector Machines}
\acrodef{AE}[AE]{Autoencoders}
Rapid high-fidelity non-demolition single-shot readout is essential for quantum computing. \sx{In the recent demonstration of a beyond-breakeven logical qubit, a significant portion of the error budget stemmed from the readout process~\cite{Acharya2023}; highlighting the impotance of improved readout for quantum error correction.} For solid-state quantum processors, readout is typically performed by probing a component, such as a harmonic oscillator, that couples to the qubit. The response signal carries information that can be used to determine the state of the qubit. One example is dispersive readout~\cite{PhysRevA.98.023849,PhysRevA.69.062320}, where the resonator exhibits a frequency shift depending on the state of the qubit. Dispersive readout has been demonstrated across various solid-state qubit architectures, including spin-qubits~\cite{Crippa2019,PhysRevB.100.245427,PhysRevB.109.155304}, quantum dots~\cite{PhysRevLett.110.046805,10.1063/1.4984224, PhysRevB.94.195305} and superconducting qubits~\cite{PhysRevLett.95.060501, PhysRevApplied.7.054020,PhysRevLett.115.203601}. 
\\
\indent The conventional state discrimination method integrates the time-domain readout signal and yields a single complex-valued data point that reflects the resonator response at the probing frequency~\cite{10.1063/1.5089550}. This method assumes the qubit state remains unchanged during readout, and neglects \sx{qubit} relaxation~\cite{PhysRevX.10.011001,2305.10508} and measurement-induced state transitions~\cite{PhysRevApplied.20.054008, PhysRevLett.117.190503,dumas2024unified} that may occur during the readout. On the other hand, the dispersive readout signal captures a continuous record of the readout resonator response, \ps{potentially enabling the qubit state to be tracked throughout the entire readout process}. Previous studies have applied statistical learning techniques to analyze the continuous readout signal records, such as \ac{LDA}, \ac{QDA}, and \ac{SVMs}~\cite{PhysRevLett.114.200501}. In addition, machine learning models such as the \ac{HMM}~\cite{PhysRevA.102.062426}, \ac{FFNN}~\cite{PhysRevApplied.17.014024}, and \ac{AE}~\cite{PhysRevApplied.20.014045} have been applied to improve the readout fidelity by taking the time-series data of the continuous measurement record as inputs. \sx{Although these methods can improve upon the conventional approach, they have their own drawbacks. To satisfy the Markovian condition, the HMM approach assumes that the \ps{mean} received readout signal corresponding to a given qubit state remains constant over time. This makes it ill-suited for scenarios in which the transient response of the readout resonator dominates the measurement record. In contrast, FFNN- and AE-based classifiers do not impose such constraints but often require extensive hyperparameter optimization.}
\\
\indent Additionally, for algorithms requiring feedback, it is important that measurements are quantum non-demolition (QND). Although the state discriminator may accurately identify the state at the beginning of the measurement, a state transition that \sx{occurs} during the \sx{measurement} will \sx{cause a QND error}. However, if these mid-measurement state transitions can be detected and classified, \sx{it introduces the possibility of correcting QND errors through active feedback or decoding.} 
\\
\indent This study proposes a feature engineering method~\cite{Hastie2009-jt} for capturing information about mid-measurement state transitions by employing the stochastic time series tool ``path signature'' to \sx{provide an expressive representation of the readout measurement record}~\cite{MR3727607, chevyrev2016primer, 10041999}. \sx{In contrast to linear methods such as aggregation and Fourier analysis, the signature offers a mathematically principled representation of paths by capturing their nonlinear characteristics. It has proven effective in applications including handwriting recognition \cite{Xie2016LearningSC}, sepsis prediction \cite{Cohen2024}, and quantitative finance \cite{10.1145/2640087.2644157}.} The implementation process is shown in Fig.~\ref{fig:procedure}. \sx{First, qubit readout is performed and the time-trace measurement record is collected. Next, the path-signature of this measurement record is evaluated. Finally, the path signature is fed to a classifier algorithm in order to assign the most probable qubit state at the start and end of the measurement.} \ps{We} show that the path-signature \ps{feature set} captures information about mid-measurement state transitions, and as a result, it enhances the assignment fidelity and enables the \sx{detection and classification of mid-measurement state transitions. In addition, the path signature can in principle be evaluated during the data integration time~\cite{MR3727607, chevyrev2016primer, 10041999}, allowing this feature engineering step to be completed without adding a time overhead.} 
\\
\indent \sx{We consider the case in which the resonator’s response to an external drive is conditioned on the state of the qubit.} The response signal is represented as a complex-valued time series, where each data point in the time series provides incremental information about the state of the qubit, denoted as vector \(\mathrm{d}X(t) = \{I(t), Q(t)\}\mathrm{d}t\). Here, \(t\in [0,T_{r}]\) represents time with readout pulse width $T_r$, and the real and imaginary parts are the in-phase channel $I(t)$ and the quadrature channel $Q(t)$ of the response signal, respectively. For dispersive readout, the response function \(\mathrm{d}X(t)\) is typically nontrivial~\cite{10.1063/1.5089550, RevModPhys.93.025005}, and here we model it as an arbitrary continuous function that is differentiable on some much finer \ps{timescale} \sx{compared} to the demodulated time\ps{-}slice \ps{interval}. \sx{State discrimination is conventionally performed by computing an optimally weighted integral of the measurement signal, using a weighting function $w(t)$, to obtain the complex-valued scalar  $\Tilde{R} = \int_0^{T_r}  w(t)\mathrm{d}X(t)$, which is then used as a feature set to assign the qubit state~\cite{PhysRevLett.95.060501, PhysRevApplied.7.054020, PhysRevApplied.10.034040}.} Although this method has the benefit of simplicity, it discards valuable information about qubit state transitions \sx{that may occur} during the readout process.
\\
\indent Instead of taking this approach, we define a path $X(t) \equiv \int_0^{t} w(\tau) {\mathrm{d}}X(\tau)$. A set of example paths collected on the Oxford qutrit hardware~\cite{2210.04857, 2303.04796} are shown in Fig.~\ref{fig:state_transition}(a). Given a multi-dimensional path \( X(t) \) defined over time, the degree-\( N \) path signature \( S^{(N)} \) is a collection of all iterated integrals of \( X(t) \), up to \( N \) iterations. The depth-\( N \) signature is then the collection of all the signatures up to degree $N$. These integrals are defined in the tensor algebra and are constructed by taking combinations of the components of \( X(t) \)~\cite{MR3727607, chevyrev2016primer, 10041999}. For example, the degree-1 signature is given by $S^{(1)}(X) = \{X(T_r) - X(0)\}$. Since $X(0)$ is always zero in our scenario, and $X(t)$ is the integral of the received signal, $S^{(1)}(F)$ falls back to the conventional approach of \ps{optimally weighted integration} of the signal. The degree-2 signature is related to the L\'{e}vy area~\cite{levy1940mouvement,levy1951wiener} \(A\)  of \(X(t)\), which is a measure of the signed area enclosed by the stochastic process and encodes useful information about its behavior. \reviewerA{Taking the notation \( S(X)^{I,Q} \) to represent the degree-2 iterated integral \( \int_{0 \leq t_1 \leq t_2 \leq T_r} dI(t_1) \, dQ(t_2) \), and \( S(X)^{Q,I} \) to represent \( \int_{0 \leq t_1 \leq t_2 \leq T_r} dQ(t_1) \, dI(t_2) \), the Lévy area can be written as $A = \frac{1}{2}\left( S(X)^{I,Q} - S(X)^{Q,I} \right)$.} An example of these two integrals is shown in Fig.~\ref{fig:state_transition}(b). The higher-order signatures generalize the integral to higher dimensional volumes. \ps{See the \appendixref{\ref{app:signature}} for more details on the higher-order signatures.} Since the same dimension can appear in the integral multiple times, for a path with $m$ dimensions, the depth-$N$ signature of the path forms a vector with $\sum_{k=1}^N m^k$ elements. This higher-dimensional object contains more useful information than $\Tilde{R}$, \sx{which we claim can be utilized} to improve assignment fidelity and detect measurement-induced state transitions. 
\\
\indent \ps{To illustrate this, we consider the paths shown in Fig.~\ref{fig:state_transition}(a)}. While most paths exhibit a random walk drifting in a certain direction, a few paths, notably the highlighted green path, show a direction change halfway, indicating a state transition event. If only the last point of the path is used for classification, which \sx{is equivalent to the conventional method that performs classification using the feature set $\tilde{R}$}, the highlighted path will be mistakenly classified into the $\ket{1}$ state. Analyzing the degree-2 signature of such paths can effectively capture this state transition event, \ps{as evidenced by the signed areas} in Fig. \ref{fig:state_transition}(b).
\\
\indent To \ps{further demonstrate the ability of the path signature feature set to detect mid-measurement state transitions}, we evaluate the path signature for a qutrit dataset (AQT Qt) on a device reported in a previous study~\cite{PRXQuantum.6.010307,Kreikebaum2020}, which experienced significant measurement-induced state transitions when measured in the $\ket{1}$ state. Fig.~\ref{fig:visualization}(a) shows the distribution of the conventional weighted-integration \ps{feature set,} and Fig.~\ref{fig:visualization}(b) shows the projected distribution of the path signature \ps{feature set}. We include the time as the third dimension of the path, usually referred to as ``time argumentation'', and evaluate the depth-5 path signature to obtain 363 features. Then we apply LDA to project the signatures into two-dimensional vectors, allowing us to visualize their distribution. We observe that the distributions of the projected signatures are better separated than the distributions of the optimally-weighted signal integrals. We also verify that these distributions remain stable over time (see \appendixref{\ref{app:stability}}).
\\
\indent \ps{To benchmark the path signature for quantum state classification}, we collected time-domain \ps{measurement record} datasets from \ps{five} different superconducting circuits experimental setups with different device parameters and quantum-limited amplifiers. The details about these setups are provided in the \appendixref{\ref{app:OXF_Qt} \ref{app:AQT_Qt} \ref{app:OXF_Simo} \ref{app:RQC}}. \ps{We then trained and tested three different state classification methods on these datasets. In the first method, a Gaussian-Mixture Model (GMM) classifier is applied to the optimally weighted integral of the measurement record. In the second method, a Random Forest (RF) classifier is applied to the optimally weighted measurement record data. In the third method, which we refer to as ``Sig+RF", the same RF classifier is applied to the depth-5 path signature of the measurement record data.} Further details about the GMM and RF machine learning methods are provided in the \appendixref{\ref{app:ml_methods}}. The performance of each method is \ps{characterized} by the assignment fidelity \(F \equiv \left[\sum_{i} P(i|i)\right]/N\), where \(P(a|b)\) represents the probability that a qubit prepared in the \(\ket{b}\) state is measured in the \(\ket{a}\) state. The indices \(i\) refer to arbitrary basis states, such as \(\ket{0}, \ket{1}, \ldots\), and \(N\) denotes the total number of basis states considered in the computation. The resulting assignment infidelities found using the different discrimination methods are reported in Tab.~\ref{tab:detailed_comparison}.
\\
\indent \ps{The results show} that the Sig+RF method consistently outperforms the other methods across all of the datasets, reducing infidelity by an average of $39(4)\%$, $26(12)\%$, $13(4)\%$ and $20.6(35)\%$ for the Oxford qutrit, Oxford qubits, RQC qubits and AQT qutrit datasets, respectively, compared to the GMM method. \ps{Furthermore, we find that the RF method generally performs worse than the GMM method when it is provided with the weighted measurement record as opposed to the path signature, showing the effectiveness of the path signature \ps{as a feature set for quantum} state discrimination.} In order to balance the computational cost, we utilized the signatures up to degree 5. The training time of the RF model was under one minute \footnote{The training time depends on several factors, including the size of the selected dataset, the choice of signature depth, the number of states, and the performance of the computer, so that it can vary under different conditions. Here, we report the training time for the signature with an RF model, which uses 5,000 traces for each state, resulting in a total of 15,000 traces. Of these, 3,000 were set aside for testing accuracy and were not involved in the training process, 2,400 were used for validation to optimize hyperparameters, and the remaining 9,600 were used for training. The complete training process, including hyperparameter optimization, takes an average of 46 seconds on an Intel Xeon w7-2495X CPU.}.
\begin{table*}[ht!]
    \centering
    \begin{ruledtabular}
        \begin{tabular}{c|ccc|ccc|c|c|c|c|c|c}
        Dataset & \multicolumn{3}{c|}{$(1-F)\times 10^2$} & \multicolumn{3}{c|}{$(1-F_{\mathrm{EOM}})\times 10^2$} & Mux/Amp & $T_r$ & $T_1$ & $T_{2r}$ & $\kappa/2\pi$ & $2\chi/2\pi$  \\
         & GMM & RF & Sig+RF & Baseline & RF & Sig+RF &  & $(\mu s)$ & $(\mu s)$ & $(\mu s)$ & (MHz) & (MHz) \\
        \hline
        OXF Qt & 13.16(44) & 11.56(50) & 8.05(40) & 20.49 & 13.61(33) & 12.02(26) & N/H & 10 & 189 & 102 & 0.5 & -0.29 \\
        \hline
        OXF Q1 & 1.23(27) & 1.91(29) & 1.05(28) & N/A & N/A & N/A & Y/H & 2 & 47 & 19 & 0.8 & -1.7 \\
        OXF Q2 & 1.17(27) & 2.17(32) & 1.05(29) & N/A & N/A & N/A & Y/H & 2 & 50 & 27 & 0.9 & -1.3 \\
        OXF Q3 & 5.70(37) & 6.64(52) & 2.69(24) & N/A & N/A & N/A & Y/H & 2 & 52 & 21 & 1.6 & -1.5 \\
        OXF Q4 & 1.80(20) & 2.69(35) & 1.30(24) & N/A & N/A & N/A & Y/H & 2 & 43 & 26 & 1.5 & -1.3 \\
        \hline
        RQC Q1 & 1.28(9) & 1.68(12) & 1.10(7) & 3.16 & 2.92(7) & 2.70(8) & N/J+H & 0.5 & 15 & 23 & 6.9 & -0.46 \\
        RQC Q2 & 2.74(17) & 2.94(21) & 2.50(18) & 5.94 & 5.20(10) & 4.64(12) & N/J+H & 0.5 & 13 & 10 & 8.0 & -0.69 \\
        RQC Q3 & 1.57(12) & 1.68(12) & 1.44(10) & 4.07 & 3.82(9) & 3.50(9) & N/J+H & 0.5 & 18 & 6 & 6.6 & -0.66 \\
        RQC Q4 & 1.52(11) & 1.54(11) & 1.33(10) & 4.52 & 4.27(8) & 3.66(7) & N/J+H & 0.5 & 17 & 11 & 5.5 & -0.73 \\
        \hline
        RQC Q5 & 1.66(15) & 1.95(18) & 1.41(11) & 2.40 & 1.95(4) & 1.58(4) & Y/H & 0.536 & 13 & 21 & 5.9 & -0.95 \\
        RQC Q6 & 1.05(12) & 0.90(11) & 0.88(12) & 2.33 & 1.92(7) & 1.54(7) & Y/H & 0.536 & 21 & 26 & 6.6 & -1.74 \\
        RQC Q7 & 1.28(13) & 1.44(12) & 1.04(11) & 2.29 & 1.94(8) & 1.59(8) & Y/H & 0.536 & 16 & 19 & 4.2 & -1.88 \\
        RQC Q8 & 1.72(11) & 1.90(12) & 1.50(10) & 2.30 & 1.91(8) & 1.35(8) & Y/H & 0.536 & 8 & 14 & 5.5 & -1.00 \\
        \hline
        AQT Qt & 4.01(14) & 4.19(17) & 3.18(9) & 15.17 & 7.88(14) & 4.43(14) & N/T+H & 1 & 130 & 41 & 1.3 & -0.71 \\
        \end{tabular}
    \end{ruledtabular}    
    \caption{Comparison of the assignment infidelity $(1-F)$ and the EOM infidelity $(1-F_{\mathrm{EOM}})$ using different discrimination approaches, along with the device parameters. \sx{All of the} reported infidelity values are \sx{the lowest values obtained from a sweep of the total measurement window length}. The ``Mux/Amp" column indicates whether the data was \ps{(Y) or was not (N)} collected \ps{simultaneously} using frequency-multiplexed readout, and denotes which \sx{amplifiers were utilized in the measurement chain}. \sx{Here,} ``H'', ``J+H'', and ``T+W'' correspond to HEMT only, {Josephson parametric amplifier (JPA)} with HEMT, and \sx{traveling-wave parametric amplifier (TWPA)} with HEMT, respectively. The $T_r$ value is the readout pulse width. The $T_1$ and $T_{2r}$ values correspond to the qubit energy relaxation time and the Ramsey decoherence time, respectively, and $\kappa$ and $\chi$ denote the resonator linewidth and dispersive shift. For more details on the experimental setups and pulse sequences, refer to \appendixref{\ref{app:OXF_Qt} \ref{app:AQT_Qt} \ref{app:OXF_Simo} \ref{app:RQC}}. }\label{tab:detailed_comparison}
\end{table*}
\\
\indent Finally, we investigate whether the path signature can improve the prediction of the qubit state at the end of the readout by detecting mid-measurement state transitions. Here, we introduce the ``end-of-measurement fidelity" ($F_{\mathrm{EOM}}$), defined as \( F_{\mathrm{EOM}} \equiv \left[\sum_{i} P_{\mathrm{EOM}}(i|i)\right]/N\), where \(P_{\mathrm{EOM}}(c|d)\) represents the probability that a qubit classified as being in the \(\ket{d}\) state at the end of a measurement will be classified as being in the \(\ket{c}\) state at the start of a consecutive measurement. As a result, only untracked state transitions during measurement will contribute to EOM infidelity. The EOM infidelities found using the different classification methods are summarized in Tab.~\ref{tab:detailed_comparison}. \ps{Here, the results using the RF and Sig+RF methods are compared to a ``Baseline" method, where the predicted state $\ket{d}$ is set equal to the prepared qubit state prior to the first measurement. }  
\\
\indent \ps{We find that the Sig+RF method reduces EOM infidelity across \sx{all of} the datasets. On the RQC dataset, the \ps{EOM} infidelity decreases by an average of $35.0(15)\%$ and $21.5(23)\%$ compared to the baseline and RF methods, respectively. Similarly, on the Oxford qutrit dataset, the \ps{EOM} infidelity is reduced by $41.3(13)\%$ and $11.7(29)\%$ compared to the baseline and RF methods. We would like to highlight that in the case of the AQT qutrit dataset, where significant mid-measurement state transitions occur when the qubit is prepared in the $\ket{1}$ state, the Sig+RF method reduces EOM infidelity from $15.17\%$ to $4.43(14)\%$. This represents a $70.8(9)\%$ improvement over the baseline and a $43.8(2)\%$ improvement over the RF approach.} Since the machine learning models for predicting the EOM state label are the same as those used for state assignment, they both have similar computational costs for training and inference. For more details please refer to \appendixref{\ref{app:ml_methods}}. \sx{Further comparison of the Sig+RF classification method against additional machine learning models for state discrimination is provided in the \appendixref{\ref{app:comparisons}}.}
\\
\indent In conclusion, the signature-based features for state discrimination provide the following benefits: First, \ps{using the path signature as a feature set for state classification outperforms the standard approach that uses the optimally weighted measurement record}. Secondly, \ps{the path signature feature set enables} detection and classification of mid-measurement state transitions, which in turn \ps{enables} a more accurate prediction of the qubit state at the end of the measurement. \ps{Thirdly, we have shown that the pipeline of evaluating the path signature followed by applying a basic machine-learning classifier is robust across multiple devices, and does not require careful hyperparameter optimization in order to achieve significant enhancement over the conventional approach. This ``out-of-the-box" property makes the path-signature method attractive where ease of implementation and broad applicability across quantum devices with differing readout parameters are important.} In the future, there are a few more directions \sx{to explore.} The first is to combine the signature method with multi-frequency probing, which could provide additional dimensional information~\cite{Chen2023}, potentially improving the readout fidelity further. In addition, \sx{the path signature} method shows potential in quantum trajectory studies~\cite{PhysRevX.10.011006, Dorsselaer_2000} and weak-measurement experiments~\cite{Flack_2014} for accurately analyzing data traces and tracking state changes.

\begingroup
\renewcommand{\addcontentsline}[3]{}
\acknowledgments
\endgroup
This project is supported by the Eric and Wendy Schmidt AI in Science Postdoctoral Fellowship, a Schmidt Science program. This work was supported by the U.S. Department of Energy, Office of Science, Advanced Scientific Computing Research Testbeds for Science program under Contract No. DE-AC02-05CH11231. This work was supported in part by the Ministry of Education, Culture, Sports, Science and Technology (MEXT) Quantum Leap Flagship Program (QLEAP) (Grant No. JPMXS0118068682).
S. C. was supported by Schmidt Science. 
Z. S. was supported by the EPSRC [EP/S026347/1].
J.-Q. Z. was supported by the Kennedy Trust Prize Studentship [AZT00050-AZ04].
P.S. was supported by the JSPS Grant-in-Aid for Scientific Research (KAKENHI) (Grant No. JP22H04937). 
P.L.~acknowledges support from the EPSRC b. [EP/T001062/1, EP/N015118/1, EP/M013243/1]. 
M.B. acknowledges support from the EPSRC QT Fellowship grant [EP/W027992/1]. 
T. L. was funded in part by the EPSRC [EP/S026347/1], in part by The Alan Turing Institute under the EPSRC [EP/N510129/1], the Data Centric Engineering Programme (under the Lloyd’s Register Foundation grant G0095), the Defence and Security Programme (funded by the UK Government) and in part by the Hong Kong Innovation and Technology Commission (InnoHK Project CIMDA). This material was funded in part by the U.S. Department of Energy, Office of Science, Office of Advanced Scientific Computing Research Quantum Testbed Program under contract DE-AC02-05CH11231. 
We thank Youpeng Zhong, Sam Morley, and Benjamin Walker for the insightful discussion and feedback. The authors would like to acknowledge the use of the University of Oxford Advanced Research Computing (ARC) facility in carrying out this work~\cite{zenodo.22558}.
 
\bibliography{bib}
 
\newpage
\appendix

\onecolumngrid

\newcommand{\iptm}[1]{\begin{tabular}{c} \includegraphics[width=3.5cm,height=2.2cm]{#1}\end{tabular}}

\newcommand{\iptsq}[1]{\begin{tabular}{c} \includegraphics[width=3cm,height=3cm]{#1}\end{tabular}}

\section{More details on path signatures\label{app:signature}}

\indent The path signature is a mathematical object that encodes the geometric and algebraic properties of a path. It is a tool used to differentiate paths based on their geometry, capturing both the overall structure and finer details of how the path evolves. \ps{An introduction to the path signature in the context of machine learning applications is given in Ref.~\cite{chevyrev2016primer}. An efficient implementation for computing path signatures is provided by the Python package iisignature~\cite{reizenstein2018iisignature}.} Formally speaking, the signature of a multi-dimensional time-series path is a graded, infinite collection of iterated integrals, where the signature of a path $X$ with dimension $d$ up to degree $N$ is the collection

\begin{equation}\label{eq:sig}
\begin{split}
S^N(X) := \Bigg(&\underset{0 < t_1 < \cdots < t_k < t_r}{\int\cdots\int} 
\frac{\mathrm d X_{i_1}}{\mathrm dt}(t_1) \cdot \frac{\mathrm d X_{i_2}}{\mathrm dt}(t_2) \frac{\mathrm d X_{i_k}}{\mathrm dt}(t_k) \mathrm dt_1 \cdots \mathrm dt_k \Bigg)
_{\substack{\!\!1 \leq i_1, \ldots, i_k \leq d \\ \!\!k=0, 1, 2, \ldots, N}}.
\end{split}
\end{equation}

where $\{i_1, \ldots, i_k\}$ denotes different dimensions of the collected signal. For this study, it may refer to the $I$ channel or the $Q$ channel.

\begin{figure}[h!]
    \centering
    \includegraphics[width=0.4\linewidth]{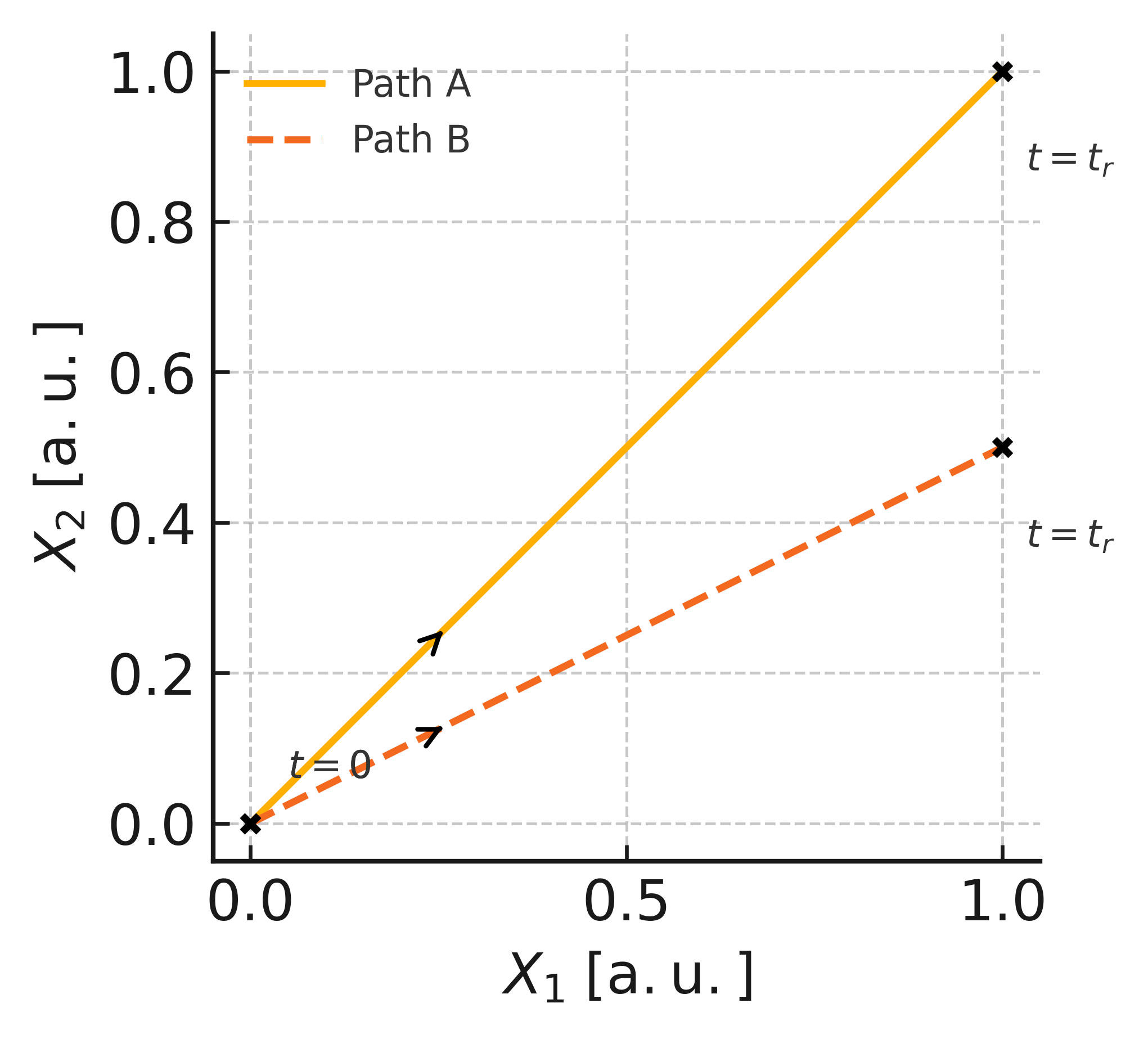}
     \caption{Two paths with different first-order signatures; their total displacements differ.}
     \label{fig:sig1}
\end{figure}

\indent As an illustrative example, let $X = (X_1, X_2)$ represent a two-dimensional path parameterized over some interval, where $X_1(t), X_2(t)$, denote the coordinates of the path at time $t$. The path signature is constructed by integrating specific combinations of the increments of these coordinates, providing a hierarchy of features that describe the path. The first-order signature of the path corresponds to its total displacement and is defined as:

\[
S^{(1)}(X) = \left( \int X_1'(t) \, dt, \int X_2'(t) \, dt \right) = \left( X_1(T) - X_1(0), X_2(T) - X_2(0) \right),
\]

where \( T \) is the terminal time of the path. This simply captures the net change in each coordinate between the start and end points of the path, ignoring the intermediate behavior. Thus the two paths shown in Fig.~\ref{fig:sig1} can be differentiated by their first order signatures, because they have different $S^1$.

\indent In contrast, the two paths shown in Fig.~5 cannot be distinguished solely by their first-order signatures, as both have the same total displacement. The second-order signatures provide additional detail by capturing the shape of the path, rather than focusing solely on its endpoint. For example, consider one of the components of the second-order signature:
 
\begin{figure}[h!]
    \centering
    \includegraphics[width=0.4\linewidth]{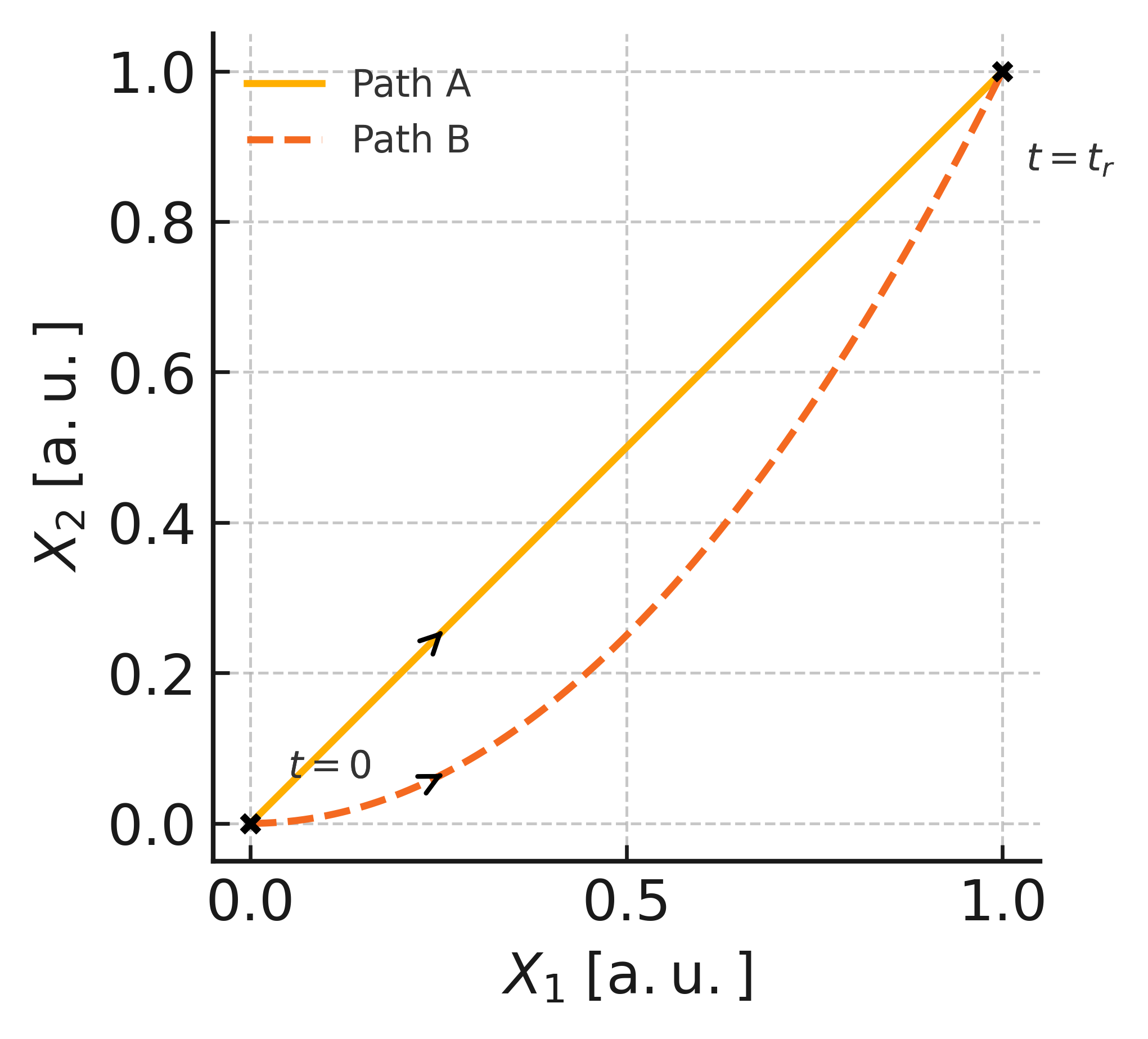}
    \caption{Two paths which share the same end-point displacement but differ smoothly in shape, yielding distinct second-order (area) signatures.}
    \label{fig:sig2}
\end{figure}

$$
S^{1,2} = \int_{t=0}^1 \int_{s=0}^t dX_1(s) dX_2(t) 
= \int_{t=0}^1 X_1(t) dX_2(t)
= \sum_j X_1(t_j) \Delta X_2(t_j).
$$
\indent The two paths in Fig. \ref{fig:sig2} can be effectively distinguished using their second-order signatures. A simple model, such as a decision tree, could easily classify the paths by splitting on the value of \( S^{1,2} \). 

\begin{figure}[h!]
    \centering
    \includegraphics[width=0.4\linewidth]{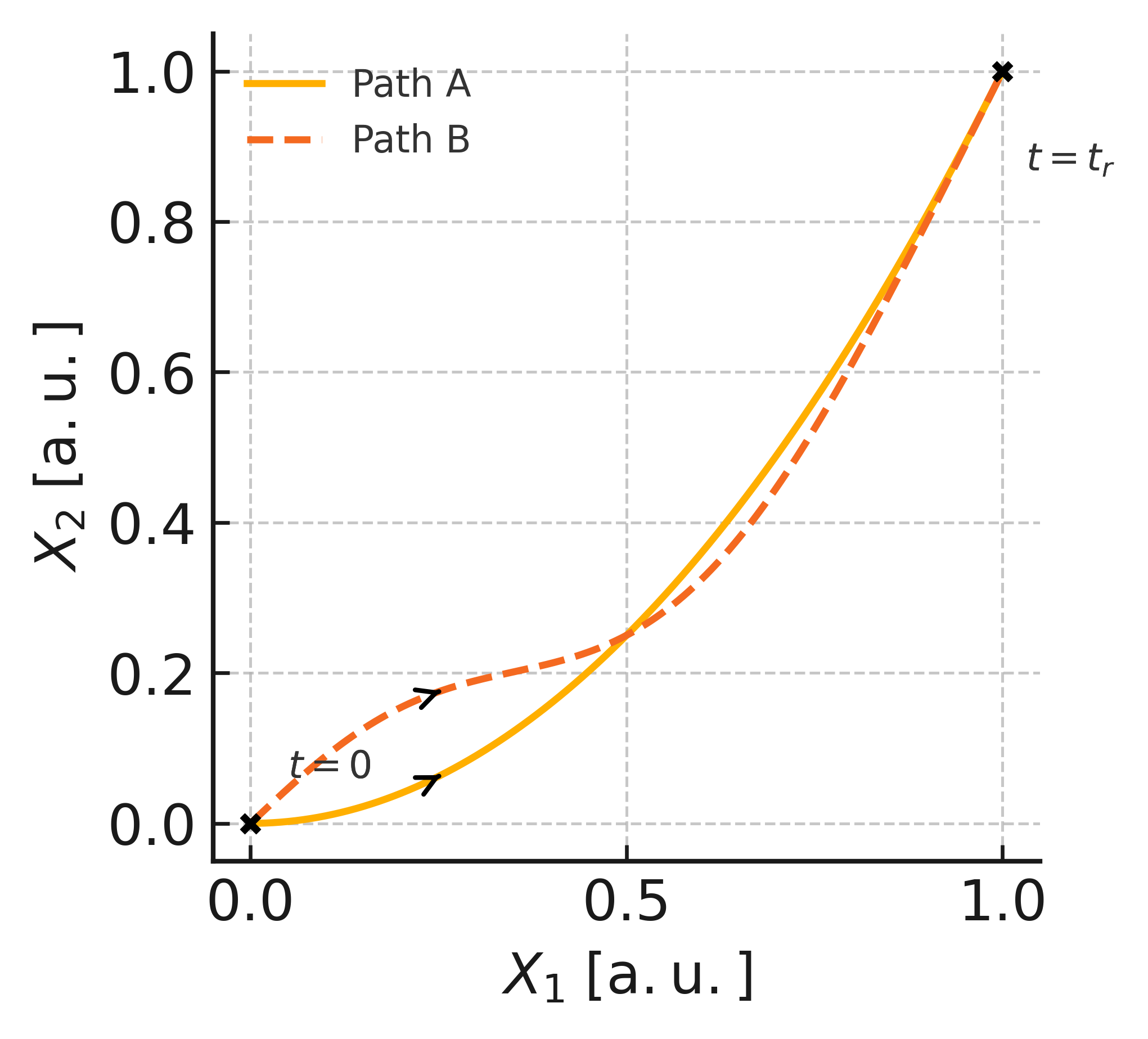}
    \caption{Two paths with matching displacement and enclosed area. The higher-order (third-order) signatures can still differentiate the paths.}
    \label{fig:sig3}
\end{figure}

\indent Higher-order signatures provide even greater discriminatory power, capturing finer geometric details. For example, in Fig.~\ref{fig:sig3}, both paths have identical start/end points and are shaped so that their signed area—the second-order signatures—are the same. However, their third-order signatures still differ, allowing for differentiation at a higher level of detail.

\section{Machine learning methods}\label{app:ml_methods}

In this appendix we explain the machine learning methods used in this experiment.

\paragraph{Gaussian Mixture Model}

The Gaussian Mixture Model (GMM) is a probabilistic machine learning algorithm that models data as a mixture of multiple Gaussian distributions. Training a GMM is the process of determining the parameters of these distributions. Each Gaussian component, denoted as $\mathcal{N}(x | \mu_k, \Sigma_k)$, is characterized by its mean $\mu_k$ and covariance $\Sigma_k$. The Probability Distribution Function of a Gaussian component is expressed as:

$$
\mathcal{N}(x | \mu_k, \Sigma_k) = \frac{1}{\sqrt{(2\pi)^d |\Sigma_k|}} \exp\left(-\frac{1}{2}(x-\mu_k)^T\Sigma_k^{-1}(x-\mu_k)\right)
$$,

where $d$ denotes the dimensionality of each data point, and $k$ denotes the index of a Gaussian component.

The variance within each Gaussian component is characterized by its covariance matrix, $\Sigma_k$, which accounts for potential correlations across different dimensions. In the context of readout signal classification, the aggregated noiseless trajectory serves as the center of the Gaussian distribution, while the noise in the readout signal defines the spread. Typically, noises in the $I$ and $Q$ channels are independent Gaussian noise, allowing us to model the covariance matrices $\Sigma_k$ as diagonal, with equal diagonal elements. This assumption corresponds to modeling the distribution as 'spherical' in the I-Q plane, where noise is isotropic and independent across dimensions. However, when quantum amplifiers operate in a regime where amplification becomes nonlinear (e.g., as seen in the RQC dataset for Q1, Q2, and Q3), this assumption may no longer hold. In such cases, the GMM model is trained using the full covariance matrix, without imposing any additional constraints.
The implementation of this Gaussian mixture model is provided by the scikit-learn library \cite{pedregosa2011scikit}.

\paragraph{Linear Support vector machine (Linear-SVM)}
A Support Vector Machine (SVM) \cite{Cortes1995} is a supervised machine learning algorithm used for classification tasks. It operates by identifying the optimal hyperplane that separates data points from different classes in a high-dimensional feature space. The goal of SVM is to maximize the margin between the nearest data points (called support vectors) of opposing classes, enhancing the model’s ability to generalize to unseen data. In this study, we used a linear SVM, which operates in the raw feature space derived from the path signature of the readout signal.

\paragraph{Random Forest (RF)} 

The Random Forest (RF) algorithm \cite{ho1995random} is an ensemble learning technique that builds multiple decision trees to improve classification performance and reduce overfitting compared to individual decision trees. The decision tree is a machine learning model that follows decision rules, where each node splits the data based on a condition (e.g., ``Is the first sample of the received I-channel signal $>$ 0.01?") to split the data into branches, continuing until a final decision or classification is reached. A key advantage of Random Forest is its ability to model non-linear relationships, making it effective in handling complex data patterns that linear models might miss. 

\paragraph{Hyperparameter search of the RF}

The hyperparameter search in this experiment follows standard practices for training RF. It focuses on several key parameters: the number of trees in the forest, which was tested over the range of values 50, 60, 70, 80, 90, 100, 110, 120, 130, 140, 150, and the maximum depth of the trees, which was varied over three possible values: 10, 20, and 30. The minimum number of samples required to split a node was explored with values of 2, 5, and 10, while the minimum number of samples required to be present at a leaf node was tested at values of 1, 2, and 4. In this experiment, the randomized search was chosen to explore the hyperparameter space efficiently. This method selects a fixed number of random combinations to test, reducing the computational costs and allowing for the exploration of a wider parameter space.

The hyperparameter search method employs StratifiedKFold cross-validation to ensure a robust evaluation of the parameter combinations. StratifiedKFold was used with five splits to maintain the proportion of classes across both training and validation datasets during cross-validation. This approach is crucial in imbalanced classification tasks, as it preserves the original class distribution in each fold, ensuring that no class is over- or under-represented in either the training or validation set. In each fold, the data is split into five subsets, and in each iteration of cross-validation, four of these subsets are used for training, while one is used for validation. This process is repeated five times, such that each subset serves as the validation set once, and the remaining subsets are used for training. This ensures a reliable estimate of model performance.

\paragraph{Evaluating the path signature} 

In this work, we use the ``sktime'' package \cite{loning2019sktime,sktime} to compute the path signature. Given that the GMM method performs best with the aggregated value of the path signature, we follow the same approach here, first calculating the weighted trajectory before evaluating the signature. For qubit datasets, the weighting is determined by averaging the traces and computing the difference between the ground and excited states. For qutrit datasets, we compute the differences between the averaged traces of each of the three states, and then use the average of these differences as the weights. We finally computed the signature directly from the weighted entire trace. 

\paragraph{Performance evaluation}

For training the GMM model, we randomly split the dataset into training and testing sets, trained the model on the training set, and reported the accuracy on the testing set. For the SVM and RF models, we selected a random subset of samples from the dataset and divided it into training, validation, and testing sets. The training set was used to train the models with various hyperparameters, while the validation set was used to assess each model’s performance and select the best one. 
For Random Forest, key hyperparameters included the number of decision trees, maximum depth, and the minimum samples required for splits and leaf nodes. 
For SVM, the regularization parameter was optimized to balance error minimization and complexity.
After training and hyperparameter optimization, the test set was used to evaluate the final model performance. The test set was not involved in training or hyperparameter tuning. For more details on splitting ratios and sample sizes, please refer to the appendix section of each dataset.

\section{Supplementary information for Dataset OXF Qt\label{app:OXF_Qt}}

\begin{figure}[h!]
     \centering
        \includegraphics[width=.8\linewidth]{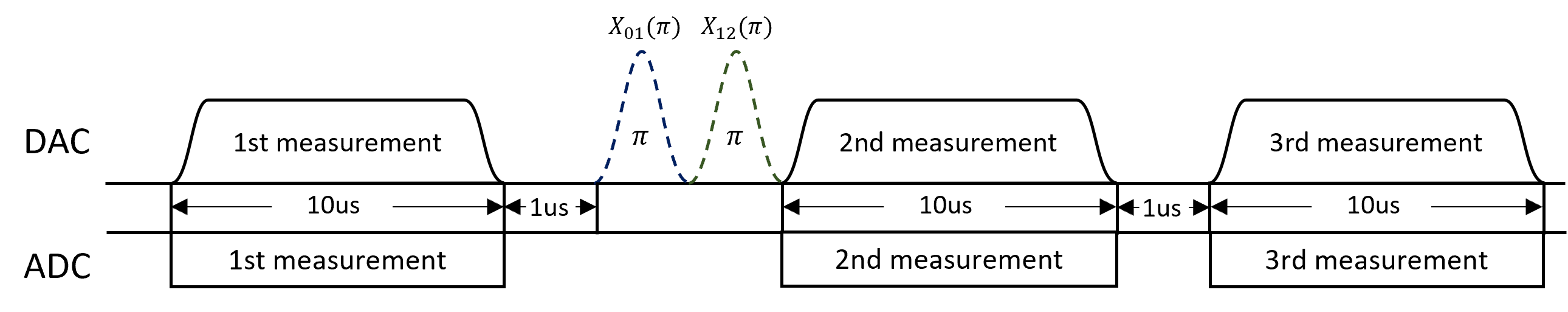}
        \caption{Experiment pulse scheme for the Oxford \ps{q}utrit dataset. There are three measurement pulses involved in the experiment. The first measurement is used to implement post-selection, ensuring the initial state is in the ground state. The second measurement pulse is analyzed using the signature approach, while the third measurement is used to detect if a state transition event occurred during the second measurement.}
        \label{fig:pulse_scheme_oxford_qutrit}
\end{figure}

The experimental pulse scheme is depicted in Fig.\ref{fig:pulse_scheme_oxford_qutrit}. It involves three measurements. The qubit state is initially determined using the conventional GMM method, based on data from the first measurement. The traces are then post-selected to ensure that the initial state is the ground state. Subsequently, the qubit is prepared into the $\ket{0}$, $\ket{1}$, and $\ket{2}$ states by applying $\pi$ pulses for the transitions $\ket{0}\rightarrow\ket{1}$ and $\ket{1}\rightarrow\ket{2}$. The gate fidelity is approximately $99.7\%$. A second measurement is conducted afterward, and the traces are recorded for analysis. Following the second measurement, a third measurement is immediately performed. The third measurement aims to identify any state transitions that occur during the second measurement. Transition events are considered to have occurred if the third measurement yields a state different from the intended preparation state. The state discrimination is implemented with the conventional approach, by applying the GMM model on integrated signals.

\begin{table}[]
    \centering
      \begin{ruledtabular}
        \begin{tabular}{lllc}
        Parameter& & OXF Qt\\\colrule 
        Resonator frequency &$f_{Res}$ (MHz)& 8783   \\
        Resonator line width &$\kappa/2\pi$ (MHz)& 0.524   \\
            $\ket{0}, \ket{1}$ Transition frequency &$f_{01}$ (MHz) & 4134.33  \\
            $\ket{0}, \ket{1}$ Relaxation time  &$T_1^{(01)}$ (us)  & $221 (30)$ \\
            $\ket{0}, \ket{1}$ Hahn decoherence time &$T_2^{(01)}$ Echo (us) & $ 126(15)$  \\
            $\ket{0}, \ket{1}$ Ramsey decoherence time &$T_2^{(01)}$ Ramsey (us) & $ 96(10) $  \\
            $\ket{1}, \ket{2}$ Transition frequency &$f_{12}$ (MHz) & 3937.66  \\
            $\ket{1}, \ket{2}$ Relaxation time &$T_1^{(12)}$ (us)      & $119(20)$  \\
            $\ket{1}, \ket{2}$ Hahn decoherence time &$T_2^{(12)}$ Echo (us) & $76(27)$ \\
            $\ket{1}, \ket{2}$ Ramsey decoherence time &$T_2^{(12)}$ Ramsey (us) & $52(4) $ \\
        \end{tabular}
        \end{ruledtabular}
    
    \caption{Summary of device parameters \ps{for the Oxford qutrit.} \label{tab:device_parameters}}
\end{table}

\begin{figure}[h!]
     \centering
        \includegraphics[width=.8\linewidth]{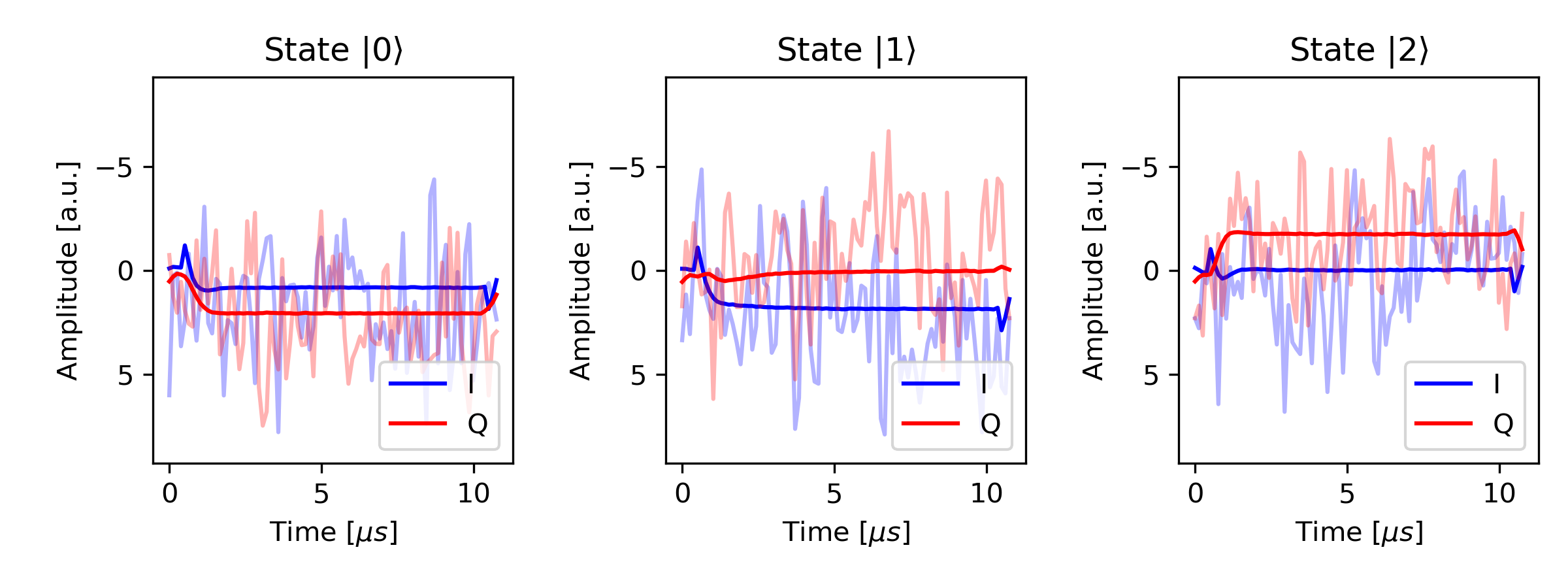}
        \caption{Signal obtained from the Oxford qutrit device by probing the resonator when the transmon qubit is in states $\ket{0}$, $\ket{1}$, and $\ket{2}$, respectively. The solid line represents the average result, and the translucent line denotes a single-shot example trace. Blue and red colors correspond to the I and Q channels of the signal, respectively.}
        \label{fig:traces}
\end{figure}

Each trace was acquired using two analog-digital converters with a sampling rate of 1 Gsps each. The traces have two distinct dimensions (I and Q). The recorded signal data has a carrier frequency of 125 MHz. A short-term Fourier transformation was applied to segments of 256 samples to demodulate the signal at this frequency. Samples of the collected traces are shown in Fig.{\ref{fig:traces}}.

Using the described experimental scheme, a database was established containing 70,000 traces for each targeted state, culminating in a total of 210,000 traces. The time spent collecting these traces was approximately two hours. The statistics of the post-selection process are presented in the Table \ref{tab:oxf_qt_statistics}.

\begin{table}[h!]
\begin{tabular}{c|ccc}
Prepared state & $N_i$    & $\Tilde{N}_i$ & $\Tilde{N}_i/N_i$ \\
\hline
$\ket{0}$      & $59,768$ & $58,273$      & $97.50\%$         \\
$\ket{1}$      & $52,630$ & $32,414$      & $61.59\%$         \\
$\ket{2}$      & $55,997$ & $32,339$      & $57.75\%$        
\end{tabular}
\caption{\sx{The statistics of the post-selection process. Here, $N_i$ represents the number of traces intended for state preparation $\ket{i}$ that pass the initial post-selection, ensuring the initial state is the ground state. $\Tilde{N}_i$ denotes the number of traces that pass both the initial post-selection and the final post-selection, where the third measurement identifies the state at $\ket{i}$. The ratio $N_i/\Tilde{N}_i$ gives the fraction of measurements with no state transition during the second measurement.}}\label{tab:oxf_qt_statistics}
\end{table}

For each machine learning classification experiment, 2,000 traces per state were randomly selected from the database, resulting in a total of 6,000 traces per experiment. These traces were split into a training set of 4,800 traces and a testing set of 1,200 traces, with the reported accuracies based on the testing set. The 4,800 training traces were further divided into 3,840 traces for training and 960 for validation. The above process is repeated 10 times, each using a different random seed for data selection and splitting to determine the confidence of the accuracies.

\begin{figure}[h!]
     \centering
        \includegraphics[width=.7\linewidth]{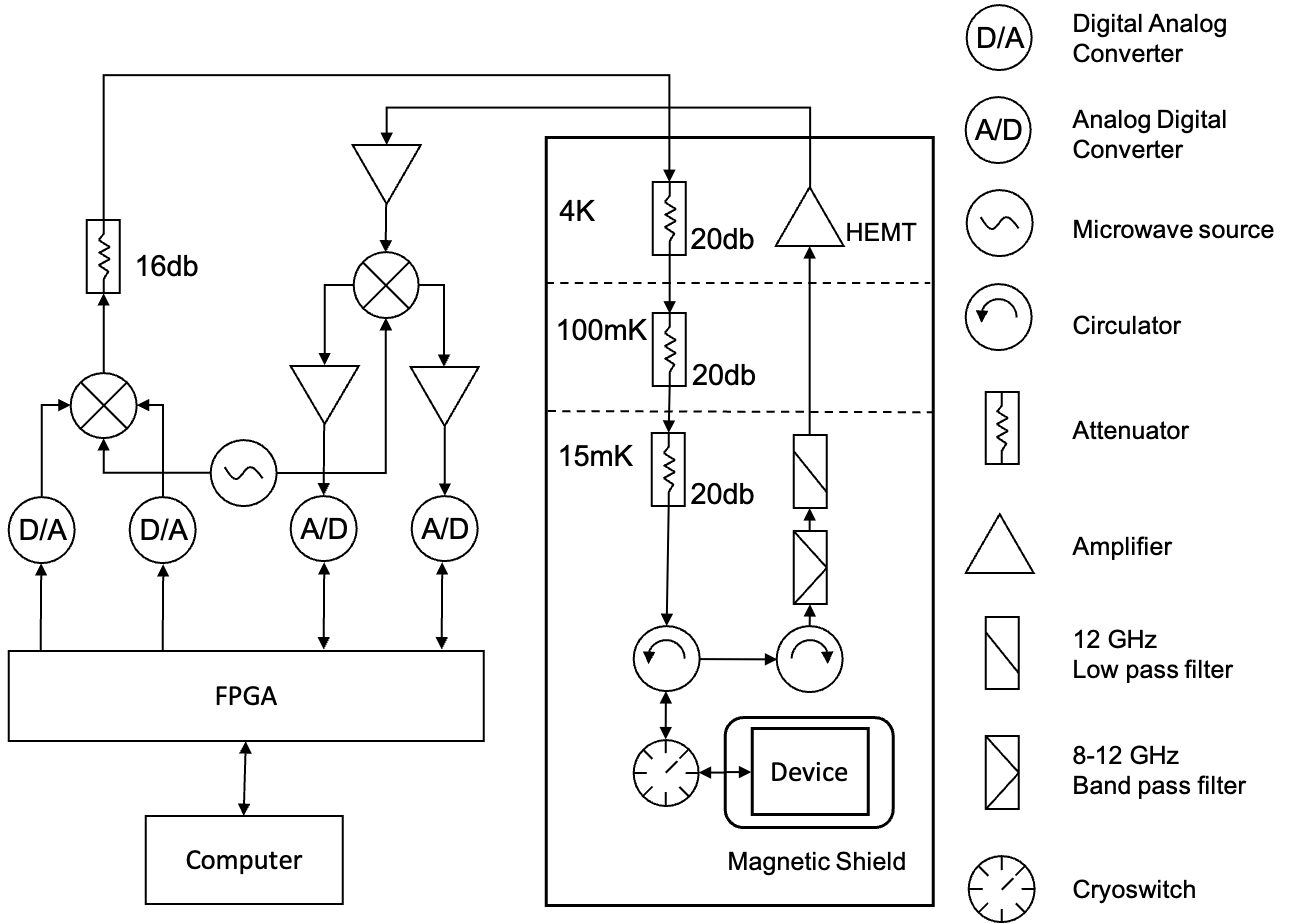}
        \caption{The schematics of the readout chain of the experiment setup \ps{for the Oxford qutrit}.}
        \label{fig:readout_chain}
\end{figure}

The detailed readout chain is described in Fig.\ref{fig:readout_chain}.

\begin{figure}[h!]
    \centering
    \includegraphics[width=\linewidth]{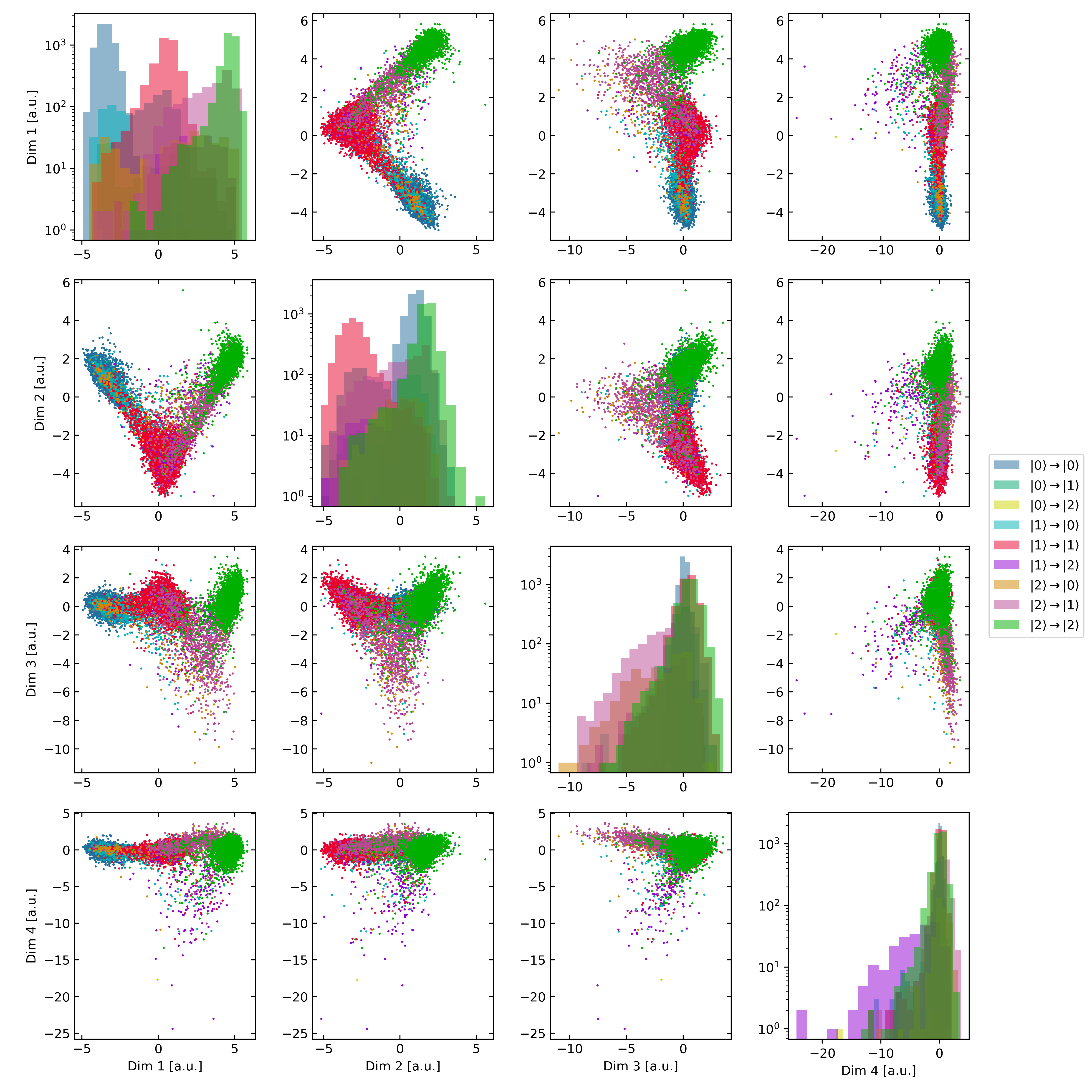}
    \caption{Pair plot of the distribution of a depth-5 signature projection from the Oxford qutrit experimental dataset, based on traces from the second measurement. The projection direction is determined using Linear Discriminant Analysis (LDA). Different colors indicate the states of the second and third measurement results.}
    \label{fig:pair_plot_square_signature}
\end{figure}

We selected a linear \ac{SVC} and \ac{RF} for implementing the classification of the signature features. Notably, the RF performs better than linear \ac{SVC}, which indicates that the signature features between classes are not linearly separable. This is likely because the readout signal follows a complex path, making linear separability difficult when analyzing the transition event. Although the signature method aims to map such complex paths into a higher-dimensional space to enable linear separation \cite{10041999,MR3727607}, the readout signal may require an exceptionally high-dimensional representation.

\begin{figure}[h!]
     \centering
        \includegraphics[width=.8\linewidth]{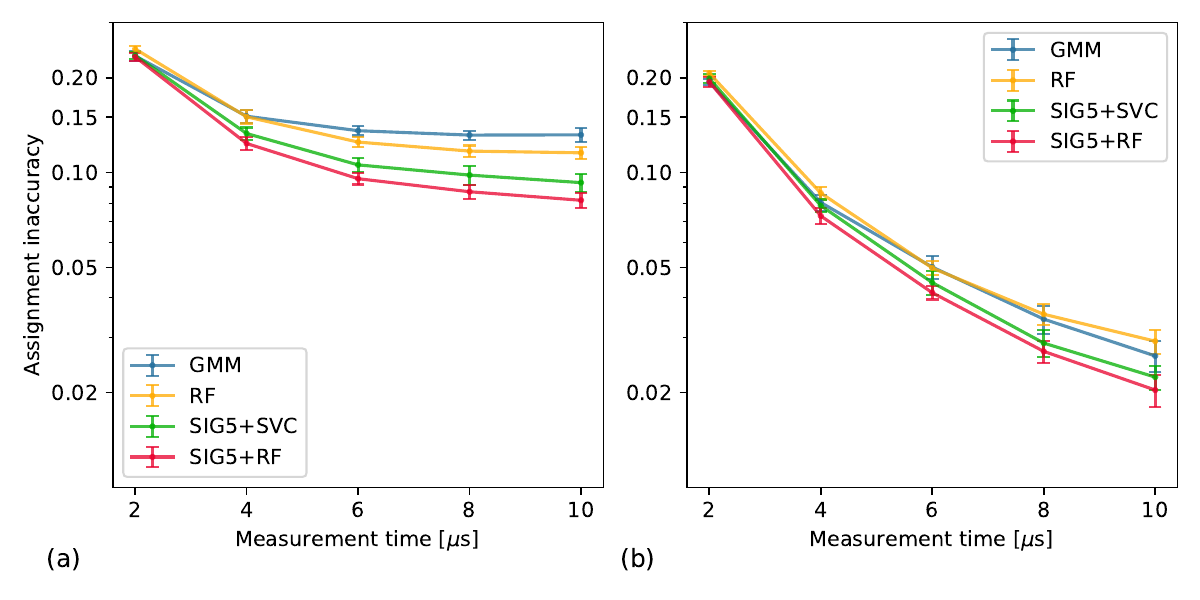}
        \caption{(a) Classification accuracy for the Oxford qutrit dataset as a function of measurement length, compared across various classification methods. The tested classification approaches are Gaussian Mixture Model (GMM), Random Forest (RF), Linear support vector classifier on path signature of depth $5$ (SIG$5$+SVC), Random forest on path signature of depth $5$ (SIG$5$+RF) (b) Classification accuracy excluding state transitions, determined by implementing an extra post-selection measurement.}
        \label{fig:classification}
\end{figure}

In addition, we \sx{here investigate the effect of incorporating} an \sx{additional} post-selection in the \sx{OXF Qt} dataset, to remove the traces where state transition likely occurred during the measurement. This is done by conducting another measurement immediately after the measurement traces are collected for classification analysis, then applying the traditional integration and the GMM method for readout on the collected signals. We kept only those traces where the last measurement agreed on the state we intended to prepare. When we applied the signature method to this post-selected dataset, we noted an improvement in accuracy at shorter measurement times. However, this enhancement diminished with longer measurement durations, see Fig.\ref{fig:classification}(b). This outcome leads further evidence to the claim that the signature approach is effective by capturing state transitions occurring during the measurement process. \sx{Note that all values reported in the main text and other appendices use only a single post-selection to reduce state preparation error; they do not include this additional post-selection.}

\begin{table}[h!]
    \centering
    \begin{tabular}{c|cccc}
        & $\ket{0}$ & $\ket{1}$ & $\ket{2}$ & Overall \\ \hline
        Baseline & 1.79 & 25.96 & 33.71 & 20.49\\ 
        RF & 1.79  & 18.71(60) & 20.34(77) & 13.61(33)\\ 
        Sig+RF & 1.79  & 16.23(45) & 18.05(61) & 12.02(26)\\ 
    \end{tabular}
    \caption{Table with mean and standard deviation values for Baseline, RF, and Sig+RF categories.}
\end{table}

\clearpage

\section{Supplementary information for the AQT dataset (AQT Qt)\label{app:AQT_Qt}}

\begin{figure}[h!]
     \centering
        \includegraphics[width=0.9\linewidth]{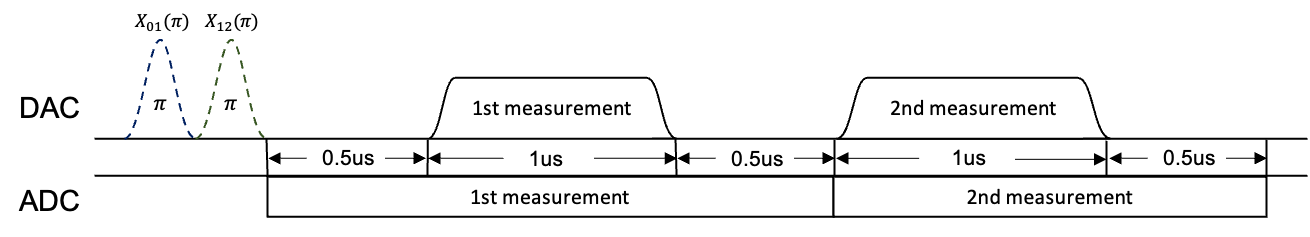}
        \caption{Experimental pulse scheme for the AQT qutrit dataset. The first measurement is analyzed using the signature approach, while the second measurement is used to detect if a state transition event occurred during the first measurement.}
        \label{fig:pulse_scheme_aqt_qutrit}
\end{figure}

\begin{figure}
    \centering
    \includegraphics[width=0.8\linewidth]{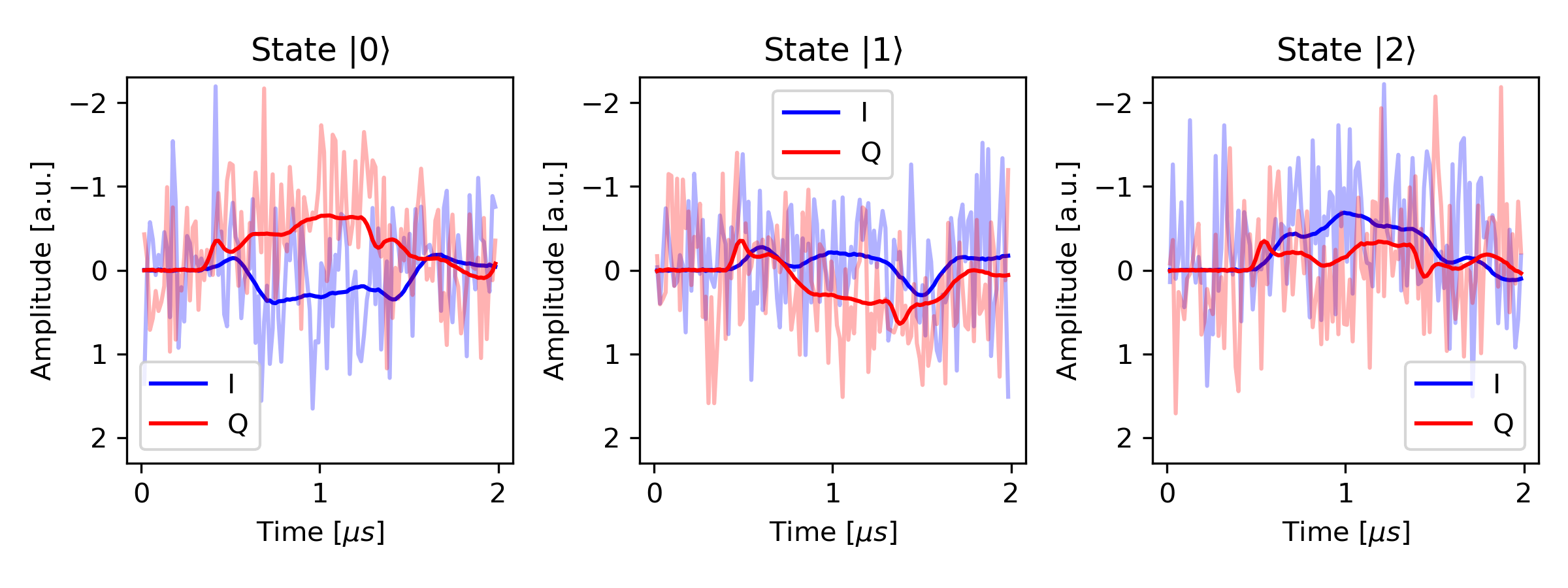}
    \caption{Signal obtained from the AQT qutrit device by probing the resonator when the transmon qubit is in states $\ket{0}$, $\ket{1}$, and $\ket{2}$, respectively. The solid line represents the average result, and the translucent line denotes a single-shot example trace. Blue and red colors correspond to the I and Q channels of the signal, respectively.}
    \label{fig:traces-aqt}
\end{figure}

\begin{figure}
    \centering
    \includegraphics[width=0.5\linewidth]{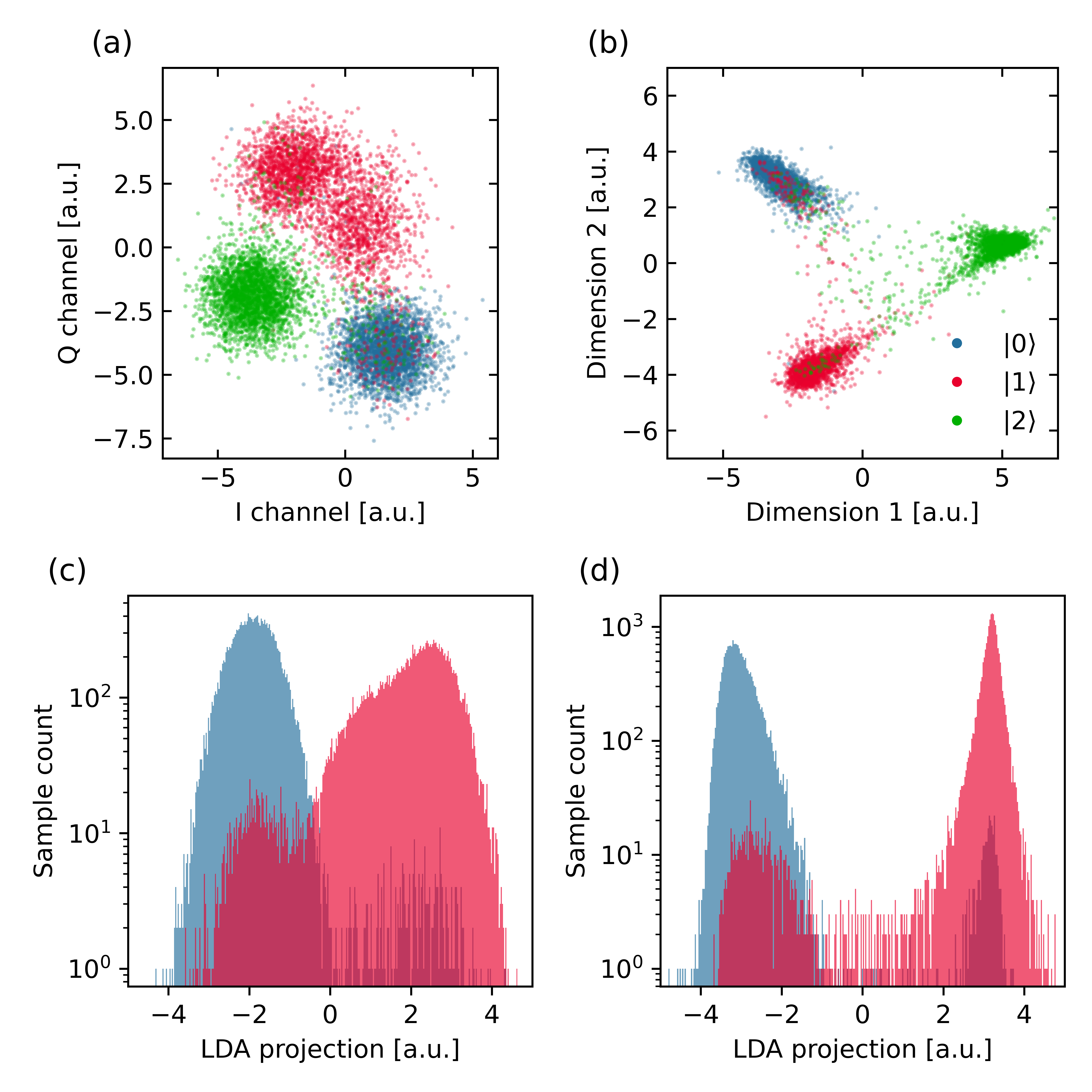}
    \caption{(a) Distribution of the AQT qutrit experimental dataset on the IQ plane using conventional integration methods. The blue, red and green denotes the state is prepared to $\ket{0}$, $\ket{1}$, and $\ket{2}$ respectively. (b) Projection of a depth-5 signature calculated from the same dataset. The projection direction is evaluated using LDA. (c) Histogram of the integration method projected linearly along the most distinguishable direction using LDA, acting on the data of $\ket{0}$ and $\ket{1}$ state only. (d) Histogram of signature features (depth=5) from the same dataset, projected similarly.}
    \label{fig:vis_sig_aqt}
\end{figure}

The experimental pulse scheme is illustrated in Fig. \ref{fig:pulse_scheme_aqt_qutrit}. The signal is collected at a rate of 2 GSa/s and demodulated with segments of 8 samples, using the QubiC 2.0 system \cite{2309.10333}. Due to the hardware limit, we could not collect three consecutive measurement pulses. In this dataset, we omitted post-selection on the initial state and reported that the fidelity of the transmon being in the $\ket{0}$ state at the start of the experiment is $98.5\%$. Immediately following the first measurement pulse, a second measurement pulse is applied. This dataset comprises 60,000 traces, with 20,000 traces each for preparing the transmon in the $\ket{0}$, $\ket{1}$, and $\ket{2}$ states. The time taken to collect these traces was approximately 165 minutes. An example of these traces is depicted in Fig.\ref{fig:traces-aqt}, and the aggregated traces and the projected signature are shown in Fig. \ref{fig:vis_sig_aqt}. For additional details on this dataset, please refer to the prior study \cite{2406.18807}.

For the classification experiment, 15,000 traces per state were randomly selected from the database, resulting in a total of 45,000 traces per experiment. These traces were split into a training set of 36,000 traces and a testing set of 9000 traces, with the reported accuracies based on the testing set. The 36,000 training traces were further divided into 28,800 traces for training and 7,200 for validation. The above process is repeated 10 times, each using a different random seed for data selection and splitting to determine the confidence of the accuracies. 

\begin{figure}
    \centering
    \includegraphics[width=\linewidth]{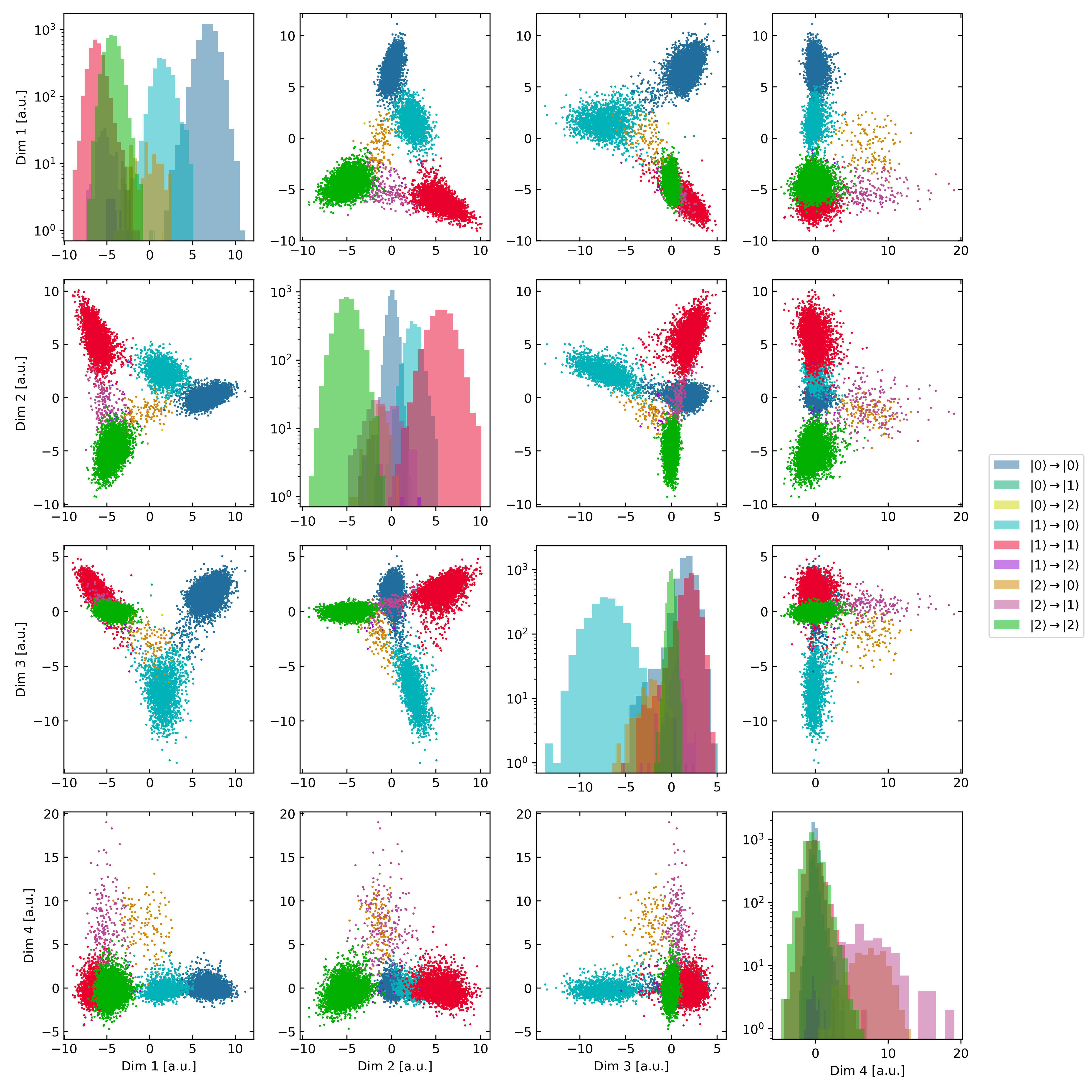}
    \caption{Pair plot of the distribution of a depth-5 signature projection from the AQT qutrit experimental dataset, based on traces from the first measurement. The projection direction is determined using Linear Discriminant Analysis (LDA). Different colors indicate the states of the first and second measurement results.}
    \label{fig:pair_plot_aqt_signature}
\end{figure}

\begin{figure}
    \centering
    \includegraphics[width=0.4\linewidth]{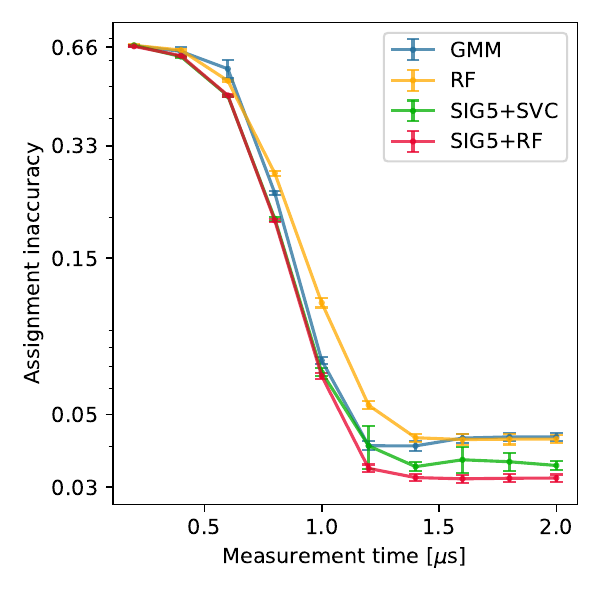}
    \caption{Classification accuracy for the AQT qutrit dataset as a function of measurement length, compared across various classification methods. The tested classification approaches are Gaussian Mixture Model (GMM), Random Forest (RF), Linear support vector classifier on path signature of depth $5$ (SIG$5$+SVC), Random forest on path signature of depth $5$ (SIG$5$+RF).}
    \label{fig:clf_sig_aqt}
\end{figure}

\begin{table}[h!]
    \centering
    \begin{tabular}{c|cccc}
        
        & $\ket{0}$ & $\ket{1}$ & $\ket{2}$ & Overall \\ \hline
        Baseline & 0.085 & 37.94 & 7.47 & 15.17 \\ 
        RF & 0.085 &  16.06(33) & 7.49(23) & 7.88(14) \\ 
        Sig+RF & 0.085  & 6.15(32) & 7.06(26) & 4.43(14) \\ 
    \end{tabular}
    \caption{Table with the mean and standard deviation values for the Baseline, RF, and Sig+RF categories.}
\end{table}

\clearpage

\section{Supplementary information for the Oxford 4Q dataset (OXF Q1-Q4)\label{app:OXF_Simo}}

\begin{figure}[h!]
     \centering
        \includegraphics[width=.7\linewidth]{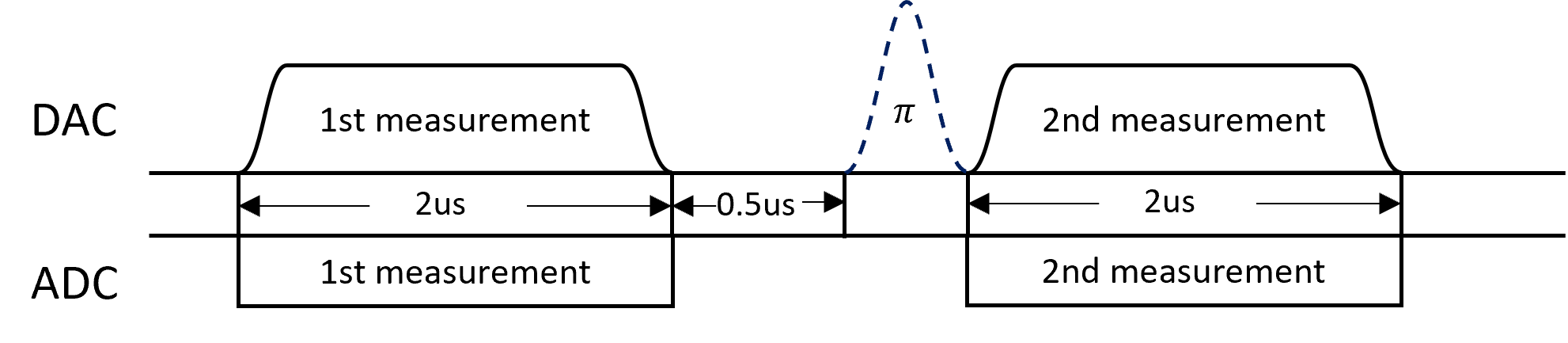}
        \caption{Experiment pulse scheme for the Oxford qubits dataset. The first measurement is used to implement post-selection, ensuring the initial state is in the ground state. The second measurement pulse is analyzed using the signature approach.}
        \label{fig:measurement_scheme_oxford_qubits}
\end{figure}

The dataset was collected from a 4-qubit multiplexed readout coaxmon device, as described in \cite{fasciati2024Complementing}. It consists of 20,000 traces, with 10,000 corresponding to ground states and 10,000 to excited states. Each trace contains 5,000 data points, sampled at a rate of 1 GSa/s. Collecting this dataset took 36 minutes and 43 seconds. The readout pulse applied to the device is a $2 \mu s$ square pulse with $10$ ns sigma Gaussian-shaped edges. During the readout, four different pulses, each at a unique frequency, are sent simultaneously to the device, and the ADC collects and demodulates the signals at each frequency using segments of 25 samples. The pulse scheme is illustrated in Fig. \ref{fig:measurement_scheme_oxford_qubits}. The first measurement ensures that the state is initialized in the ground state by performing post-selection based on the measurement yielding the $\ket{0}$ state. The post-selected initial state fidelities for qubits Q1 through Q4 are $99.30\%$, $99.18\%$, $98.92\%$, and $98.51\%$, respectively.

For each machine learning classification experiment, 2,000 traces per state were randomly selected from the database, resulting in a total of 4,000 traces per experiment. These traces were split into a training set of 3,200 traces and a testing set of 800 traces, with the reported accuracies based on the testing set. The 3,200 training traces were further divided into 2,560 traces for training and 640 for validation. The above process is repeated 10 times, each using a different random seed for data selection and splitting. The evaluated accuracies were then used for statistical analysis to produce the confidence level of the accuracies.

\begin{figure}[h]
    \centering
    \begin{tabular}{c|c|c|c|c}

Dataset & Oxford Q1 & Oxford Q2 & Oxford Q3 & Oxford Q4 \\  \hline &&& \\ 
    \begin{tabular}{c}Scatter\end{tabular} & \iptsq{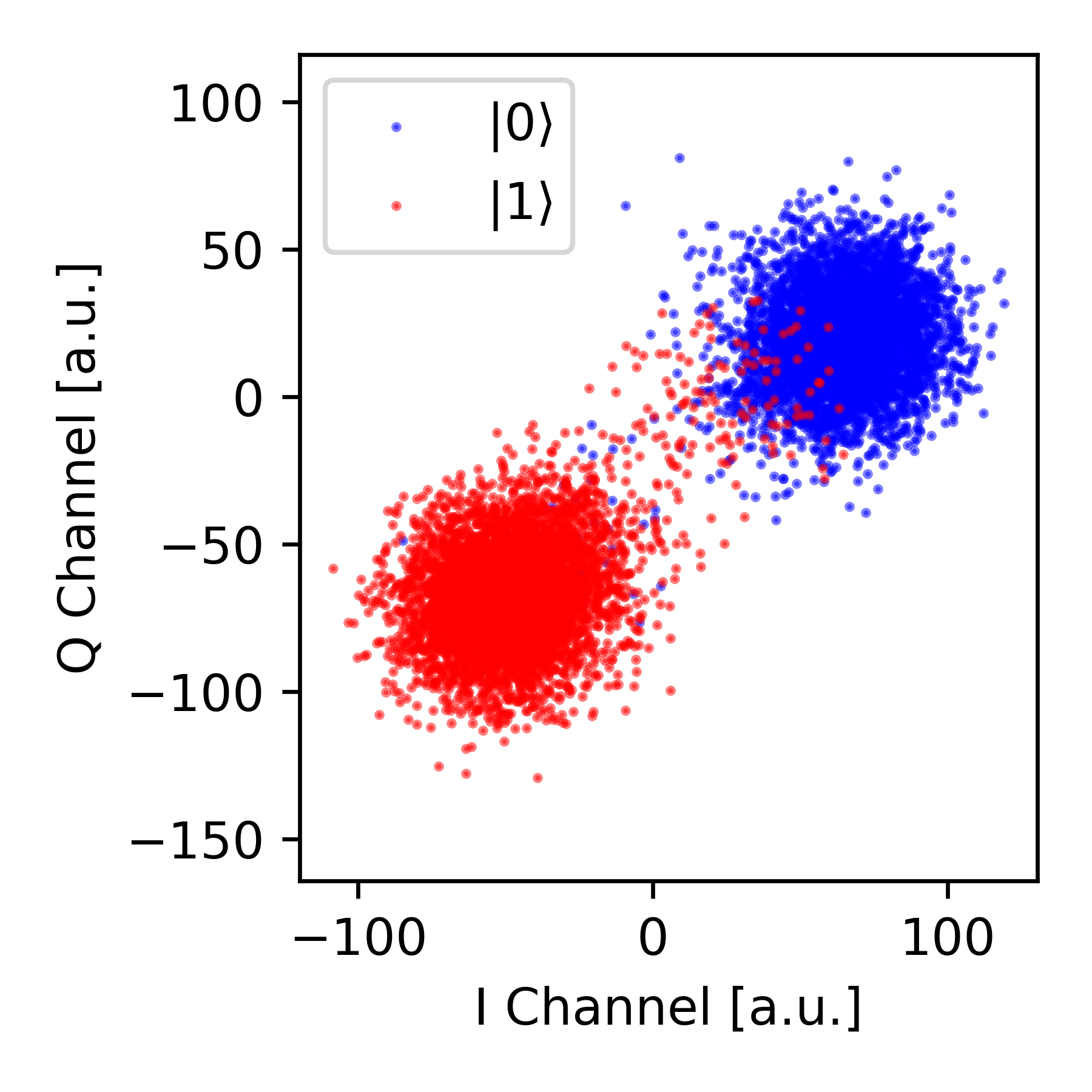}&\iptsq{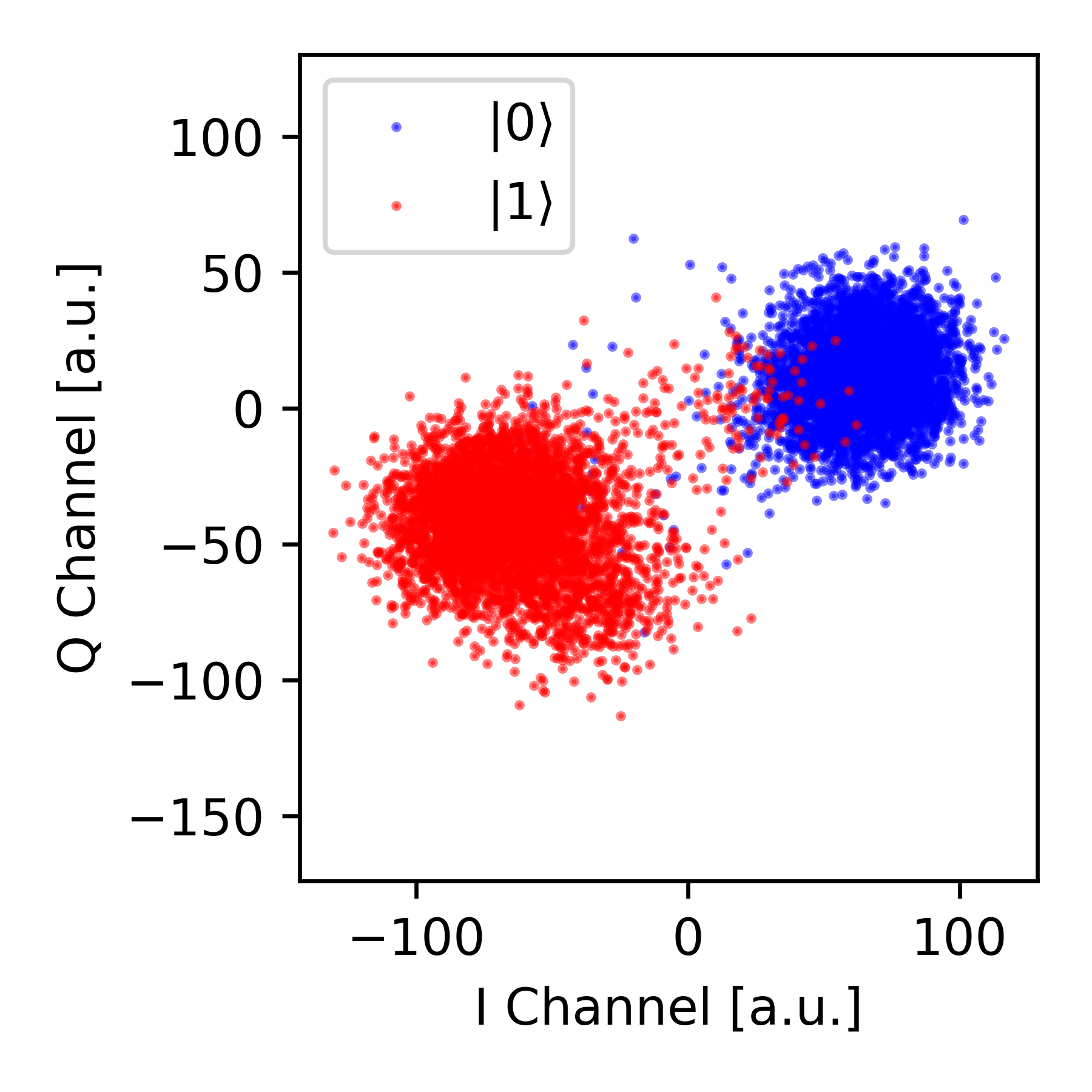}&\iptsq{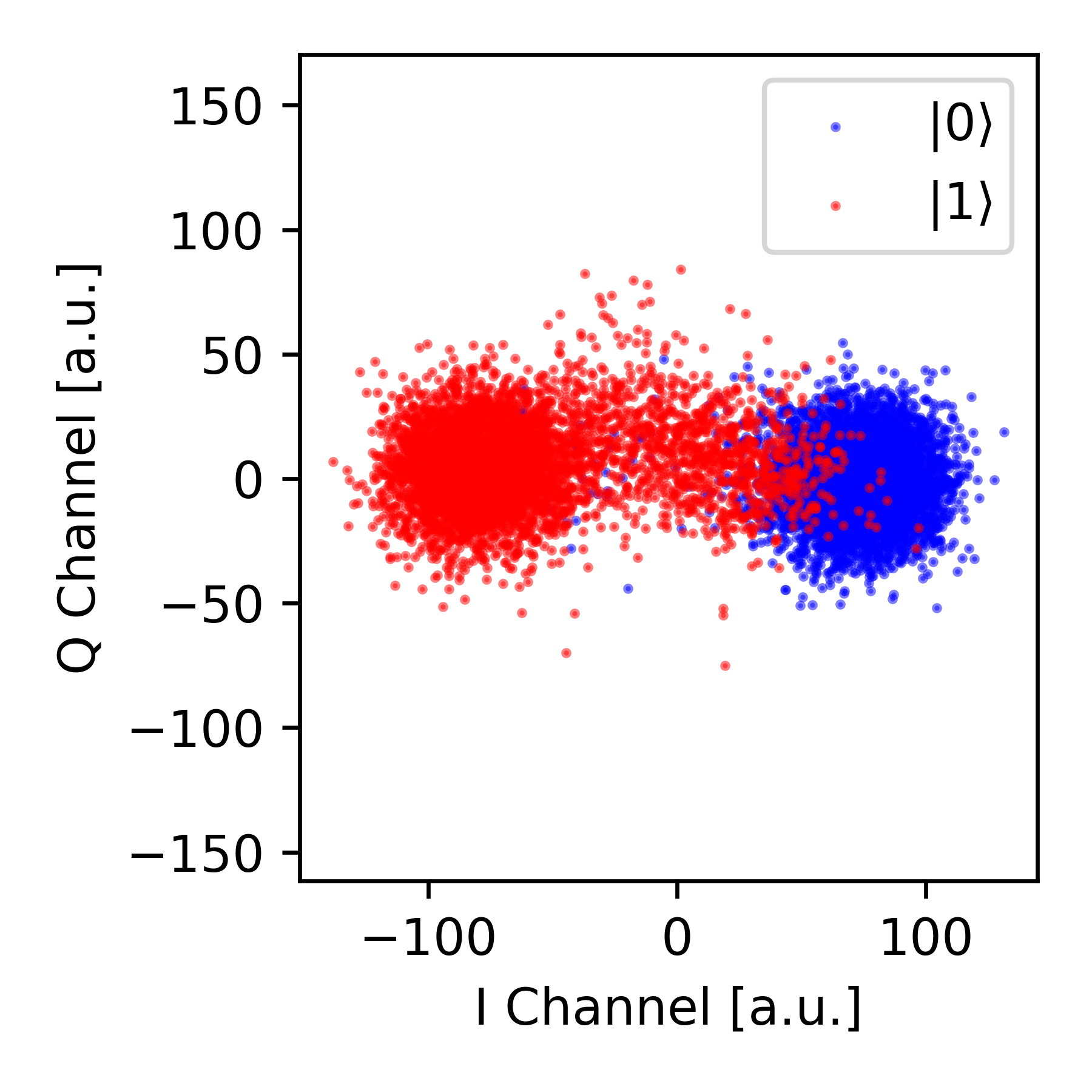}&\iptsq{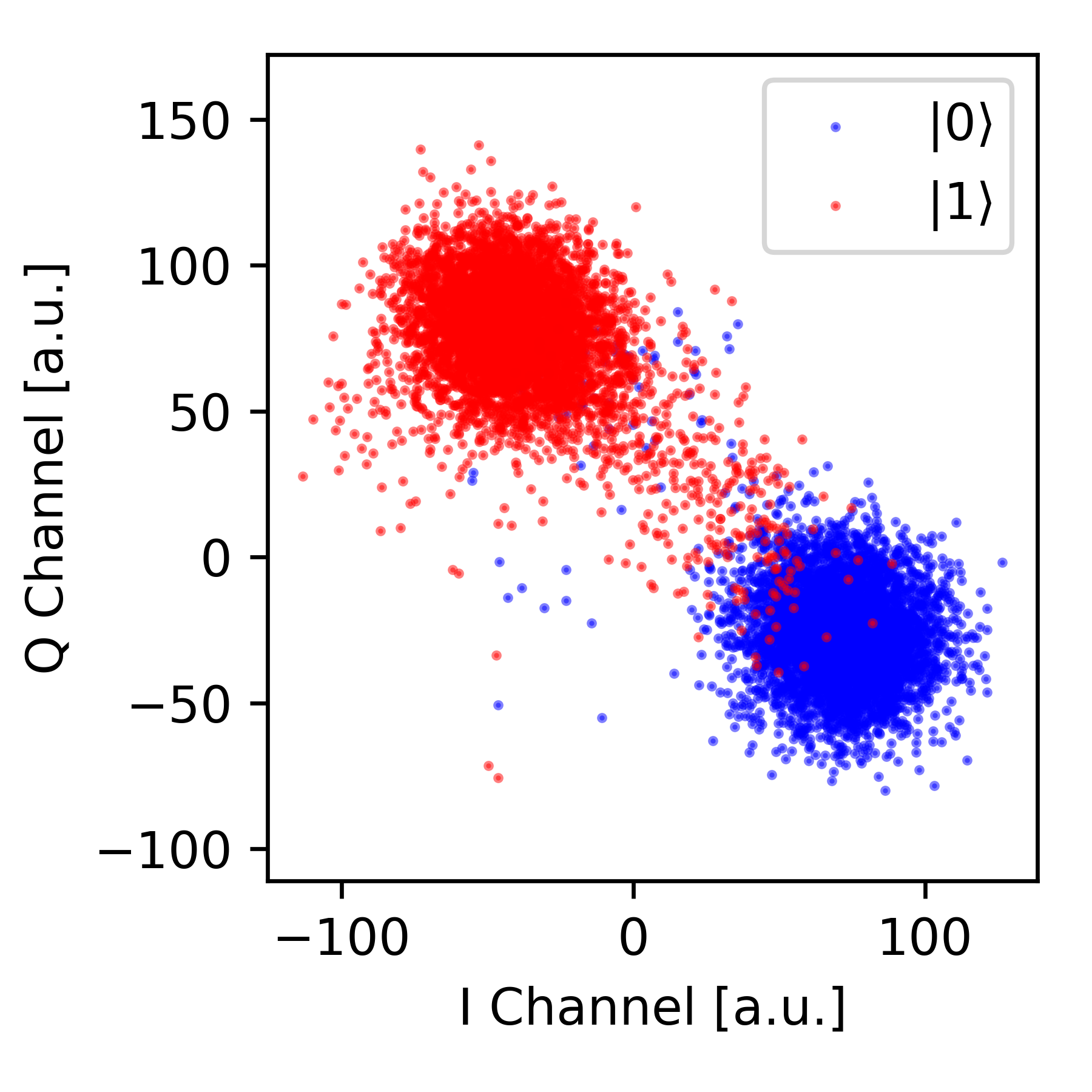}\\
    \begin{tabular}{c}Traces\\ State $\ket{0}$\end{tabular} & \iptm{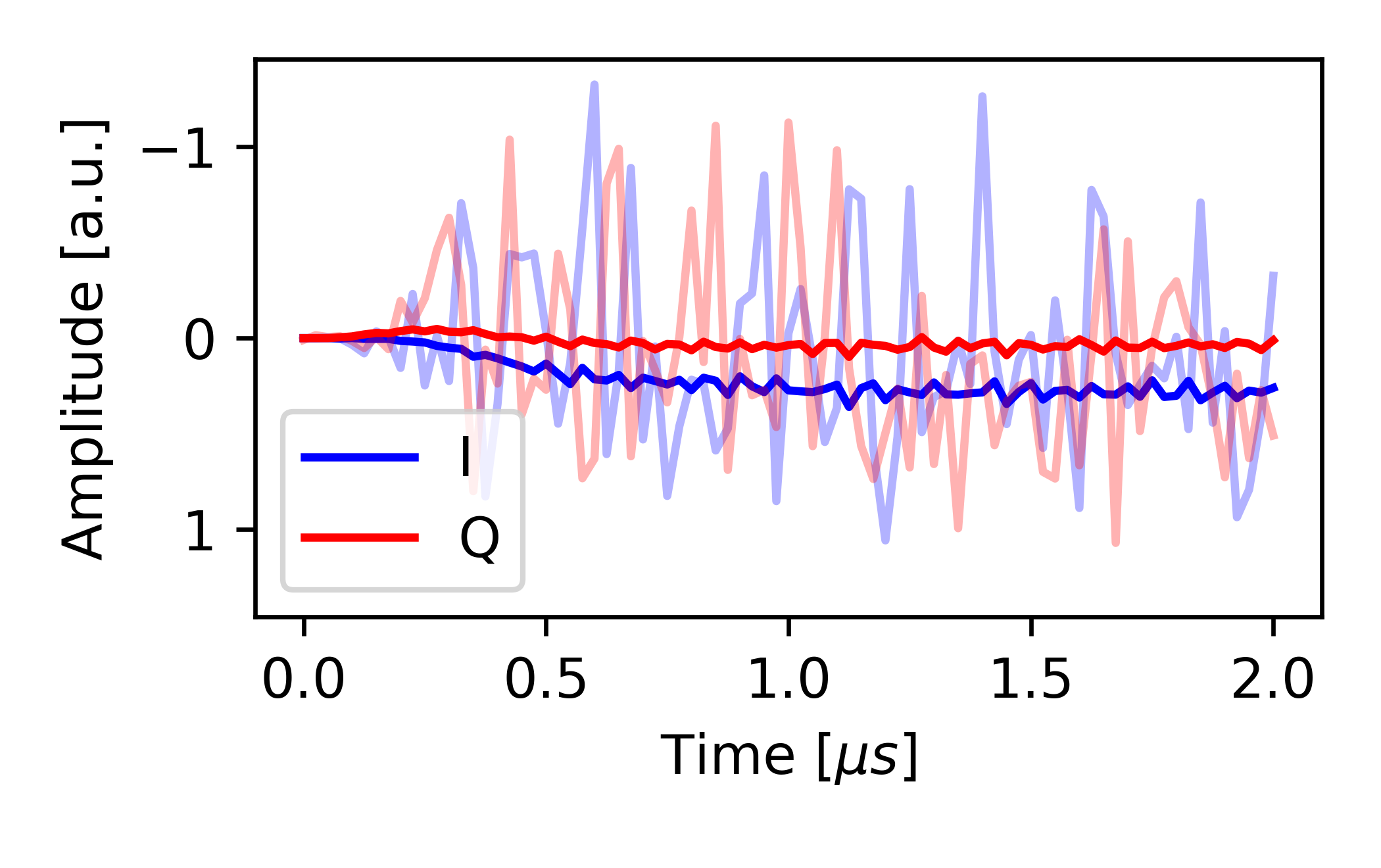}&\iptm{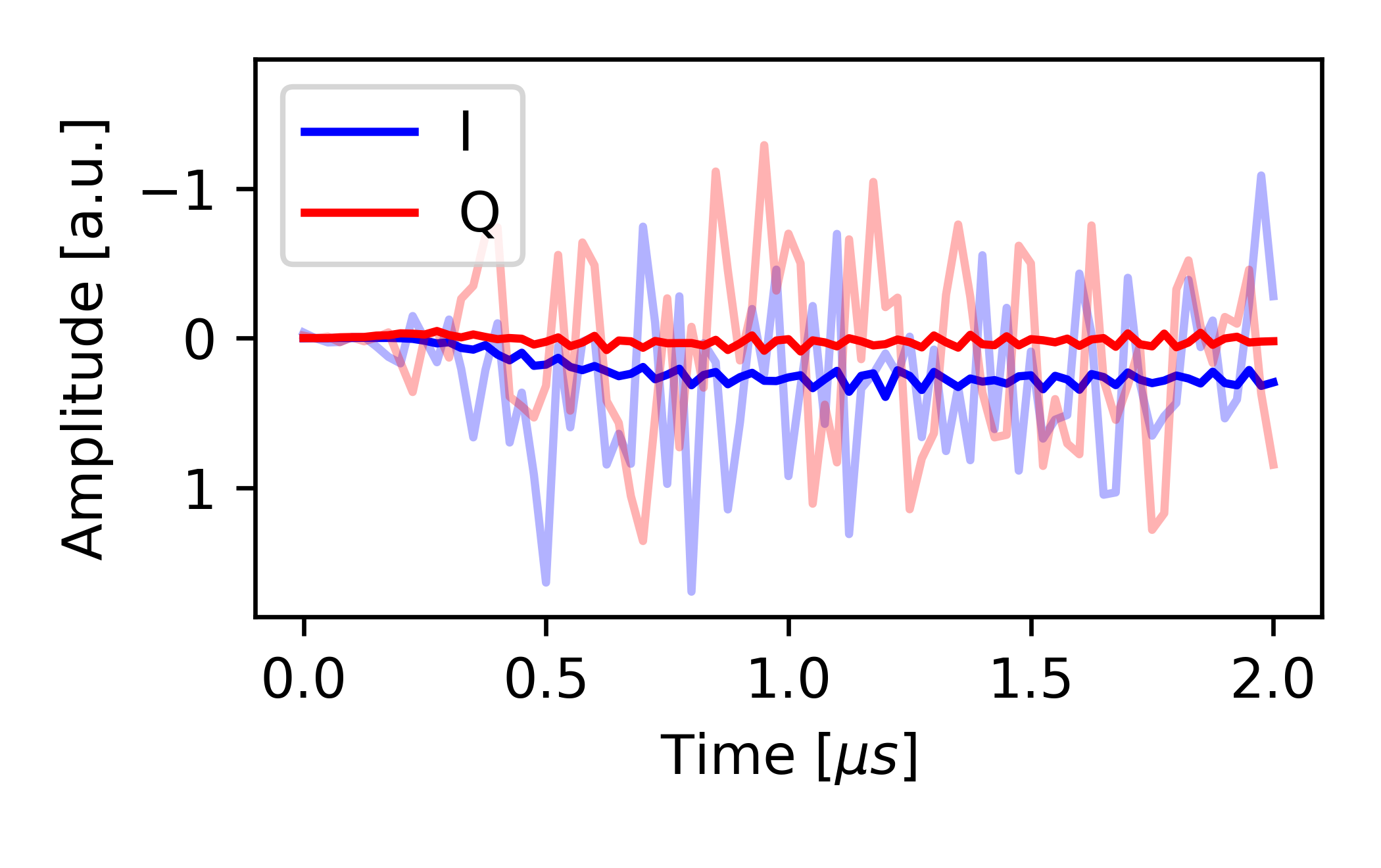}&\iptm{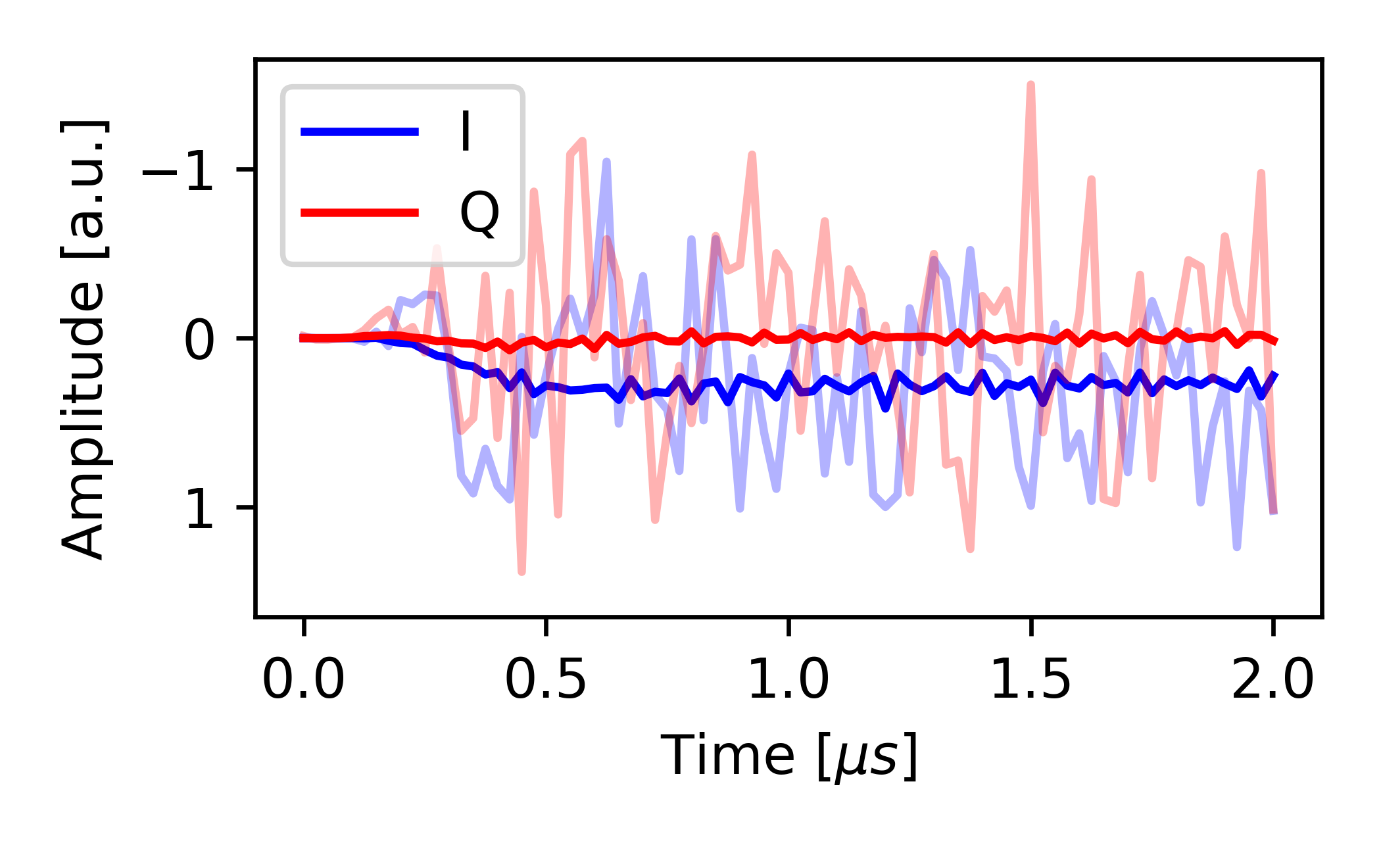}&\iptm{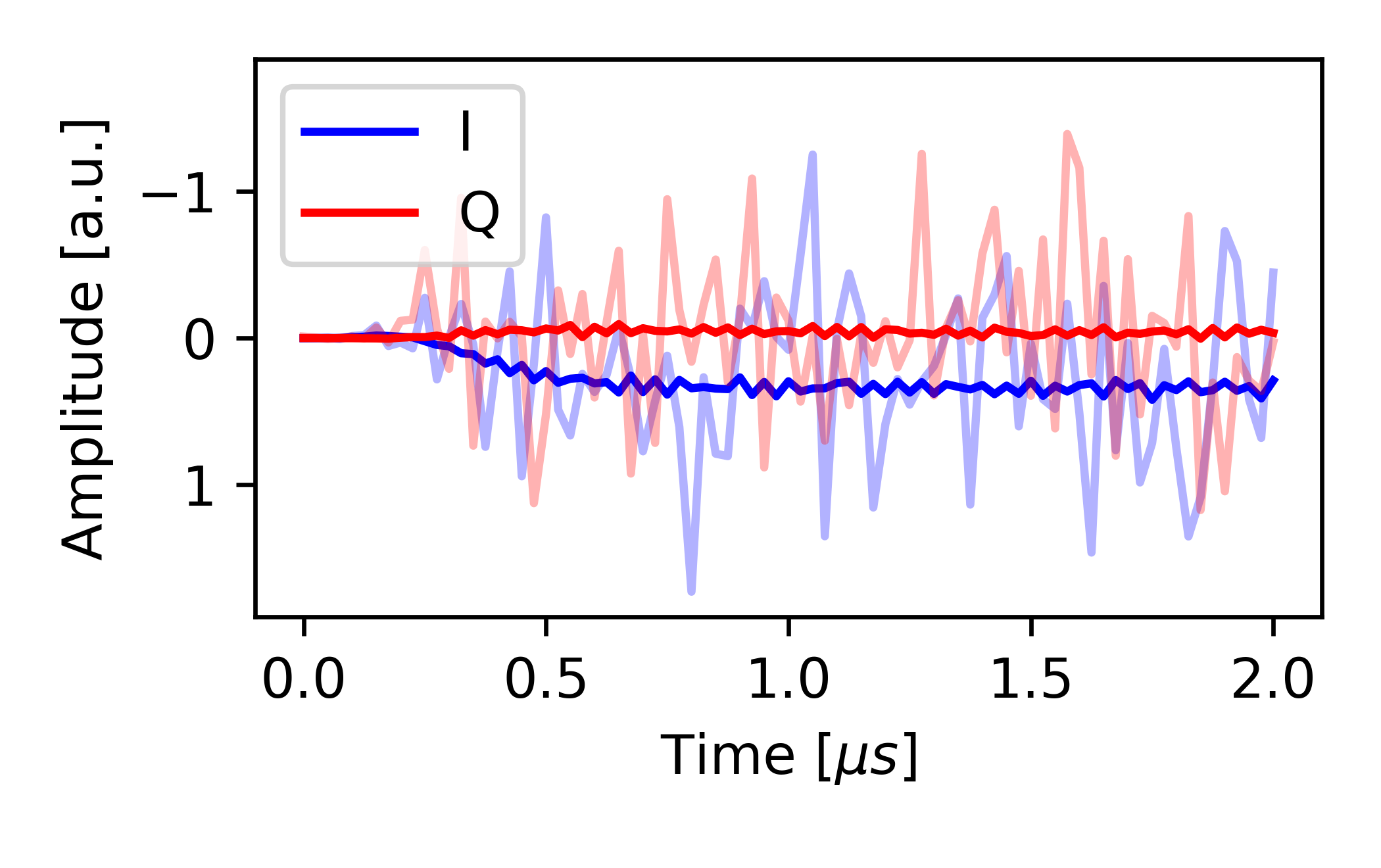}\\
    \begin{tabular}{c}Traces\\ State $\ket{1}$\end{tabular} & \iptm{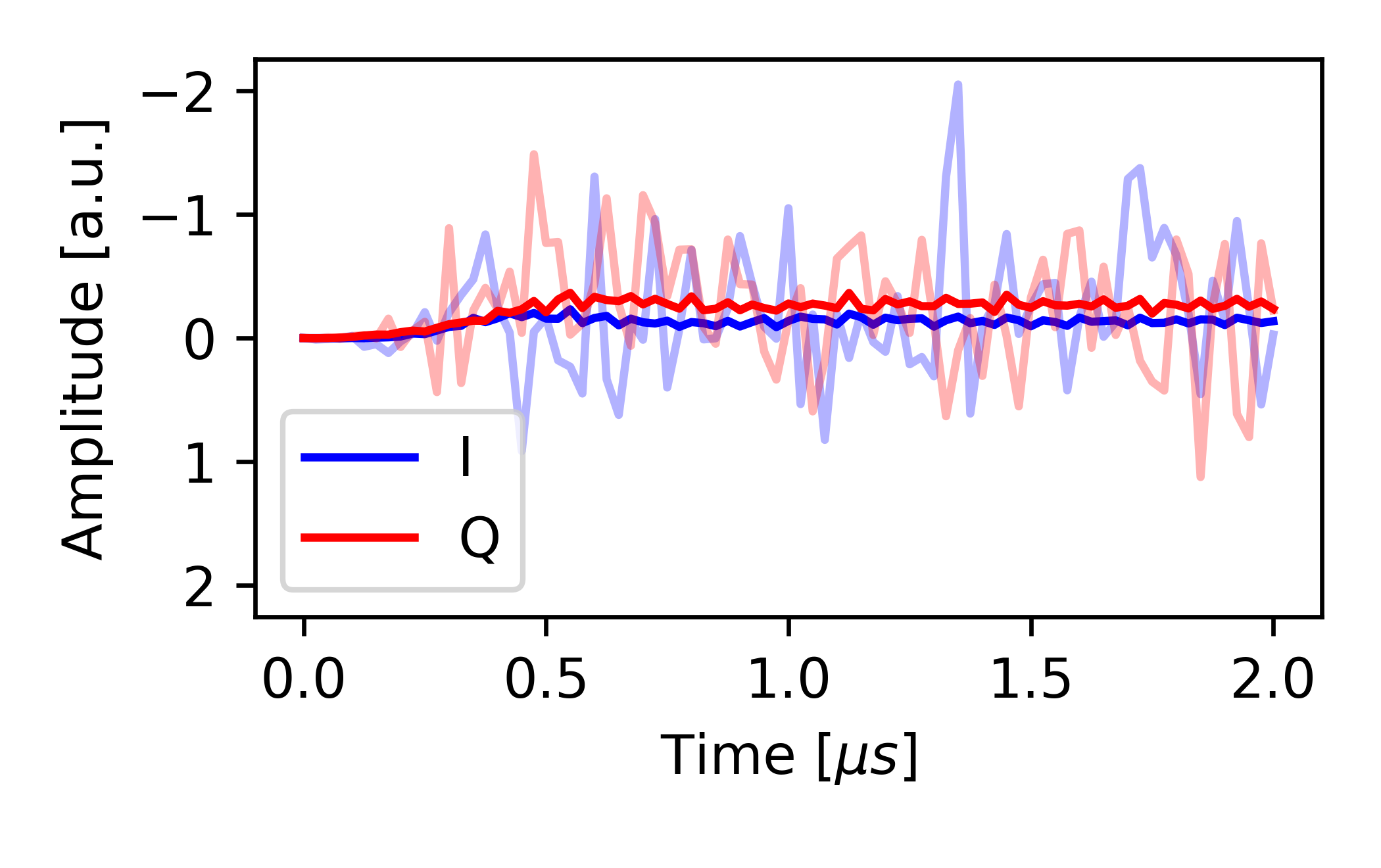}&\iptm{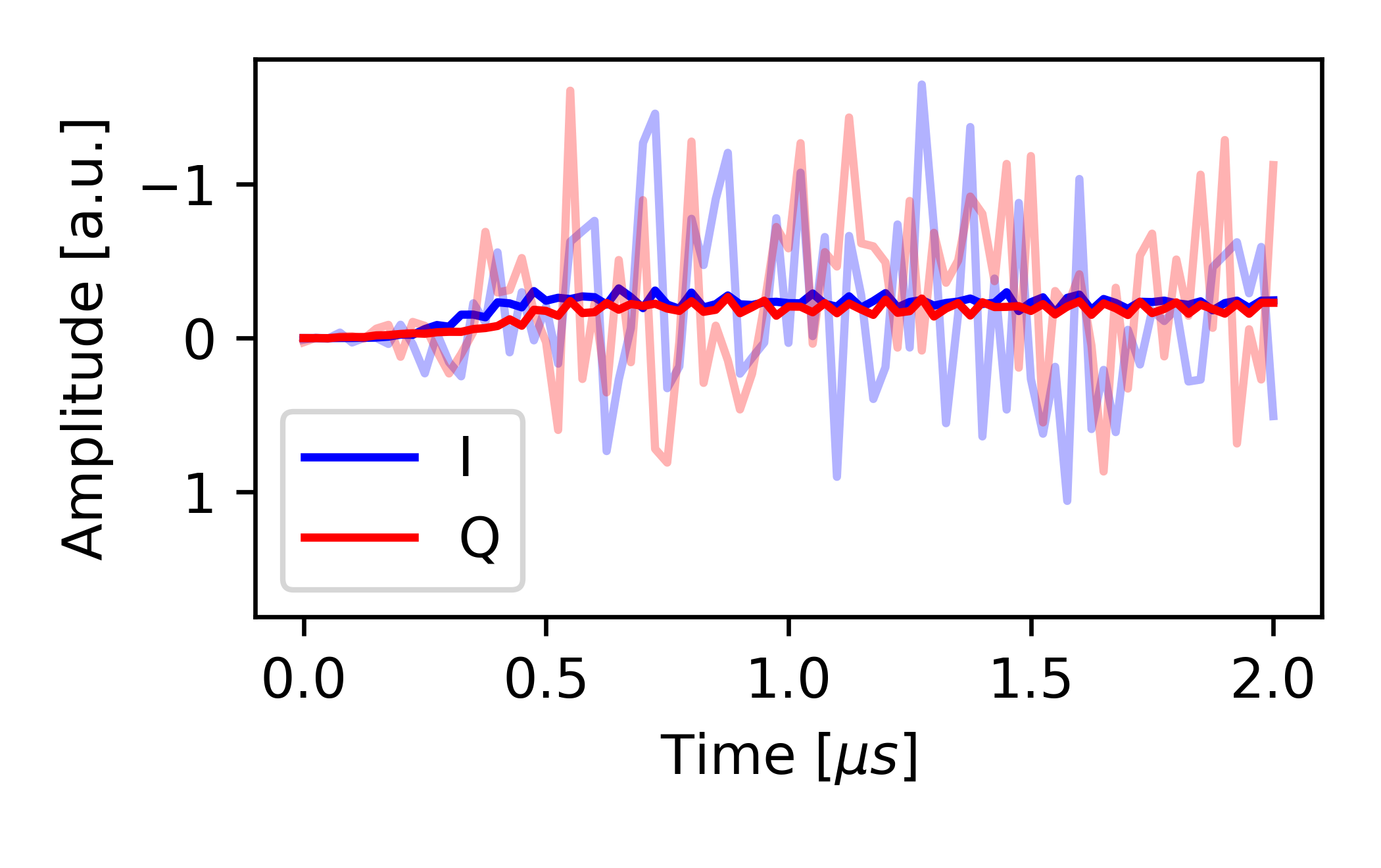}&\iptm{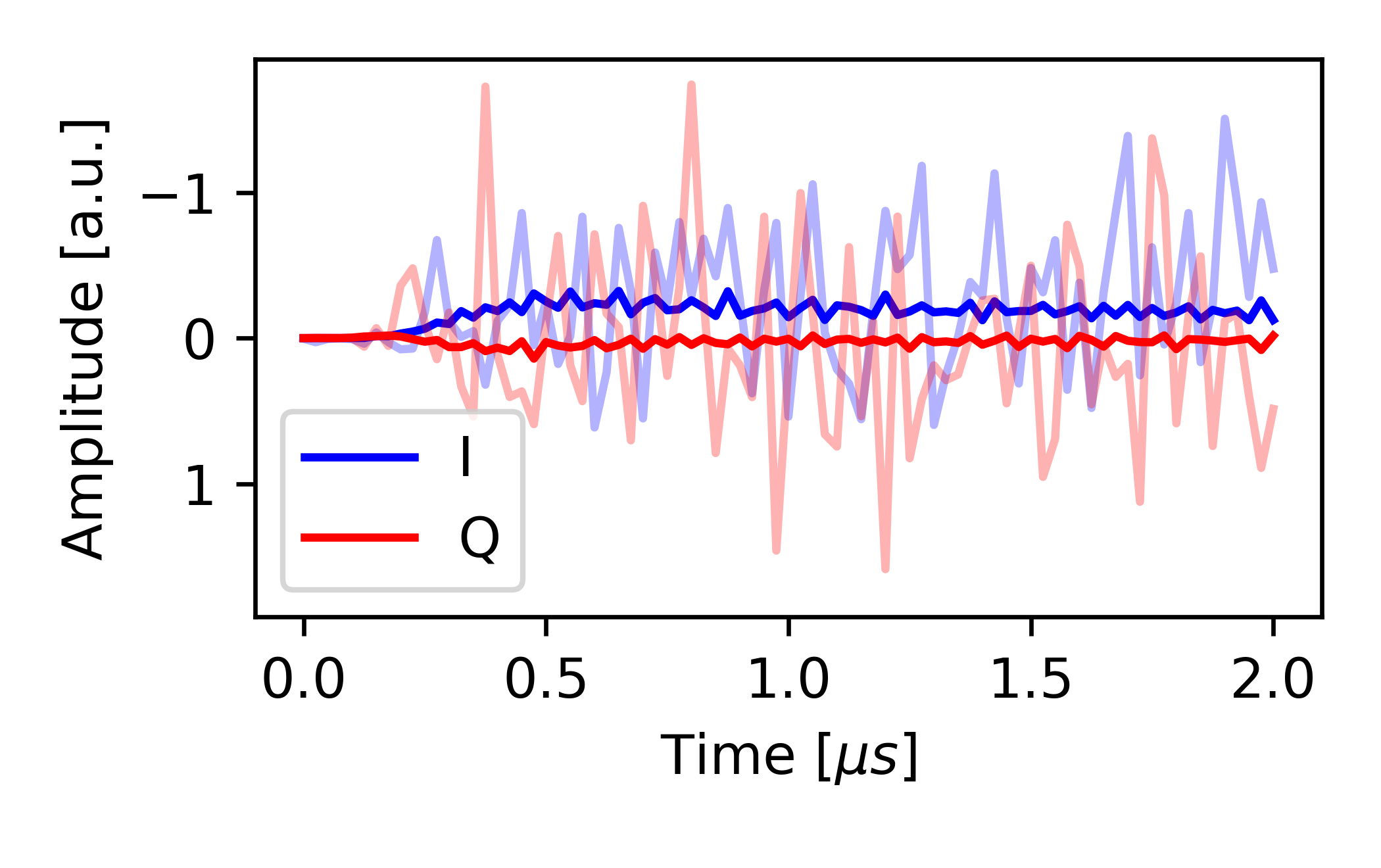}&\iptm{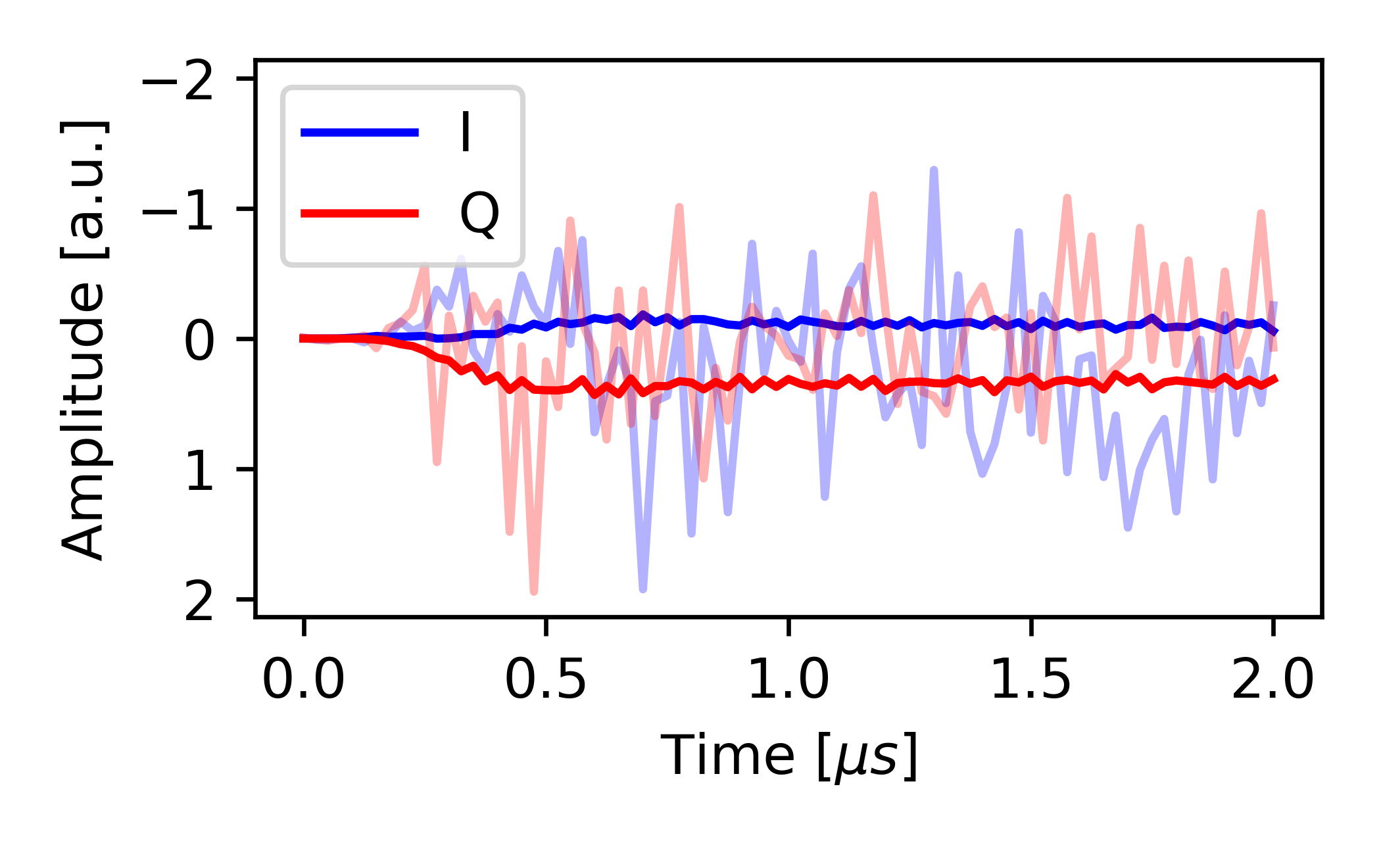}\\
    \begin{tabular}{c}Trajectory\end{tabular} & \iptsq{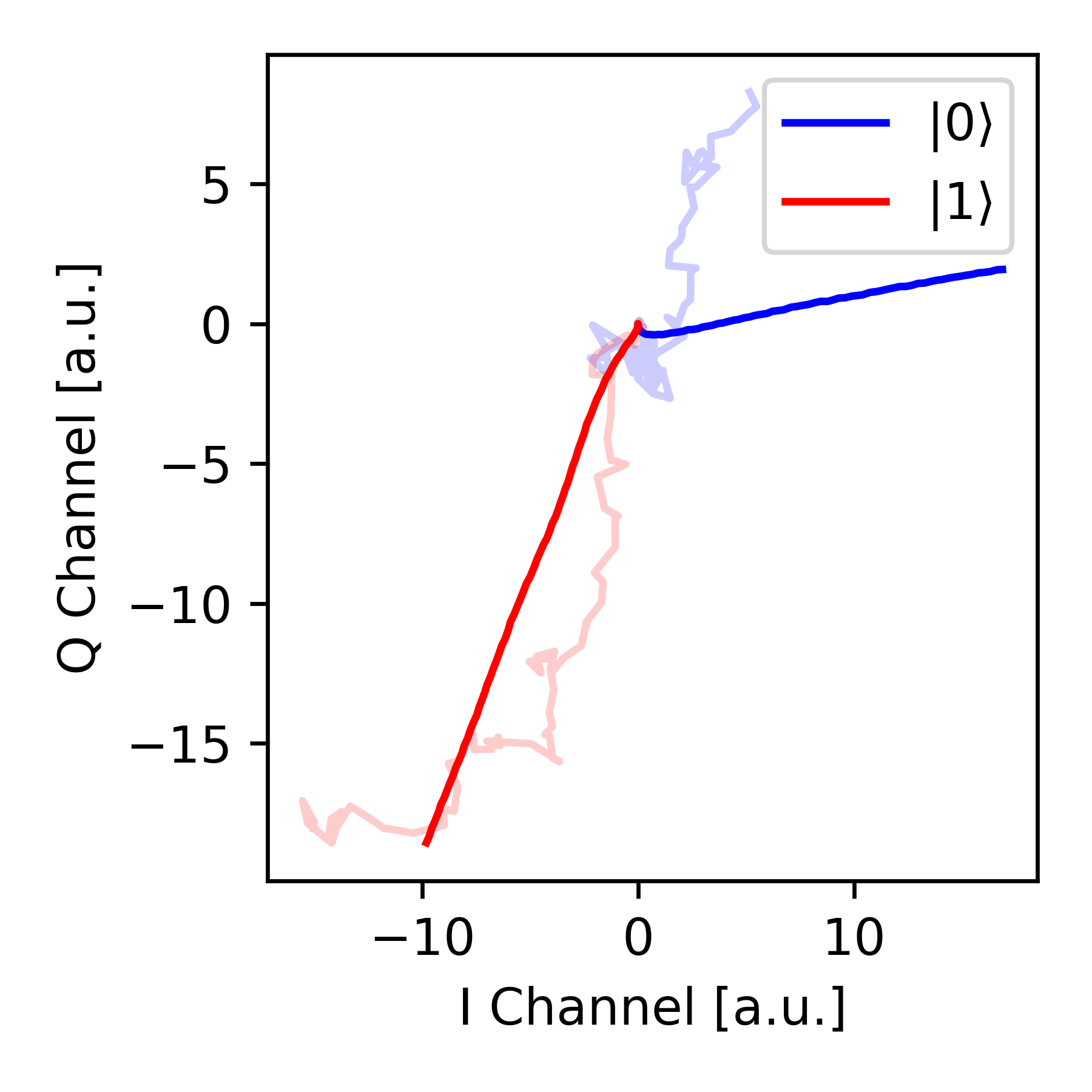}&\iptsq{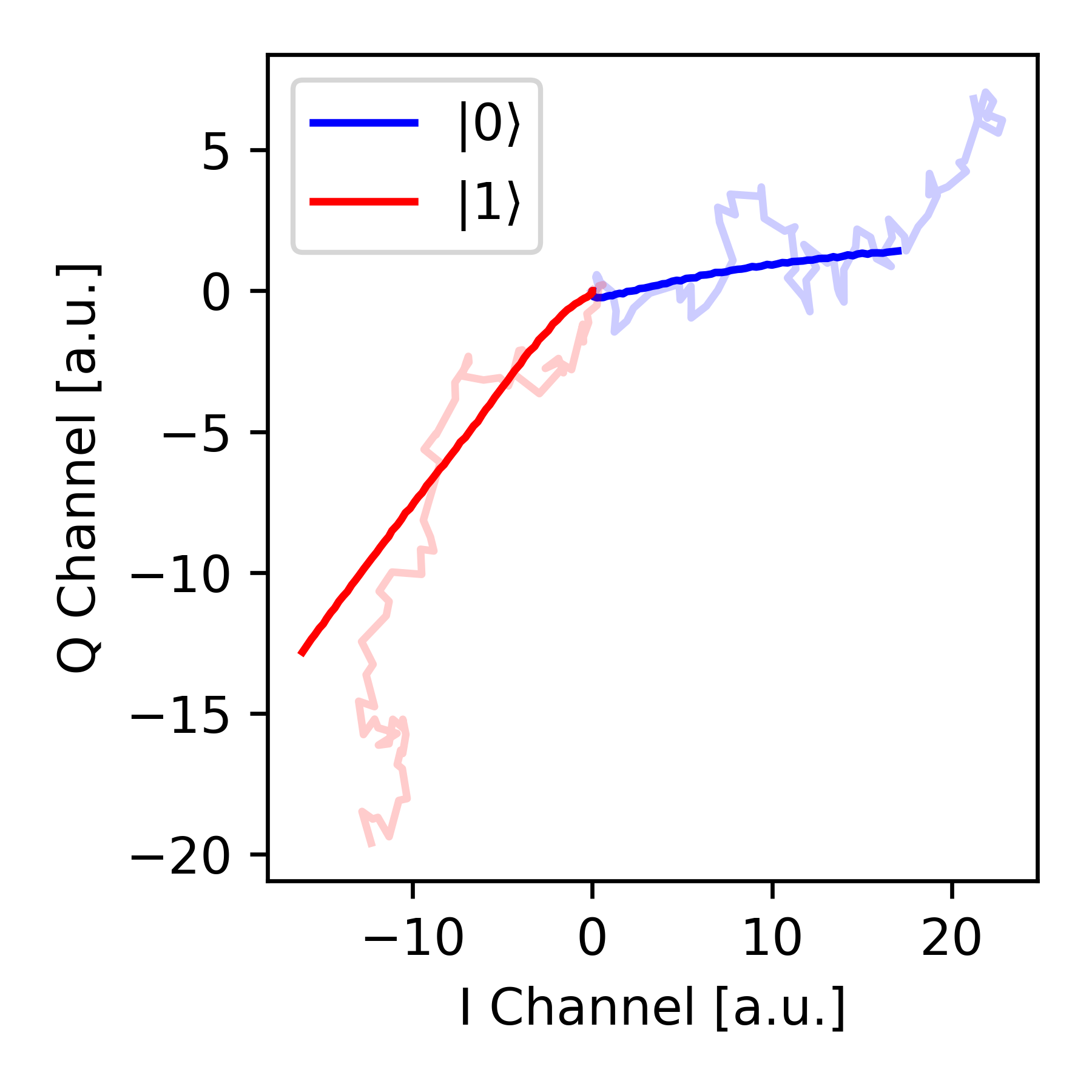}&\iptsq{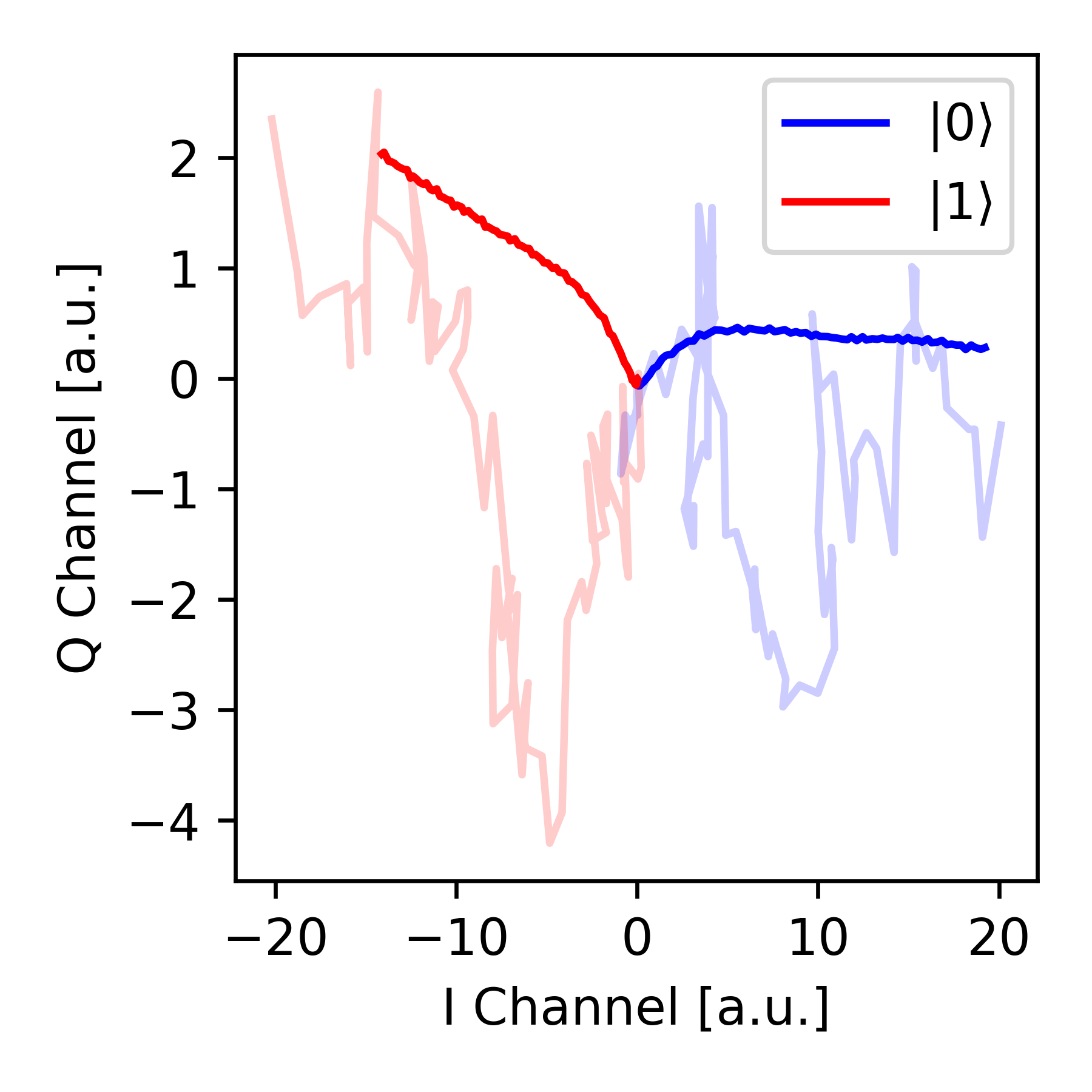}&\iptsq{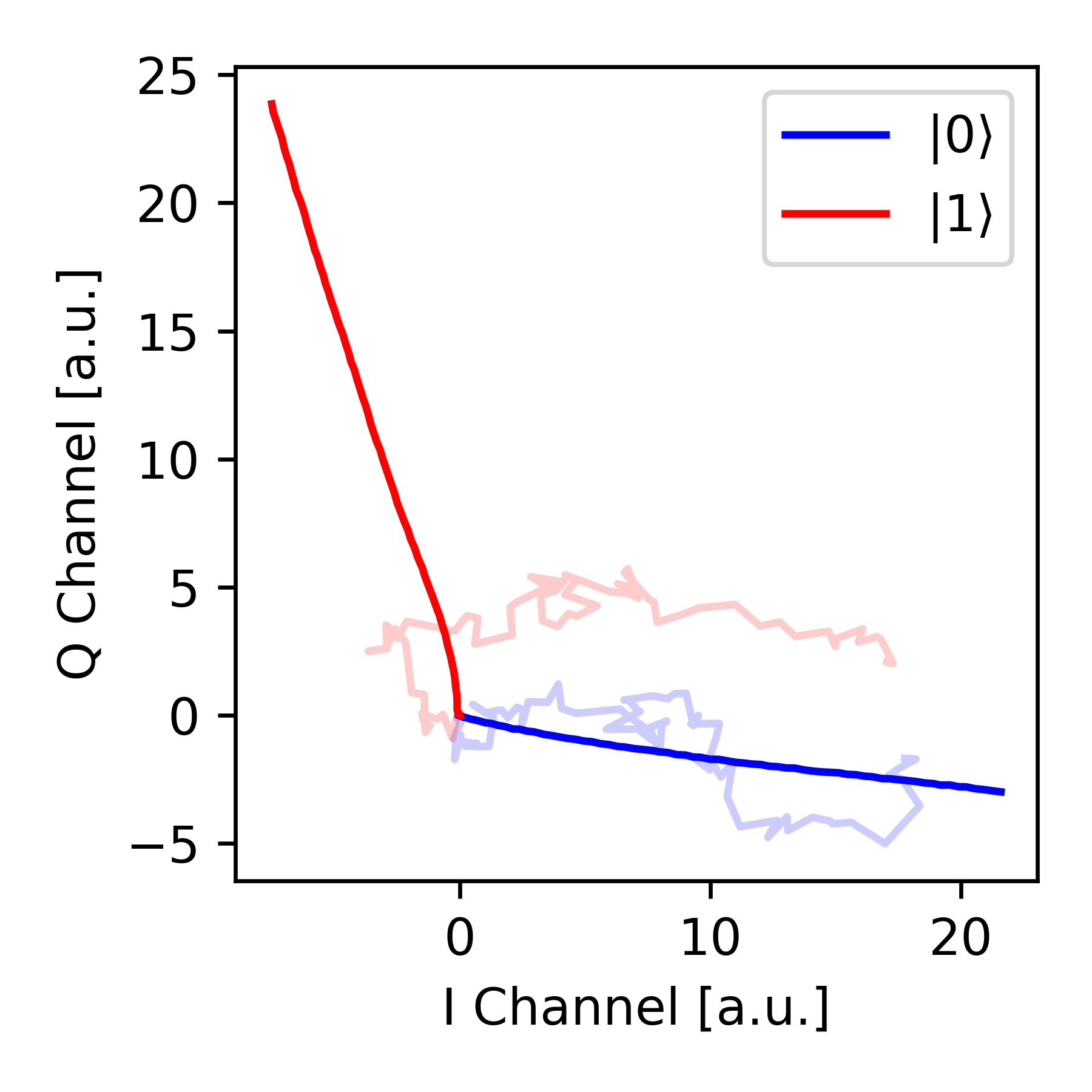}\\
    \begin{tabular}{c}Integration \\ Projection\end{tabular} & \iptm{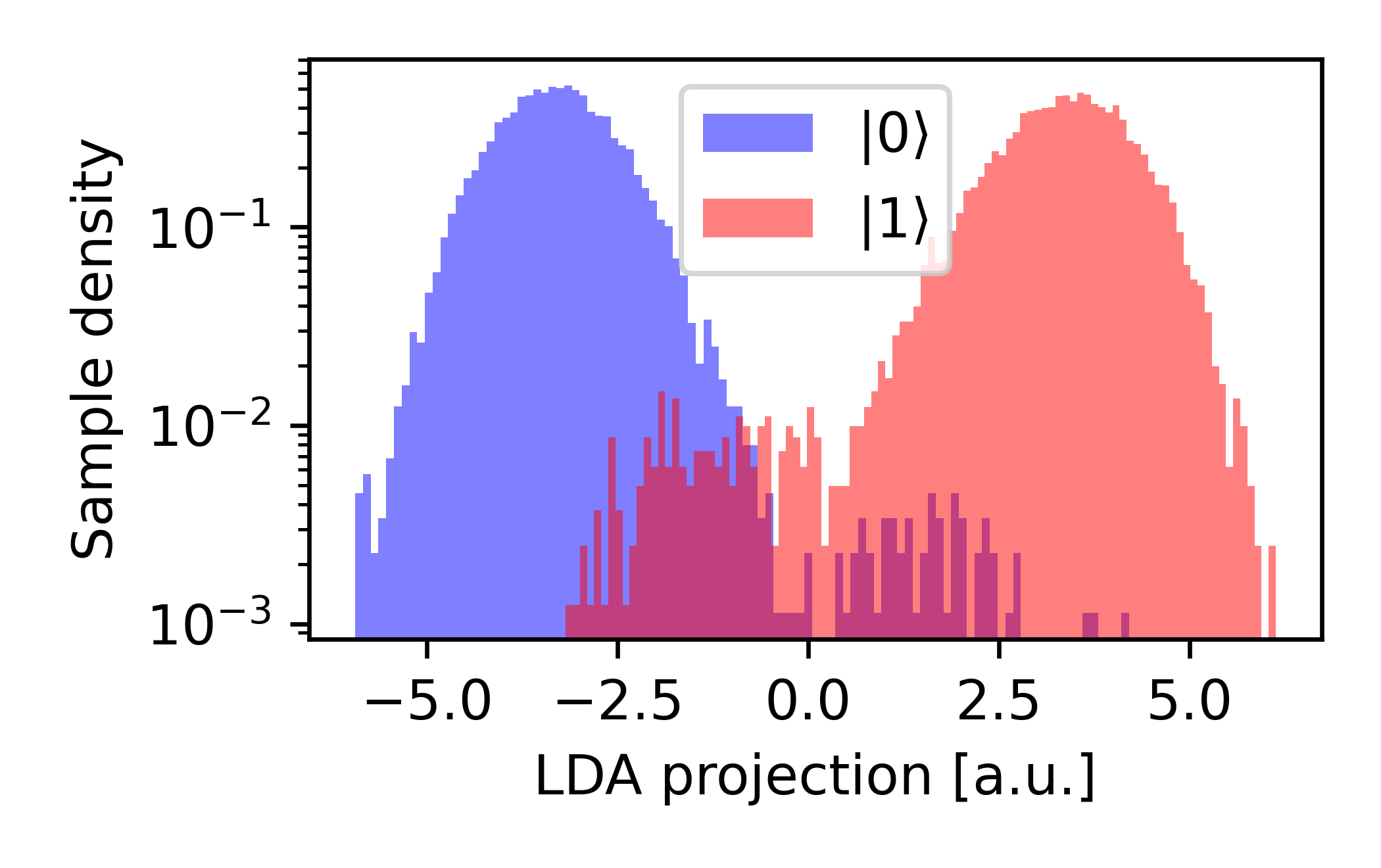}&\iptm{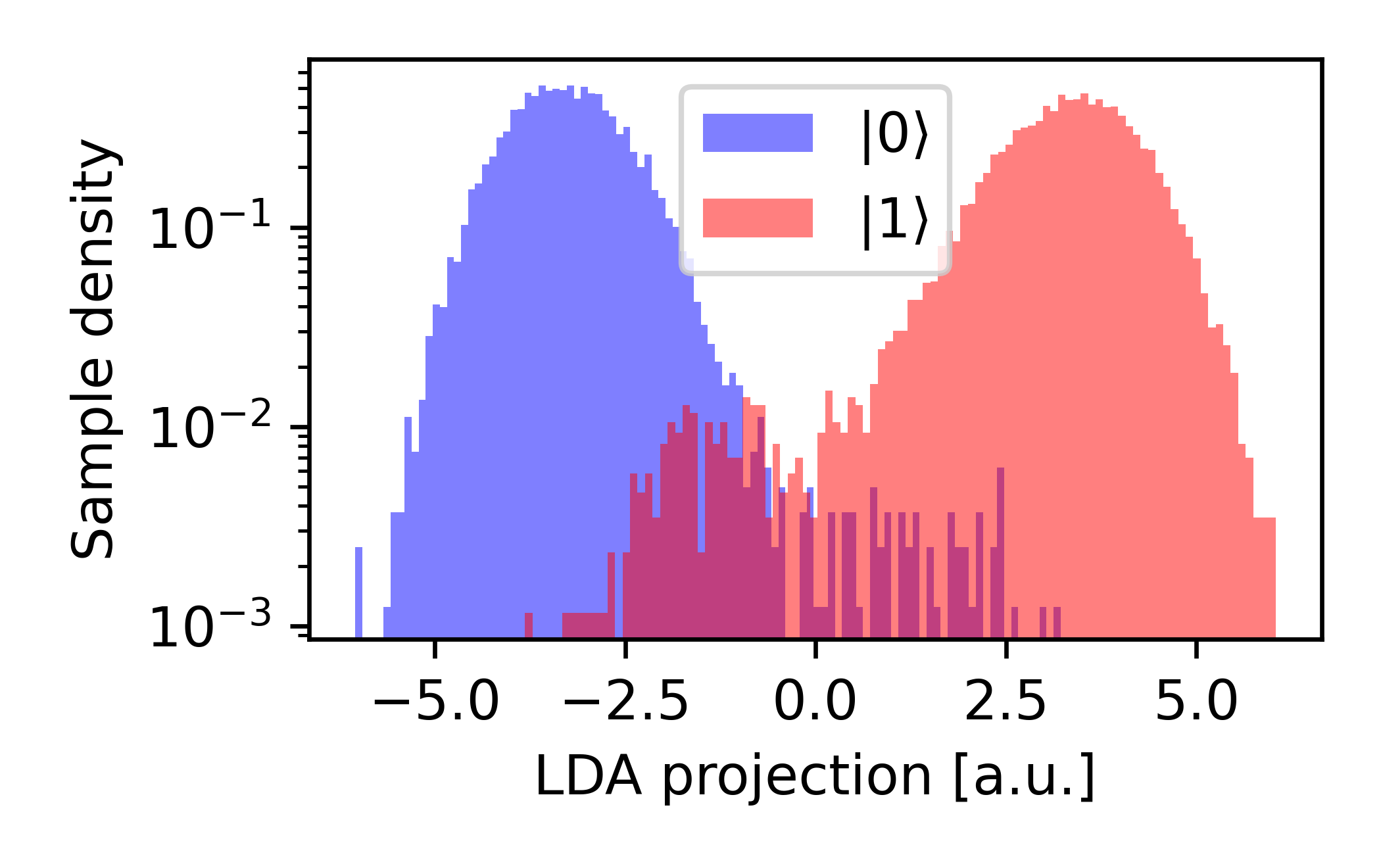}&\iptm{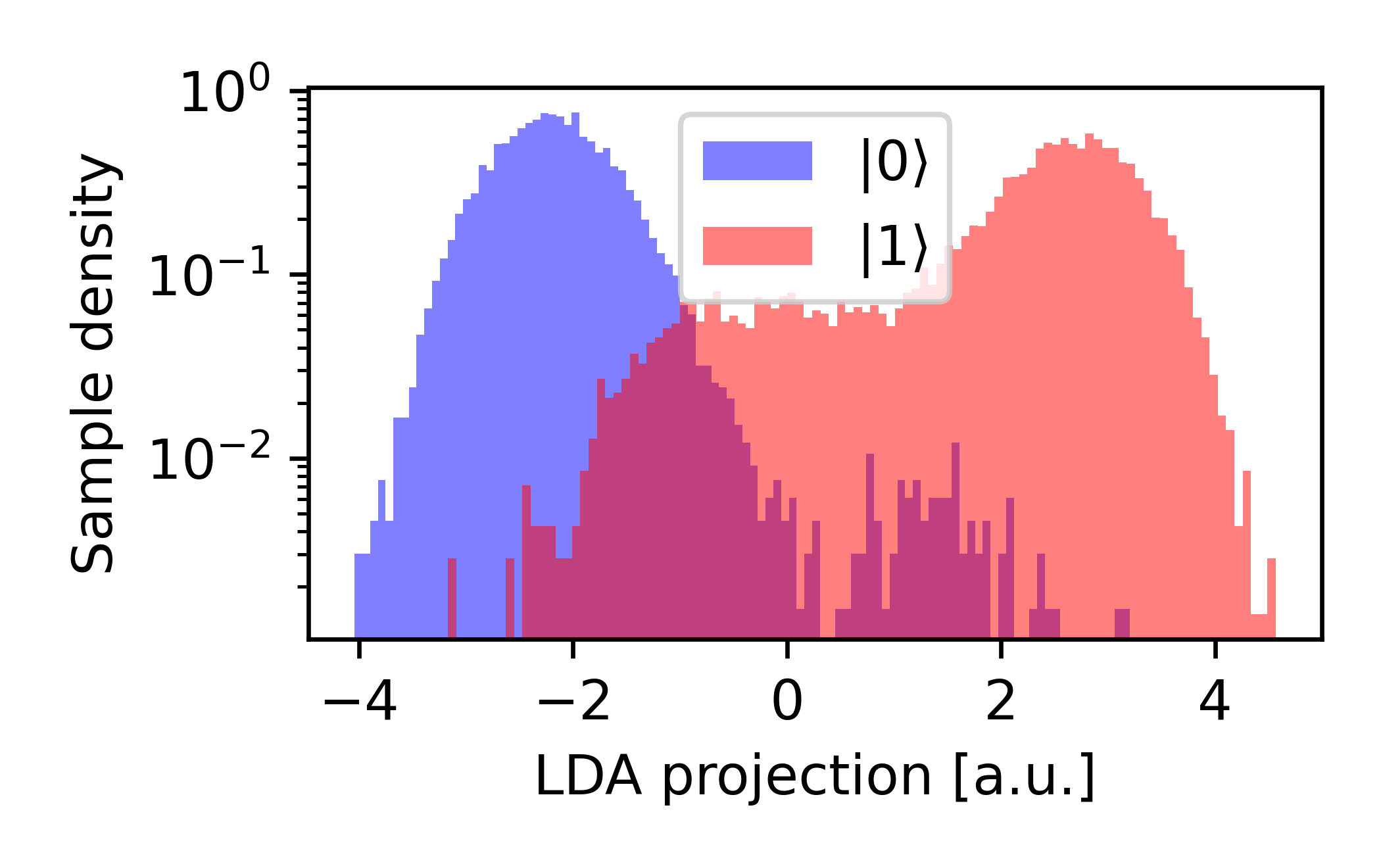}&\iptm{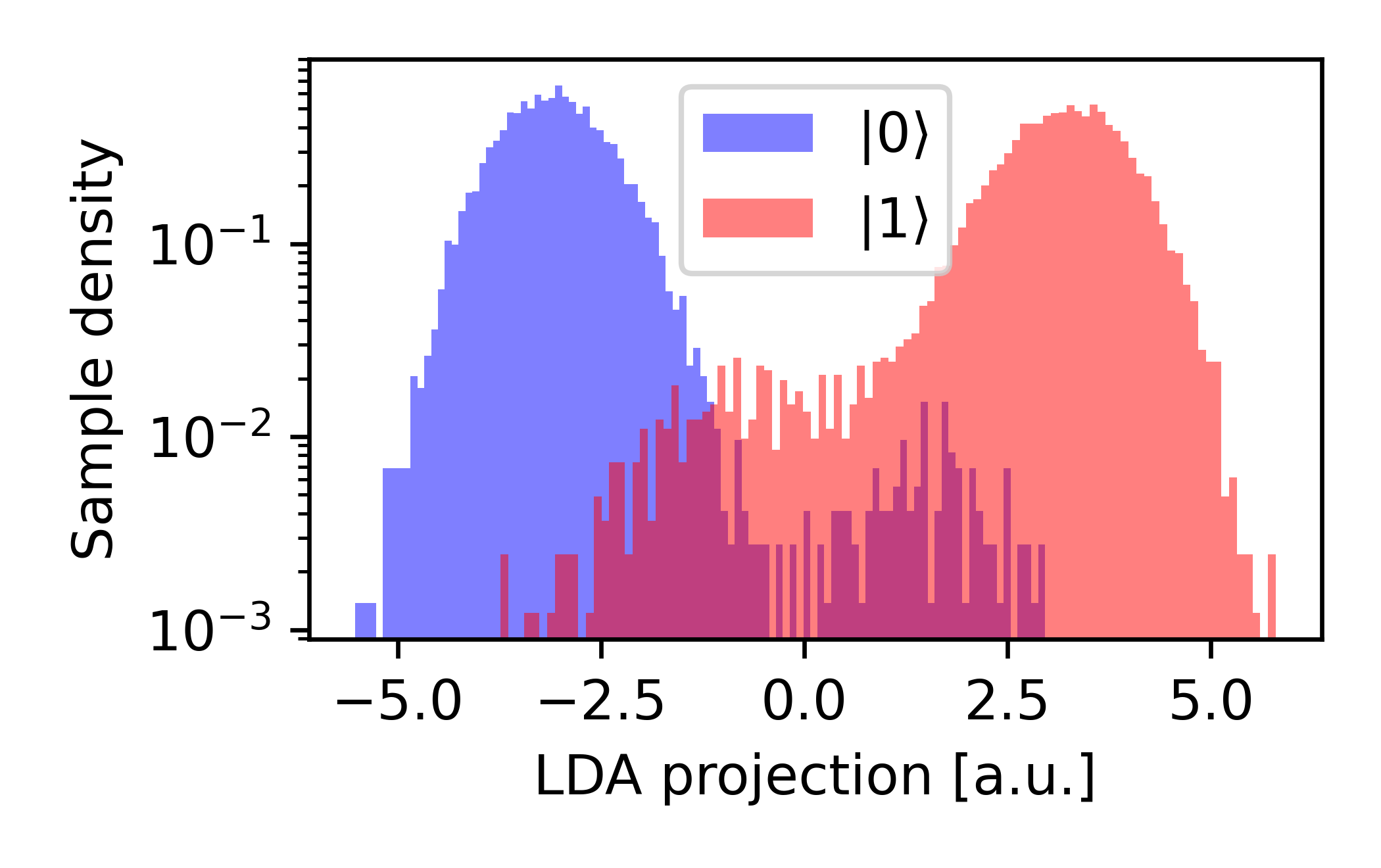}\\
    \begin{tabular}{c}Signature \\ Projection\end{tabular} & \iptm{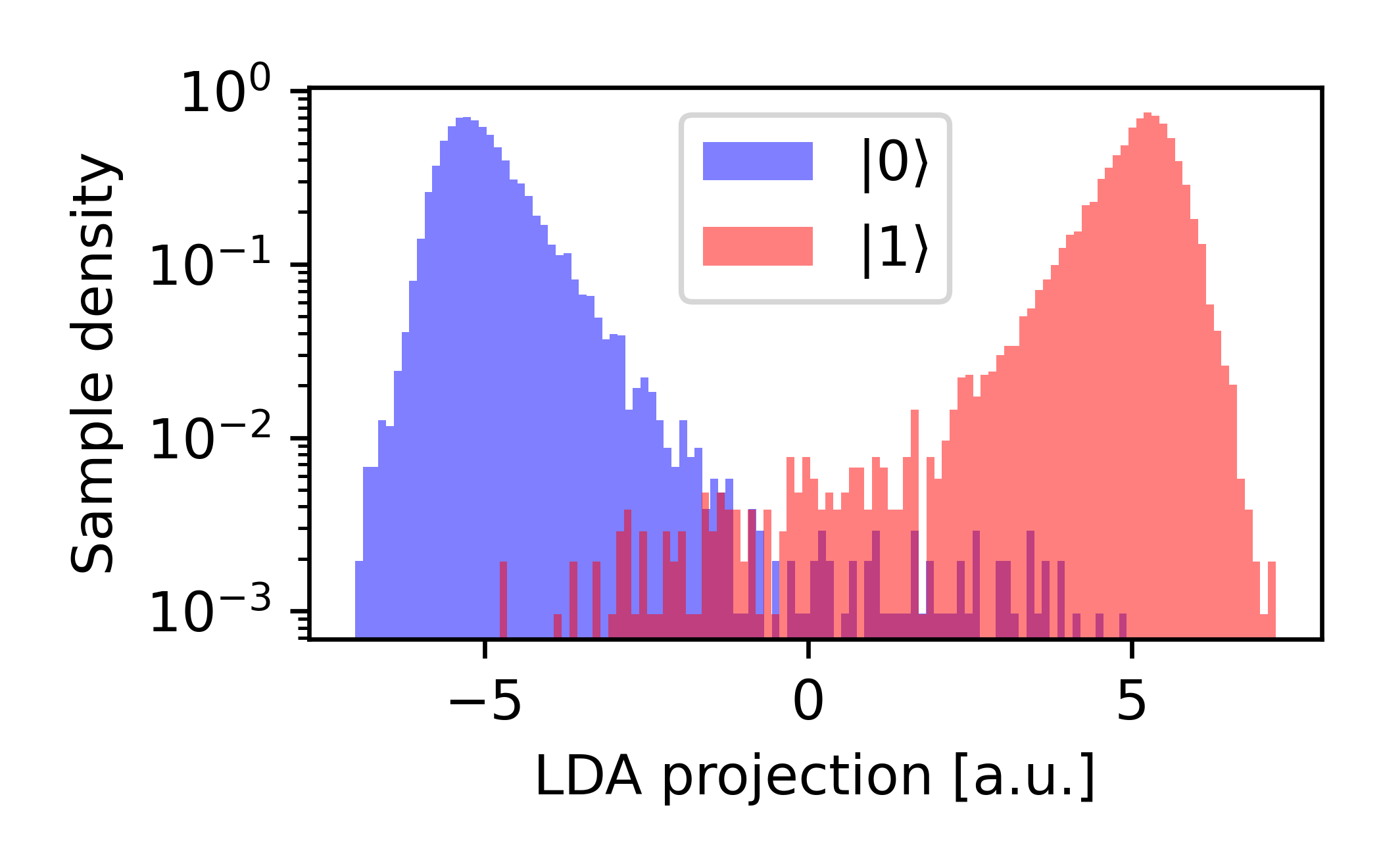}&\iptm{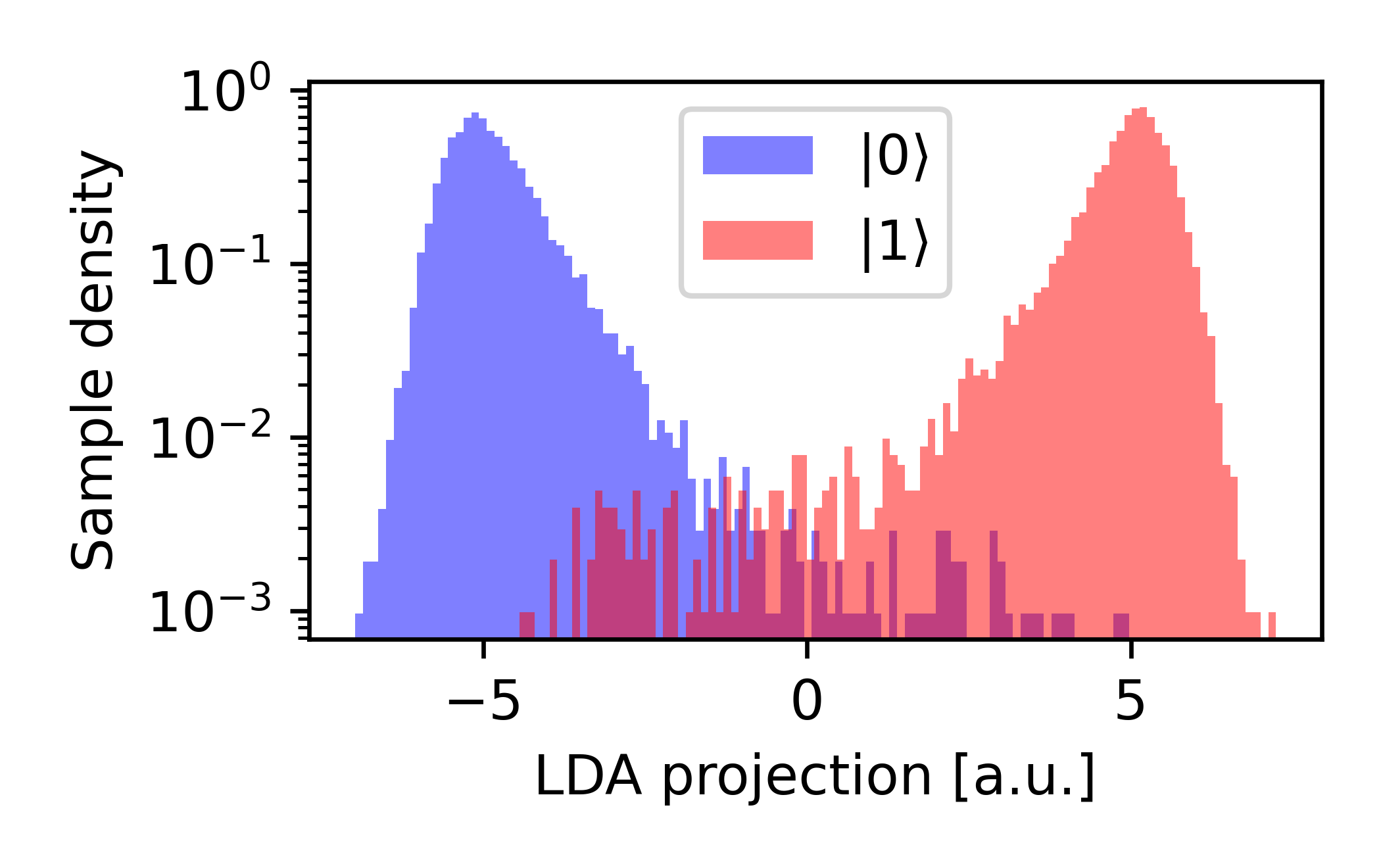}&\iptm{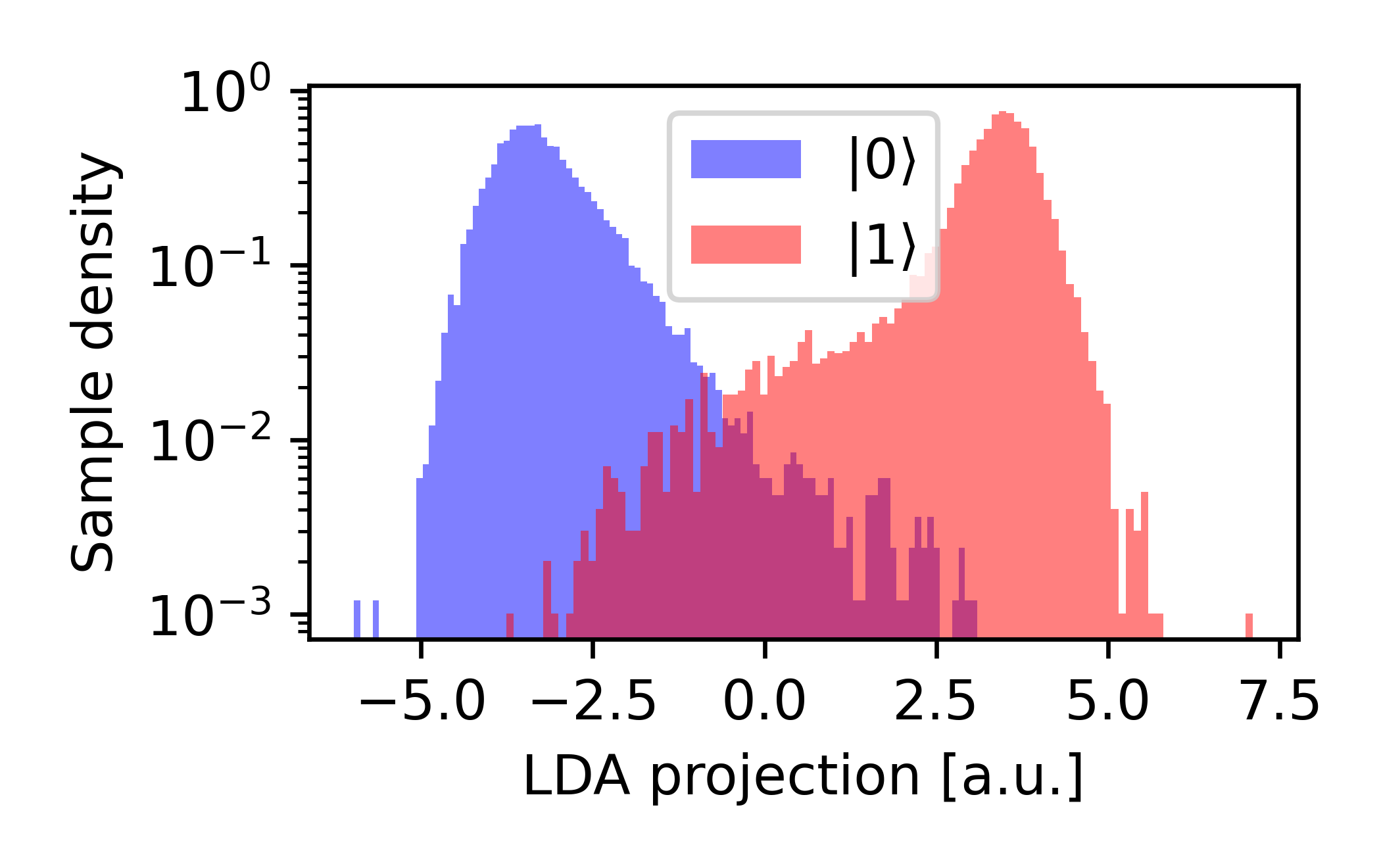}&\iptm{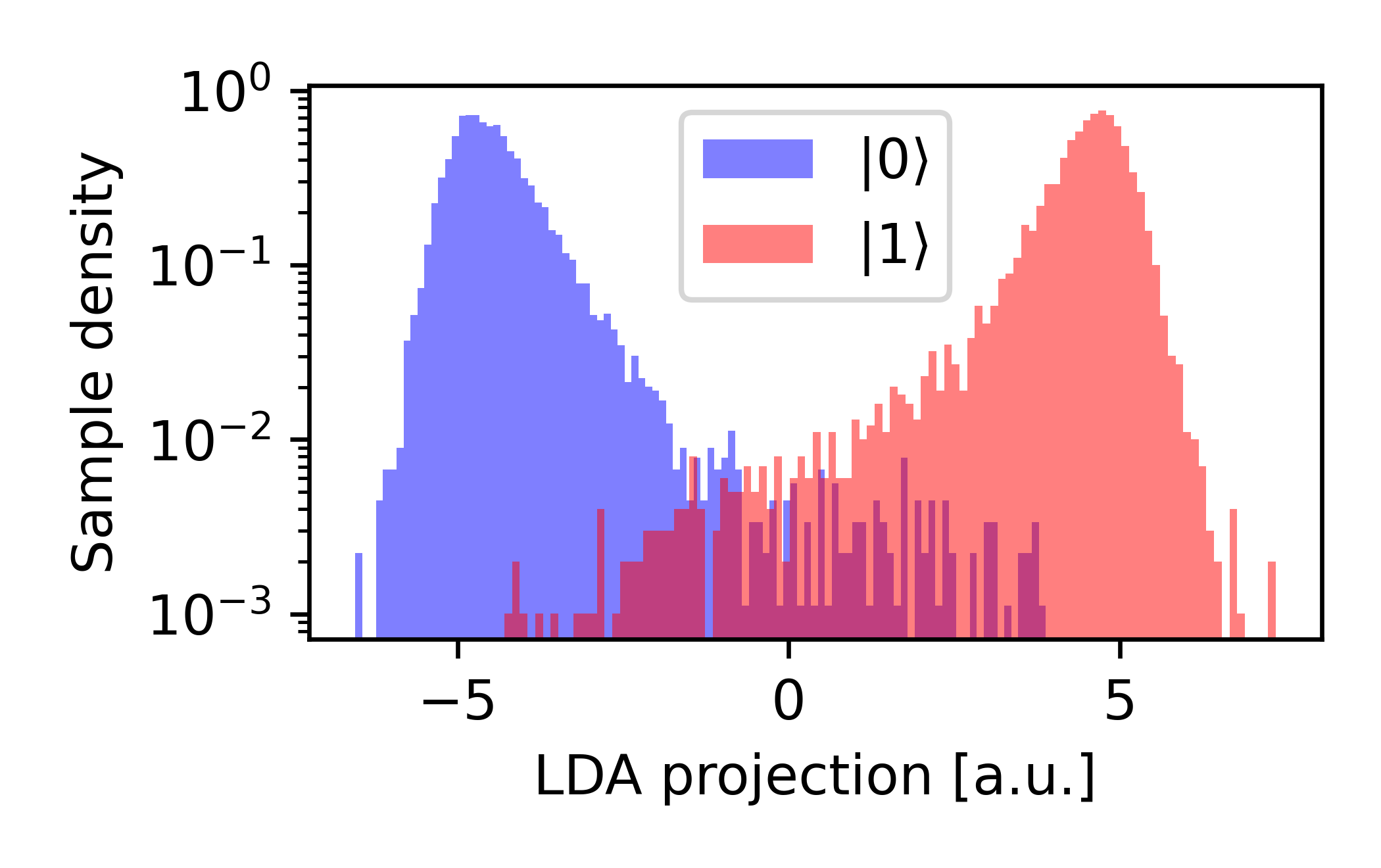}\\
    \begin{tabular}{c}Classifier \\ Performance\end{tabular} & \iptm{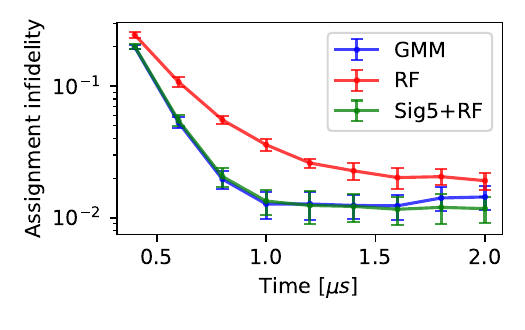}&\iptm{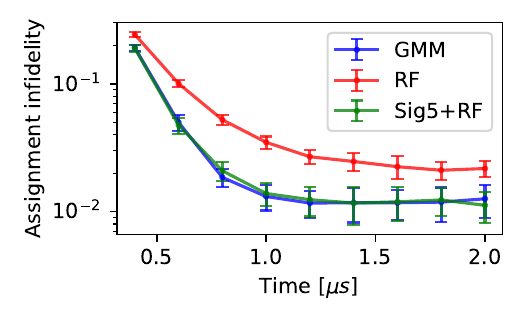}&\iptm{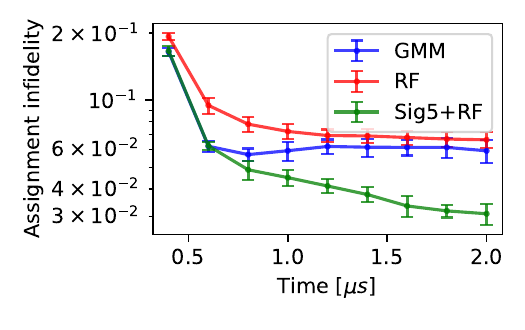}&\iptm{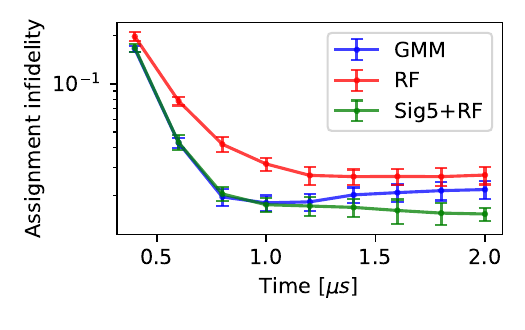}\\

         \hline
    \end{tabular}
    \caption{Statistics of the readout signal, the signal's signature, and the performance of various state discrimination approaches. These approaches and benchmarks are consistent with those reported in Section \ref{app:ml_methods} of the supplemental material.}.
    \label{tab:results_oxford_4q}
\end{figure}

\clearpage

\section{Supplementary information for RQC dataset (RQC Q1-Q8)\label{app:RQC}}

\begin{figure}[h!]
     \centering
        \includegraphics[width=.7\linewidth]{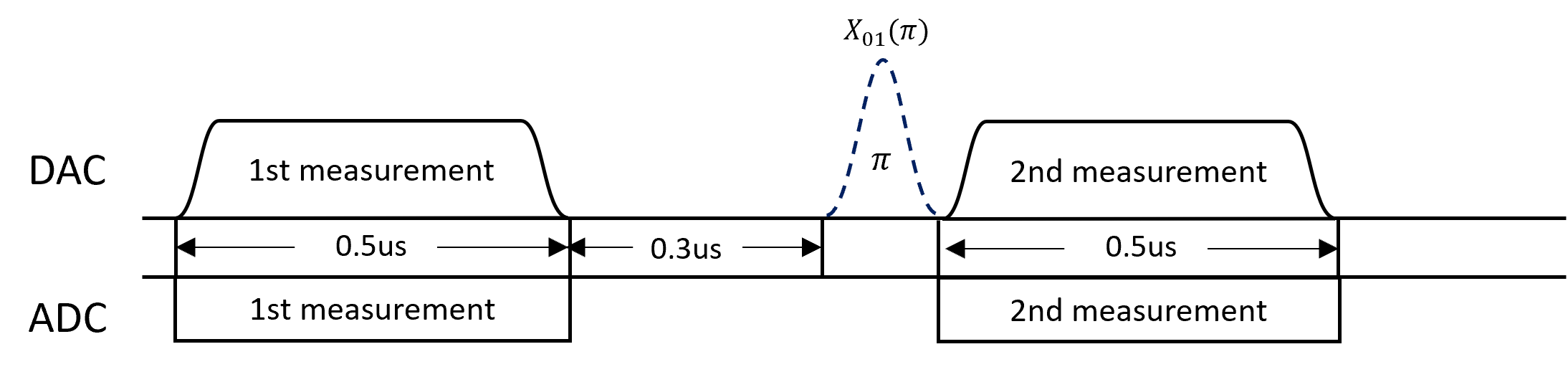}
        \caption{Experiment pulse scheme for the RQC dataset Q1 to Q4. }
        \label{fig:pulse_scheme_riken_1}
\end{figure}

\begin{figure}[h!]
     \centering
        \includegraphics[width=\linewidth]{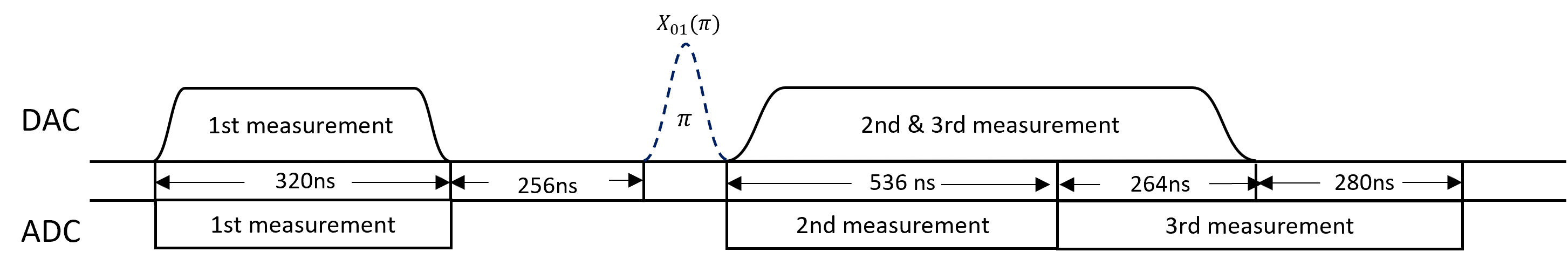}
        \caption{Experiment pulse scheme for the RQC dataset Q5 to Q8.  The second measurement pulse is longer, and it is divided into two segments, which corresponds to the second and third measurement, respectively. The first measurement is used post-selection to ensure the qubit is prepared for the ground state. The second measurement is analyzed using the signature approach, while the third measurement is used to detect if a state transition event occurred during the second measurement.}
        \label{fig:pulse_scheme_riken_2}
\end{figure}

The experimental pulse scheme is shown in Fig. \ref{fig:pulse_scheme_riken_1} for Q1 to Q4 and Fig. \ref{fig:pulse_scheme_riken_2} for Q5 to Q8, respectively. In this experiment, two consecutive readout pulses are applied. The sampling rate is 2Gsps. The dataset contains 320,000 traces for each qubit.
The initial state preparation fidelities for qubits Q1 to Q8 are measured to be $99.69\%$, $99.80\%$, $99.84\%$, $99.85\%$  
$99.28\%$, $99.33\%$, $99.21\%$, $99.43\%$, respectively. For qubits Q1 through Q4, the dataset includes an impedance-matched Josephson parametric amplifier (JPA), while qubits Q5 through Q8 do not have a JPA. The distribution of qubits Q1 to Q3 shows a non-circular shape due to the amplifier's effect on the readout signal. For the experiment setup, please refer to \cite{spring2024fast}.

Even without applying the signature method, the dataset already yields very high accuracy. To demonstrate the advantage of the signature method, we sampled 40,000 traces for each state, resulting in a total of 80,000 traces. From this dataset, we randomly selected 16,000 traces for the testing set, 51,200 traces for training, and 12,800 traces for validation to optimize the hyperparameters. This process was repeated 10 times to determine the confidence levels of the accuracies.

\begin{figure}[h]
    \centering
    \begin{tabular}{c|c|c|c|c}
Dataset & RQC Q1 & RQC Q2 & RQC Q3 & RQC Q4 \\  \hline &&& \\
    \begin{tabular}{c}Scatter\end{tabular} & \iptsq{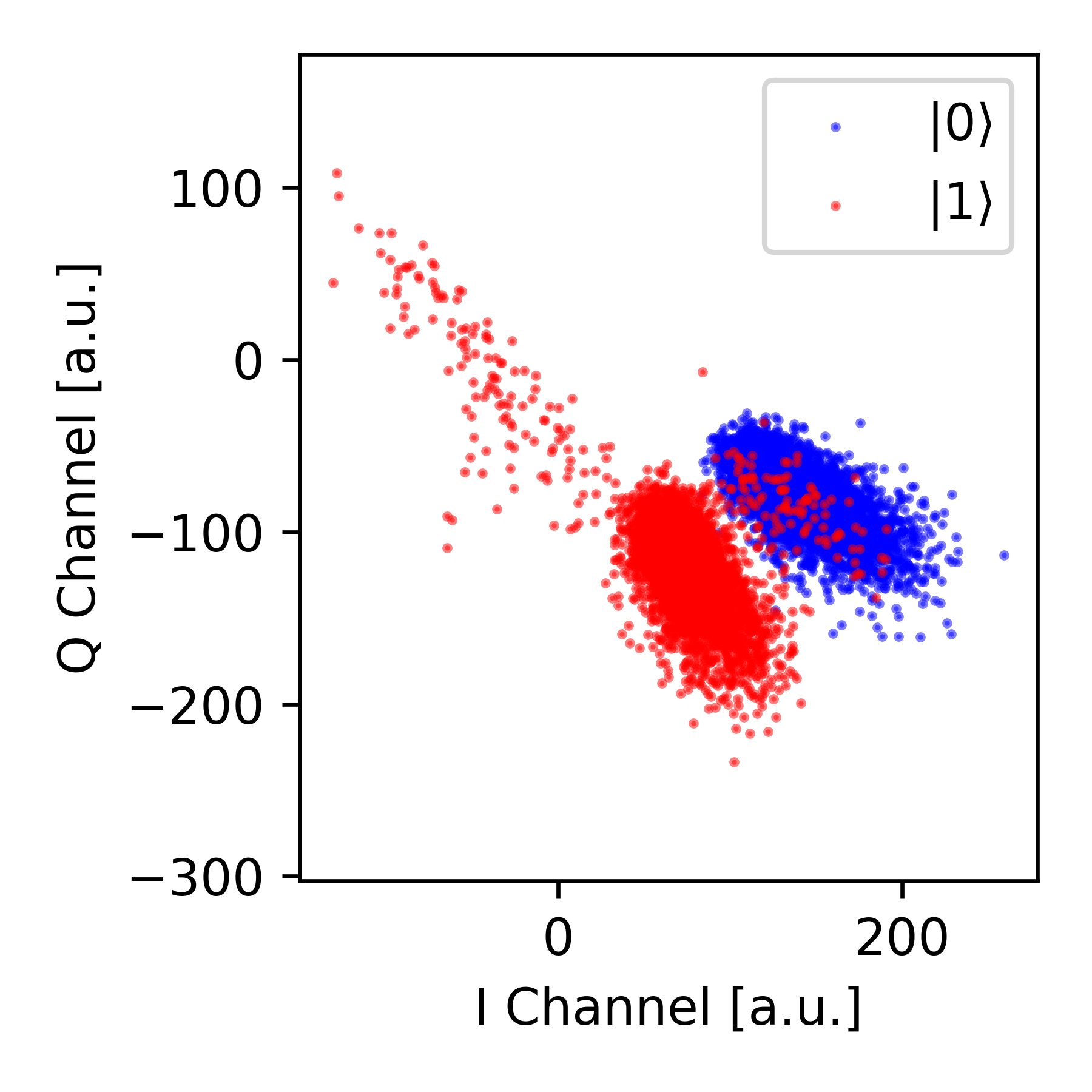}&\iptsq{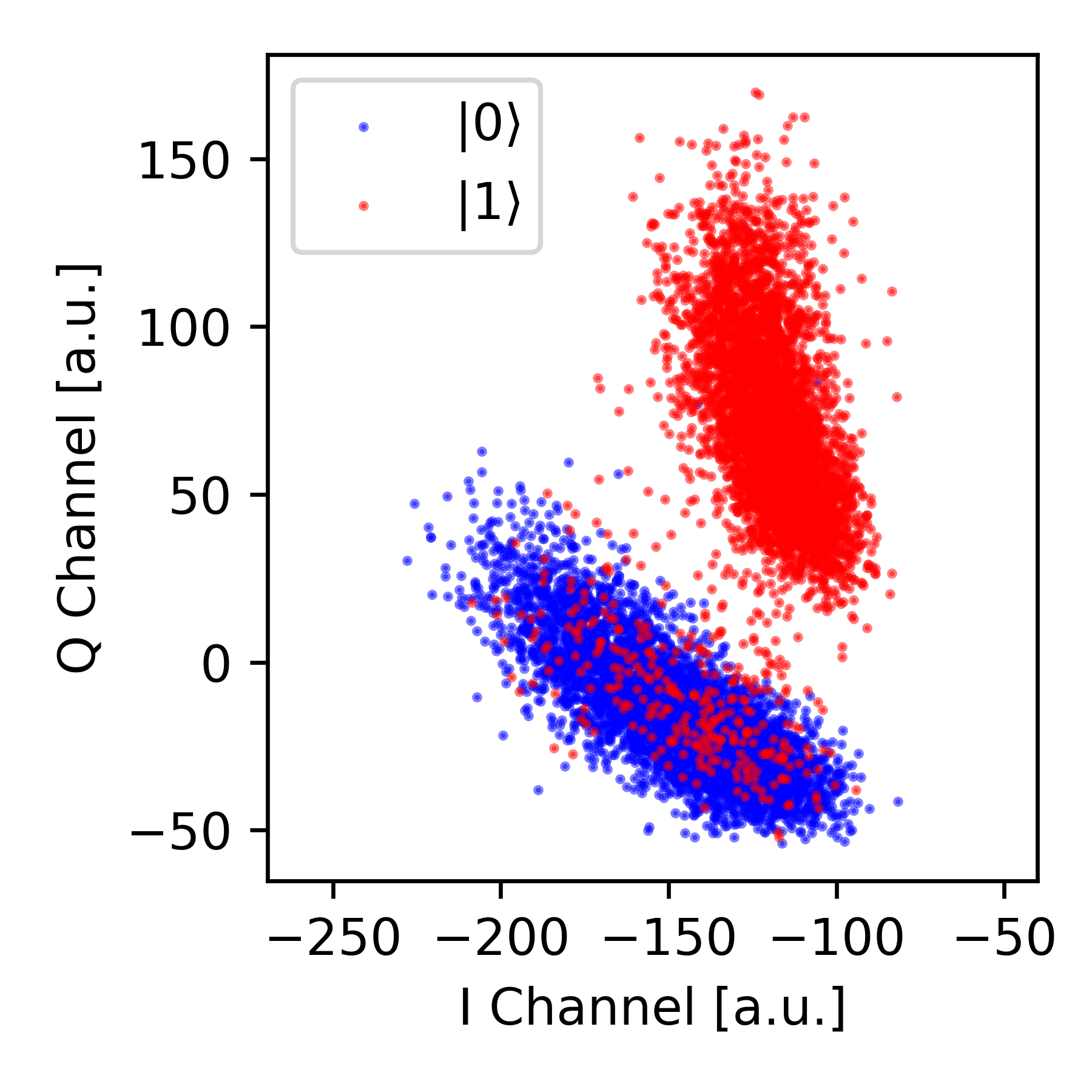}&\iptsq{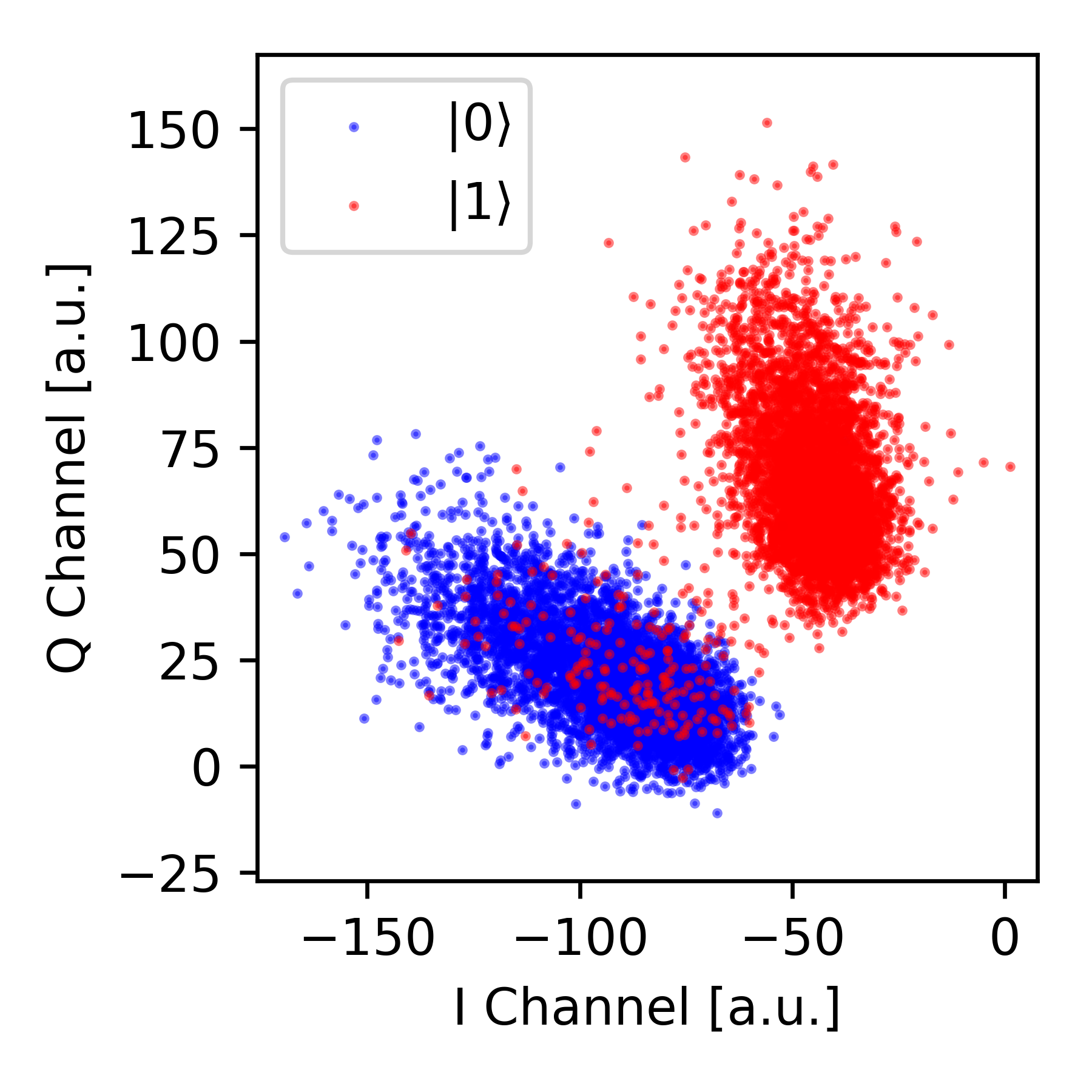}&\iptsq{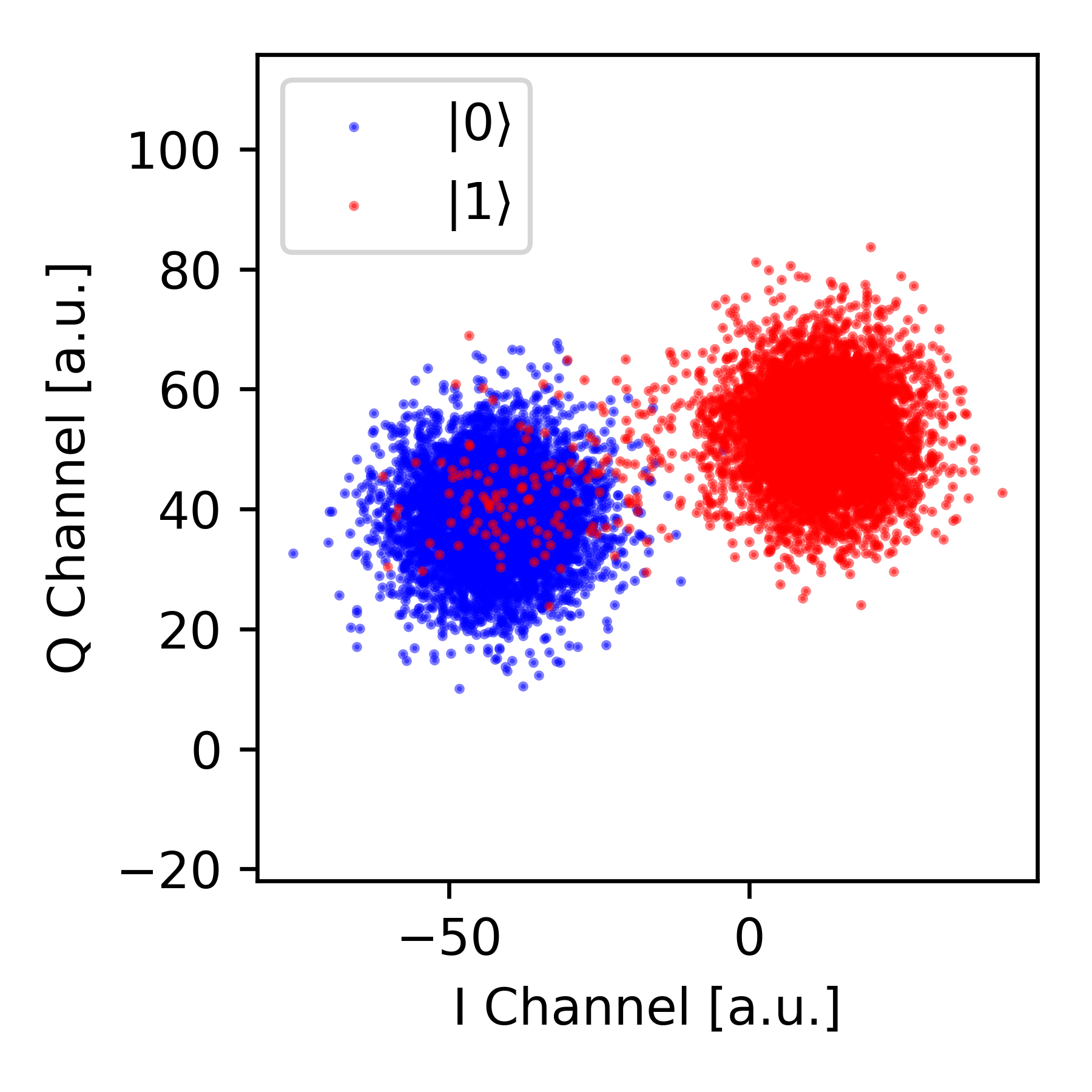}\\
    \begin{tabular}{c}Traces\\ State $\ket{0}$\end{tabular} & \iptm{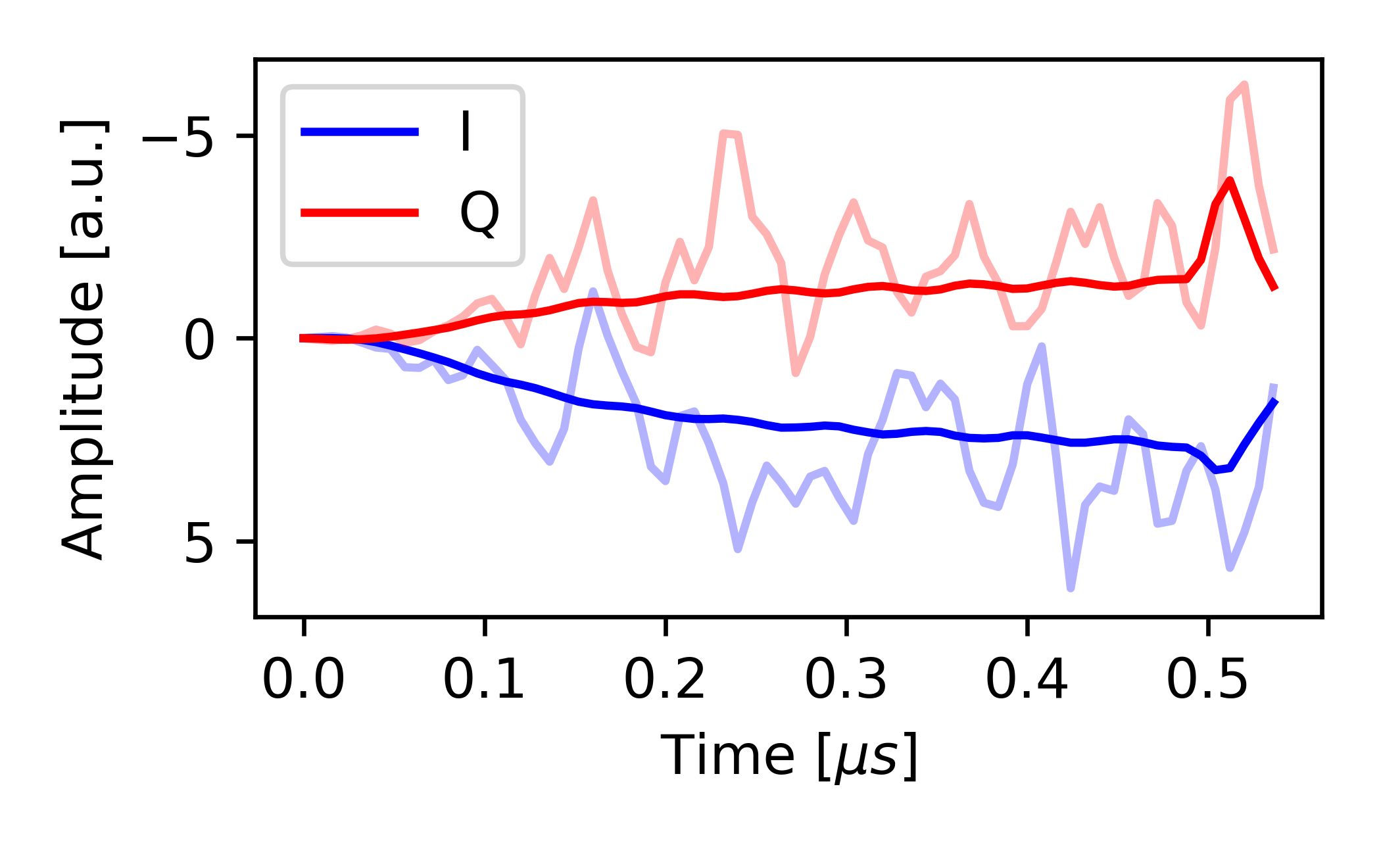}&\iptm{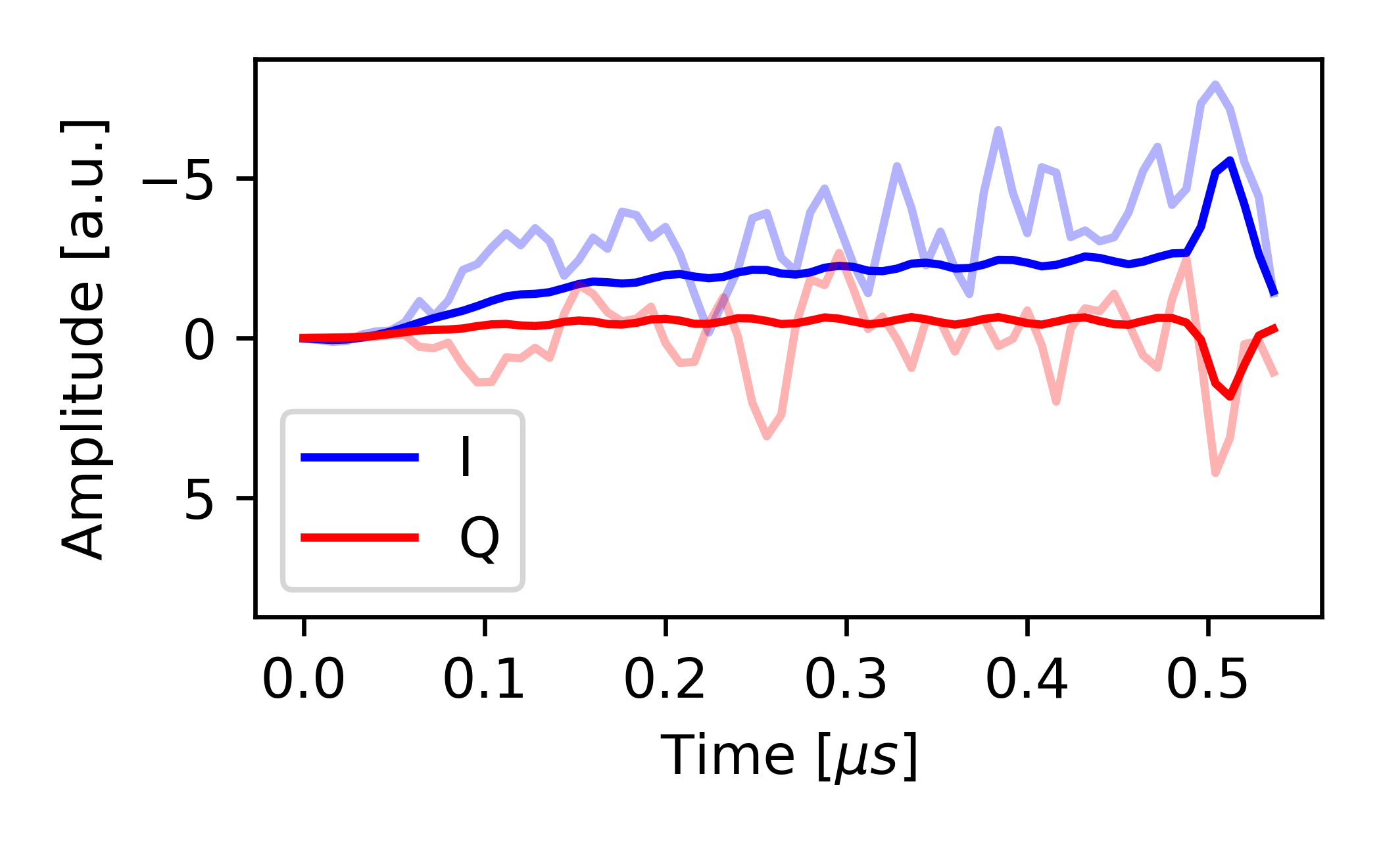}&\iptm{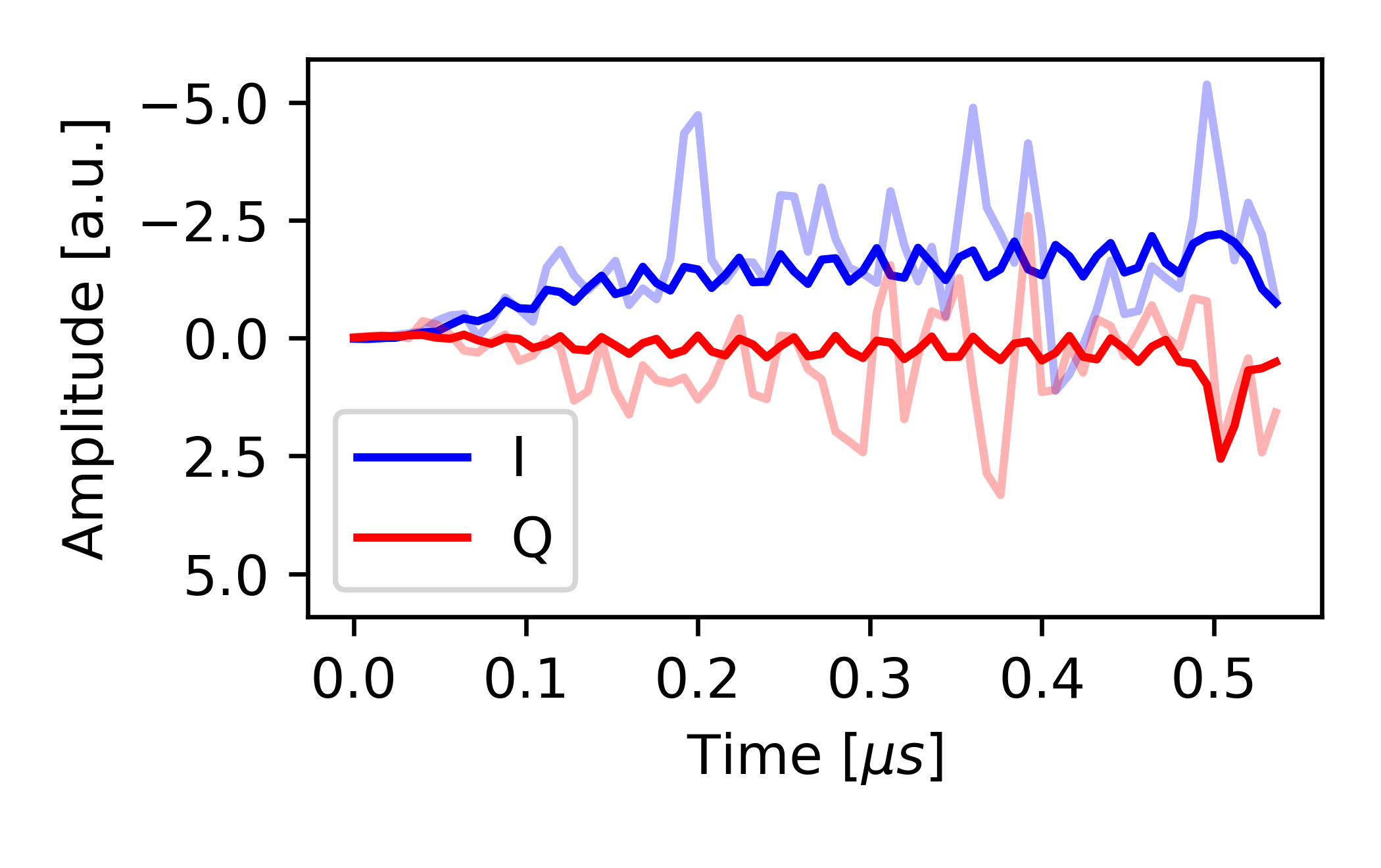}&\iptm{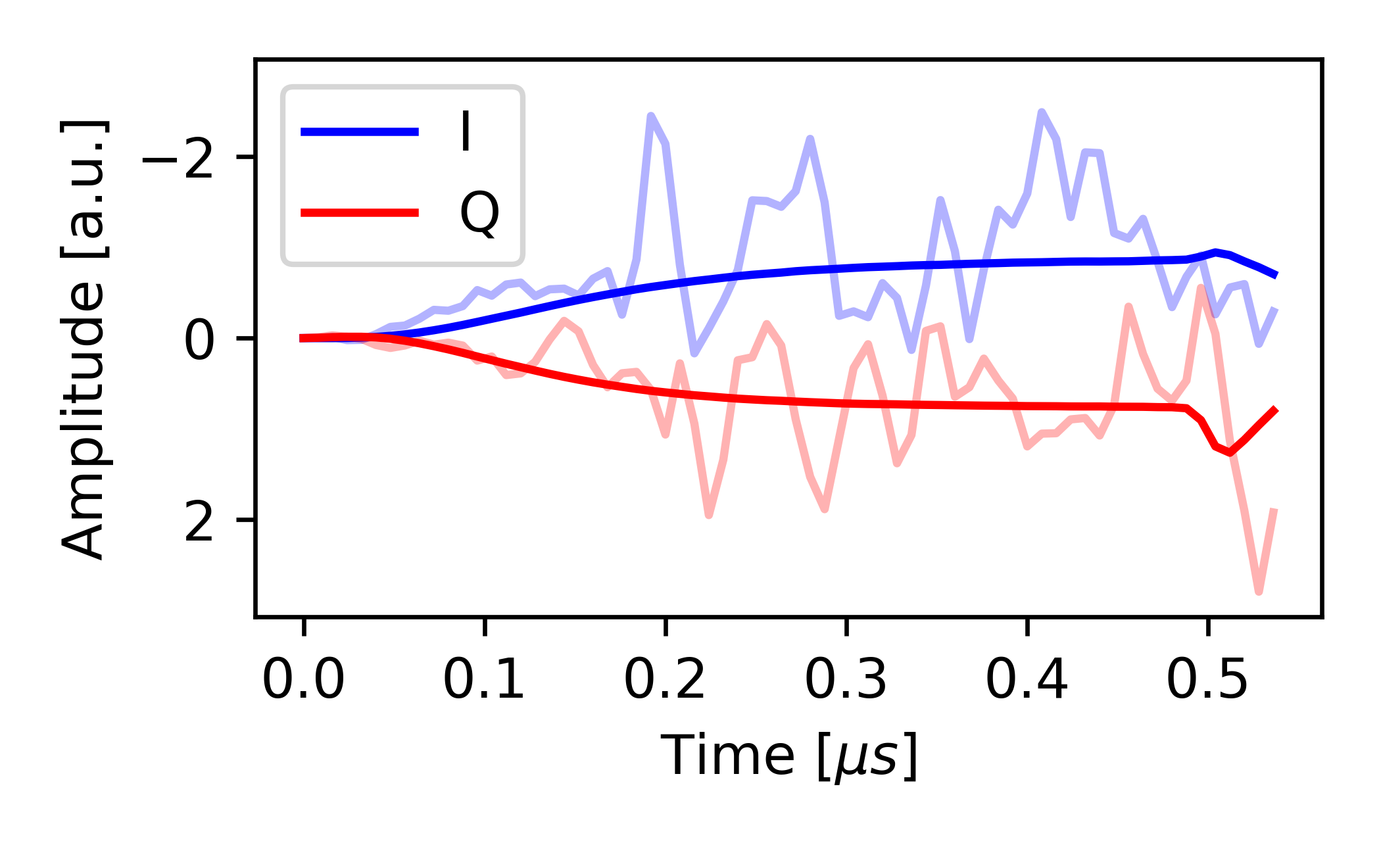}\\
    \begin{tabular}{c}Traces\\ State $\ket{1}$\end{tabular} & \iptm{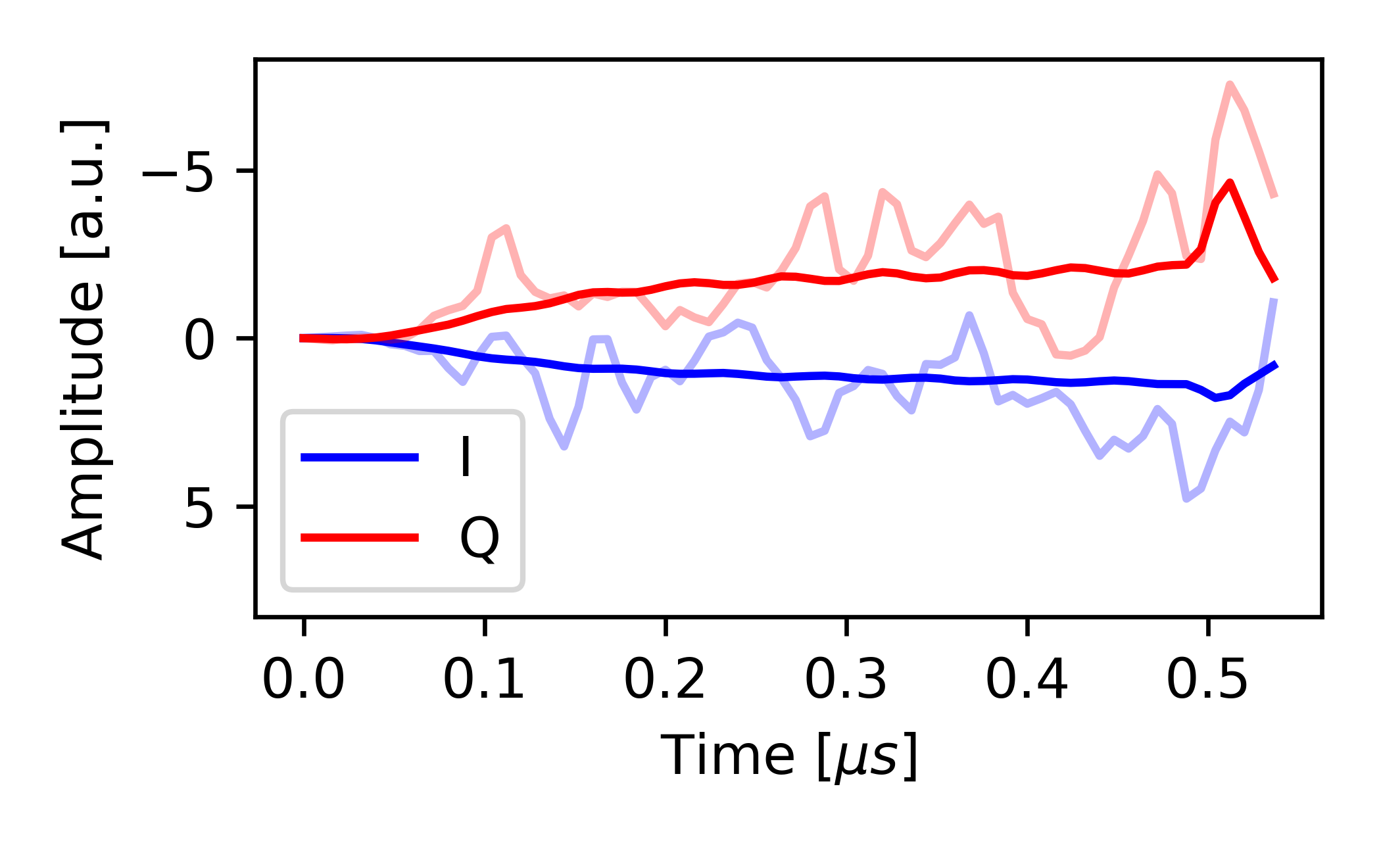}&\iptm{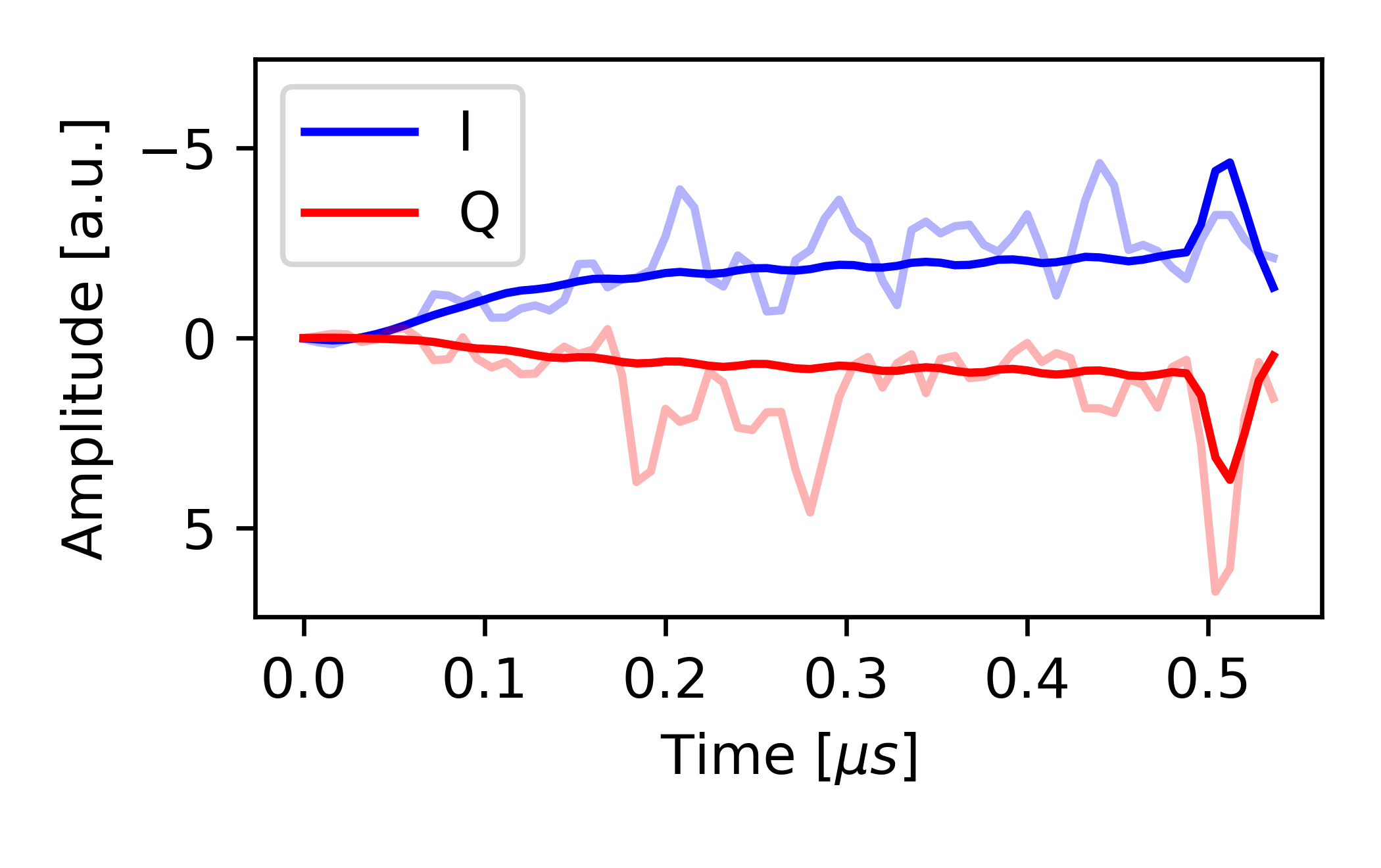}&\iptm{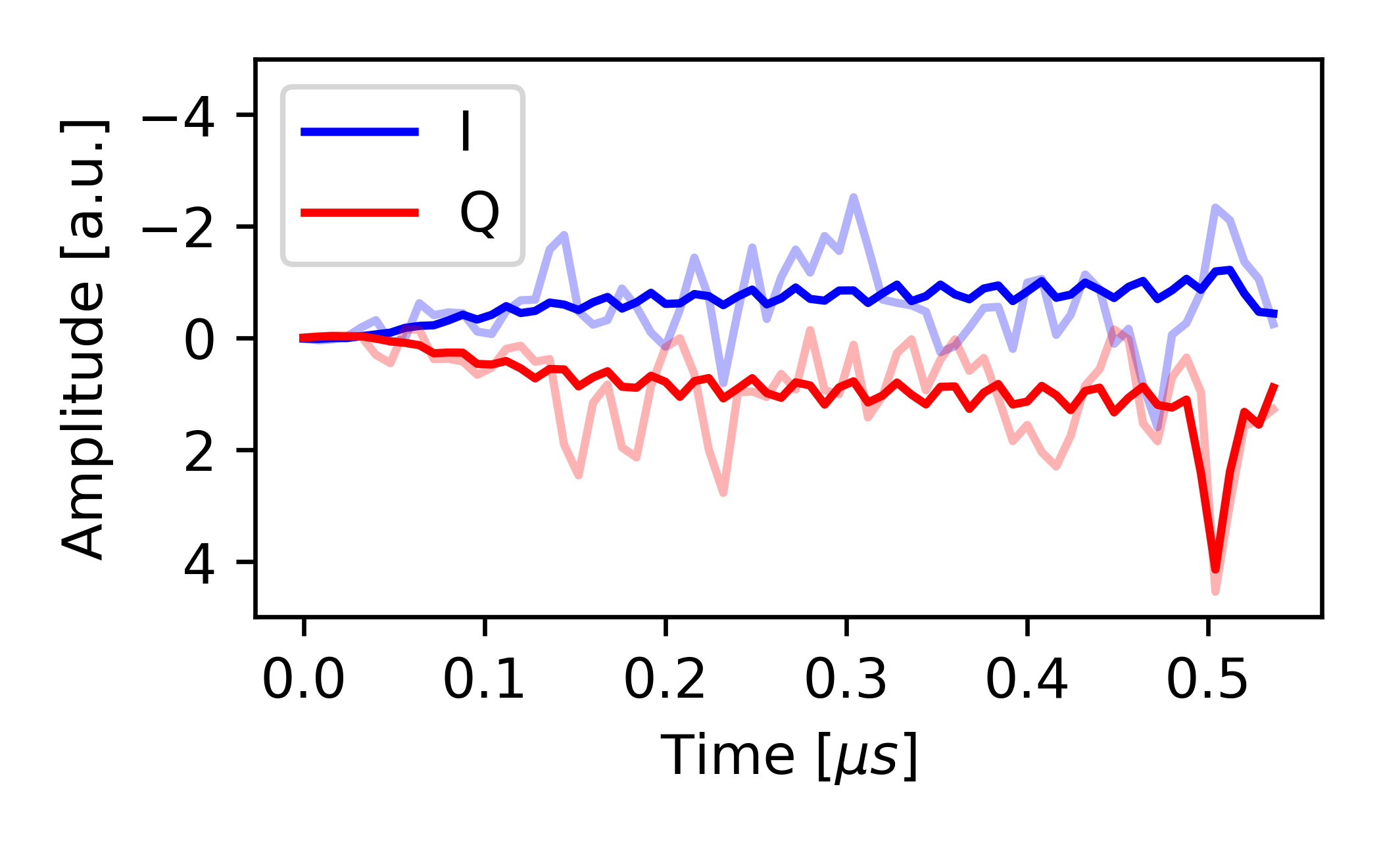}&\iptm{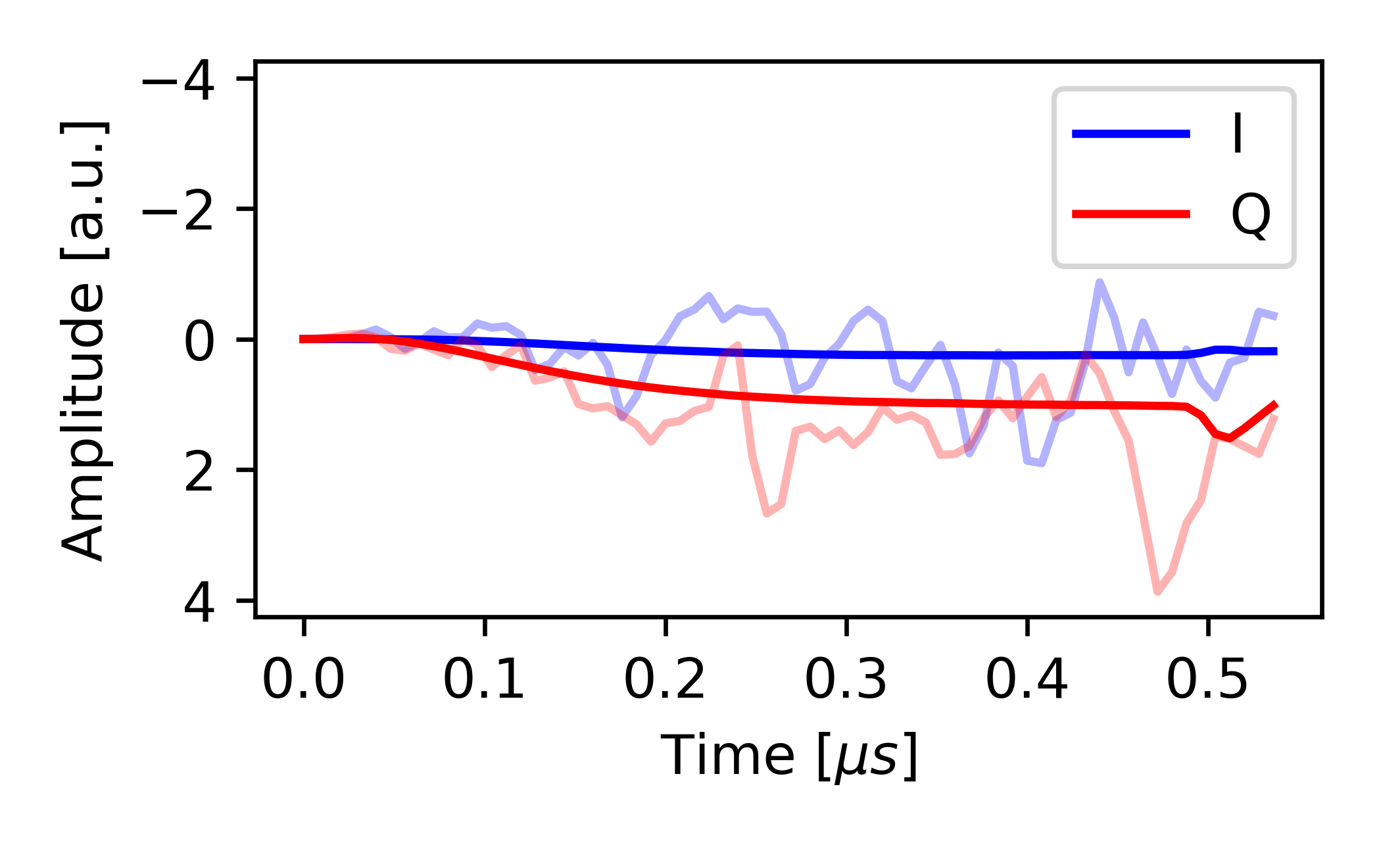}\\
    \begin{tabular}{c}Trajectory\end{tabular} & \iptsq{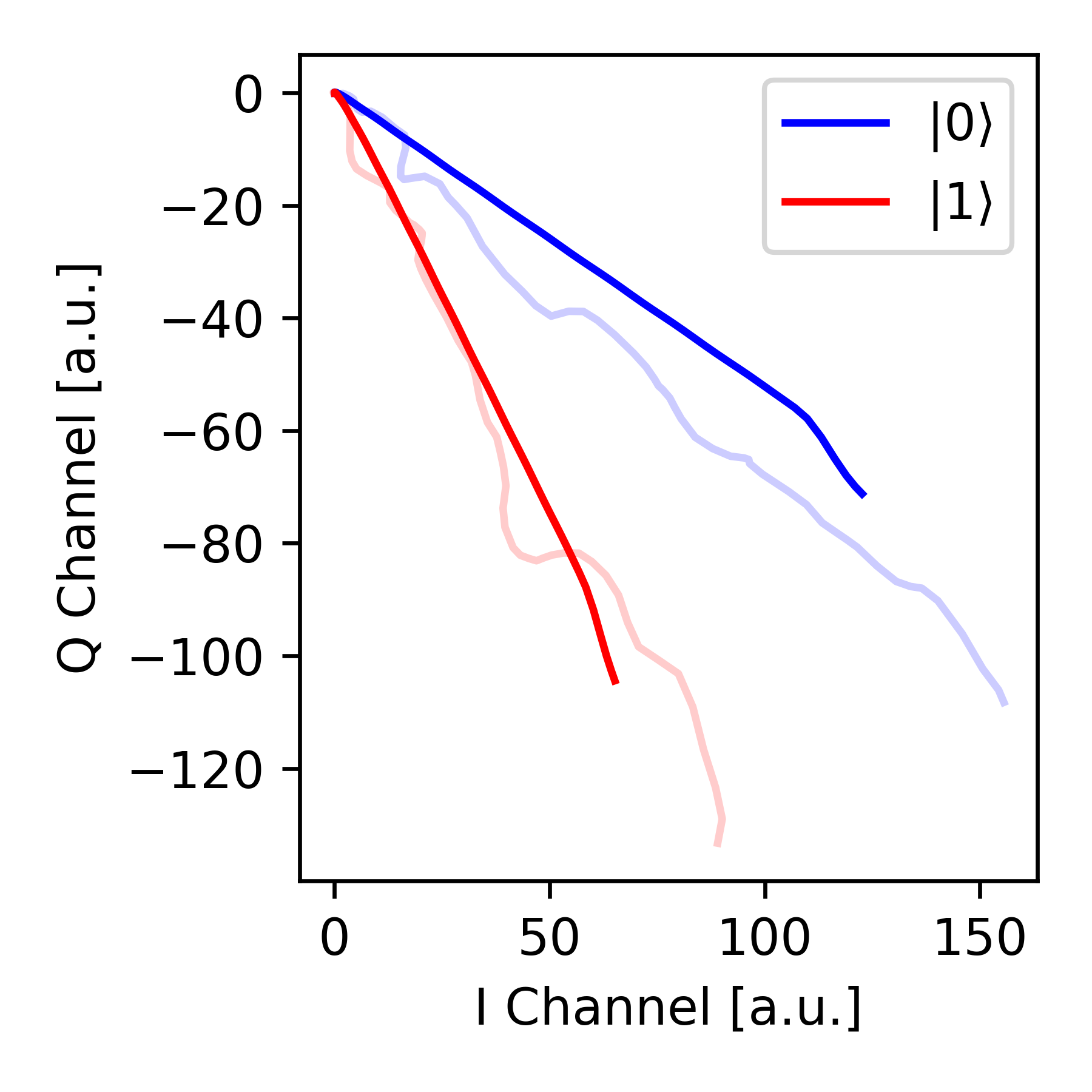}&\iptsq{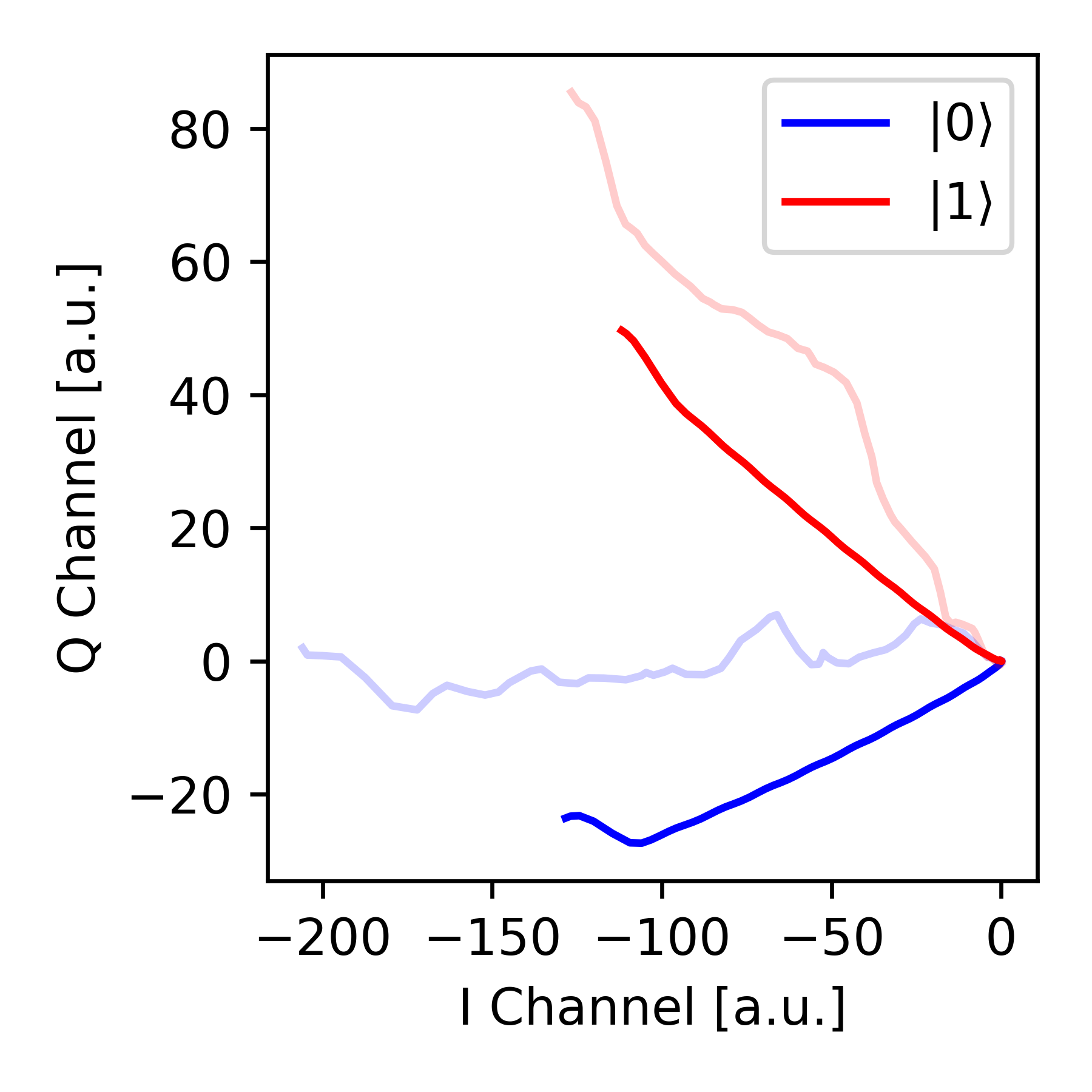}&\iptsq{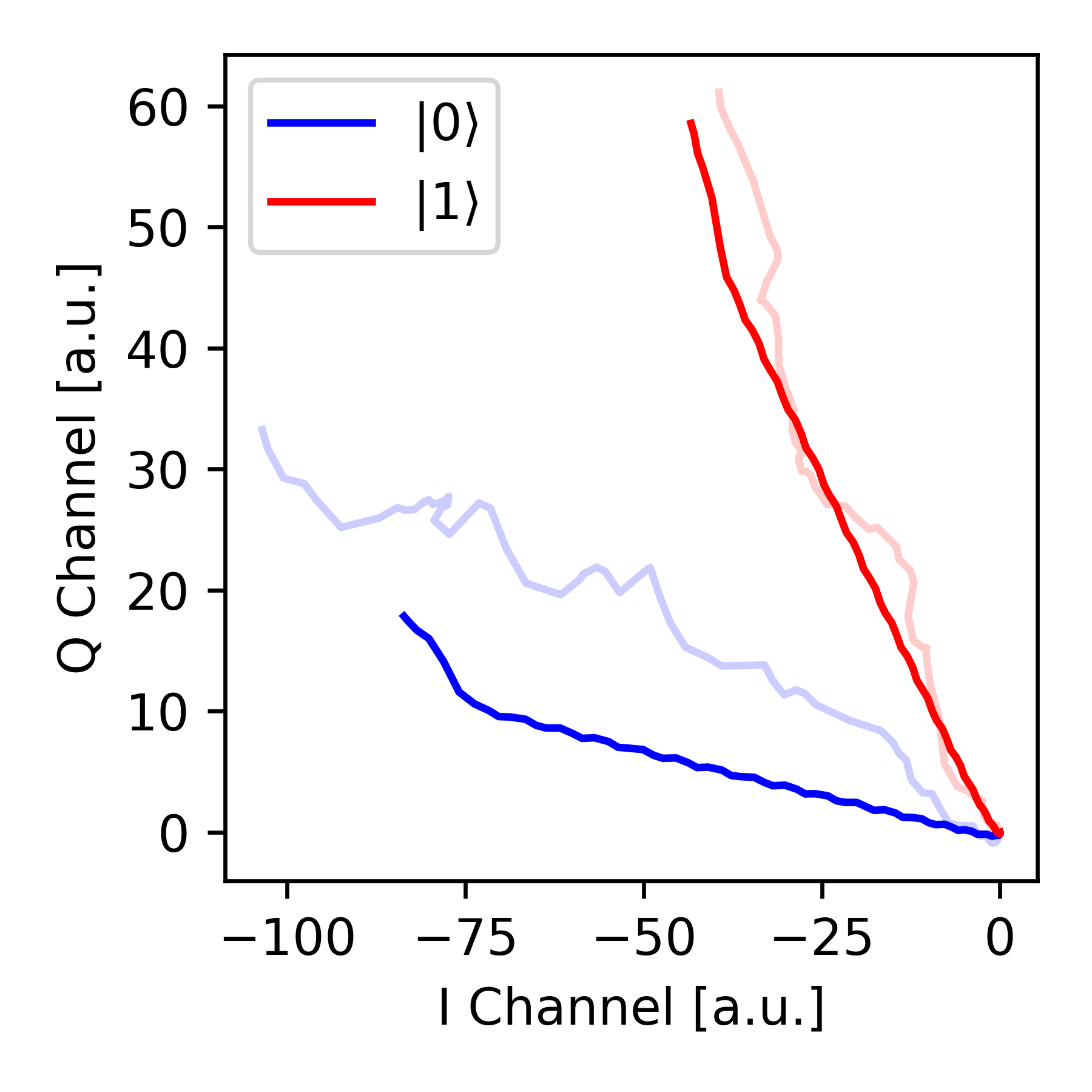}&\iptsq{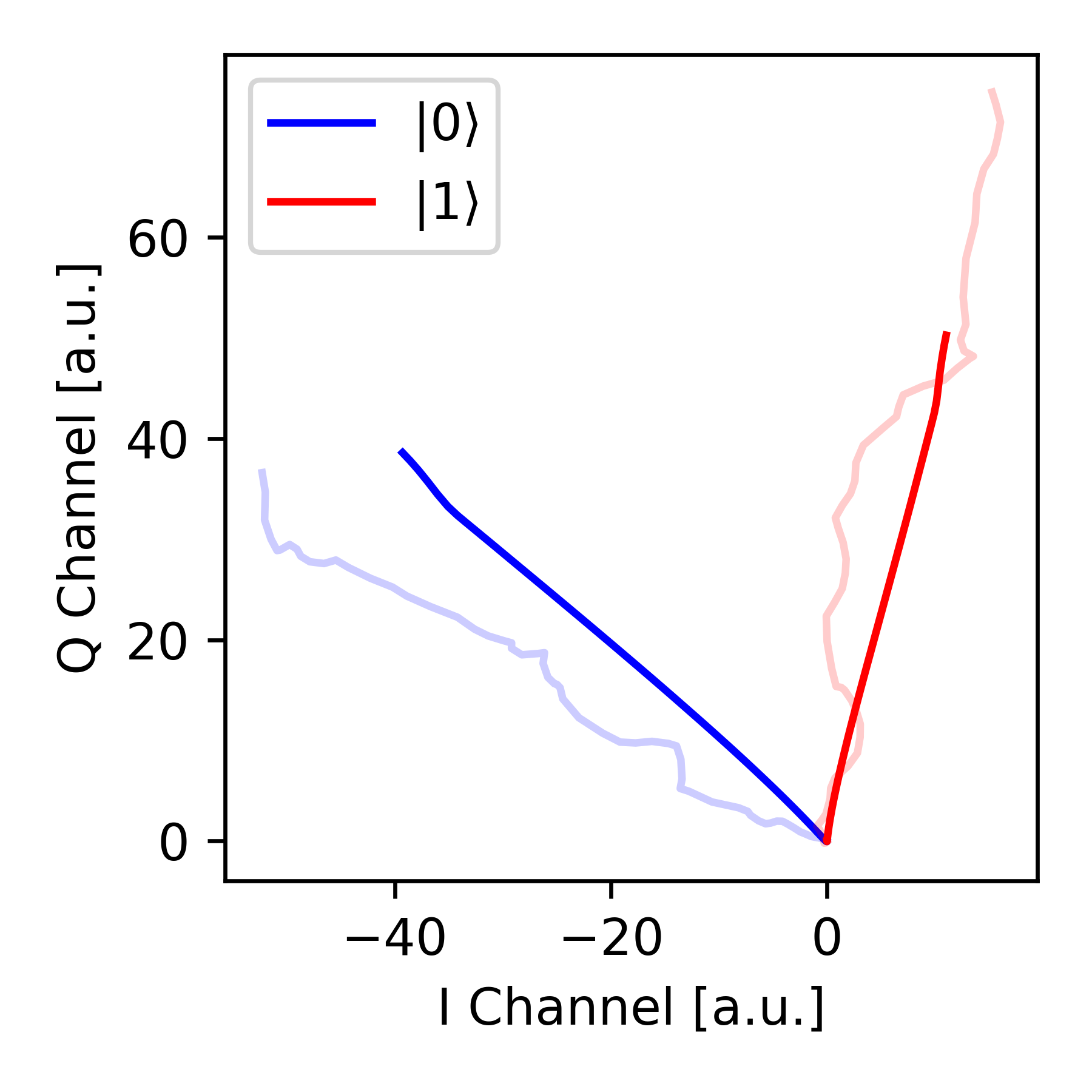}\\
    \begin{tabular}{c}Integration \\ Projection\end{tabular} & \iptm{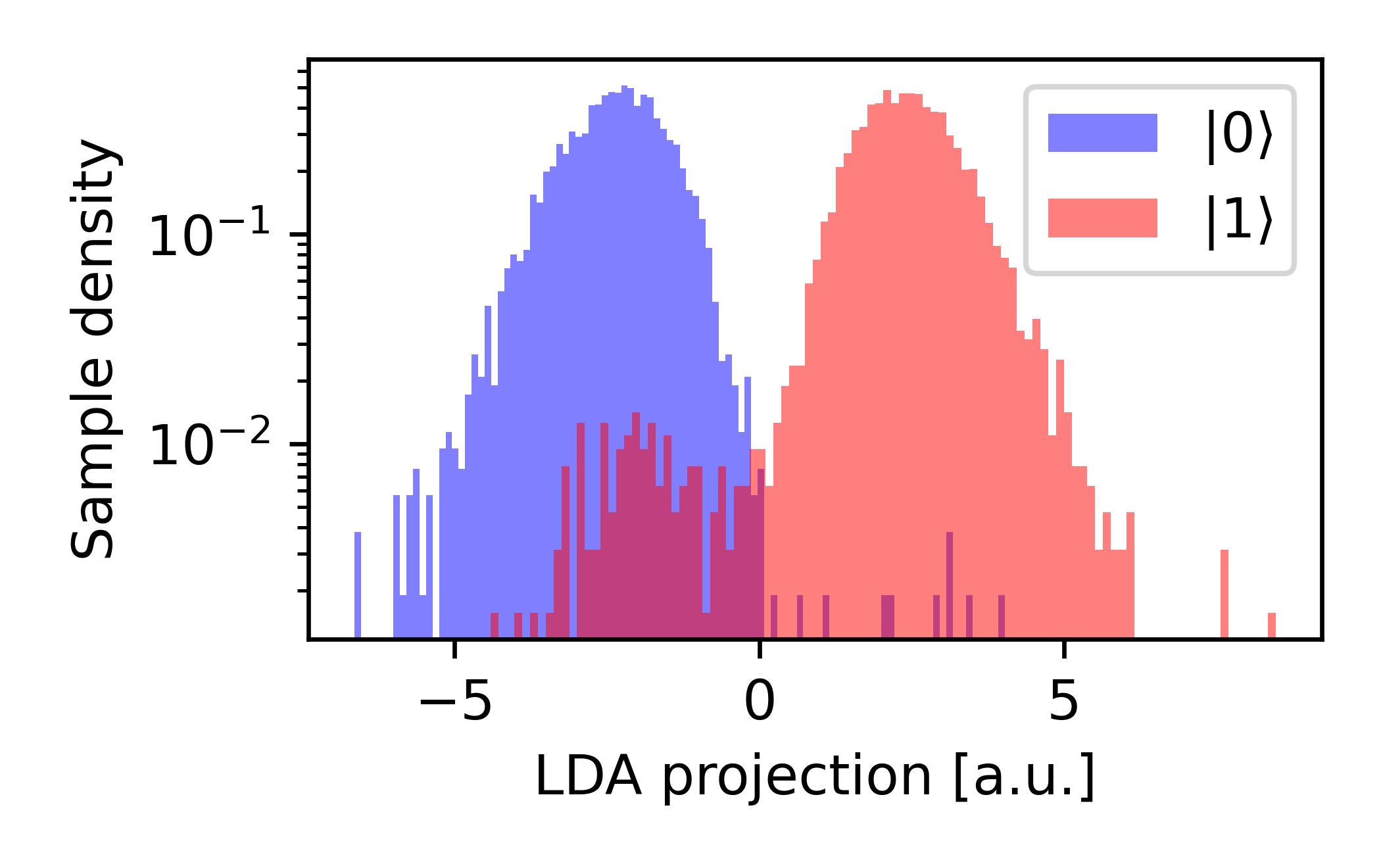}&\iptm{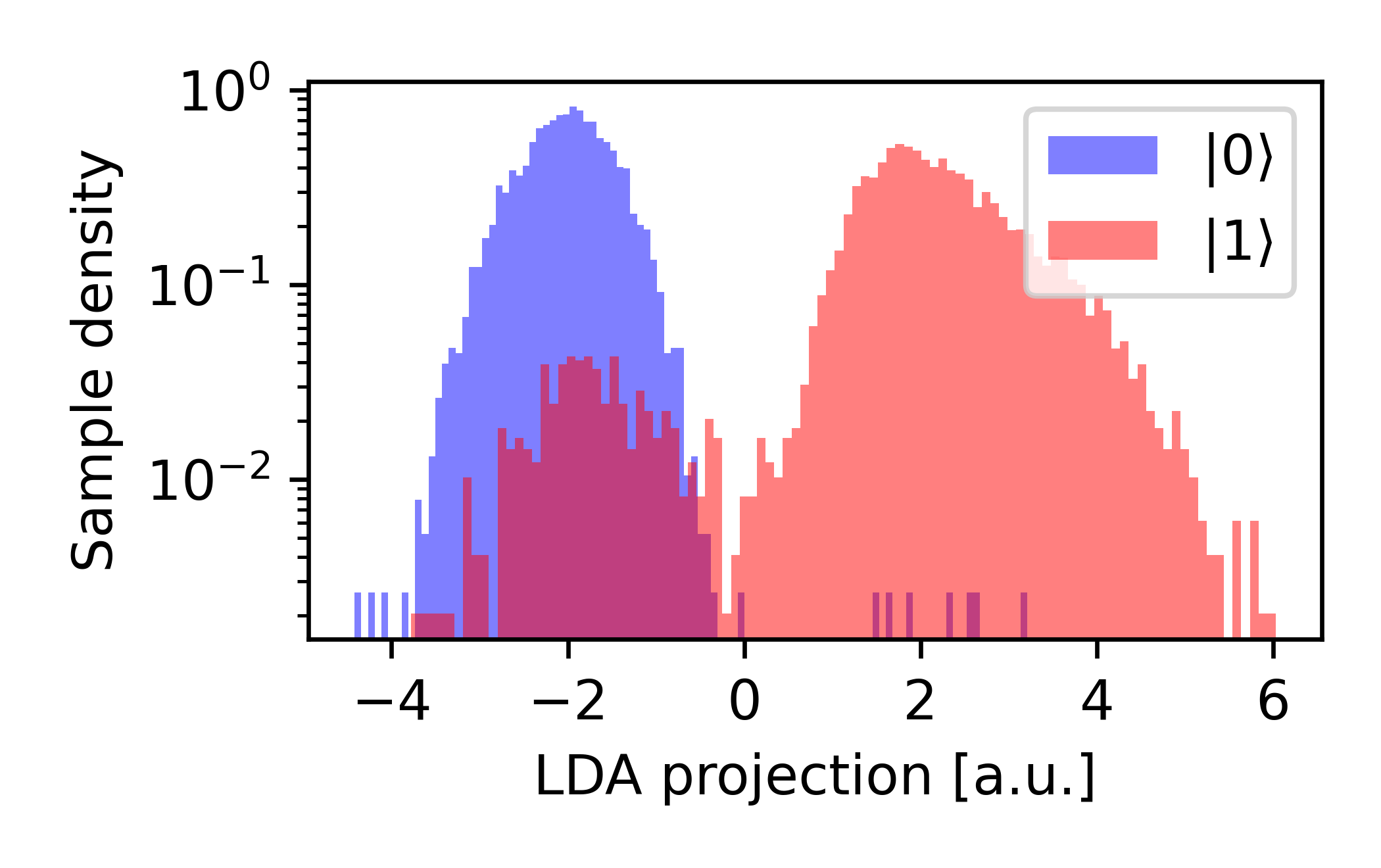}&\iptm{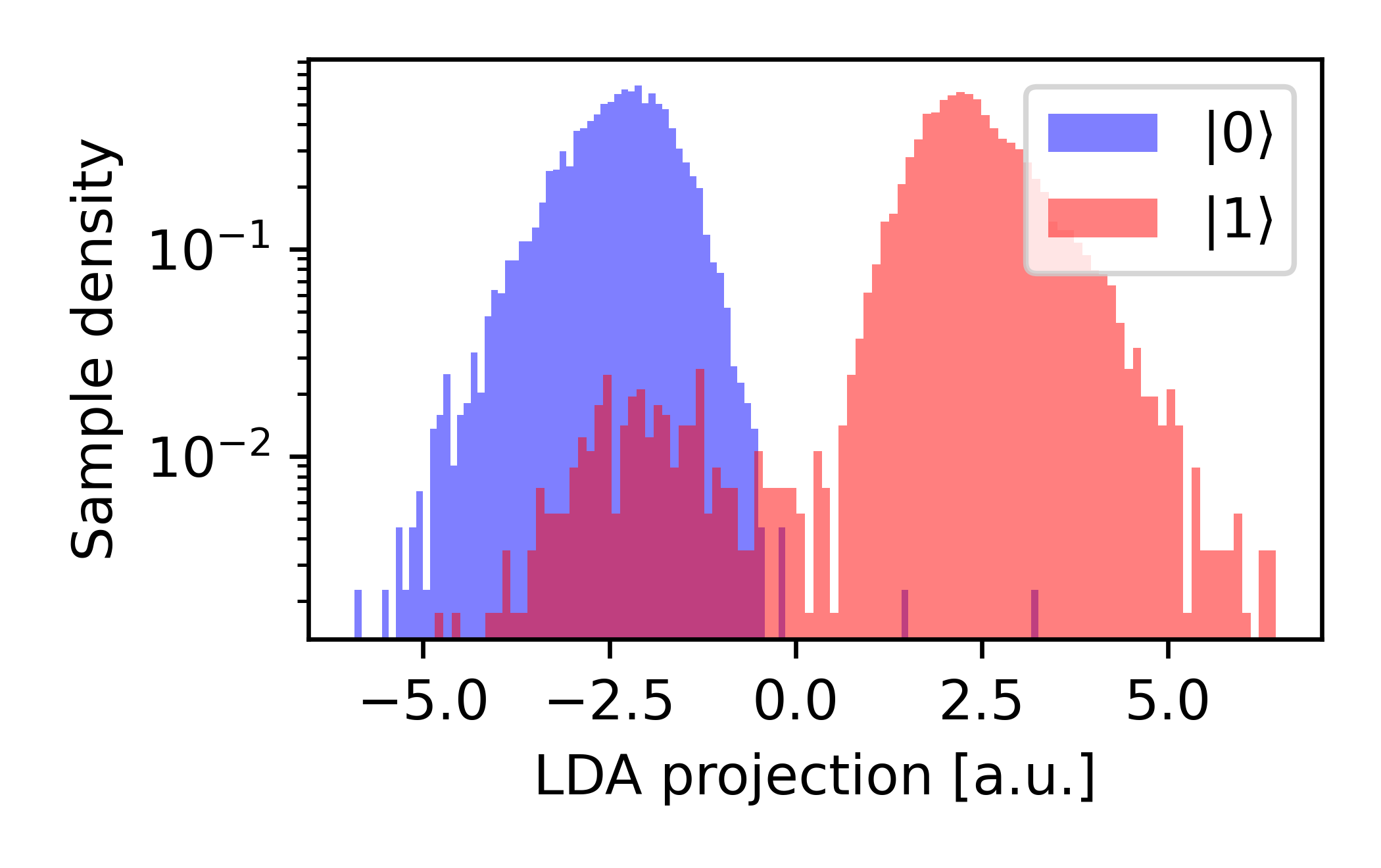}&\iptm{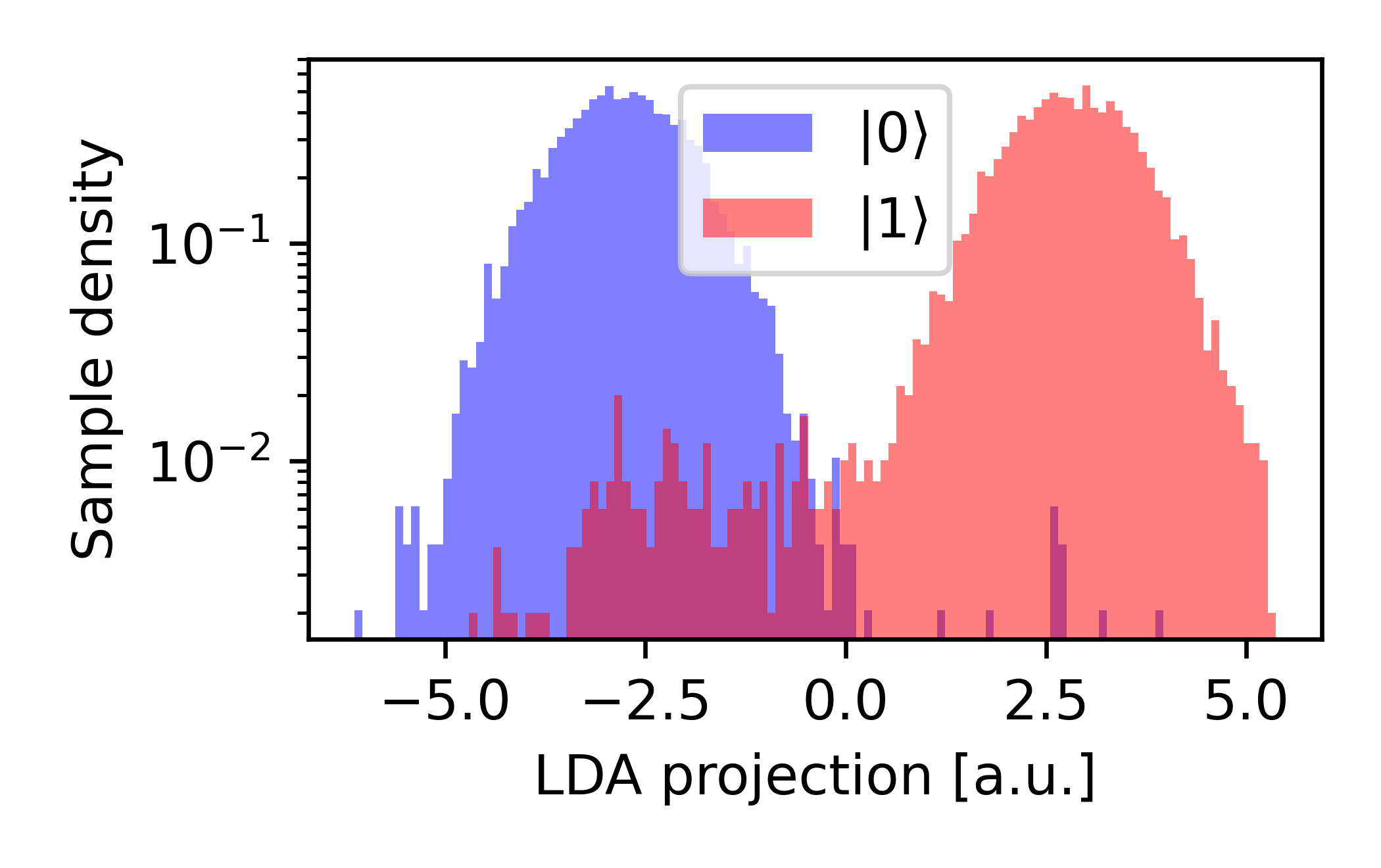}\\
    \begin{tabular}{c}Signature \\ Projection\end{tabular} & \iptm{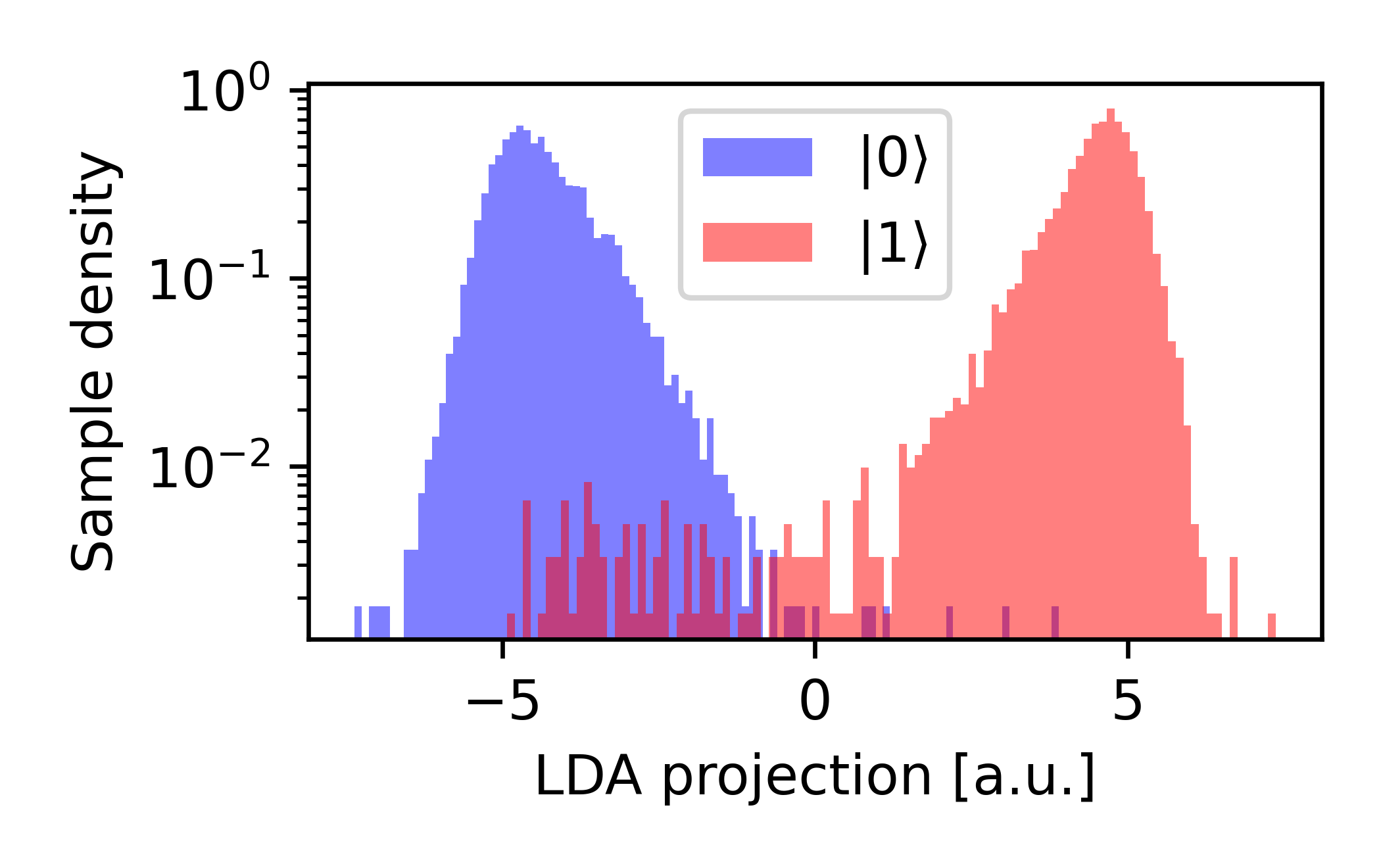}&\iptm{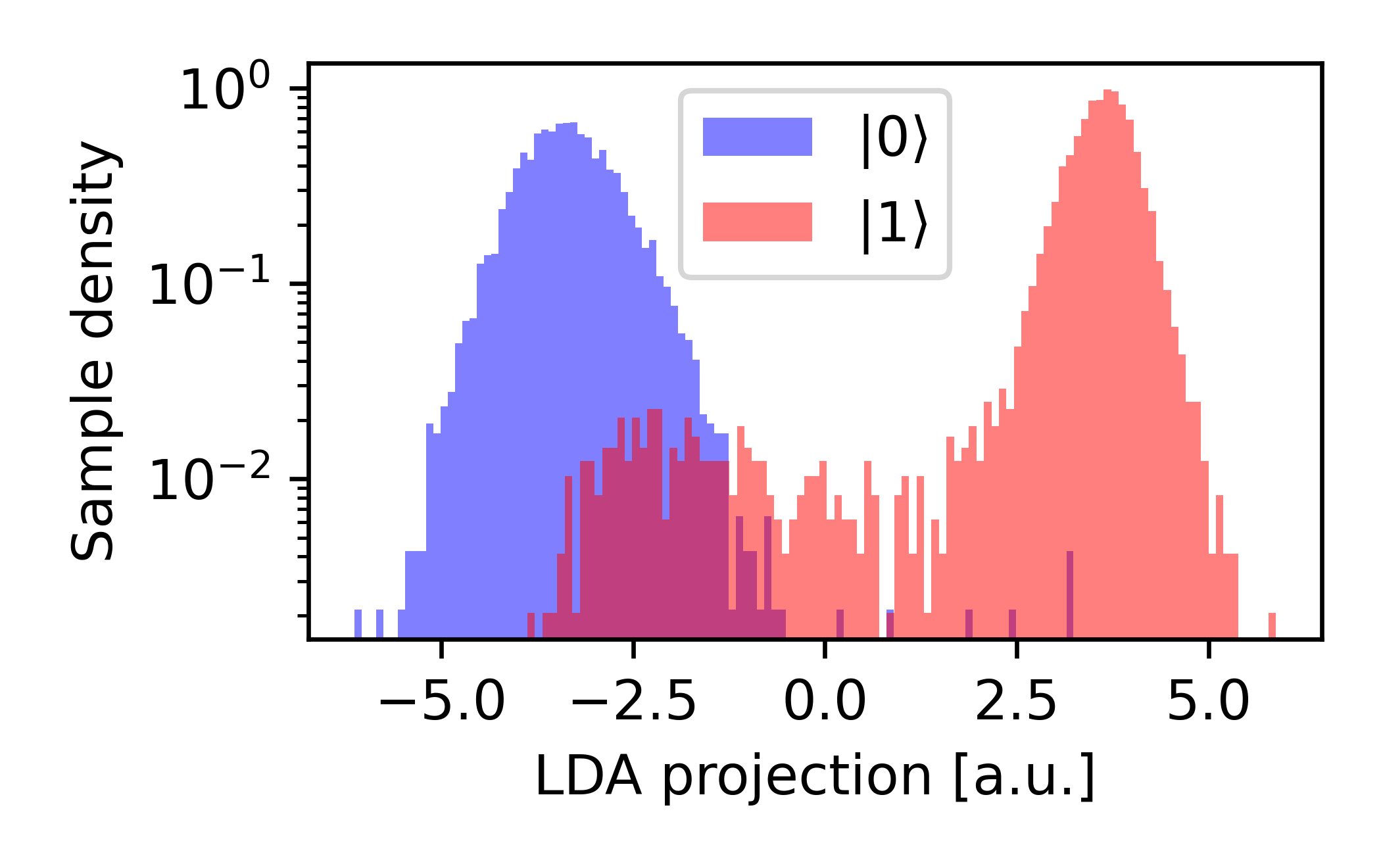}&\iptm{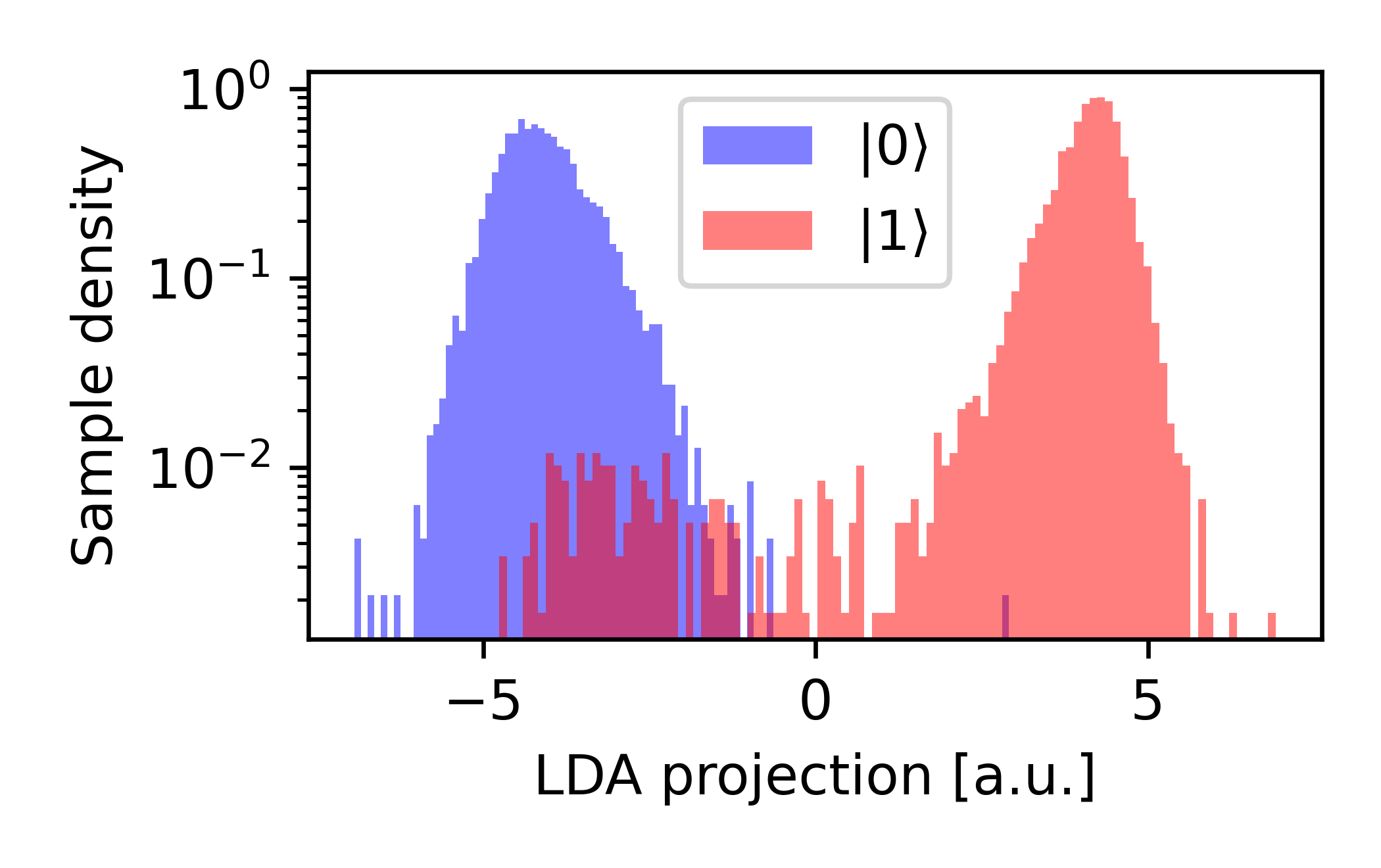}&\iptm{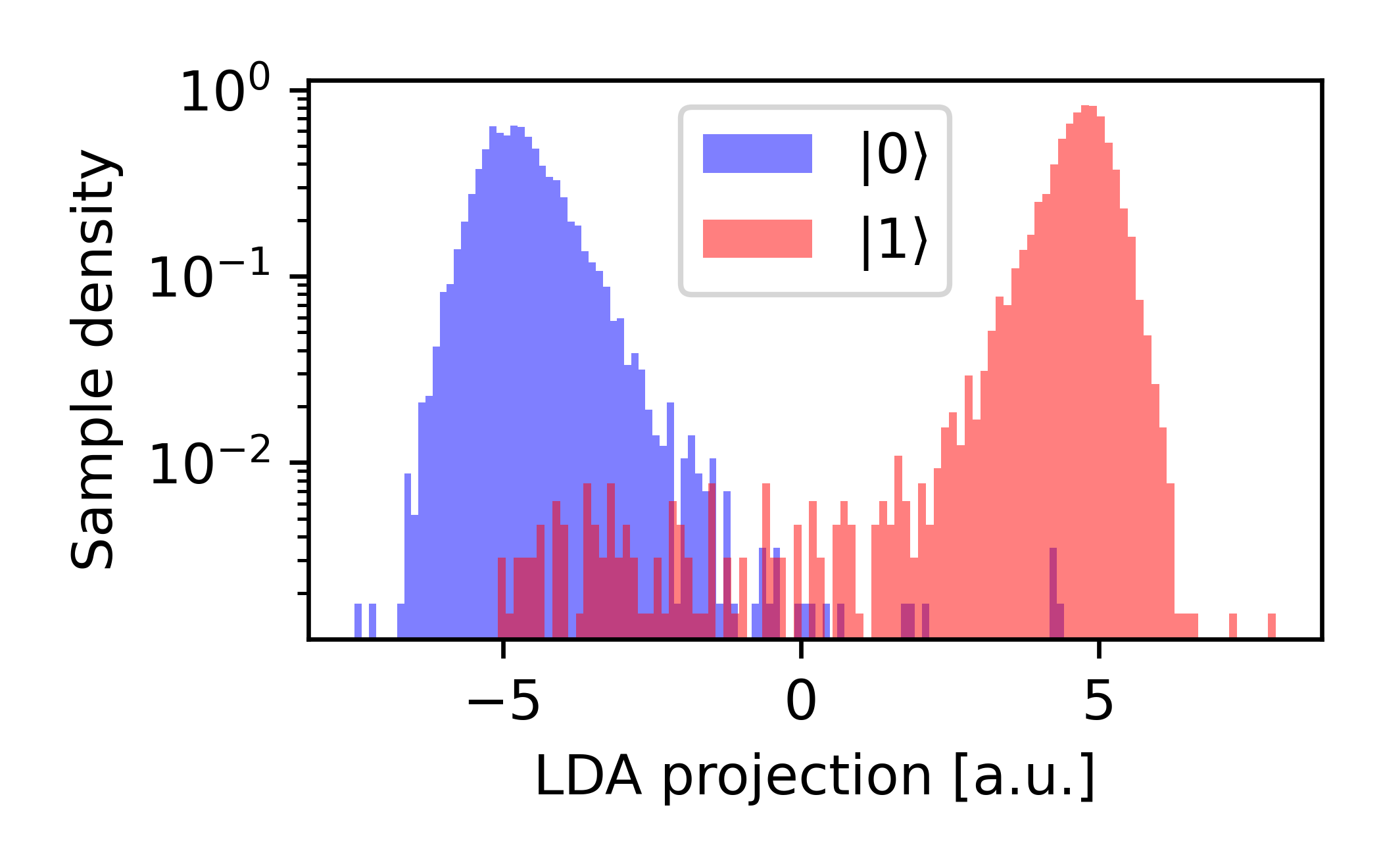}\\
    \begin{tabular}{c}Classifier \\ Performance\end{tabular} & \iptm{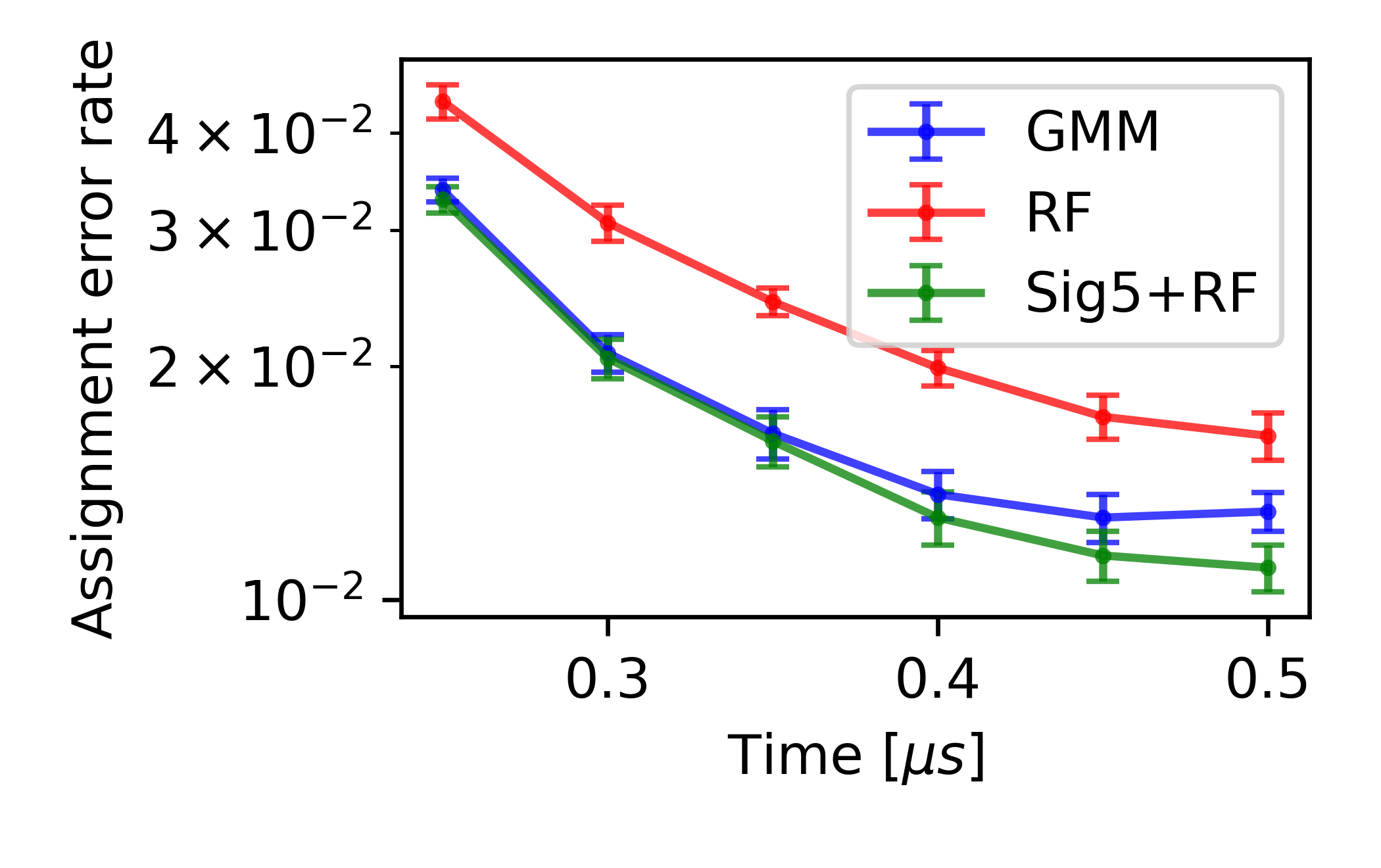}&\iptm{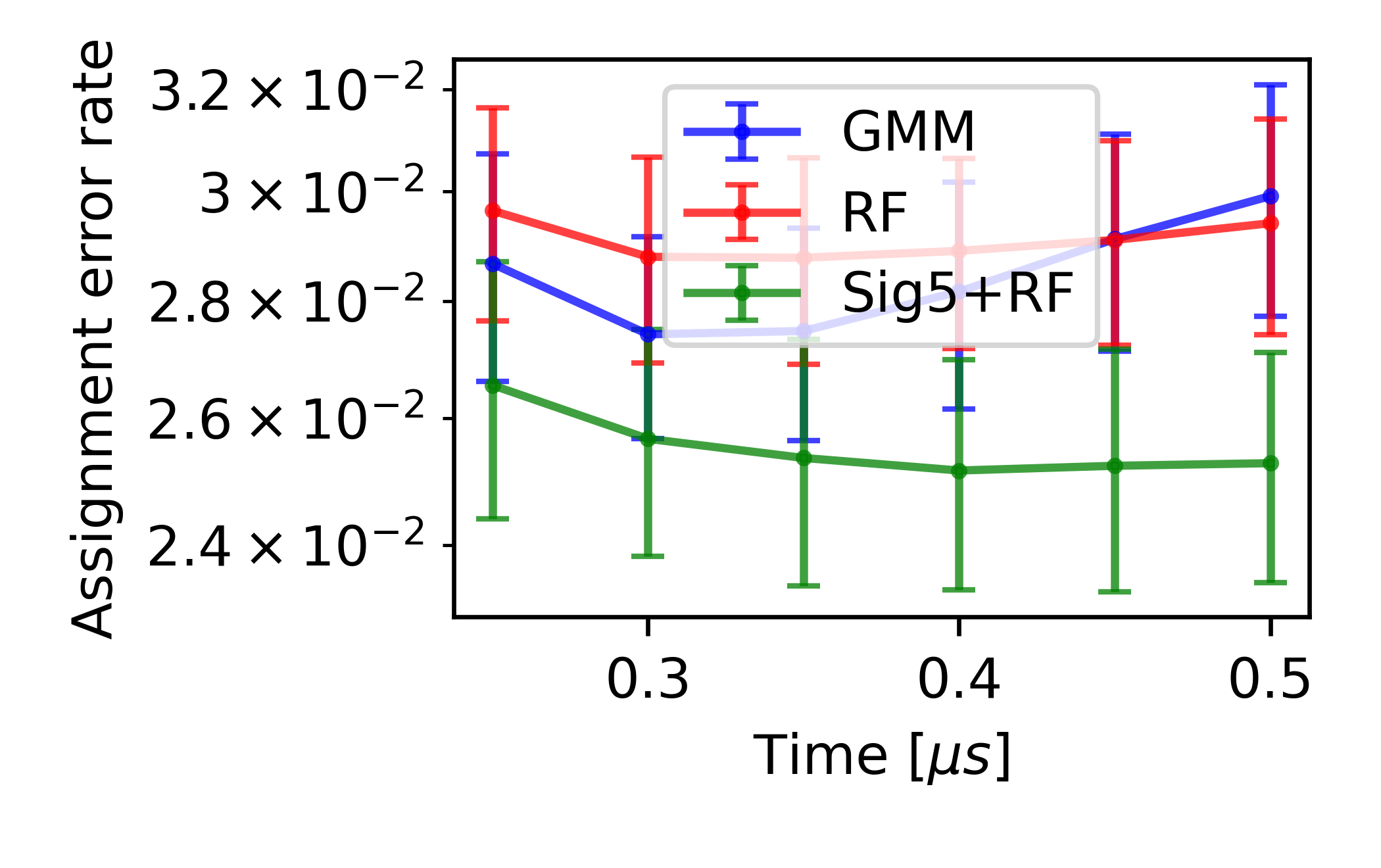}&\iptm{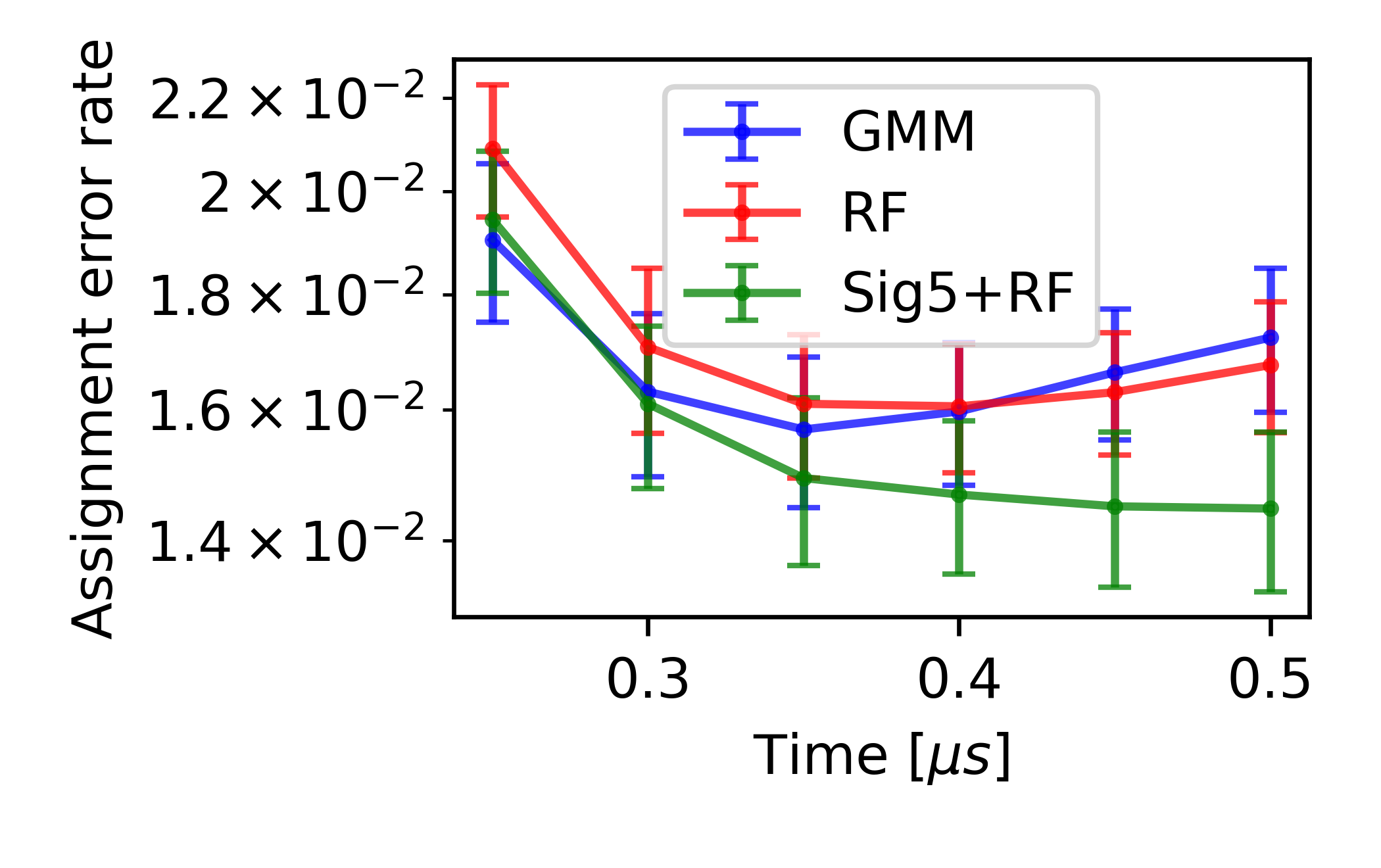}&\iptm{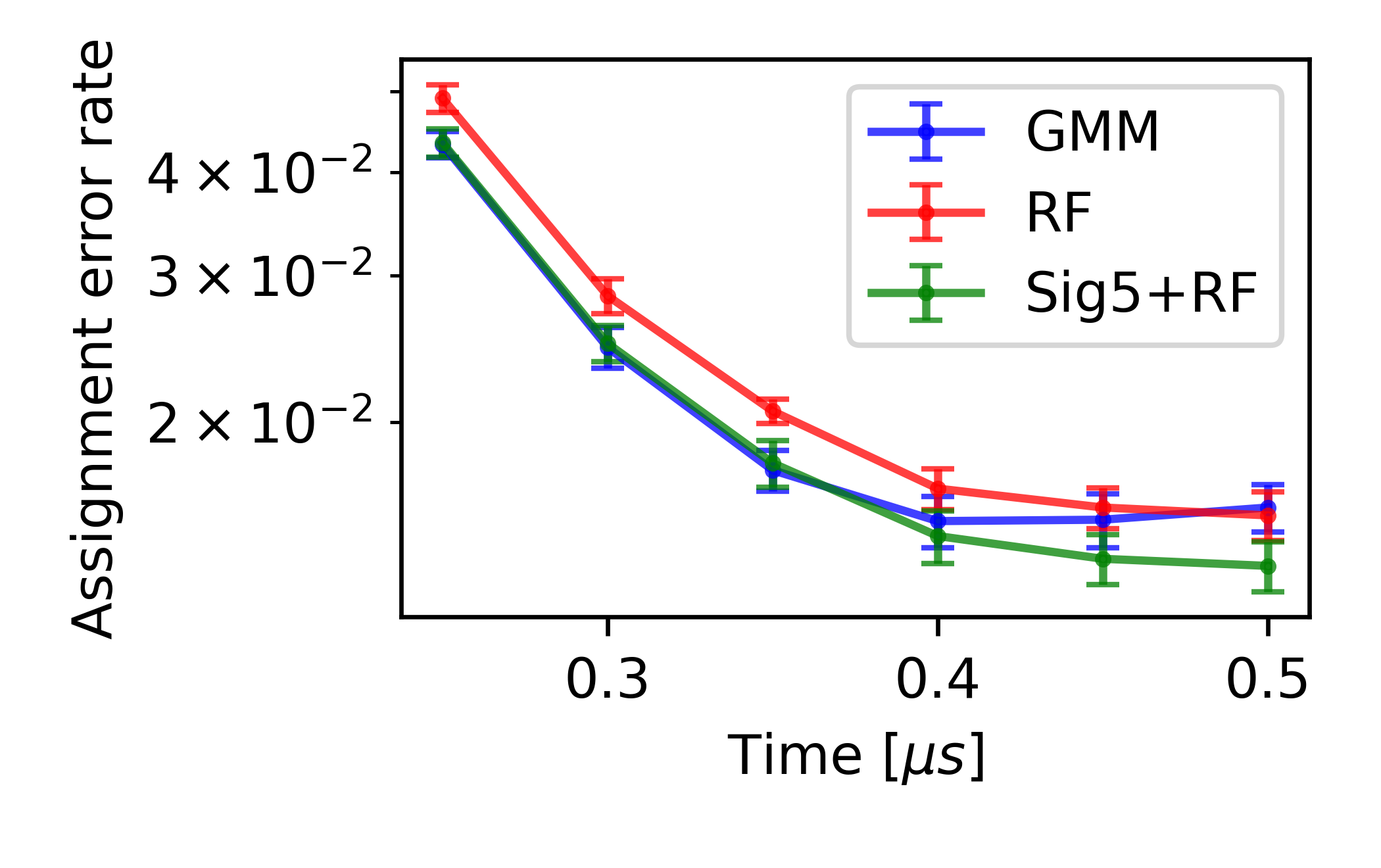}\\
    \begin{tabular}{c}Integration \\ Transition\end{tabular} & \iptm{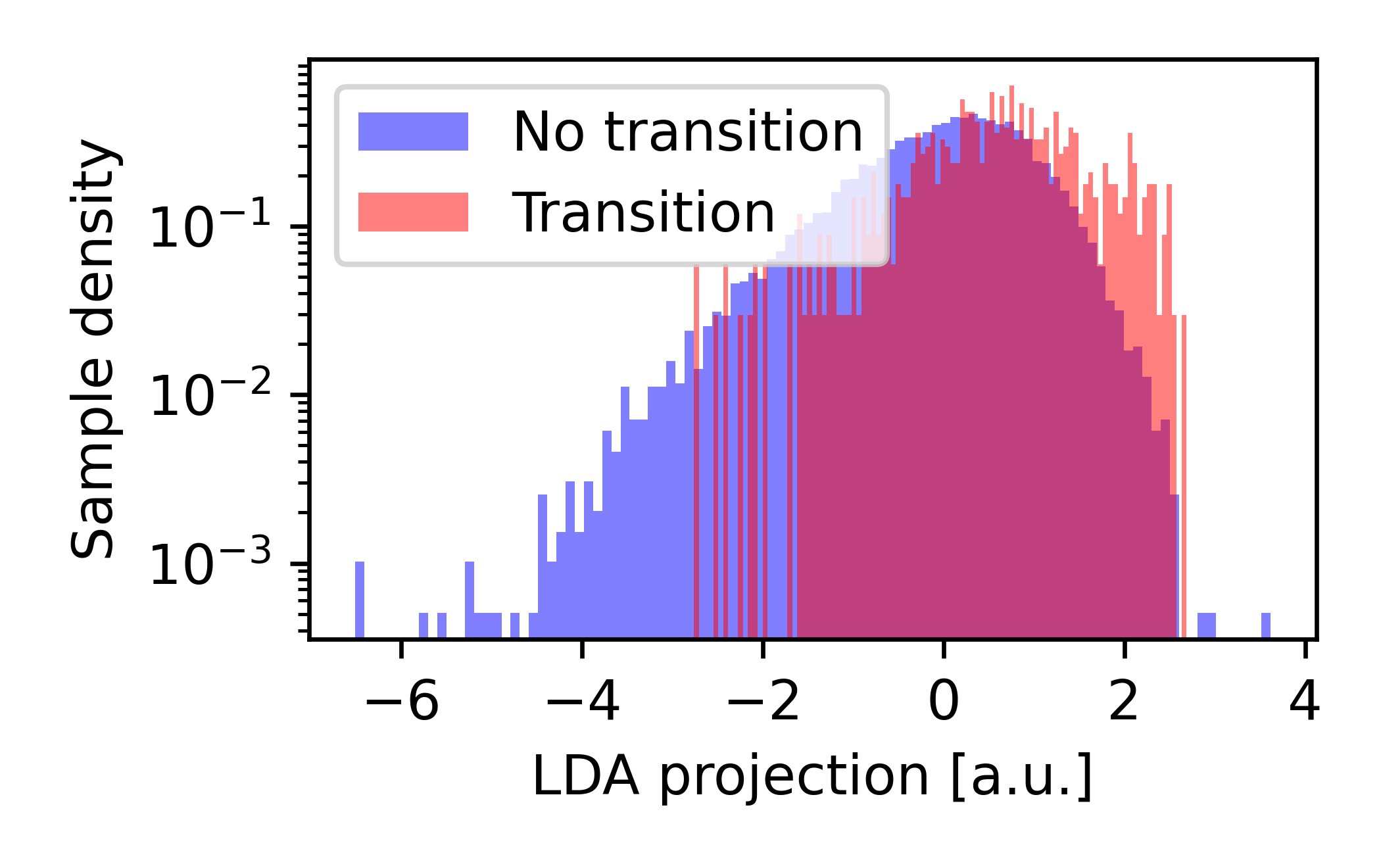}&\iptm{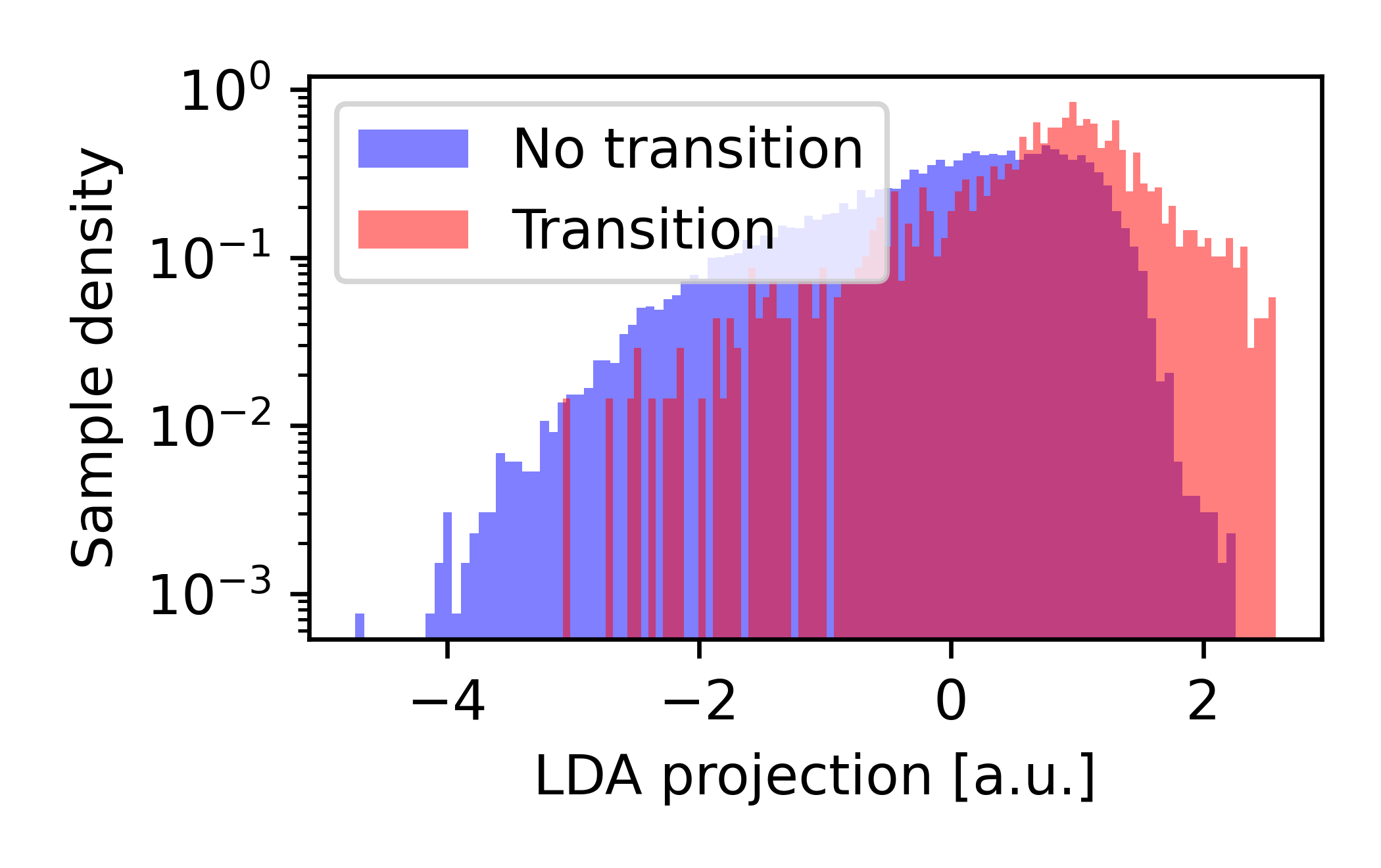}&\iptm{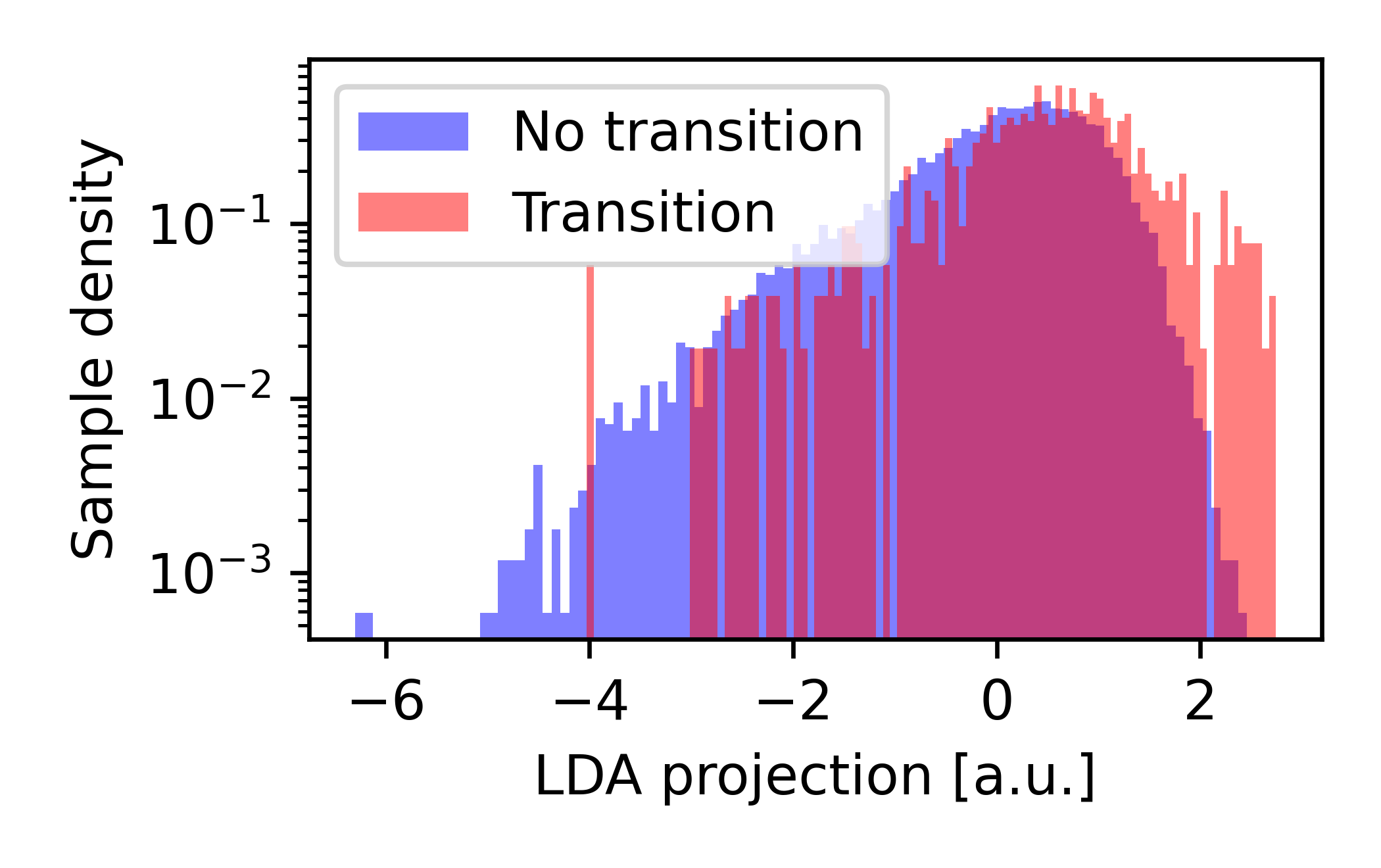}&\iptm{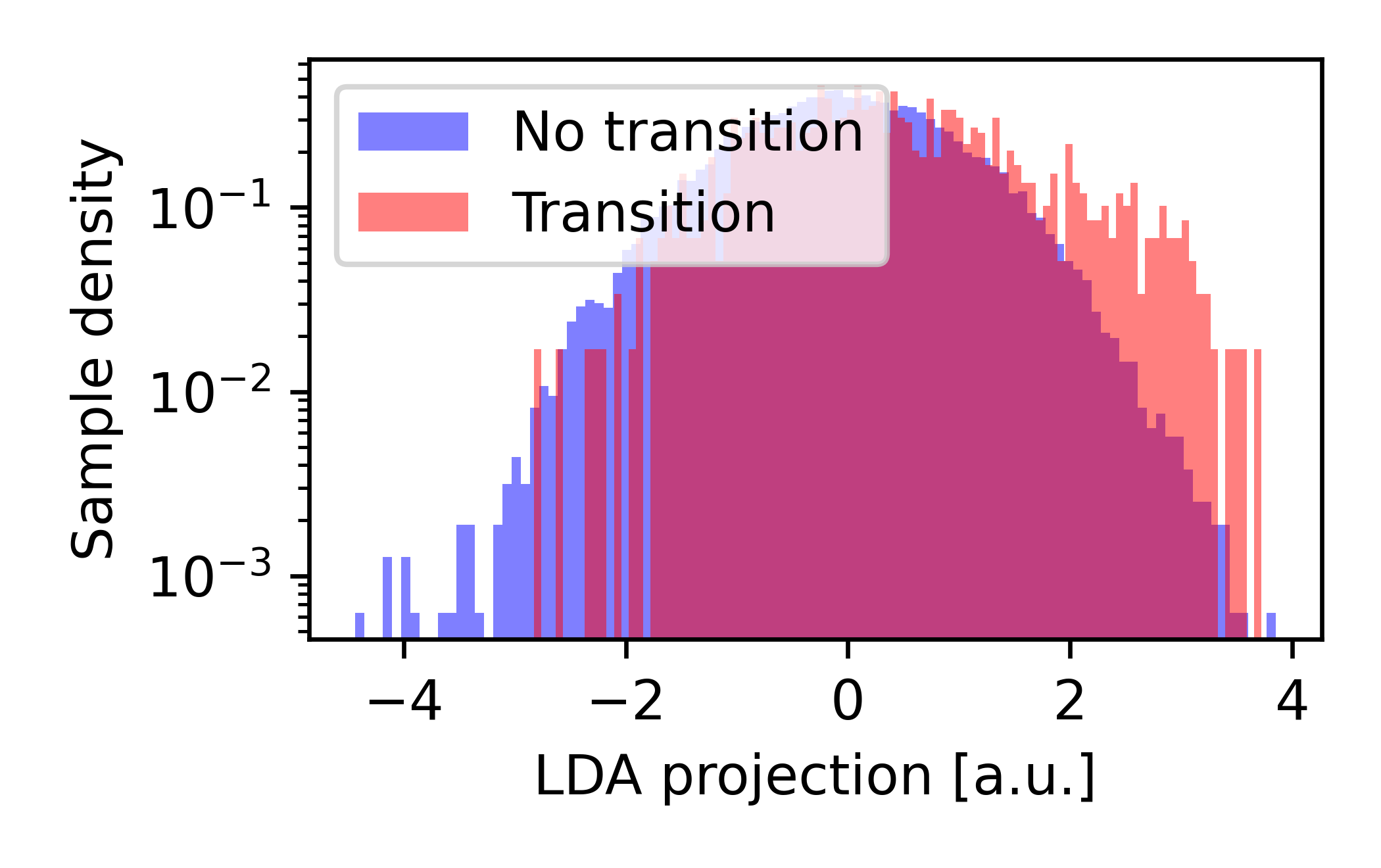}\\
    \begin{tabular}{c}Signature \\ Transition\end{tabular} & \iptm{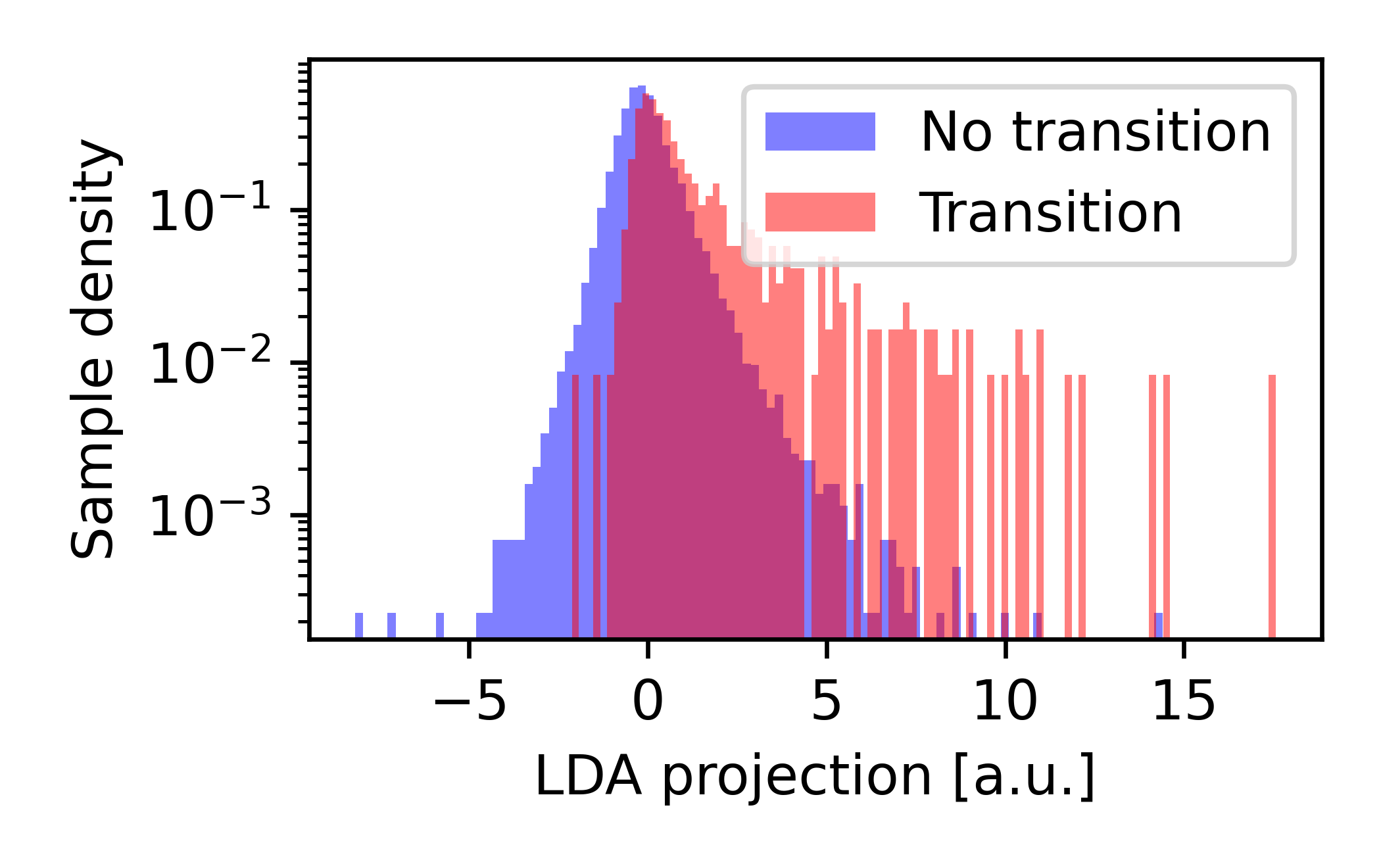}&\iptm{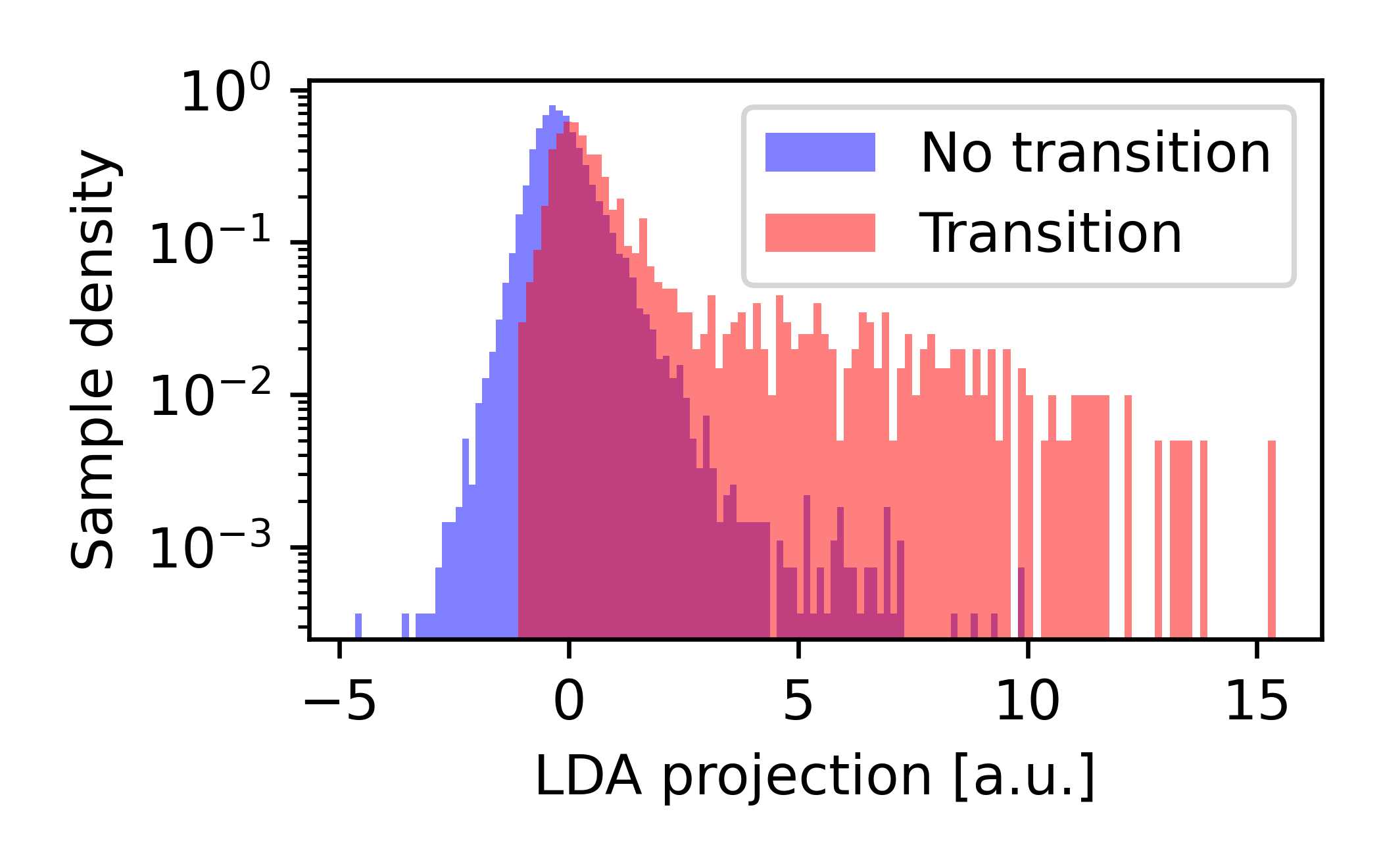}&\iptm{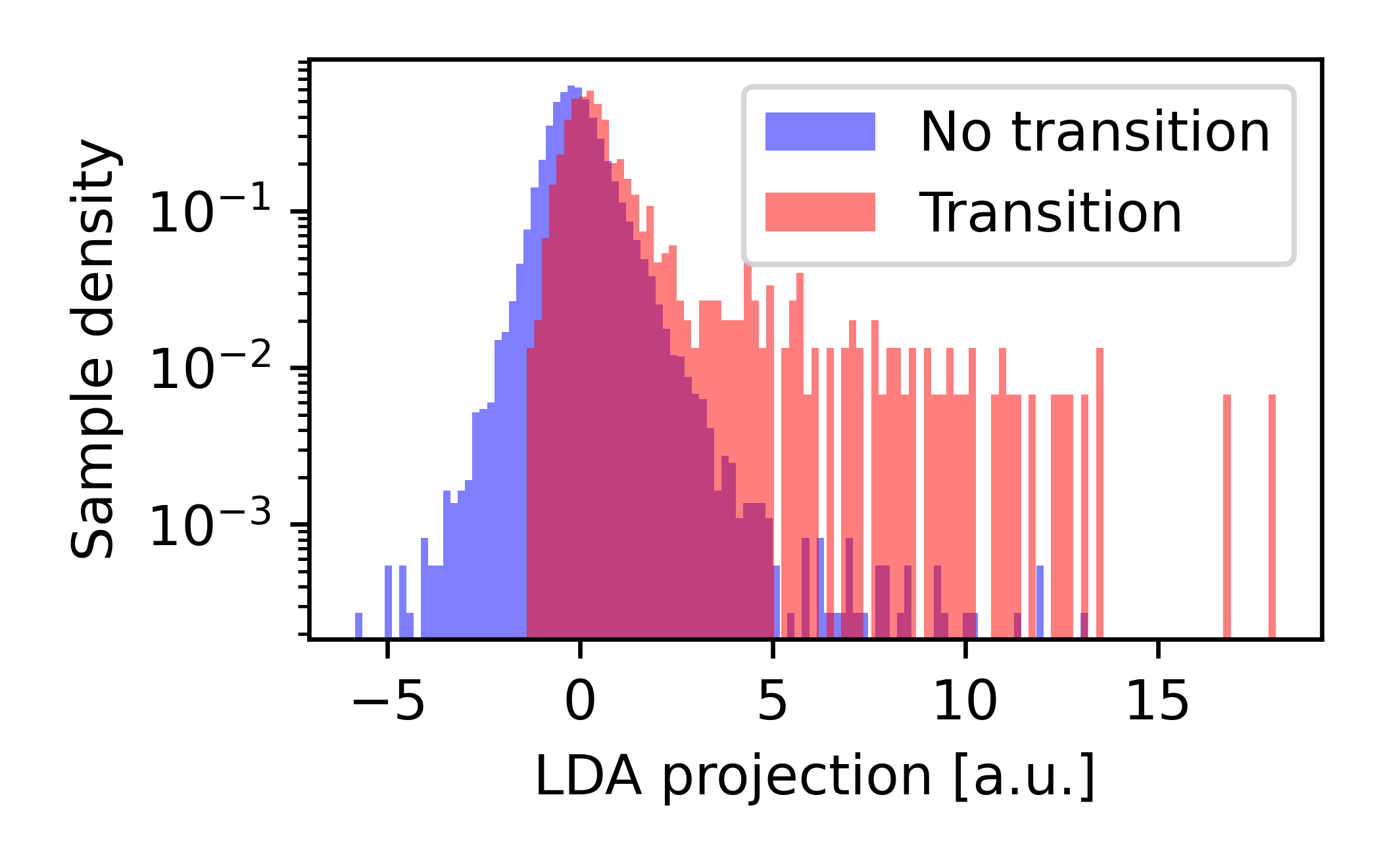}&\iptm{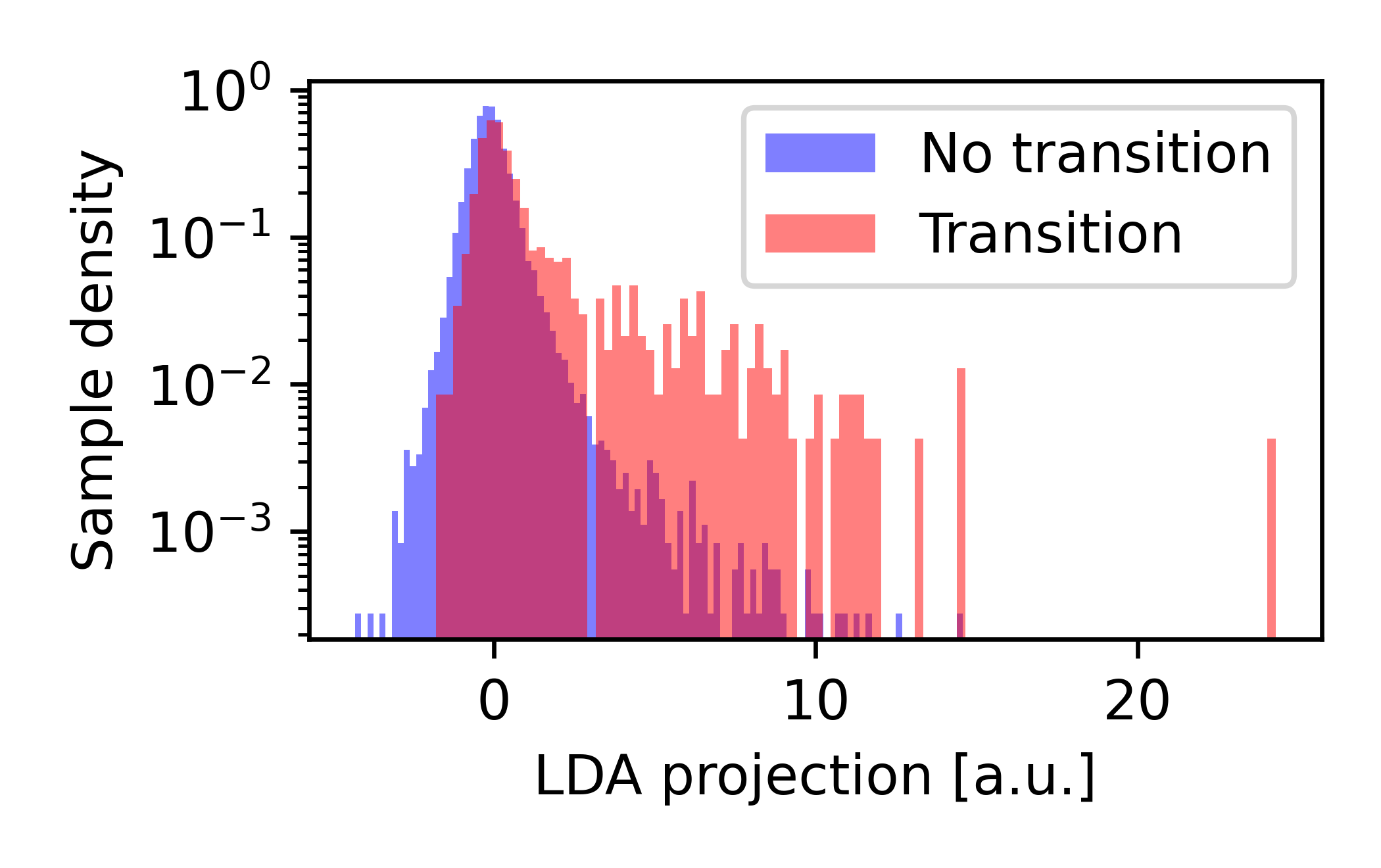}\\
    
         \hline
    \end{tabular}\caption{Statistics of the readout signal, the signal's signature, and the performance of various state discrimination approaches. These approaches and benchmarks are consistent with those reported in Section \ref{app:ml_methods} of the supplemental material.}.
    \label{tab:riken_2}
\end{figure}

There is a significant amount of leakage error in the IQ blob of Figure \ref{tab:riken_2}. We found that this leakage is due to the long integration time of 536 ns, while the optimal integration time for GMM models is 320 ns. For the signature method, the assignment error remains almost constant regardless of the integration time. In contrast, for other methods, the assignment error increases when the integration time exceeds 300 ns, indicating that the signature method demonstrates more robust performance.

\begin{table}[h!]
    \centering
    \begin{tabular}{c|ccc|ccc|ccc|ccc}
        & \multicolumn{3}{c|}{RQC Q1} & \multicolumn{3}{c|}{ RQC Q2} & \multicolumn{3}{c|}{ RQC Q3} & \multicolumn{3}{c}{ RQC Q4} \\ \hline
        & $\ket{0}$ & $\ket{1}$ & Overall & $\ket{0}$ & $\ket{1}$ & Overall & $\ket{0}$ & $\ket{1}$ & Overall & $\ket{0}$ & $\ket{1}$ & Overall \\ \hline
        Baseline & 0.33 & 5.98 & 3.16 & 0.20  & 11.68 & 5.94 & 0.20 & 7.93 & 4.07 & 0.16 & 8.87 & 4.52 \\
        RF & 0.33 & 5.50(14) & 2.92(7) & 0.20 & 10.19(19) & 5.20(10) & 0.20 & 7.43(17) & 3.82(9) & 0.16 & 8.37(15) & 4.27(8) \\
        Sig+RF & 0.33 & 5.06(15) & 2.70(8) & 0.20 & 9.08(24) & 4.64(12) & 0.20 & 6.80(17) & 3.50(9) & 0.16 & 7.16(13) & 3.66(7) \\ 
    \end{tabular}

    \vspace{1em}

    \begin{tabular}{c|ccc|ccc|ccc|ccc}
        & \multicolumn{3}{c|}{RQC Q5} & \multicolumn{3}{c|}{RQC Q6} & \multicolumn{3}{c|}{RQC Q7} & \multicolumn{3}{c}{RQC Q8} \\ \hline
        & $\ket{0}$ & $\ket{1}$ & Overall & $\ket{0}$ & $\ket{1}$ & Overall & $\ket{0}$ & $\ket{1}$ & Overall & $\ket{0}$ & $\ket{1}$ & Overall \\ \hline
        Baseline & 0.69 & 4.11    & 2.40    & 0.79 & 3.86    & 2.33      & 0.82 & 3.75     & 2.29    & 0.59& 4.01     & 2.30 \\
        RF       & 0.69    & 3.21(8) & 1.95(4) & 0.79    & 3.04(13) & 1.92(7) & 0.82    & 3.06(16) & 1.94(8) & 0.59   & 3.22(16) & 1.91(8) \\
        Sig+RF   & 0.69    & 2.47(8) & 1.58(4) & 0.79    & 2.28(13) & 1.54(7)  & 0.82    & 2.35(15) & 1.59(8) & 0.59   & 2.30(15) & 1.35(8) \\ 
    \end{tabular}
    \caption{Table \ps{of the End-Of-Measurement infidelity \((1-F_{\mathrm{EOM}})\times 10^2\) for the RQC qubits. The infidelity for each qubit state preparation is shown along with the overall infidelity.}}
\end{table}

\begin{figure}[h]
    \centering
    \begin{tabular}{c|c|c|c|c}

Dataset & RQC Q5 & RQC Q6 & RQC Q57 & RQC Q8 \\ \hline
&&& \\ 
    \begin{tabular}{c}Scatter\end{tabular} & \iptsq{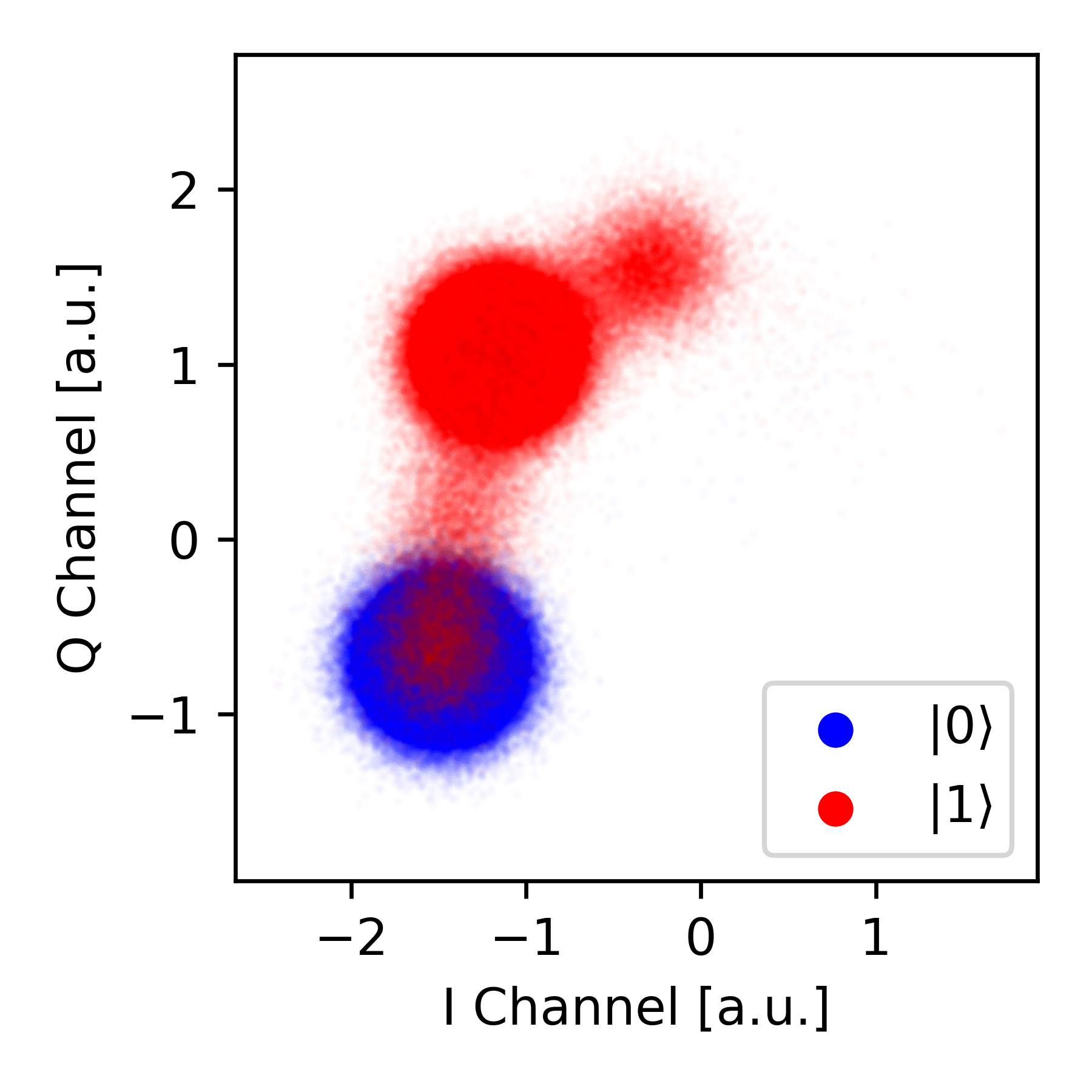}&\iptsq{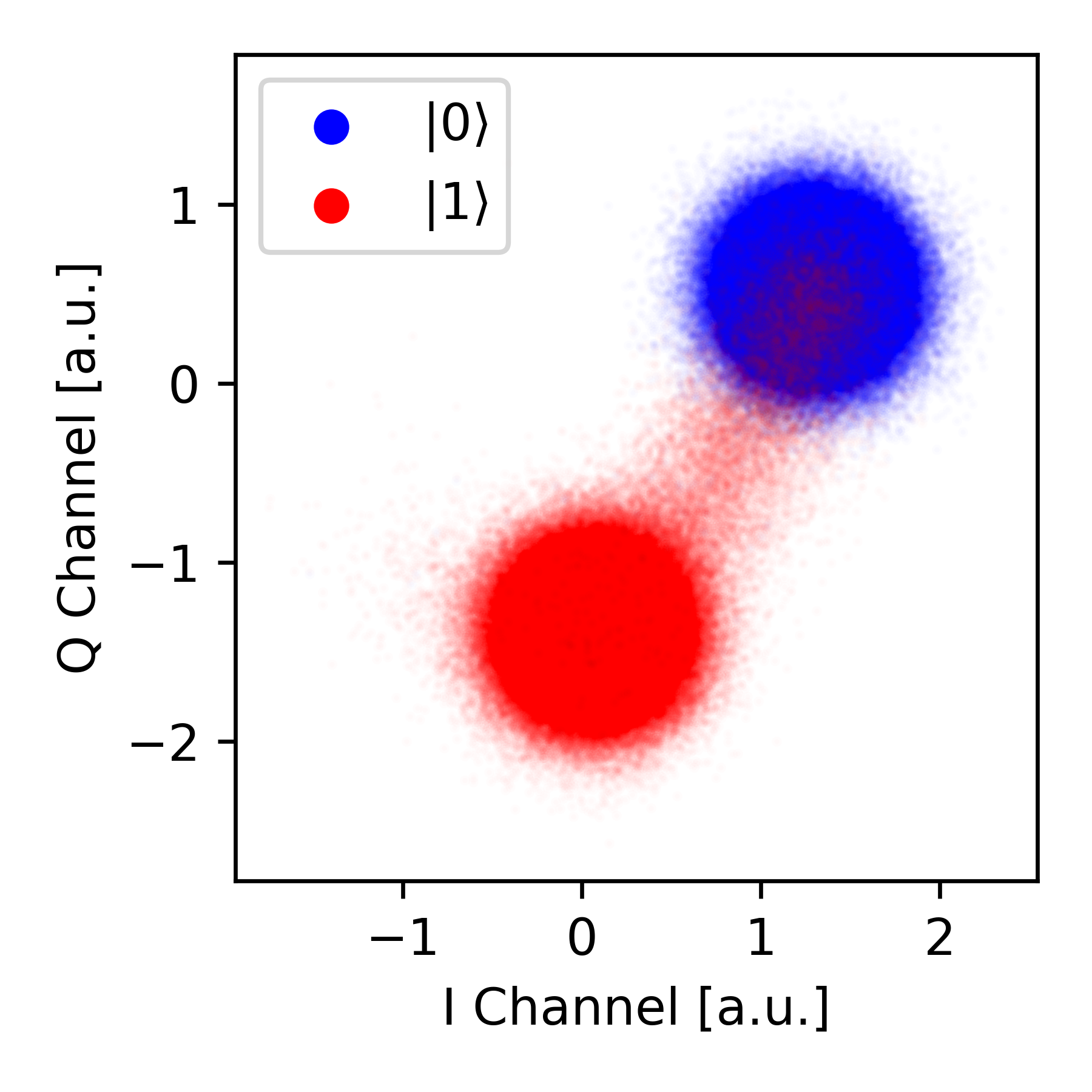}&\iptsq{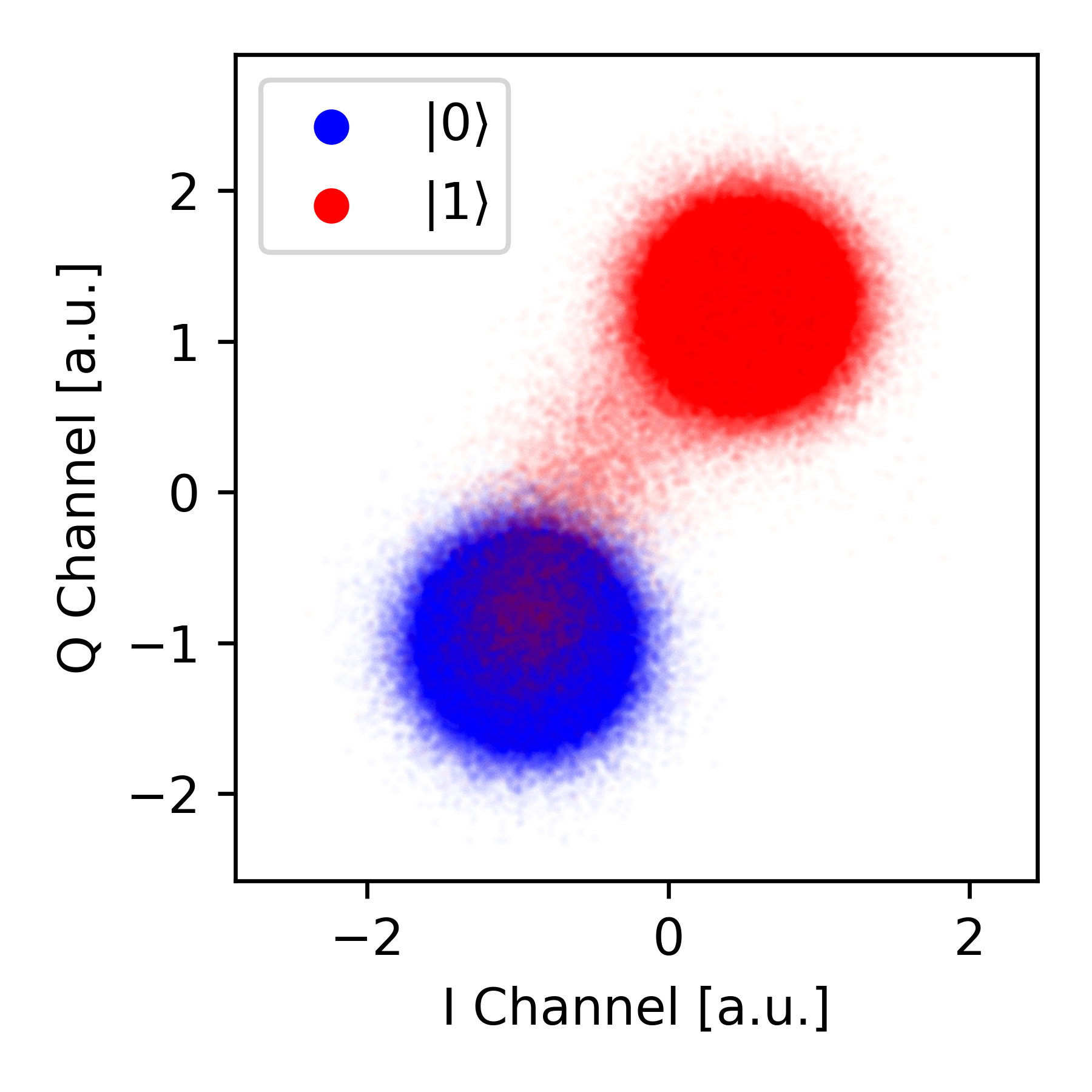}&\iptsq{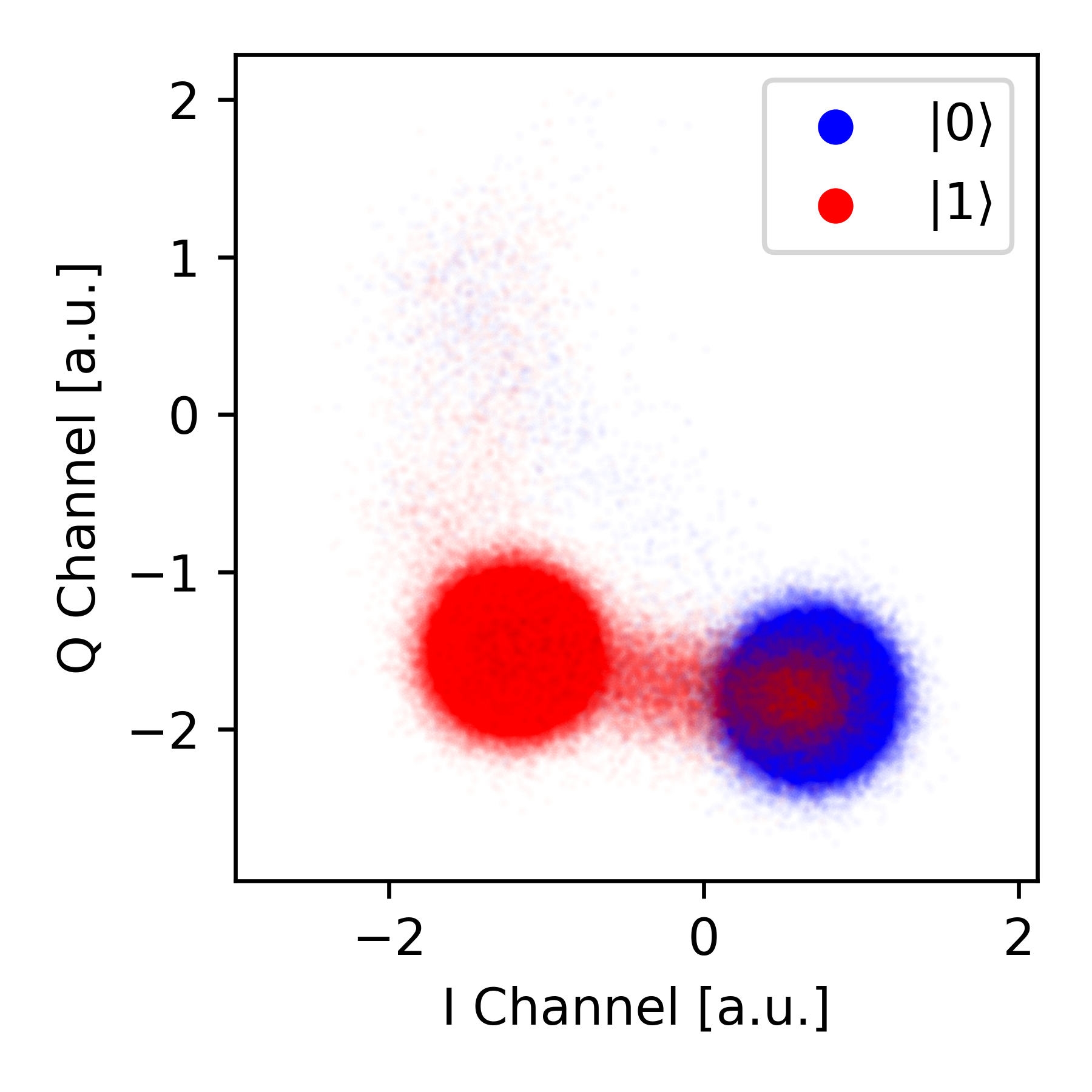}\\
    \begin{tabular}{c}Traces\\ State $\ket{0}$\end{tabular} & \iptm{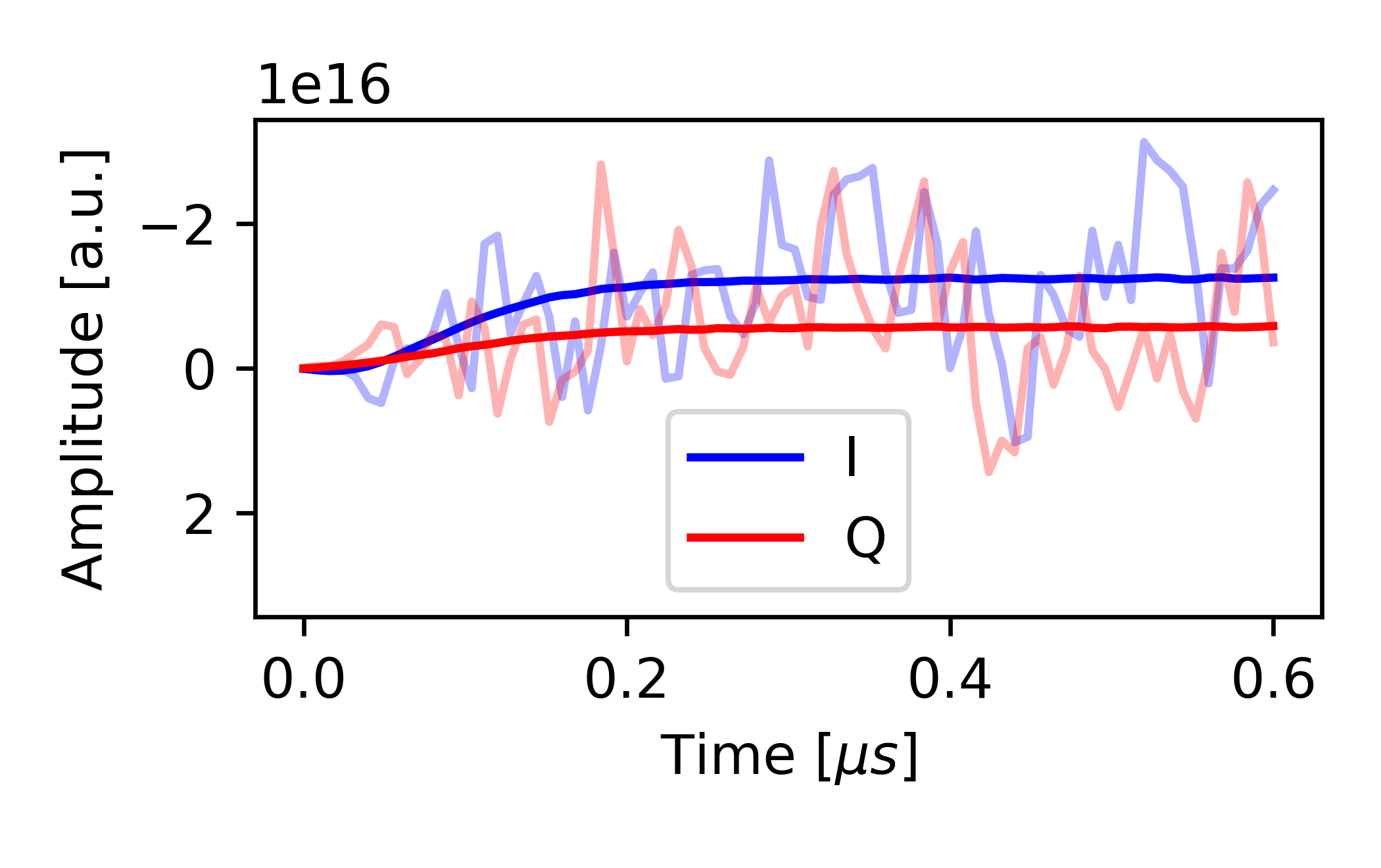}&\iptm{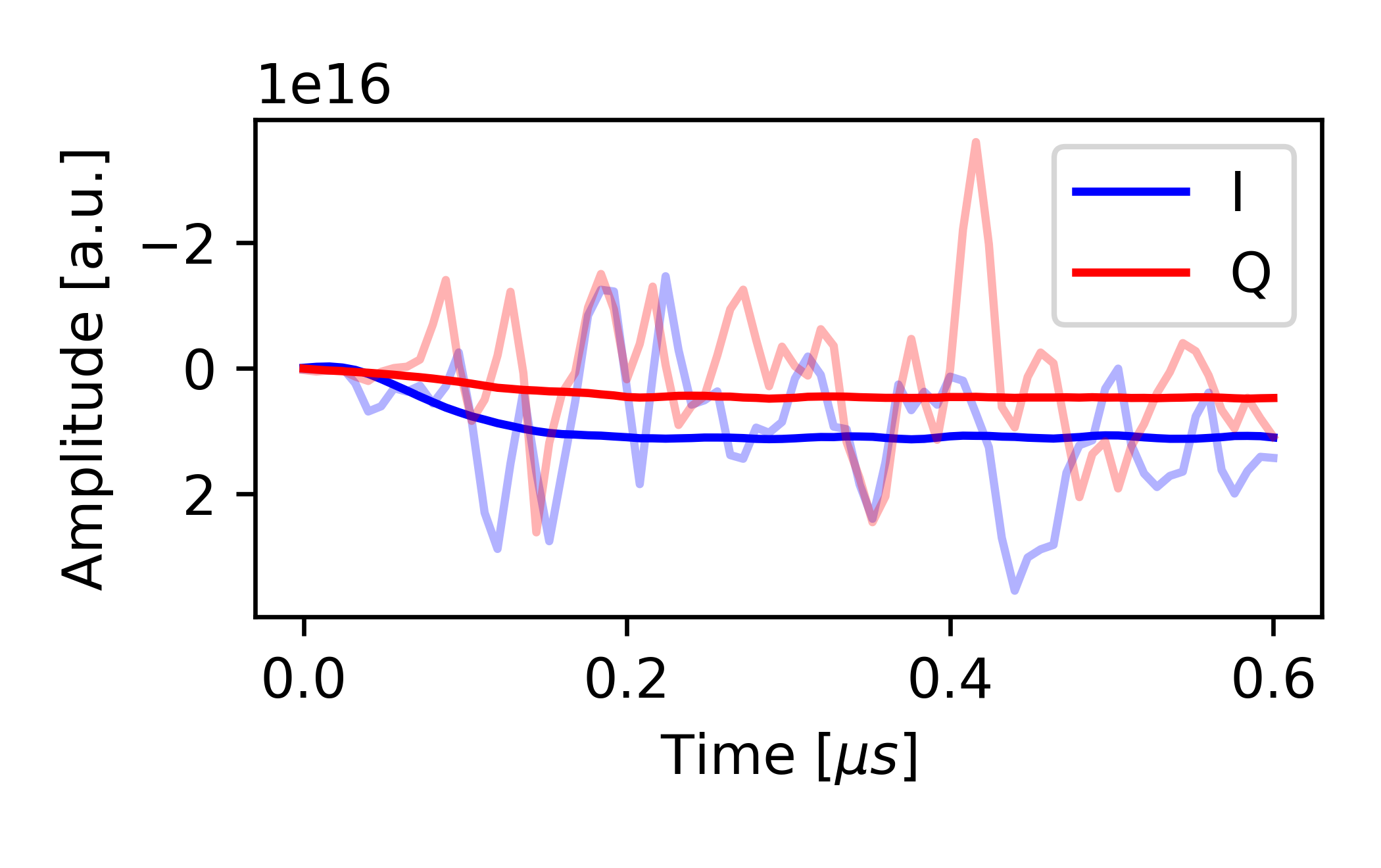}&\iptm{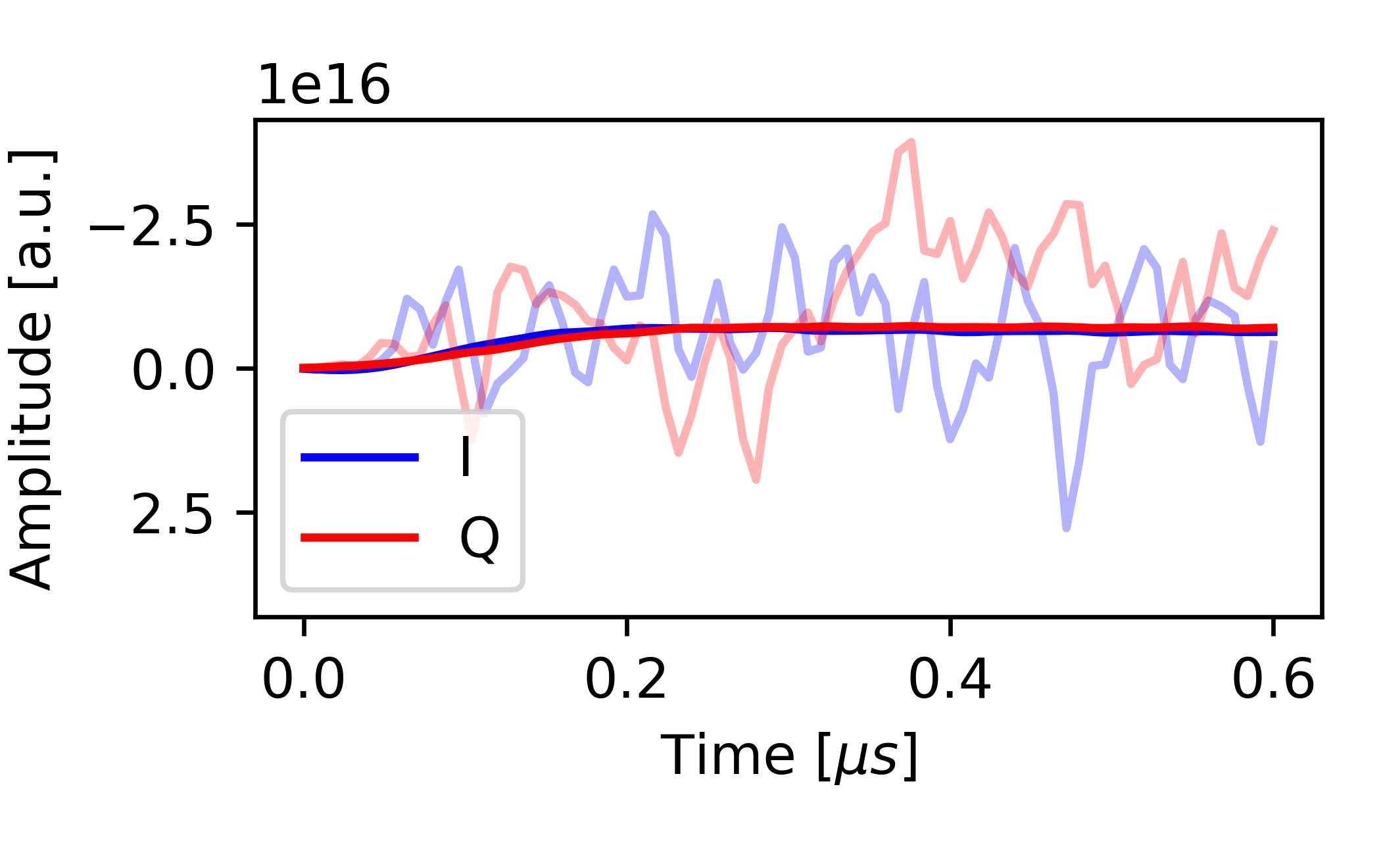}&\iptm{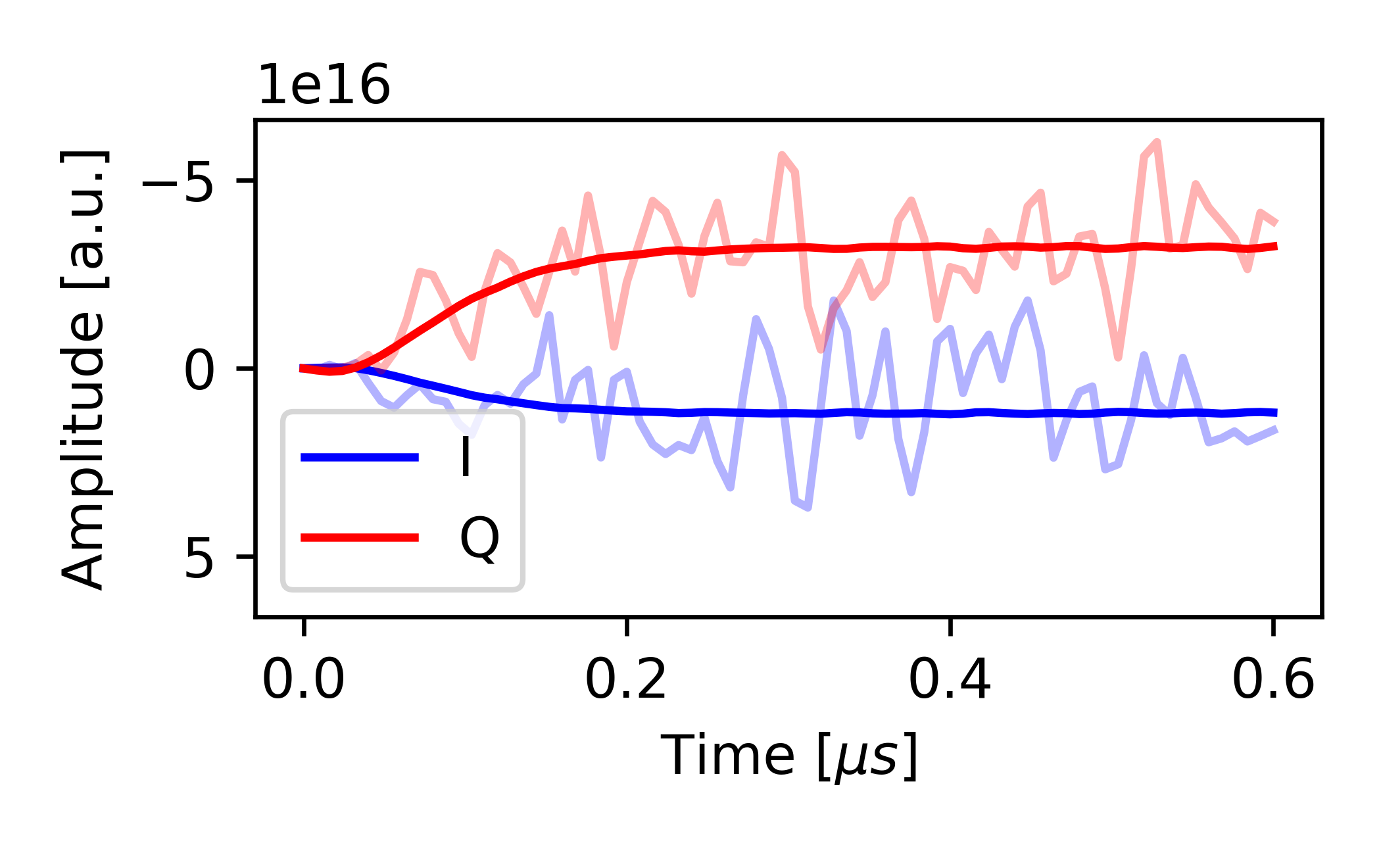}\\
    \begin{tabular}{c}Traces\\ State $\ket{1}$\end{tabular} & \iptm{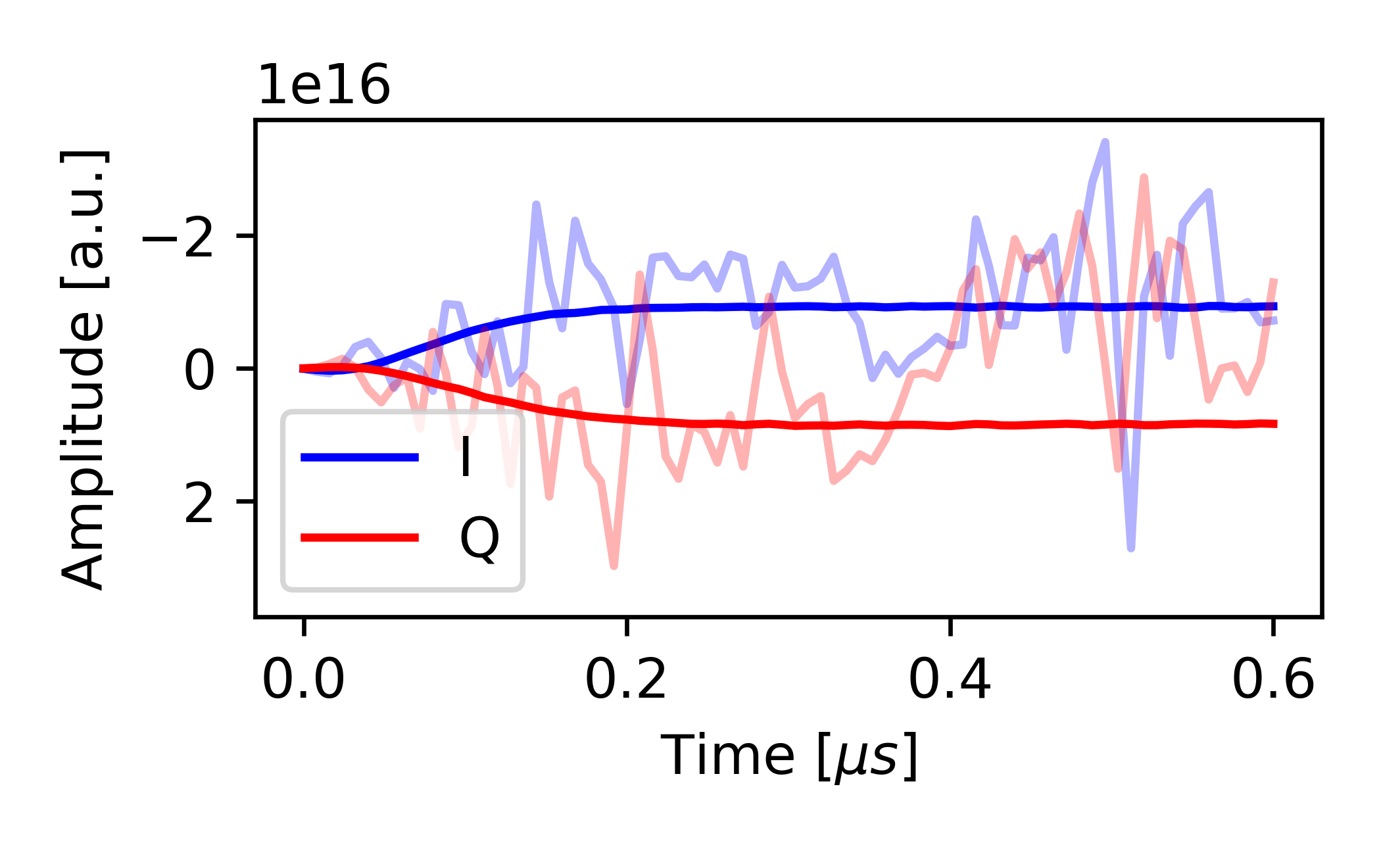}&\iptm{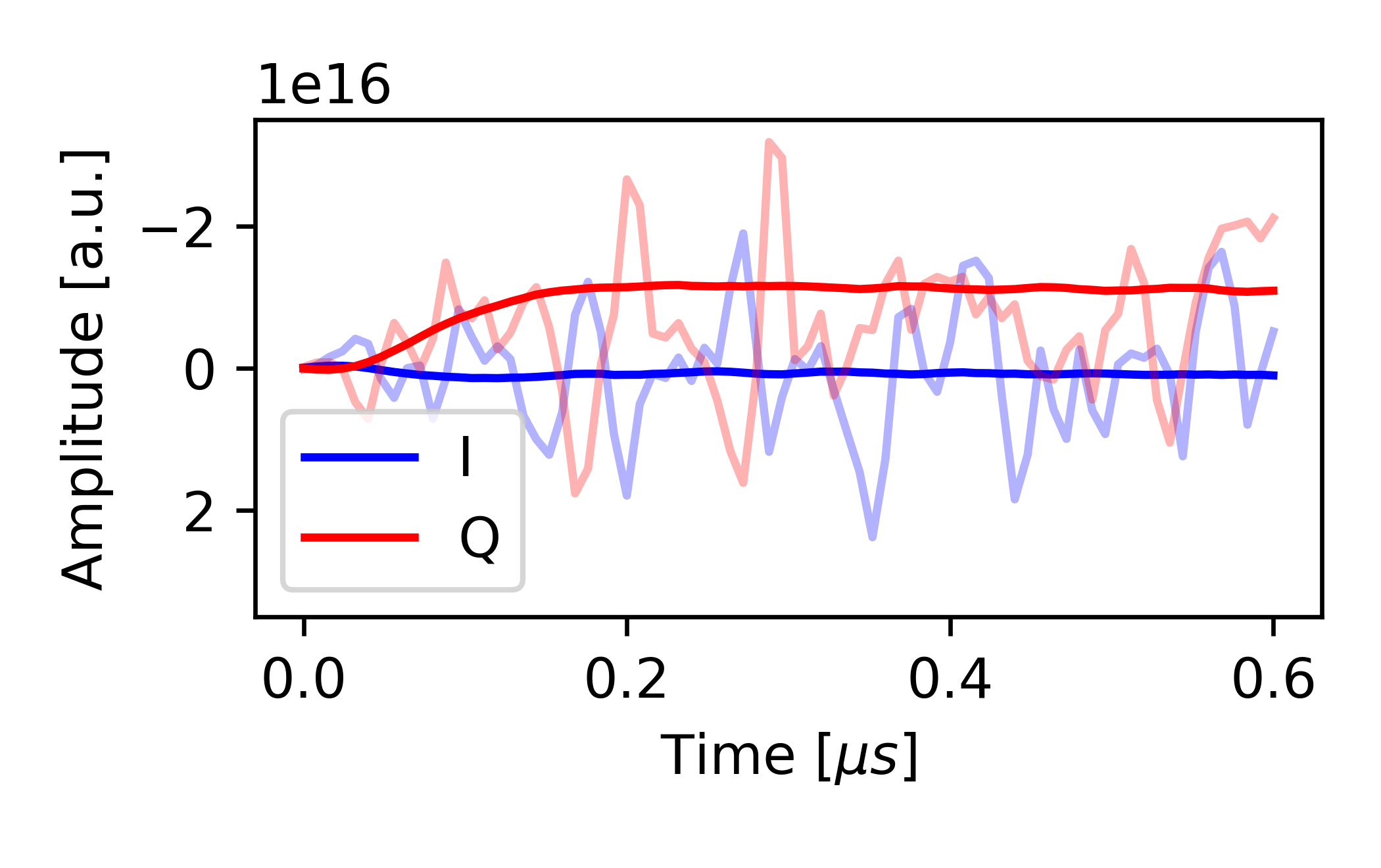}&\iptm{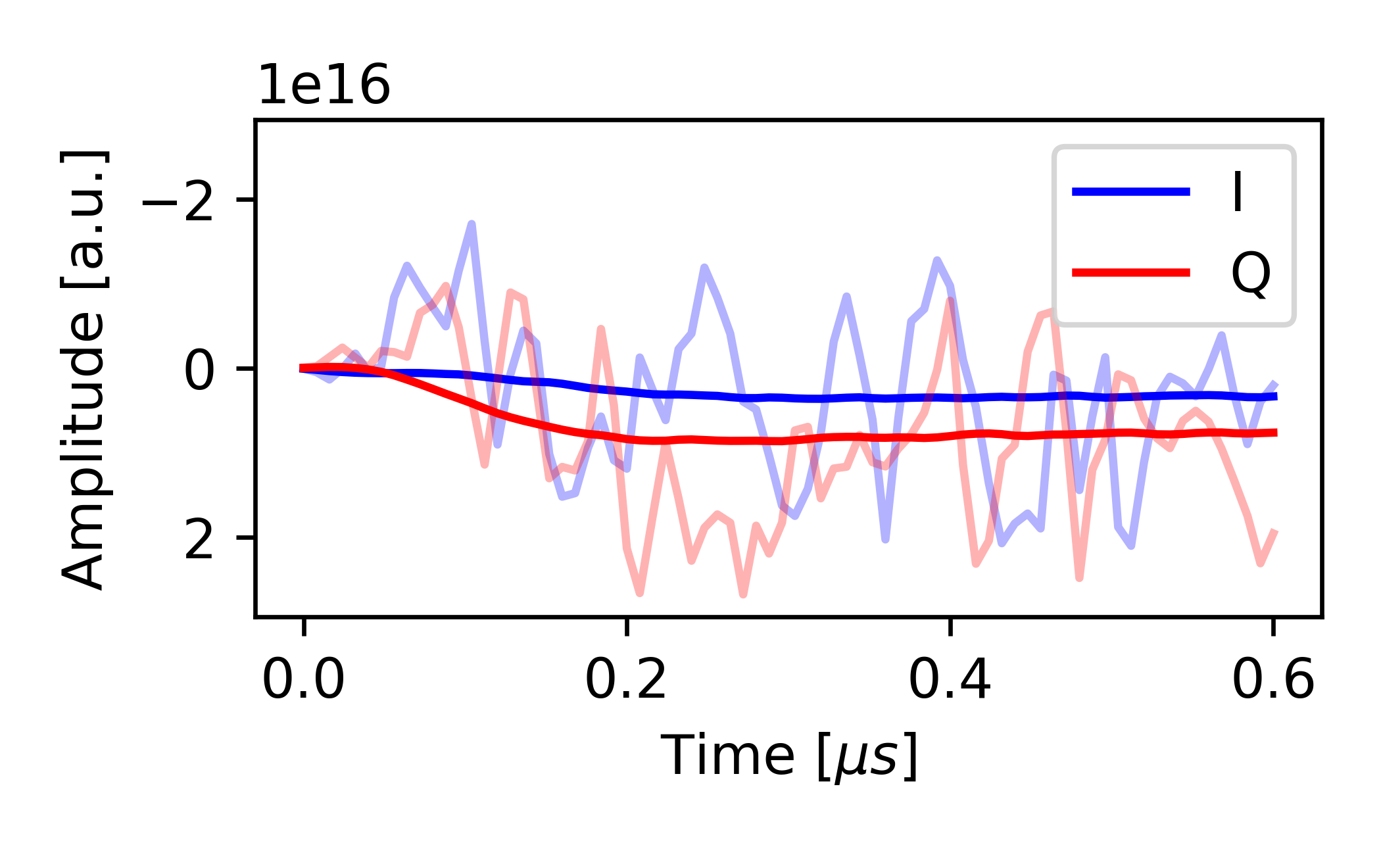}&\iptm{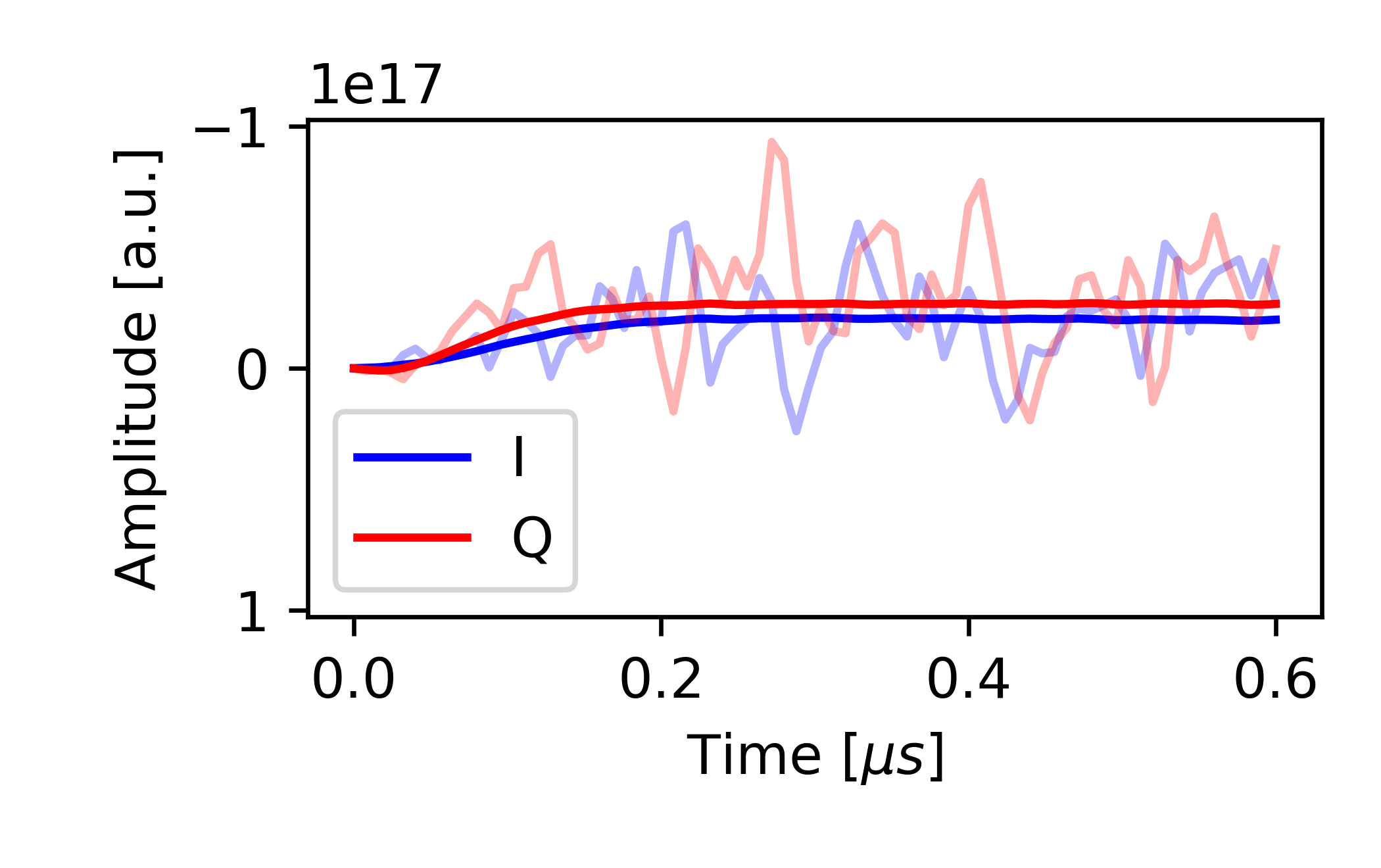}\\
    \begin{tabular}{c}Trajectory\end{tabular} & \iptsq{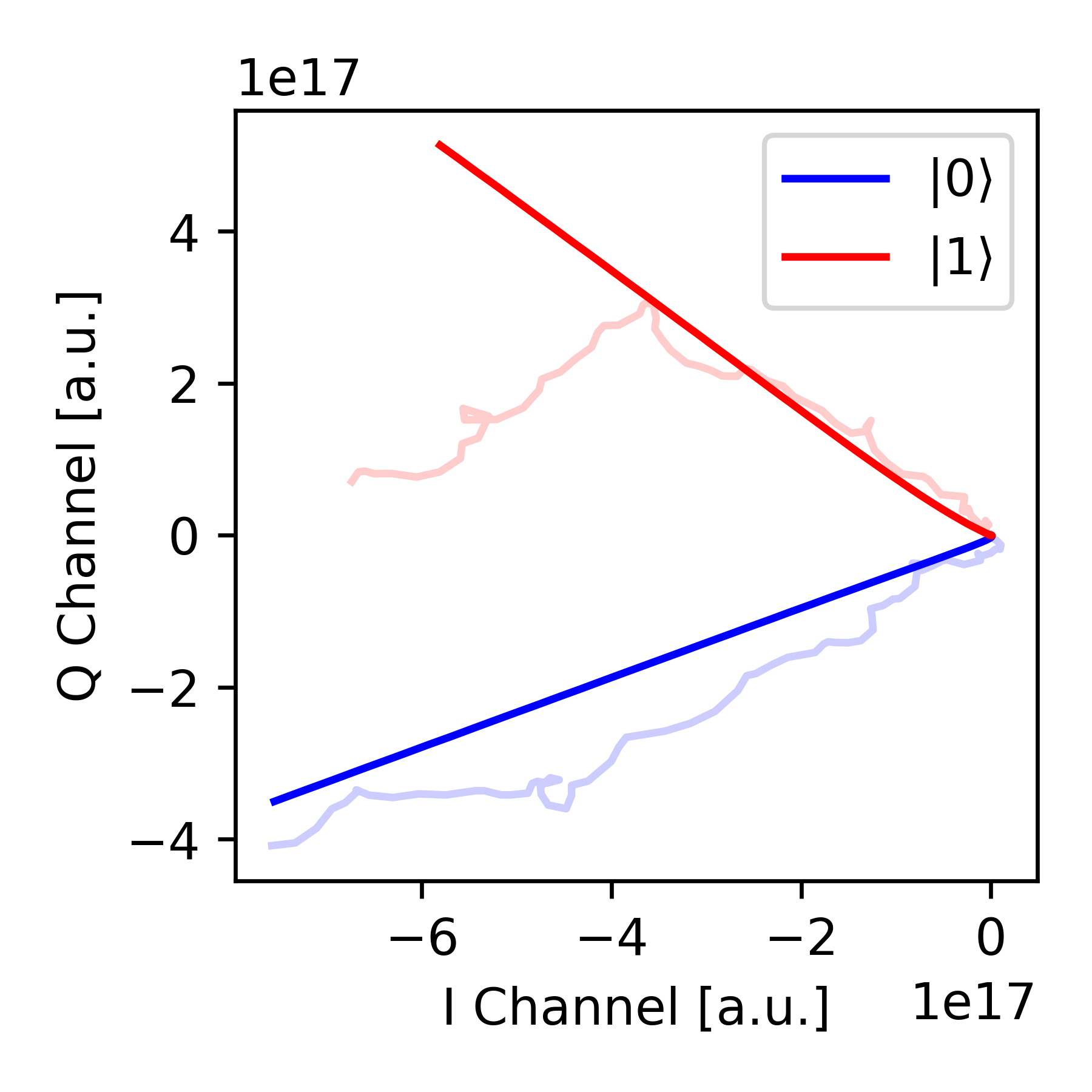}&\iptsq{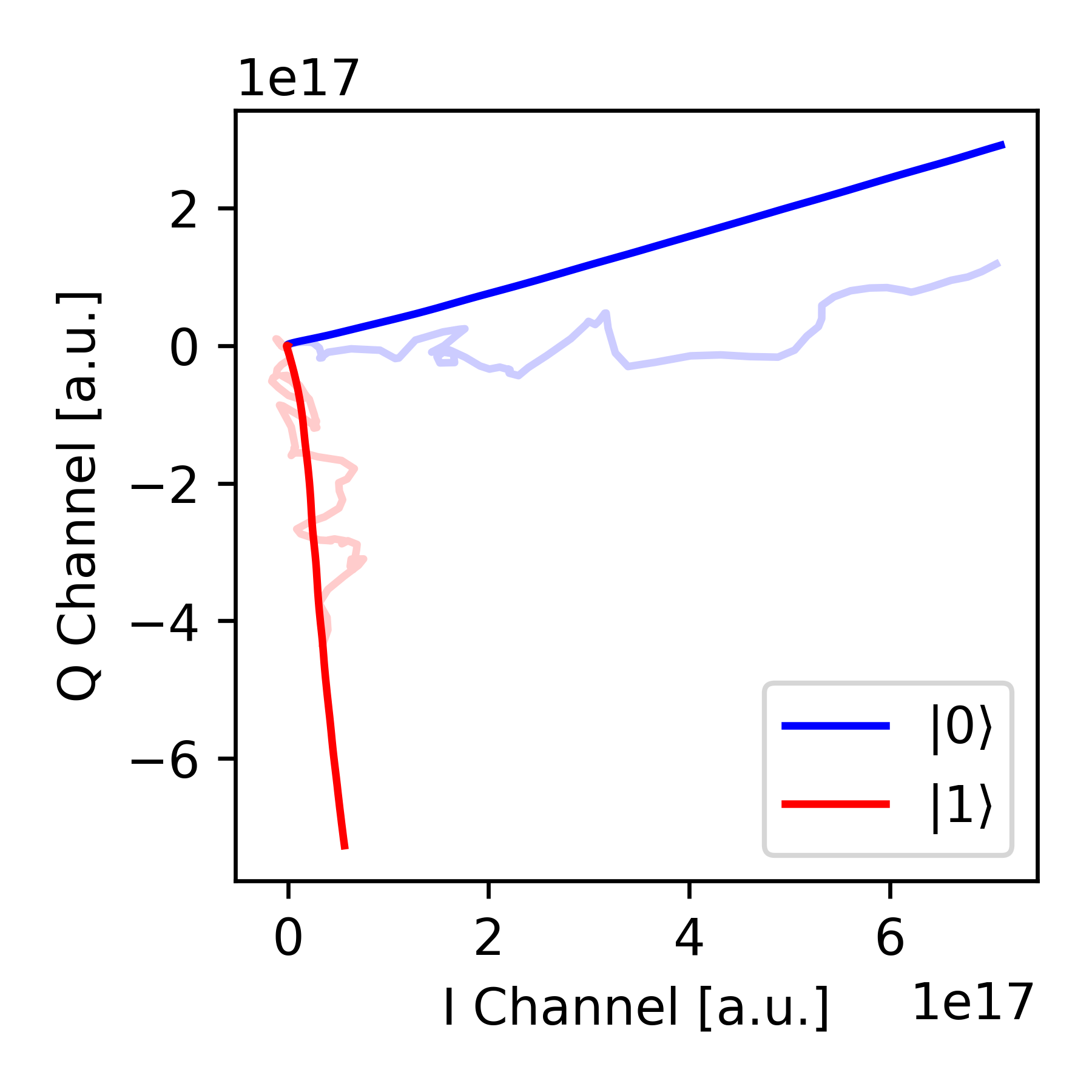}&\iptsq{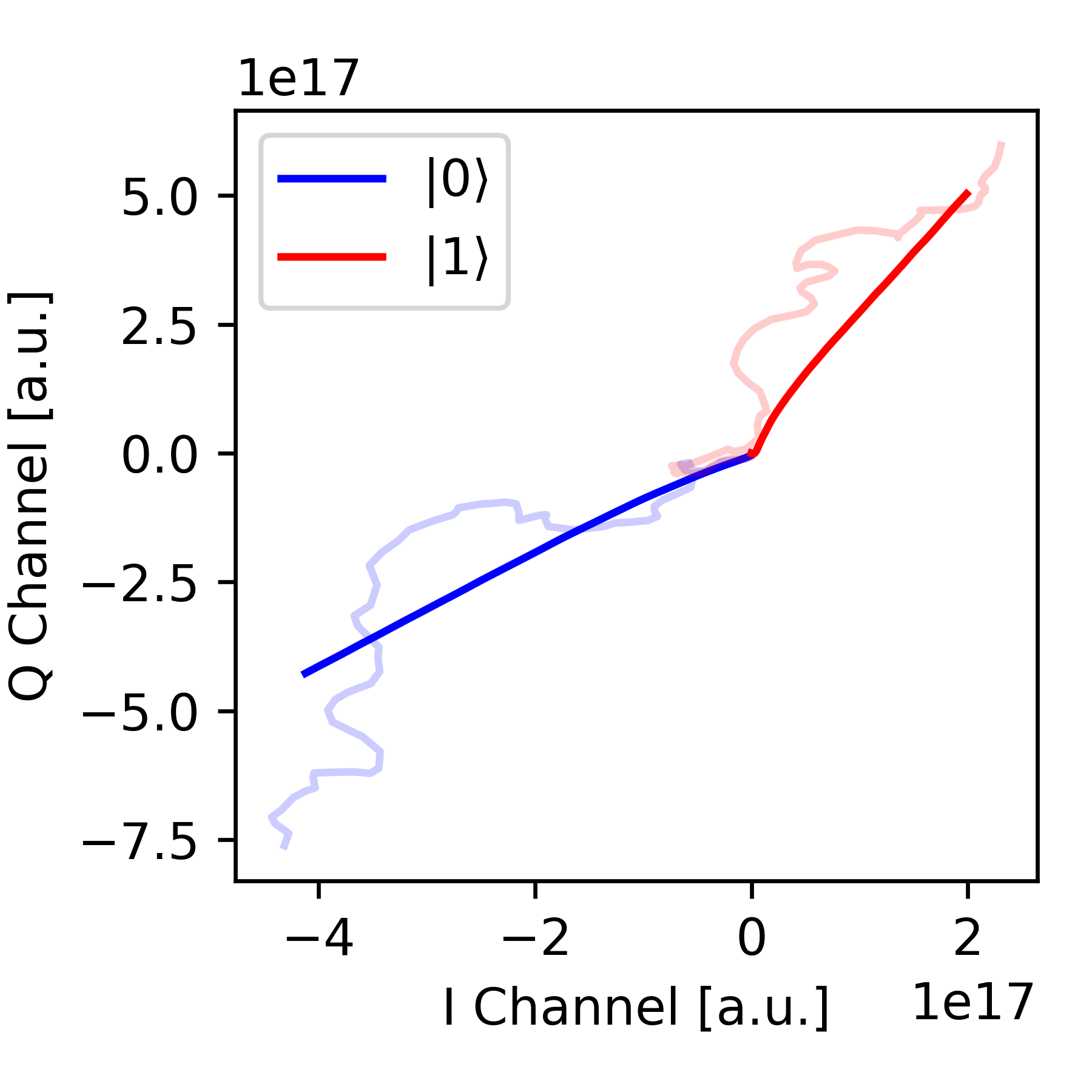}&\iptsq{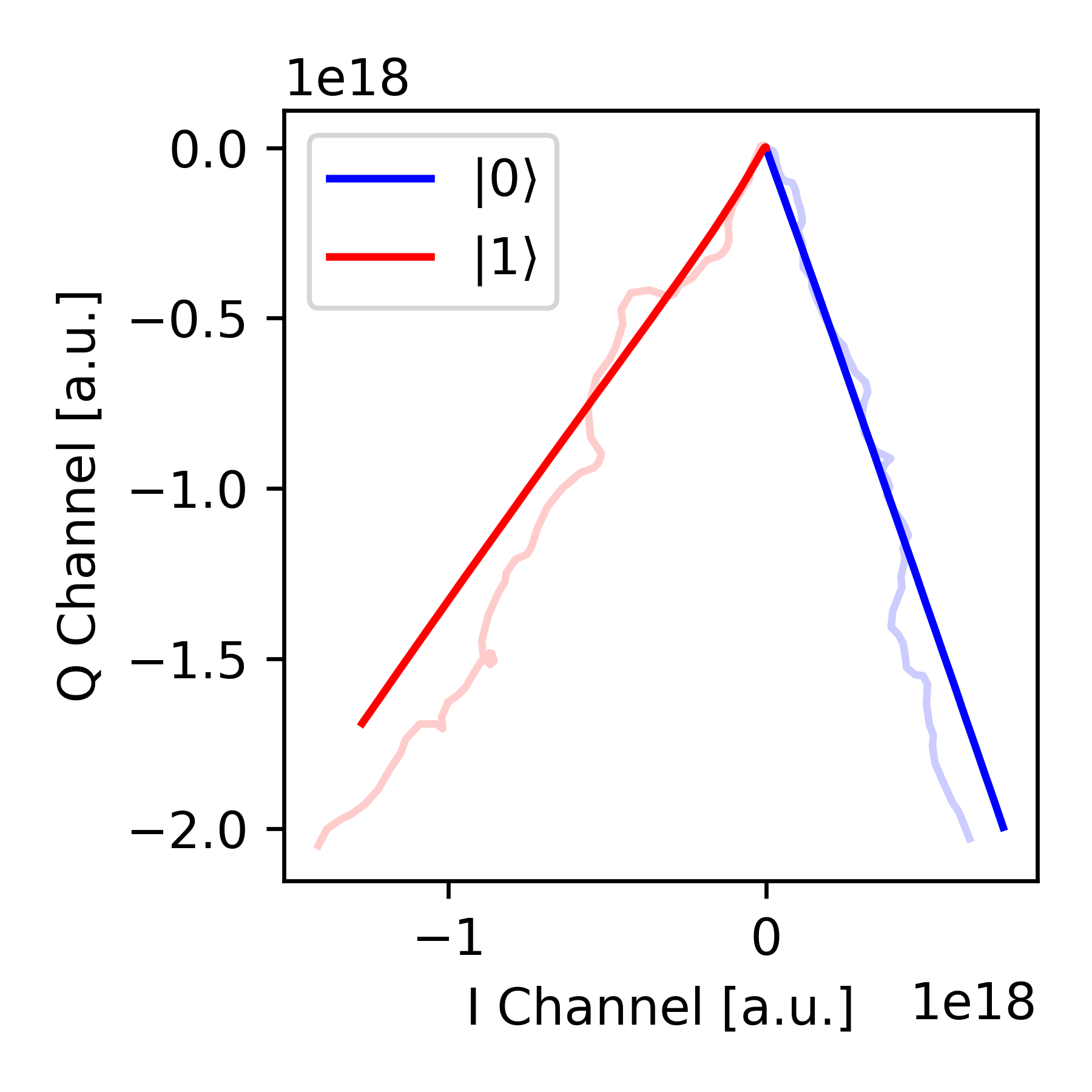}\\
    \begin{tabular}{c}Integration \\ Projection\end{tabular} & \iptm{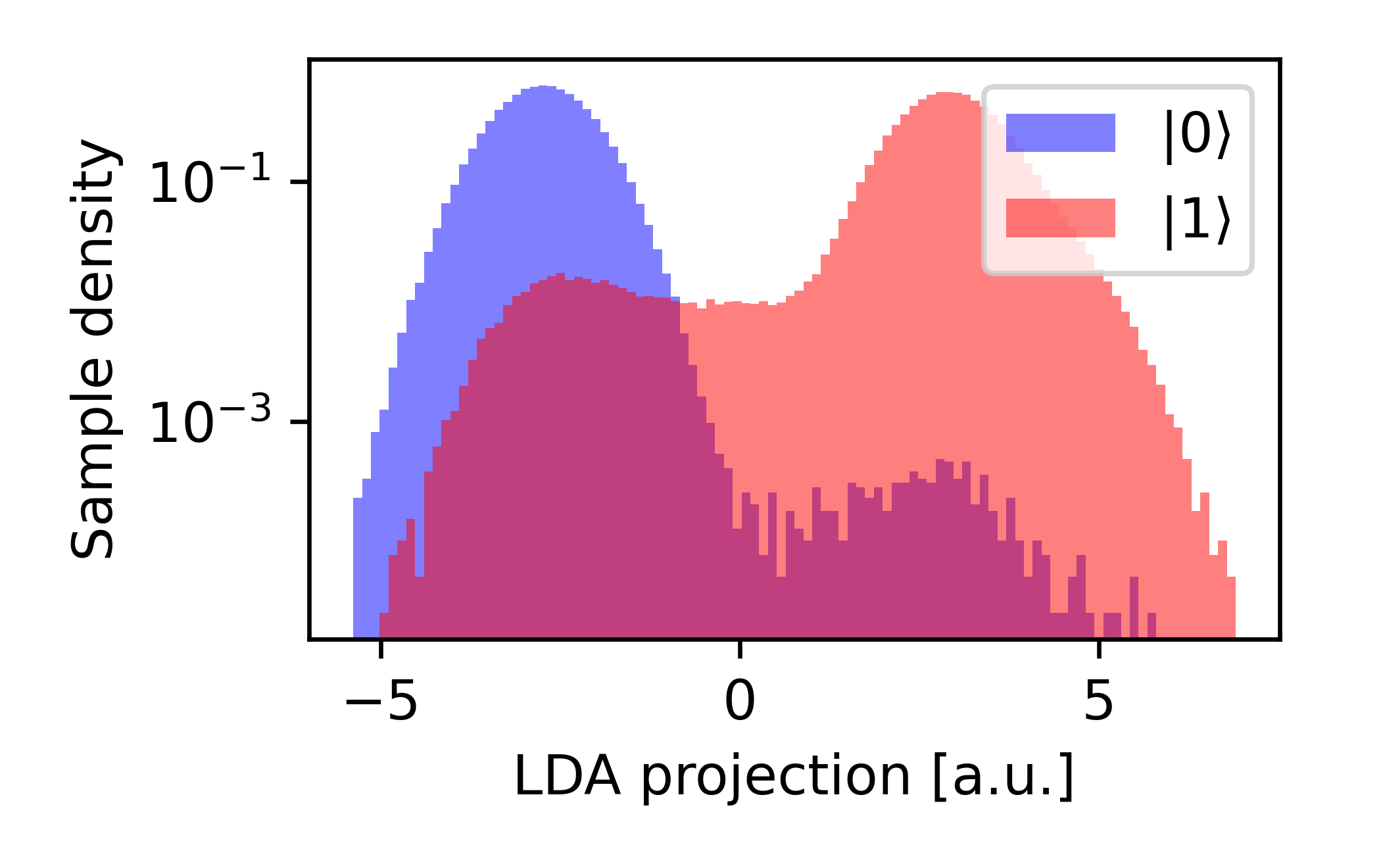}&\iptm{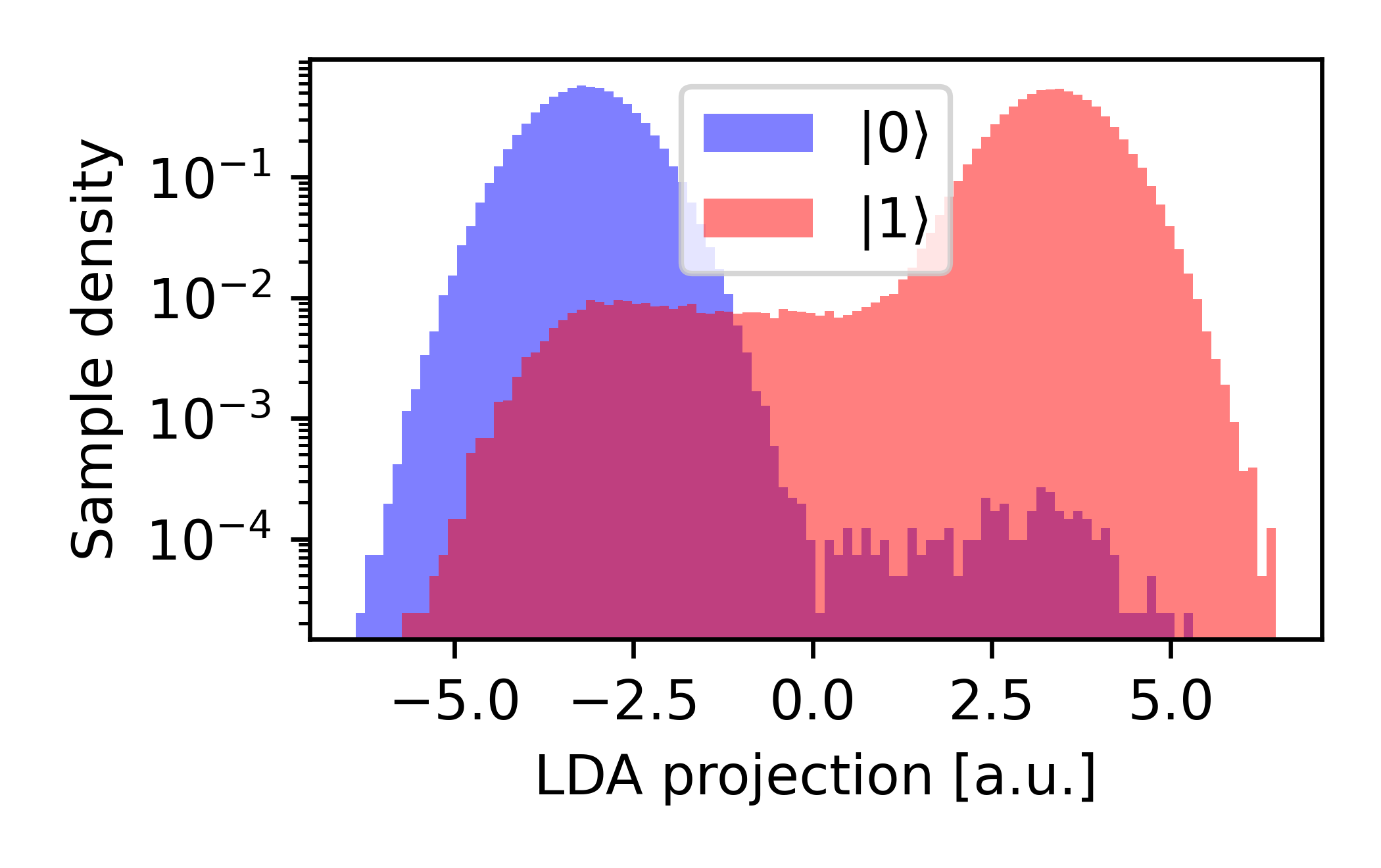}&\iptm{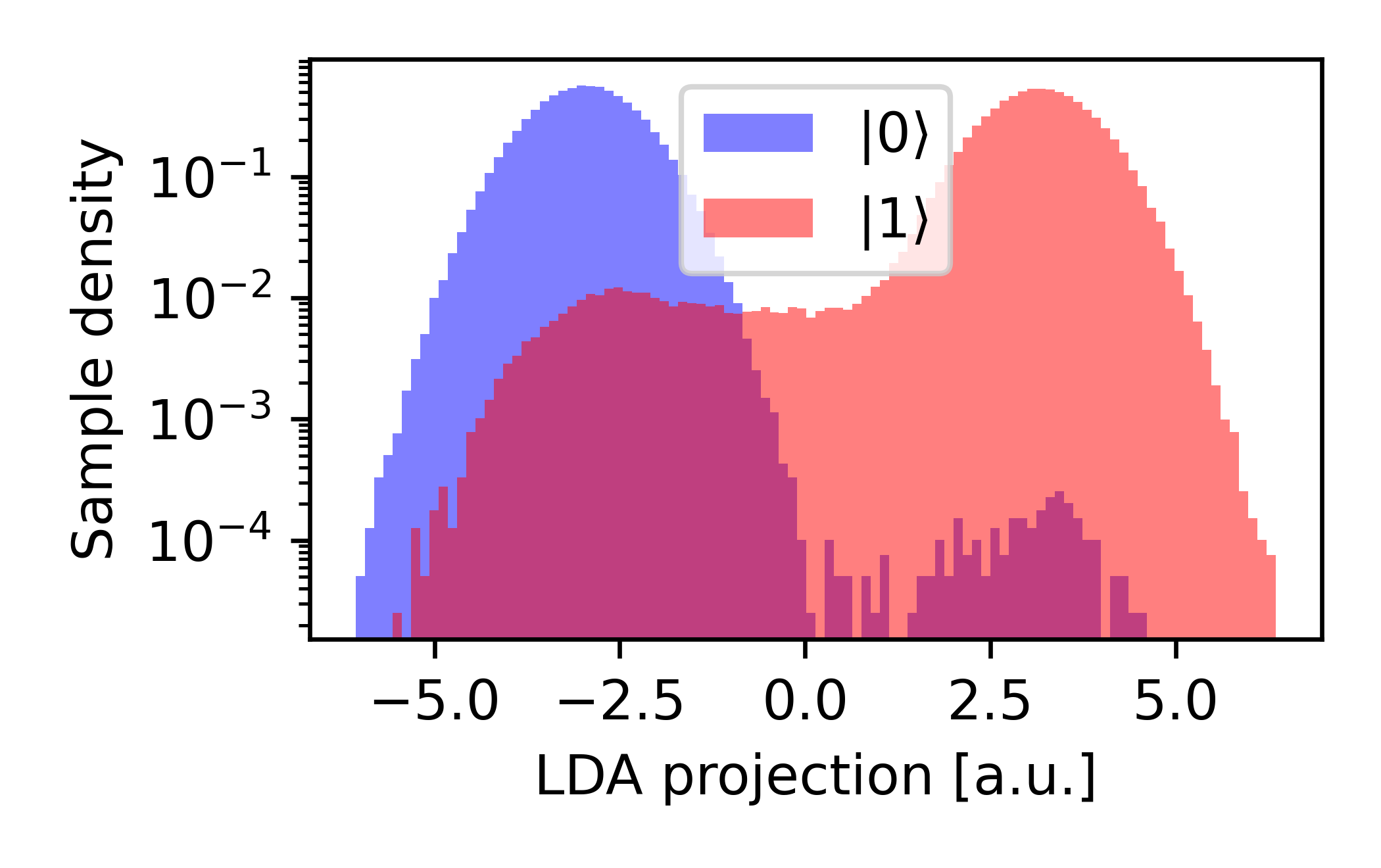}&\iptm{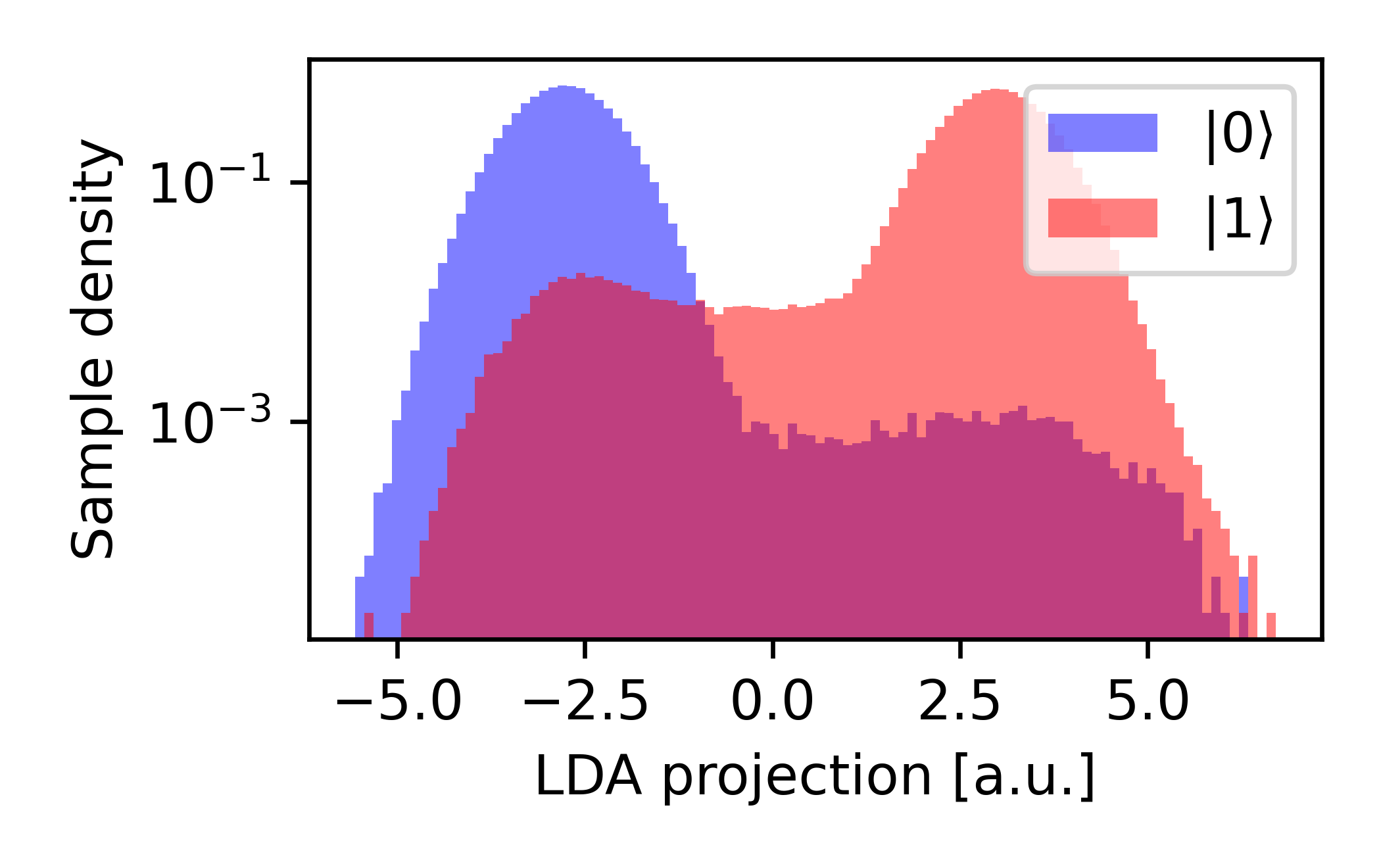}\\
    \begin{tabular}{c}Signature \\ Projection\end{tabular} & \iptm{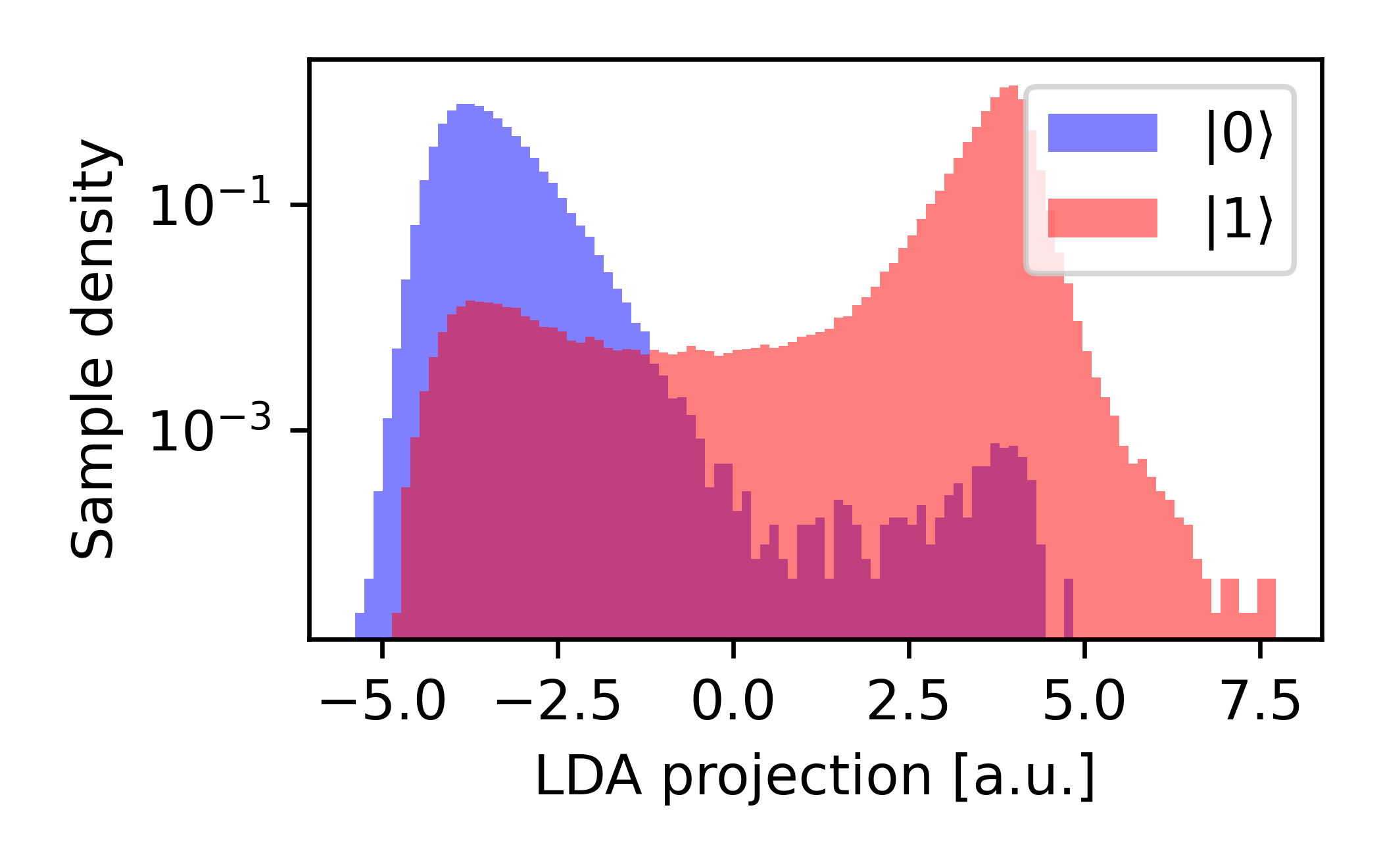}&\iptm{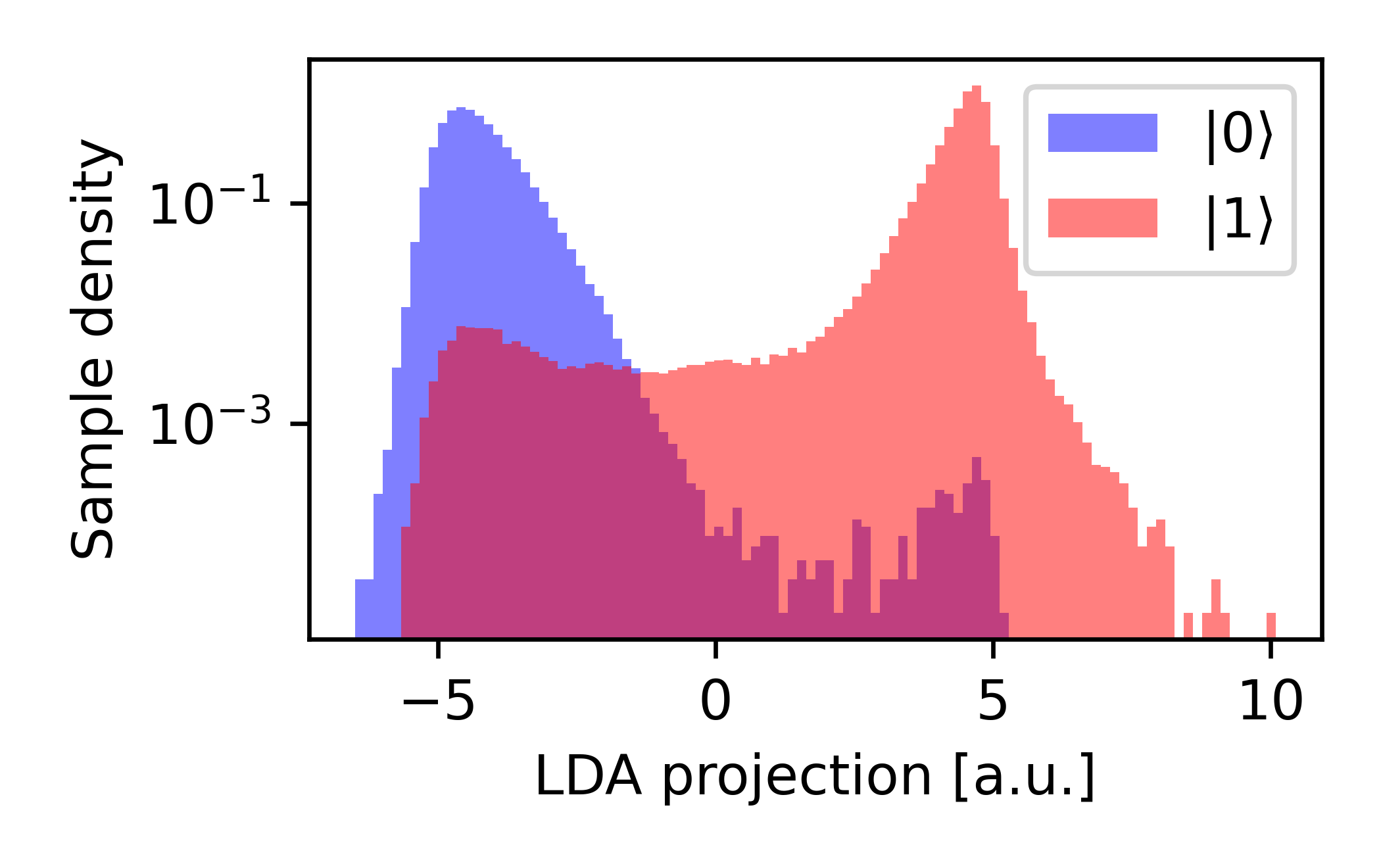}&\iptm{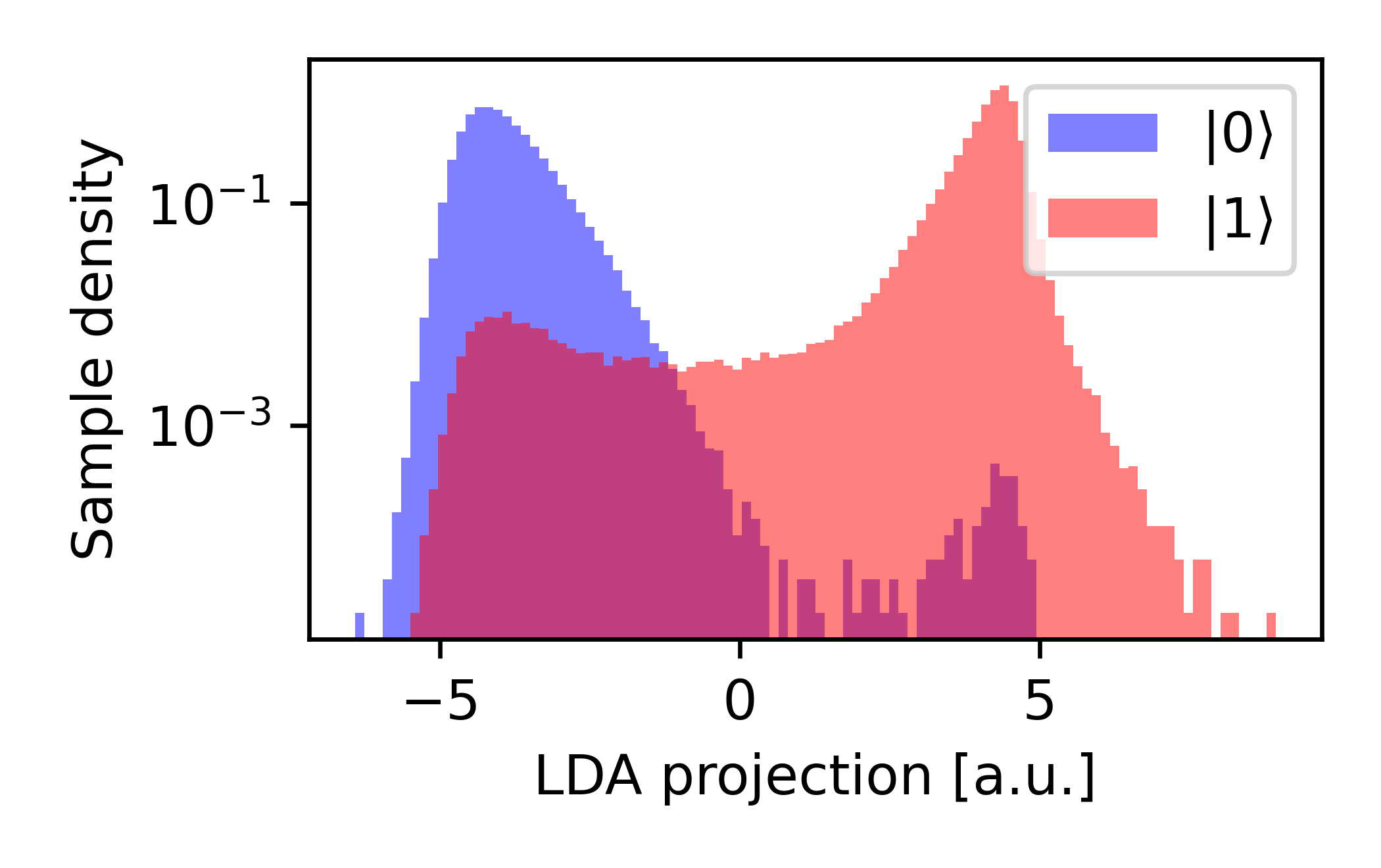}&\iptm{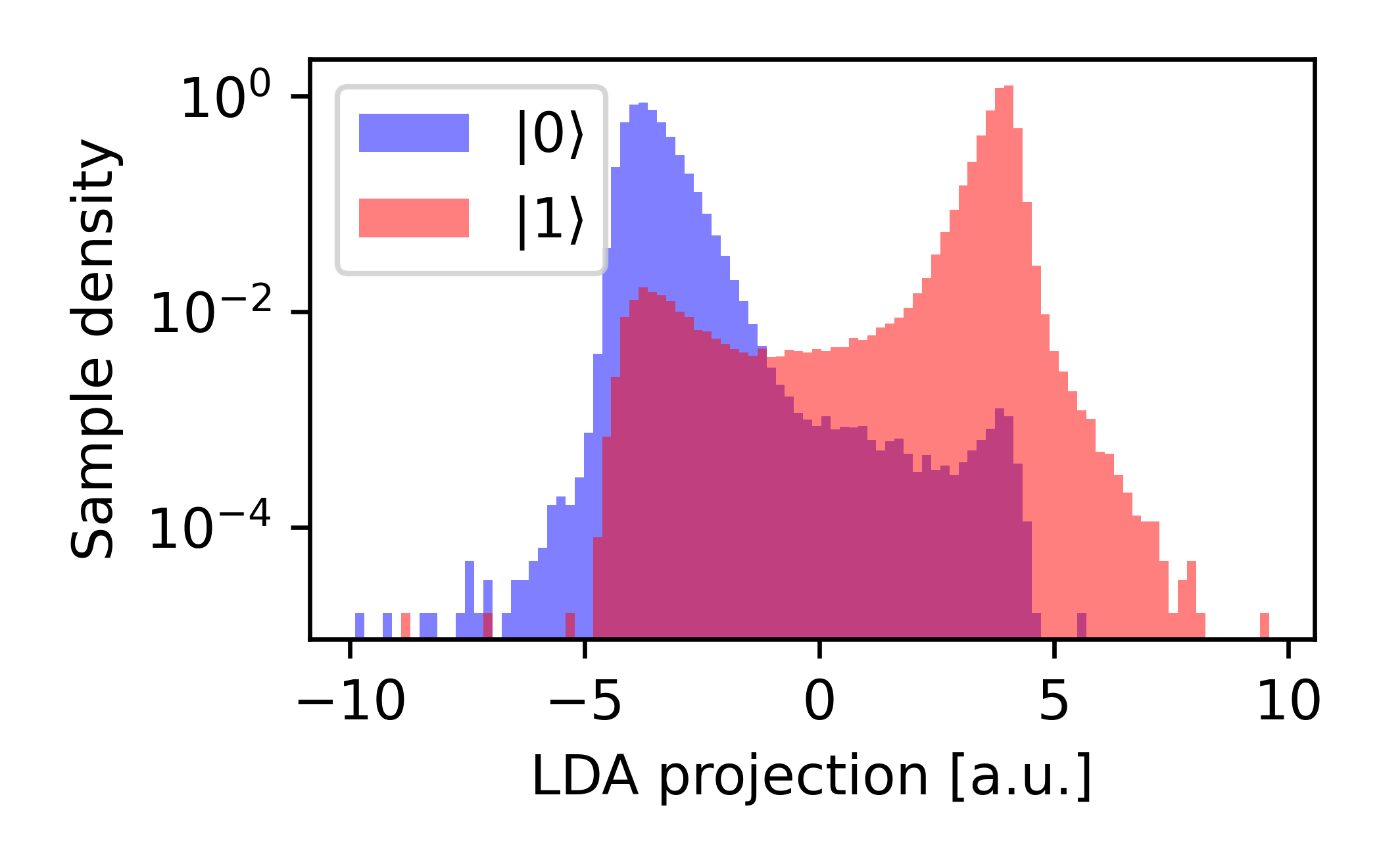}\\
    \begin{tabular}{c}Classifier \\ Performance\end{tabular} & \iptm{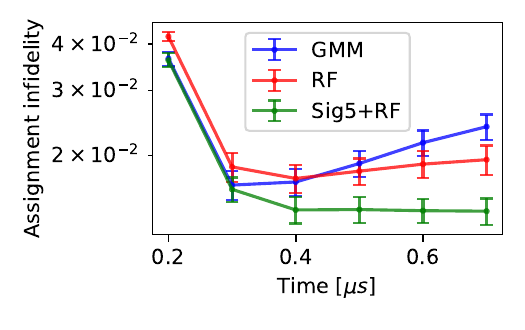}&\iptm{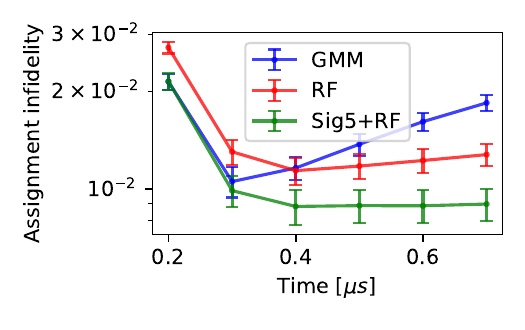}&\iptm{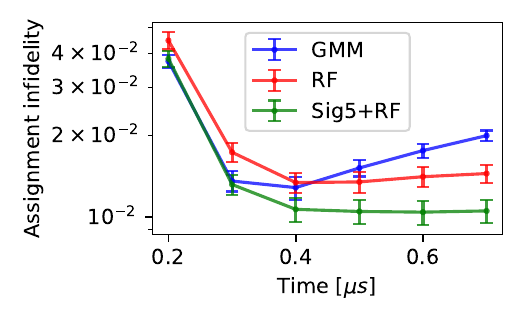}&\iptm{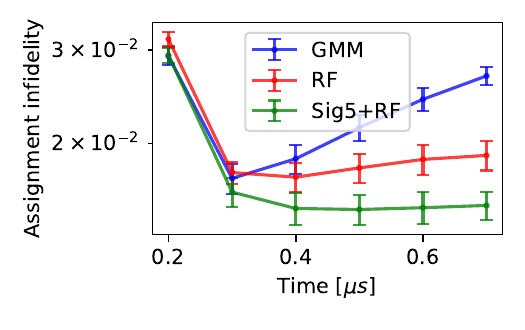}\\
    \begin{tabular}{c}Integration \\ Transition\end{tabular} & \iptm{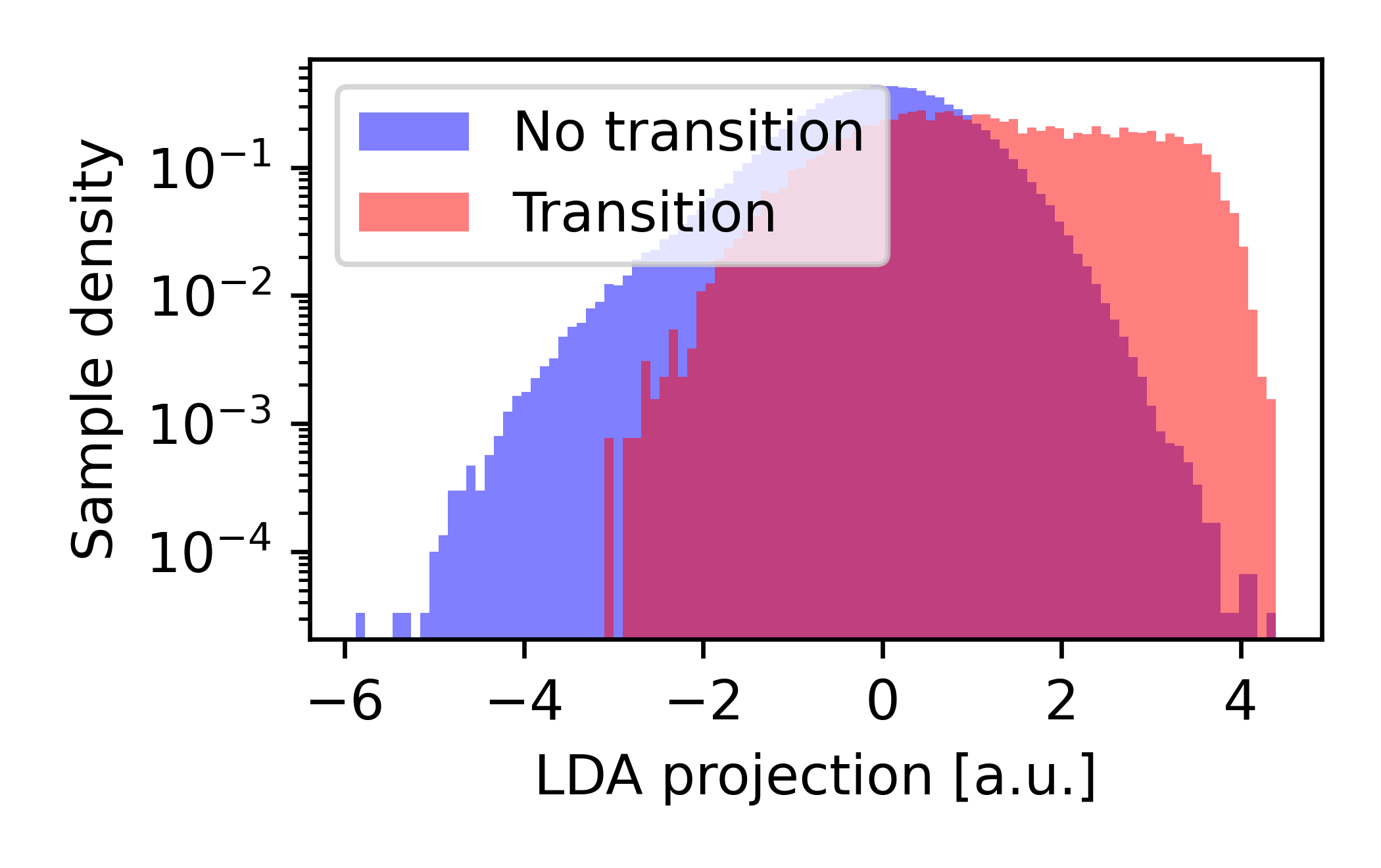}&\iptm{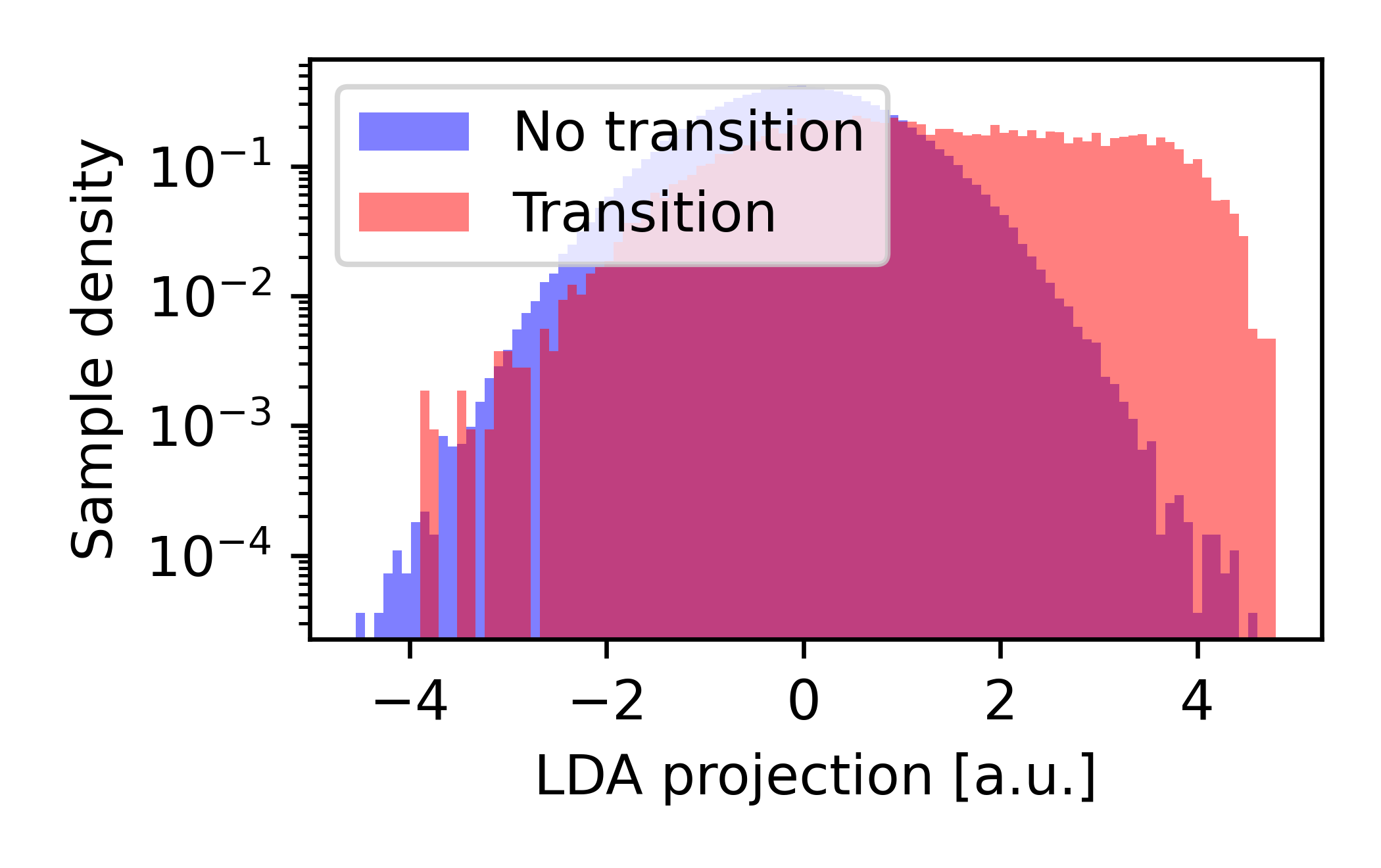}&\iptm{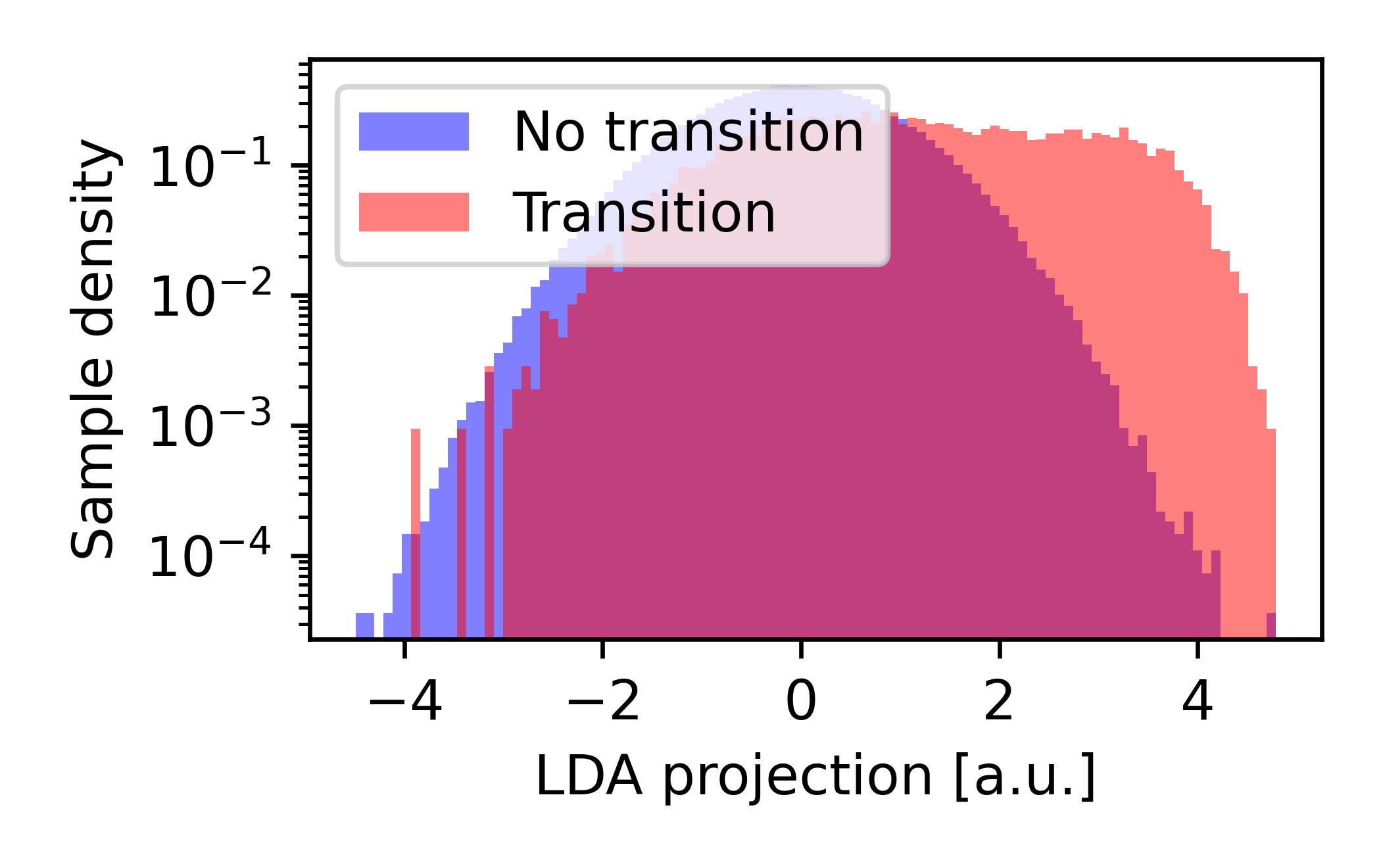}&\iptm{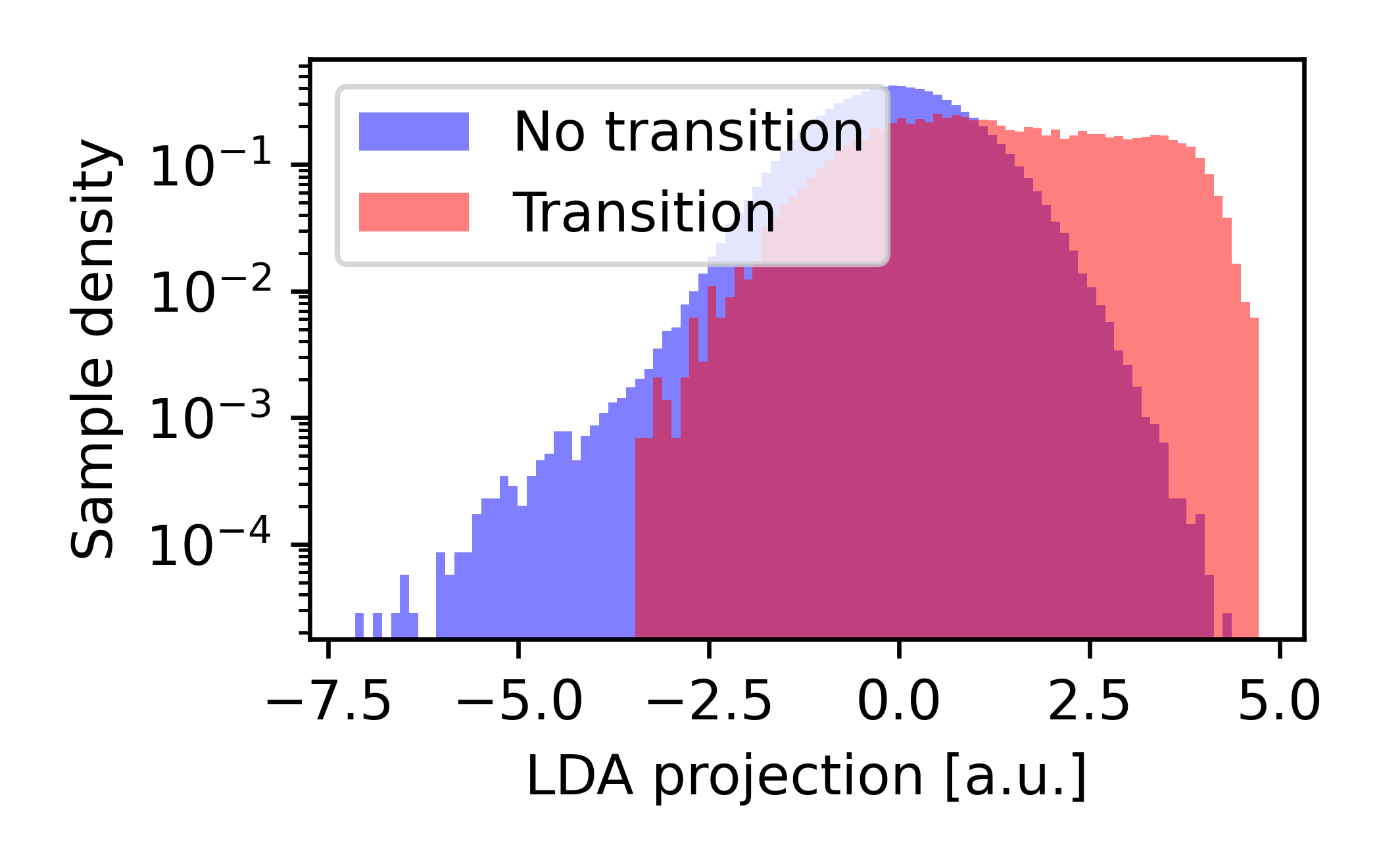}\\
    \begin{tabular}{c}Signature \\ Transition\end{tabular} & \iptm{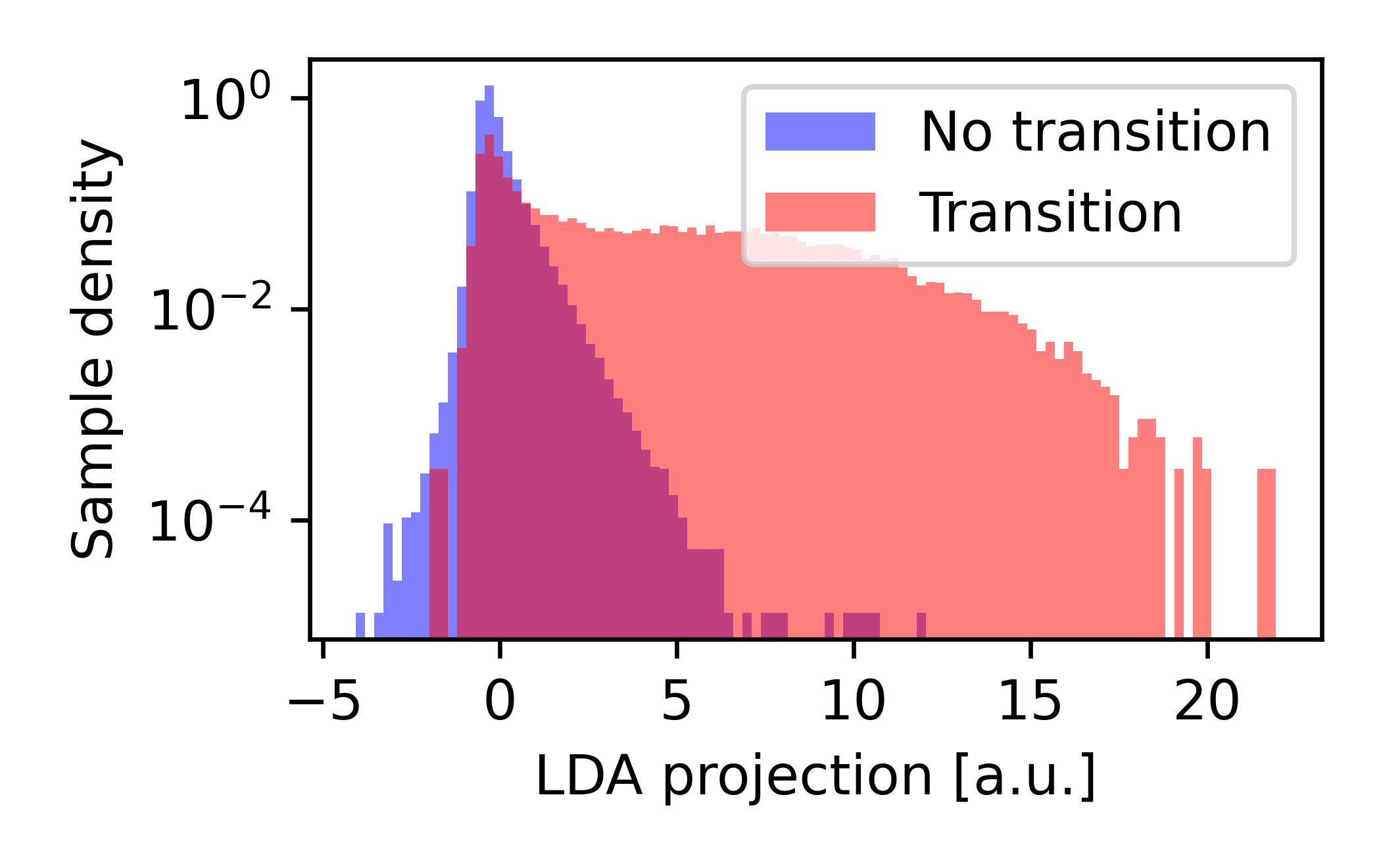}&\iptm{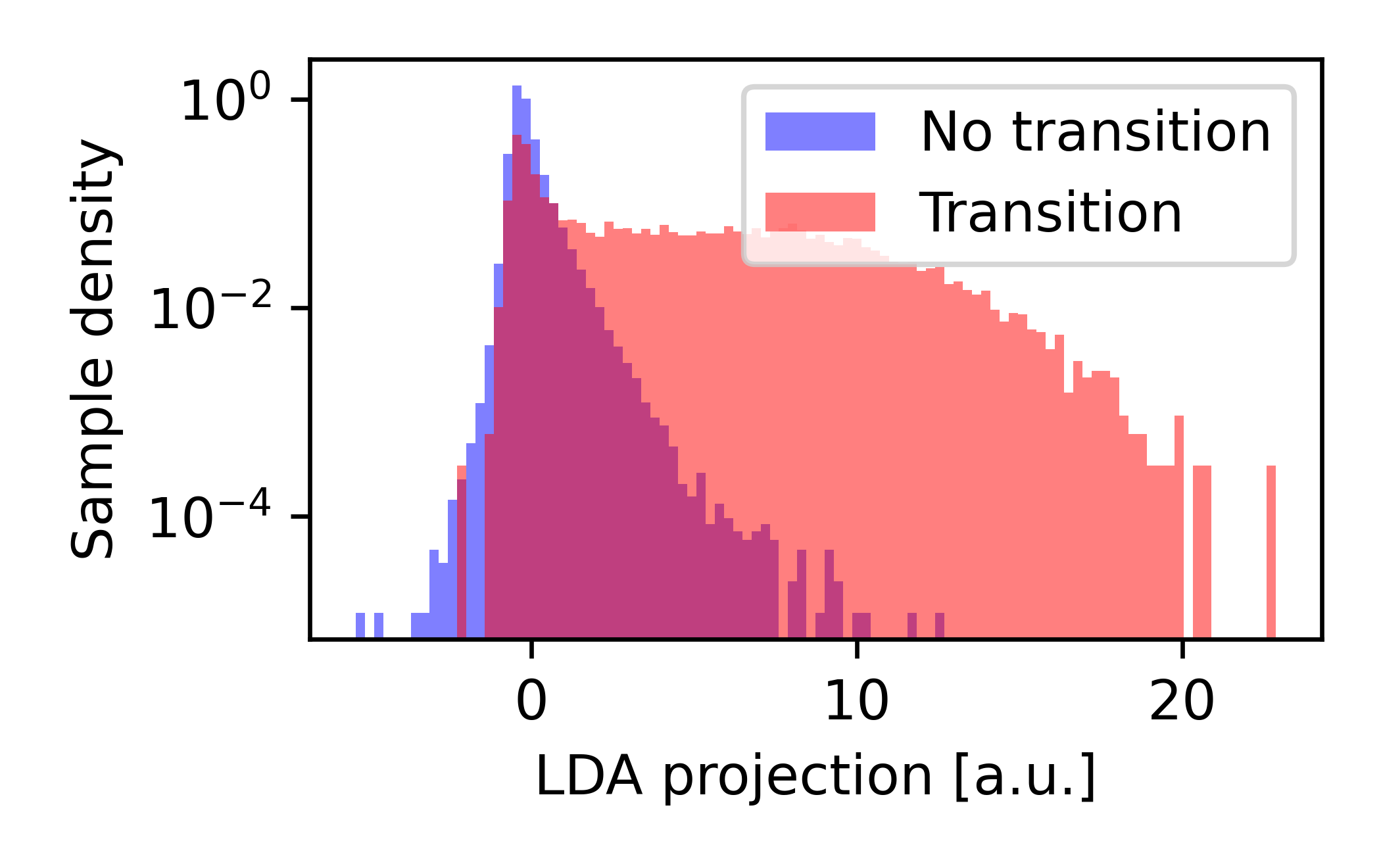}&\iptm{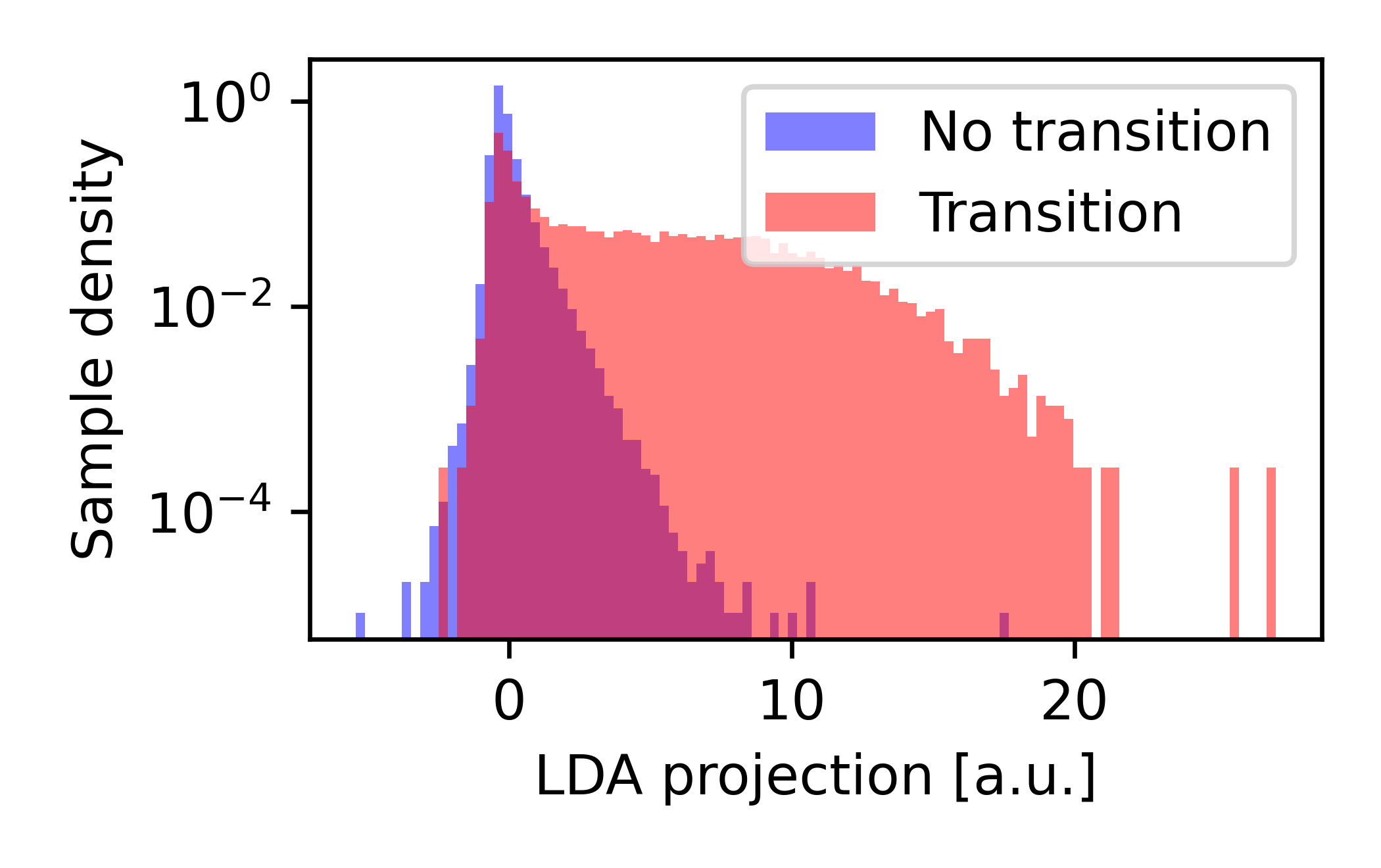}&\iptm{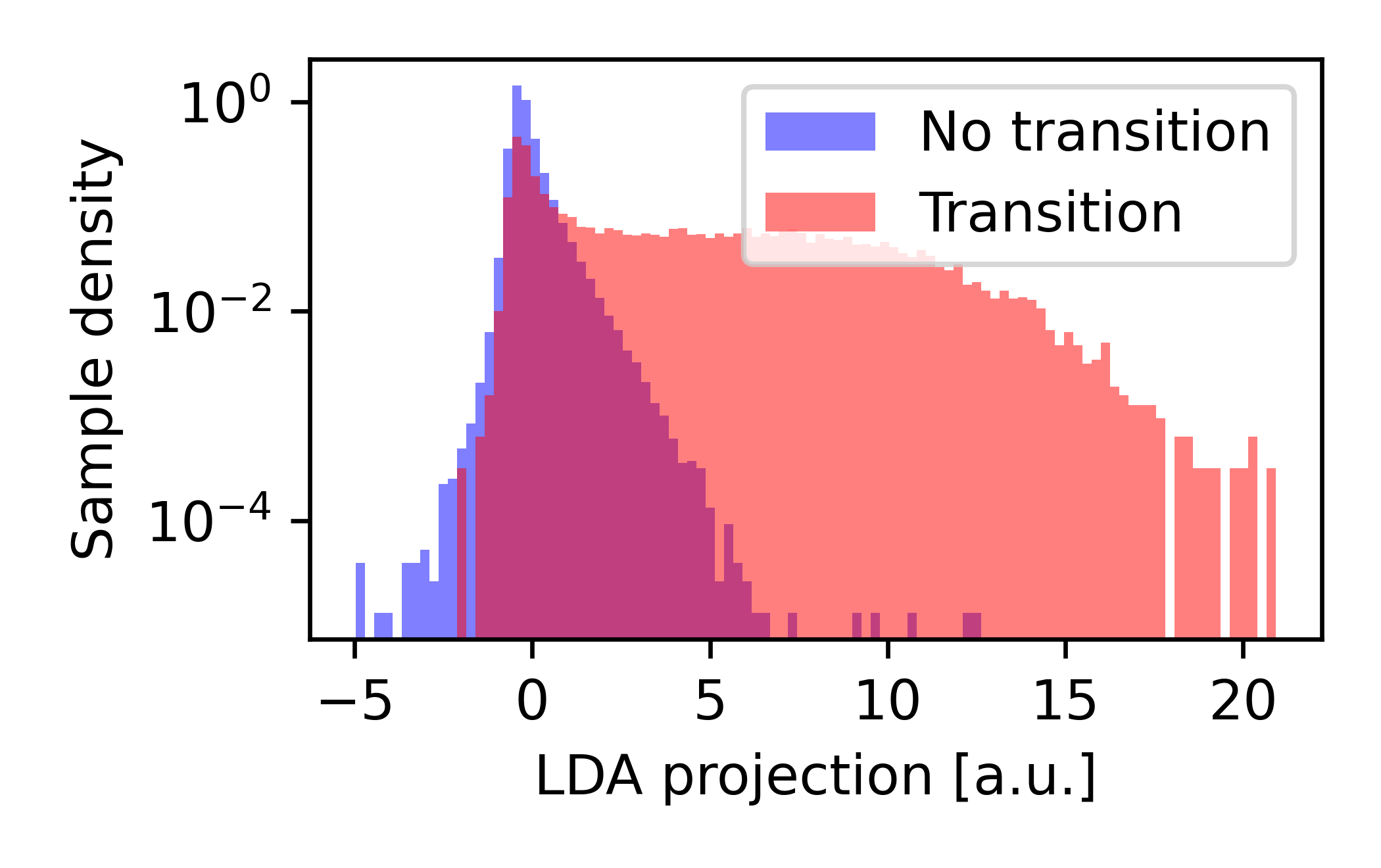}\\

         \hline
    \end{tabular}\caption{Statistics of the readout signal, the signal's signature, and the performance of various state discrimination approaches. These approaches and benchmarks are consistent with those reported in Section \ref{app:ml_methods} of the supplemental material.}.
    \label{tab:x}
\end{figure}

\clearpage

\section{Stability of the readout signal signatures} \label{app:stability}

\begin{figure}[h!]
    \centering
    \includegraphics[width=\linewidth]{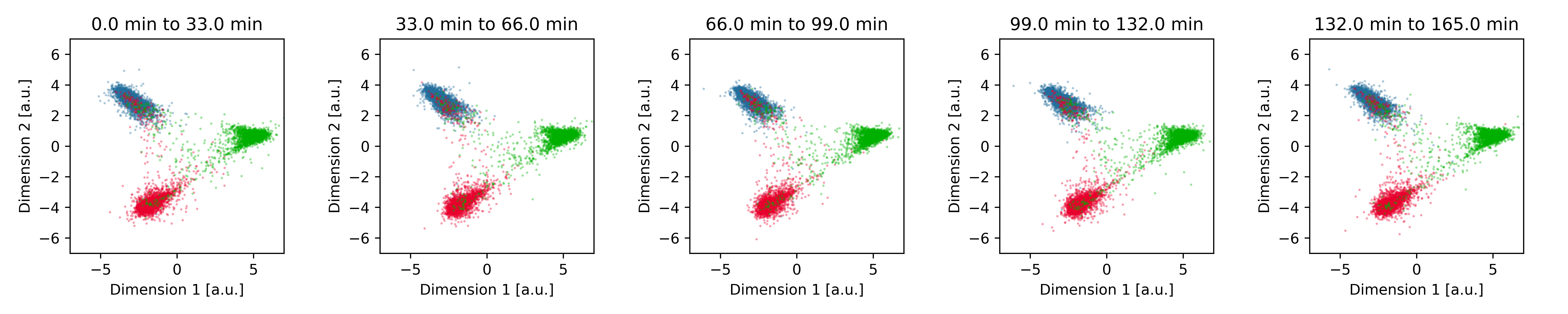}
    \caption{Projection of a depth-5 signature of the AQT qutrit dataset, collected at different time intervals, with the projection direction determined using Linear Discriminant Analysis (LDA). The blue, red, and green points represent data where the state was prepared as $\ket{0}$, $\ket{1}$, and $\ket{2}$, respectively.}
    \label{fig:stability}
\end{figure}

To assess the stability of the readout signature, we selected the AQT dataset, which had the longest collection duration of 165 minutes. We projected the depth-5 signature onto a 2D plane for the data collected over a 33-minute interval. The results show that the distribution of the signatures remains stable over time. See Fig.~\ref{fig:stability}.

To quantify the stability of the distributions, we evaluate the Hellinger Distance of the projected signature distributions between the first distribution (0-33 minutes) and the following 4 distributions. The Hellinger Distance is widely used in probability theory and statistics to quantify similarity between probability distributions. The Hellinger Distance is a measure of divergence between two probability distributions \( P \) and \( Q \) over domain $\Omega$. It is defined as:

\[
H(P, Q) = \sqrt{\frac{1}{2} \int_\Omega \left( \sqrt{P(x)} - \sqrt{Q(x)} \right)^2 \, dx}.
\]

In this analysis, we estimate the density of the distribution using a 2D histogram of the projected signature features with a total of 10,000 bins. The estimated density is then utilized to calculate the Hellinger Distance.

\begin{figure}[h!]
    \centering
    \includegraphics[width=.5\linewidth]{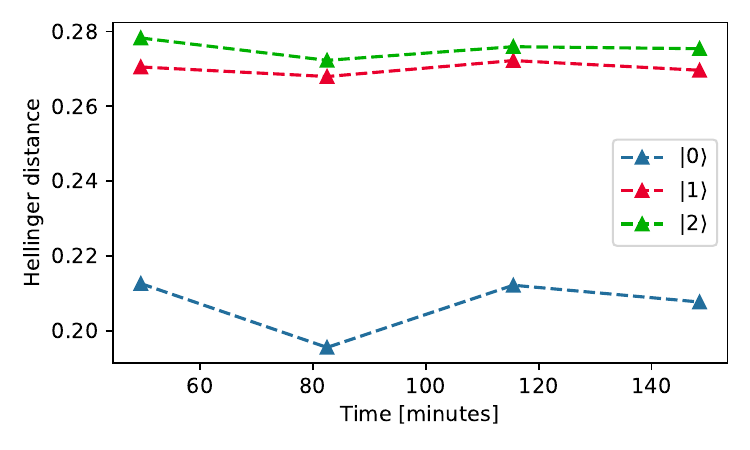}
    \caption{The Hellinger Distance was computed between the first distribution (0–33 minutes) and the subsequent four distributions, shown in Fig.~\ref{fig:stability}}
    \label{fig:hellinger}
\end{figure}

From Fig.~\ref{fig:hellinger}, we observe the Hellinger Distance is stable over time, which indicates the signature feature remains stable over time. We attribute non-zero Hellinger Distance is attributed to sampling.
\section{Fidelity comparisons between multiple methods\label{app:comparisons}}

In this appendix, we benchmark the Sig+RF classifier against a variety of established machine learning models for quantum state discrimination. We additionally evaluate the performance of these models in predicting the EOM state. Although we have made our best effort to faithfully reproduce the original models, certain hyperparameters and neural network architectures may require additional fine-tuning and optimization to achieve optimal performance. 

For each classifier model, the assignment and EOM fidelities are evaluated over 10 independent runs. In each run, 5,000 traces are randomly sampled from each dataset using a unique random seed. The data is split into 70\% for training and 30\% for testing. Fidelity values and corresponding error bars are computed from the outcomes of these runs. The resulting assignment and EOM infidelities are summarized in Tab.~\ref{tab:assignment_comparison} and \ref{tab:eom_comparison}. 

\begin{table}[h!]
\begin{ruledtabular}
\begin{tabular}{l|c|c|c|c|c|c|c|c|c|c}
Dataset & GMM & LDA & QDA & SVM(Linear) & SVM(Nonlinear) & FFNN & Autoencoder & HMM & RF & Sig+RF \\ \hline
OXF Qt & 13.16(44) & 11.08(45) & 9.54(36) & 11.64(46) & 10.00(49) & 26.66(14) & 11.06(38) & 16.64(73) & 11.56(50) & \textbf{8.05(40)} \\
OXF Q1 & 1.23(27) & 1.73(28) & 1.76(32) & 2.46(49) & 1.89(35) & 1.83(31) & 2.08(31) & 1.49(22) & 1.91(29) & \textbf{1.05(28)} \\
OXF Q2 & 1.17(27) & 1.63(32) & 1.62(34) & 2.52(43) & 1.79(44) & 1.64(27) & 1.93(34) & 1.43(34) & 2.17(32) & \textbf{1.05(29)} \\
OXF Q3 & 5.70(37) & 6.08(56) & 5.86(45) & 6.43(51) & 6.37(52) & 6.28(57) & 6.70(54) & 7.20(44) & 6.64(52) & \textbf{2.69(24)} \\
OXF Q4 & 1.80(20) & 2.49(27) & 2.52(25) & 2.77(27) & 2.75(28) & 2.48(32) & 2.80(36) & 2.41(30) & 2.69(35) & \textbf{1.30(24)} \\
RQC Q1 & 1.28(9) & \textbf{0.51(12)} & 1.05(18) & 1.00(28) & 1.07(28) & 1.07(19) & 1.32(30) & 7.27(13) & 1.68(12) & 1.10(7) \\
RQC Q2 & 2.74(17) & \textbf{0.83(16)} & 0.95(19) & 1.22(18) & 1.23(19) & 1.63(39) & 1.90(28) & 3.34(32) & 2.94(21) & 2.50(18) \\
RQC Q3 & 1.57(12) & 1.00(22) & \textbf{0.80(20)} & 0.89(18) & 1.00(27) & 1.96(59) & 2.00(27) & 1.97(37) & 1.68(12) & 1.44(10) \\
RQC Q4 & 1.52(11) & 2.33(36) & 2.65(40) & 2.77(47) & 2.51(35) & 2.76(28) & 3.26(37) & 3.25(45) & 1.54(11) & \textbf{1.33(10)} \\
RQC Q5 & 1.66(15) & 2.07(23) & 1.87(27) & 3.22(37) & 1.81(22) & 1.74(41) & 2.14(33) & 2.04(65) & 1.95(18) & \textbf{1.41(11)} \\
RQC Q6 & 1.05(12) & 1.51(28) & 1.25(28) & 2.50(42) & 1.21(25) & 1.57(35) & 1.68(33) & 1.08(21) & 0.90(11) & \textbf{0.88(12)} \\
RQC Q7 & 1.28(13) & 1.69(36) & 1.46(31) & 2.52(30) & 1.49(33) & 1.42(28) & 1.53(31) & 1.19(23) & 1.44(12) & \textbf{1.04(11)} \\
RQC Q8 & 1.72(11) & 2.21(36) & 2.09(40) & 3.30(42) & 1.79(32) & 1.83(53) & 2.67(0.) & 1.96(54) & 1.90(12) &\textbf{ 1.50(10)} \\
AQT Qt & 4.01(14) & 3.48(17) & 3.64(15) & 3.23(20) & 3.36(18) & 19.65(16) & 3.85(31) & 28.78(28) & 4.19(17) & \textbf{3.18(9)} \\
\end{tabular}
\caption{ Comparison of assignment infidelity $(1-F)\times 10^2$ across various previously reported methods. ``N/A'' indicates that the model did not properly converge (with infidelity exceeding 30\%). The best infidelities among all methods were highlighted for each dataset. \label{tab:assignment_comparison}}
\end{ruledtabular}
\end{table}

\begin{table}[h!]

\begin{ruledtabular}
\begin{tabular}{l|c|c|c|c|c|c|c|c|c|c}
 Dataset & Baseline & LDA & QDA & SVM(Linear) & SVM(Nonlinear) & FFNN & Autoencoder & HMM & RF & Sig+RF \\\hline
OXF Qt & 20.49 & \textbf{11.51(93)} & 30.49(95) & 19.95(21) & 11.55(91) & 16.32(55) & N/A & N/A & 13.61(33) & 12.02(26) \\
RQC Q1 & 3.16 & 5.19(42) & 5.48(43) & 5.15(41) & 4.76(43) & 5.23(61) & 8.53(98) & N/A & 2.92(7) & \textbf{2.70(8)} \\
RQC Q2 & 5.94 & 4.90(64) & 5.01(76) & 4.72(65) & 4.87(62) & \textbf{4.58(28)} & 6.77(75) & 5.69(13) & 5.20(10) & 4.64(12) \\
RQC Q3 & 4.07 & 4.00(57) & 4.22(64) & 4.03(63) & 4.04(59) & 4.30(77) & 8.27(0.) & 4.26(28) & 3.82(9) & \textbf{3.50(9)} \\
RQC Q4 & 4.52 & 4.93(54) & 5.20(52) & 5.12(49) & 4.95(51) & 5.33(47) & 5.41(71) & 5.39(19) & 4.27(8) & \textbf{3.66(7)} \\
RQC Q5 & 2.40 & 1.44(27) & 1.81(32) & 3.22(17) & 1.29(24) & N/A & 10.62(60) & \textbf{1.55(28)} & 1.95(4) & 1.58(4) \\
RQC Q6 & 2.33 & 1.46(30) & 1.61(27) & 3.01(93) & 1.21(32) & N/A & 3.58(16) & \textbf{1.28(15)} & 1.92(7) & 1.54(7) \\
RQC Q7 & 2.29 & 1.40(28) & 1.66(35) & 3.33(15) & 1.22(23) & N/A & 1.92(34) & \textbf{1.22(20)} & 1.94(8) & 1.59(8) \\
RQC Q8 & 2.30 & 1.68(35) & 1.81(37) & 3.67(21) & \textbf{1.33(35)} & N/A & 9.37(75) & 1.36(17) & 1.91(8) & 1.35(8) \\
AQT Qt & 15.17 & 5.21(59) & 7.97(50) & 5.05(51) & 5.24(56) & 10.30(10) & N/A & 14.62(86) & 7.88(14) & \textbf{4.43(14)} \\
\end{tabular}
\caption{ Comparison of EOM infidelity $(1-F_{\mathrm{EOM}})\times 10^2$ across various previously reported methods. ``N/A'' indicates that the model did not properly converge (with infidelity exceeding 30\%). The best infidelities among all methods were highlighted for each dataset. \label{tab:eom_comparison}}
\end{ruledtabular}
\end{table}

The results in Tab.~\ref{tab:assignment_comparison} show that the Sig+RF method provides the best assignment fidelity in 11 of the 14 datasets. We observed that for the RQC Q1, Q2, and Q3 datasets, the LDA and QDA methods statistically outperform the other approaches. These datasets were collected using an over-saturated quantum amplifier, causing non-homogeneous noise characteristics along both the I and Q directions. This may explain the better performance of LDA and QDA; however, a detailed investigation falls outside the scope of our work. Amplifier over-saturation is typically an undesirable scenario in practical applications and is generally avoided. 

The results in Tab.~\ref{tab:eom_comparison} show that the Sig+RF method achieves the highest EOM fidelity in 4 out of 10 datasets. In the remaining 6 datasets, where other models slightly outperform Sig+RF in mean EOM fidelity, the improvements are modest and not statistically significant. We further observe that the Sig+RF method demonstrates the most consistent performance across all datasets.

In our attempts to reproduce the machine learning approach (AE and FFNN), we found that the reproduced models sometimes failed to outperform even the baseline or GMM method. This does not imply the methods are invalid, but rather that the method is less robust and a successful implementation likely requires extensive hyperparameter optimization.

In the following, we summarize our implementation of the benchmarked machine learning models in Tab.~\ref{tab:assignment_comparison} and \ref{tab:eom_comparison}.

\paragraph{Linear Discriminate Analysis (LDA) \cite{PhysRevLett.114.200501}} Our experiment employs a Linear Discriminant Analysis (LDA) classifier that is trained by performing a search for hyperparameters using a 5-fold stratified cross-validation. The input data is the measurement record time series, weighted by the root mean square (RMS) of the difference between each quantum state.

\paragraph{Quadratic Discriminant Analysis (QDA) \cite{PhysRevLett.114.200501}} Our experiment implements a Quadratic Discriminant Analysis (QDA) model that is trained by performing a grid search on a hyperparameter governing regularization. Specifically, the grid search iterates over different values of the “reg\_param” parameter—ranging from 0.0 to 0.5—in order to adjust the amount of regularization applied during the estimation of the covariance matrices. 

\paragraph{Linear SVM \cite{PhysRevLett.114.200501}}The model trains a support vector machine using a linear kernel by tuning the regularization parameter C. Our implementation establishes a grid of C values ([0.1, 1, 10, 100]) and then uses GridSearchCV with a 5-fold stratified cross-validation to systematically search through these values during training. 

\paragraph{Nonlinear SVM \cite{PhysRevLett.114.200501}} The support vector machine is trained by performing an exhaustive grid search for the regularization parameters C and gamma for parametrizing the kernel function, using 5-fold stratified cross-validation. The hyperparameter C is tested with candidate values [0.1, 1, 10, 100], and the gamma parameter is evaluated using the settings ``scale'' and ``auto'', each offering different strategies to compute its value based on the input features. 

\paragraph{Feed Froward Neural Network (FFNN) \cite{PhysRevApplied.17.014024}} We implement a multi-class classifier in PyTorch using a modular architecture detailed in \cite{PhysRevApplied.17.014024}. It begins with a linear layer that maps the input dimension to a hidden dimension (h\_dim), followed by several hidden blocks—each consisting of two linear layers with ReLU activations—that progressively refine the hidden representation. The final output layer reduces the dimension to 3 which corresponding to the maximum number of qubit state labels. Then we apply a softmax function to produce class probabilities. Key hyperparameters include a learning rate of $10^{-3}$ (used by the Adam optimizer), a batch size of 64, and a total of 100 training epochs, while the architecture itself is configurable via parameters: h\_dim is set to 128, and the number of hidden blocks (block\_num) is set to 3.

\paragraph{Autoencoder \cite{PhysRevApplied.20.014045}} This training pipeline is implemented as a two-stage process that first trains an autoencoder and then a classifier on the latent representations produced by the autoencoder. The architecture is an exact reproduction of the one in \cite{PhysRevApplied.20.014045}. The autoencoder is pre-trained for 200 epochs with an MSE loss to optimize reconstruction, and the classifier is trained for 100 epochs using cross-entropy loss on the latent representations provided by the autoencoder, all optimized with the Adam optimizer at a learning rate of 1e-3 and using a batch size of 64, and the learning rate is $10^{-3}$.

\paragraph{Hidden Markov Model \cite{PhysRevA.102.062426}} We use a similar pipeline to that originally reported in \cite{PhysRevA.102.062426}. The measurement record is segmented into a set of $n$ distinct windows that are integrated in order to arrive at an array of $n$ complex IQ points. This array of IQ points is then treated as a sequence of Gaussian-distributed “emissions” that are decoded using a HMM in order to determine the most probable sequence of the true (“hidden”) qubit state. We treat the number of segments $n$ as a hyperparameter and sweep between 2 to 10 segments. In Tab.~\ref{tab:assignment_comparison} and \ref{tab:eom_comparison}, we report the best result of this sweep.  For the other hyperparameters, we set the number of hidden states to 2 for the qubits and 3 for the qutrit dataset. We choose the covariance\_type to be “full”, and the maximum iteration count n\_iter is set to 1000 for the expectation maximization algorithm.

\end{document}